%% file: fresh.tex
\RequirePackage{fix-cm}
\documentclass{UUPaper}
\usepackage{makeidx}  
\usepackage{amsmath,bbm}

\theoremstyle{definition}
\newtheorem{Example}{Example}

\usepackage[nofillcomment,ruled,lined,linesnumbered]{algorithm2e}
\usepackage{bm}
\usepackage{graphicx,color}
\usepackage{rotating}
\usepackage{xspace}

\usepackage{stmaryrd,bbding}
\usepackage{latexsym}
\usepackage{url}

\usepackage[style=base]{caption}
\usepackage{subcaption}
\usepackage{units}
\usepackage{multirow}

\usepackage{tikz}
\usetikzlibrary{arrows,positioning,fit,shapes,calc,arrows.meta}

\tikzset{blockNode/.style={draw=black!50,line width=1pt,rounded corners=2pt},
  blockEdge/.style={-latex}}

\tikzset{
  >={Latex[width=2mm,length=2mm]},
            base/.style = {rectangle, rounded corners, draw=black,
                           minimum width=4cm, minimum height=0.8cm,
                           text centered, font=\sffamily},
  activityStarts/.style = {base, fill=blue!30},
       startstop/.style = {base, fill=red!30},
    activityRuns/.style = {base, fill=green!30},
         process/.style = {base,
                           draw=black!50,line width=1pt,rounded corners=2pt,
                           minimum width=1.5cm, fill=orange!15,
                           font=\ttfamily},
}

\usepackage{floatflt}
\usepackage{csvsimple}

\newif\iftr
\trfalse

\newif\ifcomments
 \commentsfalse

\newif\ifoutline
\outlinefalse

\newcommand{\uppsat}{\textsf{UppSAT}\xspace}

\newcommand{\comment}[1]{\ifcomments{\color{blue} #1}\fi}
\newcommand{\Comment}[1]{\ifcomments\marginpar{\small\color{blue} #1}\fi}
\newcommand{\contents}[1]{\ifoutline{\color{blue}
    \begin{itemize}
    #1
    \end{itemize}
  }\fi}

\newcommand{\toAdd}[1]{}
\newcommand{\toRem}[1]{}

\newcommand{\fpa}{FPA\xspace}

\newcommand{\standard}{the IEEE-754 standard }
\newcommand{\Standard}{The IEEE-754 standard }

\newcommand{\aux}[1]{\ensuremath{\hat{#1}}}

\newcommand{\fp}[1]{\ensuremath{\mathit{FP}_{#1}}}

\newcommand{\todo}[1]{{\color{red}TODO: #1}}
 
\renewcommand{\todo}[1]{}

\newcommand{\scala}{Scala\xspace}
\newcommand{\mathsat}{MathSAT}

\usepackage{listings}

\lstdefinestyle{floating}{
    xleftmargin=10pt,
    xrightmargin=5pt,
    aboveskip=4mm,
    belowskip=4mm,
    fontadjust=true,
    columns=[c]flexible,
    keepspaces=true,
    basewidth={0.5em, 0.425em},
    tabsize=2,
    basicstyle=\ttfamily,
    commentstyle=\rm,
    keywordstyle=\bfseries,
    mathescape=true,
    captionpos=b,
    framerule=0.3pt,
    firstnumber=0,
    numbersep=1.5mm,
    numberstyle=\tiny,
    float=tbp,
    frame=tblr,
    framesep=5pt,
    framexleftmargin=3pt,
    abovecaptionskip=\smallskipamount,
    belowcaptionskip=\smallskipamount,
} 

\newcommand{\code}[1]{
   \lstinline[
               flexiblecolumns=true,
               basicstyle=\ttfamily]{#1}}

\title{Exploring Approximations for Floating-Point Arithmetic using \uppsat}
\author{ \hspace{2em}Aleksandar Zelji\'{c}$\,^1$      \and \hspace{2em}Peter Backeman$\,^1$ \and
  Christoph~M.~Wintersteiger$\,^2$ \and Philipp R\"{u}mmer$\,^1$  \and
  $^1\,$\normalsize Uppsala University \and $^2\,$\normalsize Microsoft Research}

\begin{document}
\pagestyle{fancy}

\maketitle

\begin{abstract}   
  We consider the problem of solving floating-point constraints obtained from
  software verification. We present \uppsat{} --- an new implementation of a
  systematic approximation refinement
  framework~\cite{DBLP:journals/jar/ZeljicWR17} as an abstract SMT solver.
  Provided with an approximation and a decision procedure (implemented in an
  off-the-shelf SMT solver), \uppsat yields an approximating SMT solver.
  Additionally, \uppsat yieldsincludes a library of predefined approximation
  components which can be combined and extended to define new encodings,
  orderings and solving strategies. We propose that \uppsat
  can be used as a sandbox for easy and flexible exploration of new
  approximations. To substantiate this, we explore several approximations of
  floating-point arithmetic. Approximations can be viewed as a composition of an
  encoding into a target theory, a precision ordering, and a number of
  strategies for model reconstruction and precision (or approximation)
  refinement. We present 
  encodings of floating-point arithmetic into reduced precision floating-point
  arithmetic, real-arithmetic, and fixed-point arithmetic (encoded into the
  theory of bit-vectors
  in practice). In an experimental evaluation we compare the advantages and
  disadvantages of approximating solvers obtained by
  combining various encodings and decision procedures (based on existing,
  state-of-the-art SMT solvers for floating-point, real, and bit-vector
  arithmetic).
\end{abstract}

 \section{Introduction}
\label{sec:intro}

\Comment{Motivate the research, look into jar paper / lic for motivation}

The construction of satisfying assignments of a formula, or showing
that no such assignments exist, is one of the most central tasks in
automated reasoning. Although this problem has been addressed
extensively in research fields including constraint programming, and
more recently in Satisfiability Modulo Theories (SMT), there are still
constraint languages and background theories where effective model
construction is challenging. Such theories are, in particular,
arithmetic domains such as bit-vectors, nonlinear real arithmetic (or
real-closed fields), and floating-point arithmetic; even when
decidable, the high computational complexity of such problems turns
model construction into a bottleneck in applications such as model
checking, test-case generation, or hybrid systems analysis.

In several recent papers, the notion of \emph{approximation} has been
proposed as a means to speed up the construction of (precise)
satisfying assignments. Generally speaking, approximation-based
solvers follow a two-tier strategy to find a satisfying assignment of
a formula~$\phi$. First, a simplified or \emph{approximated}
version~$\hat\phi$ of $\phi$ is solved, resulting in an approximate
solution~$\hat m$ that (hopefully) lies close to a precise
solution. Second, a \emph{reconstruction} procedure is applied to
check whether $\hat m$ can be turned into a precise solution~$m$ of
the original formula~$\phi$. If no precise solution~$m$ close to $\hat
m$ can be found, \emph{refinement} can be used to successively obtain
better, more precise, approximations.

This high-level approach opens up a large number of design choices,
some of which have been discussed in the literature. The
approximations considered have different properties; for instance, they might 
be over- or under-approx\-imations (in which case they are commonly
called \emph{abstractions}), or be non-conservative and exhibit
neither of those properties. The approximated formula~$\hat\phi$ can
be formulated in the same logic as $\phi$, or in some \emph{proxy} theory
that enables more efficient reasoning. The reconstruction of $m$ from
$\hat m$ can follow various strategies, including simple
re-evaluation, precise constraint solving on partially evaluated
formulas, or randomised optimisation. Refinement can be performed with
the help of of approximate assignments $\hat m$, using proofs or
unsatisfiable cores, or be skipped altogether.

In this paper, we aim at a uniform description and exploration of the
complete design space.  We focus on the case of (quantifier-free)
floating-point arithmetic (\fpa) constraints, a particularly
challenging domain that has been studied extensively in the SMT
context over the past few
years~\cite{bkw2009-fmcad,DBLP:conf/tacas/DSilvaHKT12,DBLP:conf/cade/ZeljicWR14,DBLP:conf/fmcad/RamachandranW16,DBLP:conf/cav/FuS16,DBLP:journals/jar/ZeljicWR17}.
To enable uniform exploration of approximation, reconstruction, and
refinement methods, as well as simple prototyping and comparative
studies, we present \uppsat as a a general framework for building
approximating solvers. \uppsat is implemented in \scala,
open-source under GPL licence, and allows the
implementation of approximation schemes in a modular and high-level
fashion, such that different components can be easily combined with
various back-ends.

With the help of the \uppsat framework, we explore several ways of
approximating SMT reasoning for floating-point arithmetic. The
contributions of the paper are:
\begin{itemize}
\item a high-level approach to design, implement, and evaluate
  approximations, presented using the case of floating-point
  arithmetic constraints;
\item a conceptual and experimental comparison of three different
  forms of \fpa approximation, based on the notions of
  \emph{reduced-precision floating-point arithmetic, fixed-point
    arithmetic,} and \emph{real arithmetic;}
\item a systematic comparison of different back-end solvers for the
  case of \emph{reduced-precision} floating-point arithmetic.
\end{itemize}

\subsection{Introductory Example}
In this paper we will use the following formula as running example to
illustrate the effects of using different approximations:
\begin{Example}
  \label{ex:running}
  Consider a floating-point formula~$\phi$ over two variables $x$ and
  $y$:
  \[ y = x + 1.75 \wedge y \geq 0 \wedge (x = 2.0 \vee x = -4.0)\]
  Note that the formula can be satisfied by the model $m = \{x \mapsto 2.0, y
  \mapsto 3.75\}$ in both single-precision and double-precision FPA
  (and a couple of further formats).
\end{Example}

We will use the formula to highlight different aspects of the
approximations discussed in this paper, in particular approximations
using reduced-precision FPA and fixed-point arithmetic. 

\subsubsection{Reduced-Precision Floating-Point Arithmetic}
\label{sec:reduced}

The first form of approximation uses floating-point operations of
reduced precision, i.e., with a reduced number of bits for the
significant and exponent. Approximations of this kind have previously
been studied
in~\cite{DBLP:conf/cade/ZeljicWR14,DBLP:journals/jar/ZeljicWR17}, and
found to be an effective way to boost the performance of
bit-blasting-based SMT solvers, since the size of FPA circuits tends
to grow quickly with the bit-width.  The change of the actual formula
lies in decreasing the number of bits used for each variable and
operator.

\begin{Example}
  \label{ex:rpfp}
  We assume reduction to the floating-point $(3,3)$ format, i.e., the
  format in which 3~bits are used for the significant, and 3~bits for
  the exponent. The approximate formula~$\hat\phi_{3,3}$ is obtained
  by replacing the variables $x$ and $y$ with re-typed variants
  $x_{3,3}, y_{3,3}$, casting all floating-point literals to the new
  format, and replacing the addition operator~$+$ and comparison predicate~$\leq$ with the
  operator~$+_{3,3}$ and the predicate~$\geq_{3,3}$ for reduced-precision arguments:
  \[y_{3,3} = x_{3,3} +_{3,3} 1.75_{3,3} \wedge y_{3,3} \geq 0_{3,3} \wedge (x_{3,3} = 2.0_{3,3} \vee x_{3,3} = -4.0_{3,3})\]
  Even though~$\hat\phi_{3,3}$ is satisfiable the models are not
  guaranteed models for the original formula, but only satisfies the
  reduced precision formula because of over/under-flows and rounding
  errors when working with only three precision and three significand
  bits. For example, $\aux m = \{x \mapsto 2.0, y \mapsto 4.0\}$), satisfies
  ~$\hat\phi_{3,3}$ because $2.0_{3,3} +_{3,3} 1.75_{3,3} = 4_{3,3}$.

  This could means that the current reduced precision does not
  allow for representation of the solutions that exists for the full
  precision formula. Therefore we need to refine the precision, and a
  simple strategy is to increase the precision of every node by the
  same amount, yielding:
  \[y_{5,5} = x_{5,5} +_{5,5} 1.75_{5,5} \wedge y_{5,5} \geq 0_{5,5} \wedge (x_{5,5} = 2.0_{5,5} \vee x_{5,5} = -4.0_{5,5})\] 
  which will have model~$\{x \mapsto 2.0, y \mapsto 3.75\}$ which is also a model for the original problem.
\end{Example}

\subsubsection{Fixed-Point Arithmetic}
\label{sec:fixedpointexample}

As a second relevant case, we consider the use of fixed-point
arithmetic as an approximation of FPA.  This is done by choosing a
fixed number of \emph{integral bits} and a fixed number of
\emph{fractional bits,} defining the applied fixed-point format, and
then recasting each floating-point constraint as a bit-vector constraint:
each floating-point operation is replaced with a set of bit-vector
operations implementing the corresponding computation over fixed-point
numbers.

\begin{Example}
  \Comment{we should also show a FP format where rounding is
    necessary; e.g., (5,1)?} For our example, we can initially choose a
  representation with 5 integral and 5 fractional bits, i.e., in the
  $(5, 5)$ fixed-point format. We can note that fixed-point $(5,
  5)$-addition is exactly implemented by bit-vector
  addition~$\oplus_{10}$ over 10~bits, and fixed-point
  comparison~$\geq$ by signed bit-vector comparison~$\geq_s$ over
  10~bits, so that the translation becomes relatively straightforward,
  resulting in the formula~$\hat\phi^F_{5,5}$:
  \begin{align*}
    &y_{10} = x_{10} \oplus_{10} 00001\,11000_2 \wedge
    y_{10} \geq_s 00000\,00000_2 \wedge{}\\
    &\quad (x_{10} = 00010\,00000_2 \vee x_{10} = 11100\,00000_2)
  \end{align*}
  Constants are interpreted as 2's complement numbers with 5
  fractional and 5 integral bits, e.g., $11100\,00000_2$ represents
  the binary number $-100.00000$, which is $-4.0$ in decimal notation.

  It can easily be seen that the constraint~$\hat\phi^F_{5,5}$ is
  satisfied by the model $\aux m = \{x_{10} \mapsto 00010\,00000_2, y_{10}
  \mapsto 00011\,11000_2\}$, which corresponds to the fixed-point
  solution $x = 2.0$ and $y = 3.75$, and to the floating-point
  solution given in Example~\ref{ex:running}.
\end{Example}

\iffalse
\subsubsection{Real Arithmetic}

The third approach presented in this paper is using real
arithmetic. In principle, we translate all floating point operations
and predicates to their real counterparts and floating point variables
are replaced by real ones.

\begin{Example}
  \[y = x + 1.75 \wedge y \geq 0 \wedge (x = 2.0 \vee x = -4.0)\]  
\begin{lstlisting}
    (assert (or (= x two) (= x minusfour)))
    (assert (>= y 0))
    (assert (= y (+ x onesevenfive)))
    (assert (= two 2))
    (assert (= minusfour (- 4)))
    (assert (= onesevenfive 1.75))
  \end{lstlisting}
\end{Example}
    \fi

\contents{
\item
find a simple example to illustrate the overall approach of
  approximation-based solving. not as extensive as in the JAR and
  IJCAR paper, but we need to show how approximation, reconstruction
  and refinement work}

 \section{\uppsat\ --- An Abstract Approximating SMT Solver}
\label{sec:uppsat}
\Comment{PR: this section needs further polishing ...}
\Comment{Should we talk about lifted formulas? Encoded formulas?}
\uppsat is an implementation of the systematic approximation refinement
framework~\cite{DBLP:journals/jar/ZeljicWR17} as an abstract SMT solver. It
takes an approximation and a back-end SMT solver to yield an approximating SMT
solver. \uppsat can implement a broad range of approximations in a simple and
modular way, and it easily integrates off-the-shelf SMT solvers.

\begin{figure}[tb]
  \begin{center}
    \clearpage{}\small
\begin{tikzpicture}[node distance=1.5cm,
    every node/.style={fill=white, font=\sffamily}, align=center]
  \node (Encoder)     [process]          {$\mathit{encode}$};
  \node (Backend)      [process, below of=Encoder, yshift=-1cm]   {$\mathit{checkSAT}$};
  \node (Decoder)     [process, below of=Backend, yshift=-1cm]   {$\mathit{decode}$};
  \node (Reconstructor)      [process, below of=Decoder, yshift=-1cm] {$\mathit{reconstruct}$};

  \node (ProofRefiner)    [process, right of=Backend, xshift=2cm] {$\mathit{refineProof}$};
  
  \node (ModelRefiner)    [process, left of=Decoder, xshift=-2cm] {$\mathit{refineModel}$};

  \draw[->]     (Encoder) -- node {Encoded Formula} (Backend);
  \draw[->]		      (Backend) -- node {Model} (Decoder);
  \draw[->]     (Decoder) -- node {Approximate Model} (Reconstructor);

  \draw[->]		      (Backend.east) -- node [above=3pt] {Proof} (ProofRefiner.west);
  \draw[->]      (ProofRefiner) |- node [above=5pt] {Precision} (Encoder);

  \draw[->]		      (Reconstructor.west) -| node [below=6pt] {Reconstructed Model} (ModelRefiner.south);  
  \draw[->]      (ModelRefiner) |- node [above=5pt] {Precision} (Encoder);

\end{tikzpicture}
\clearpage{}
  \end{center}

  \caption{The main components needed to implement approximations in \uppsat,
    and the flow of data in between them.}
  \label{fig:dataflow}
\end{figure}
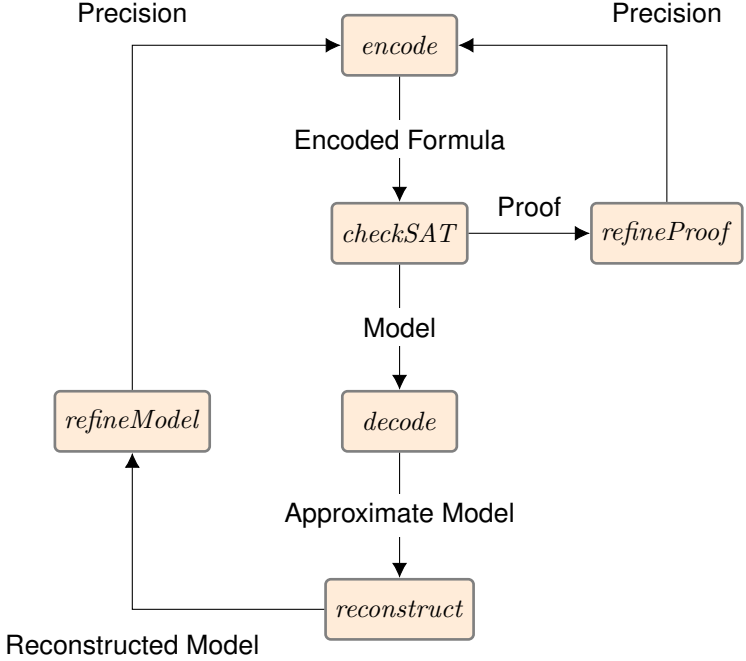

The theoretical framework~\cite{DBLP:journals/jar/ZeljicWR17} is
defined in terms of a monolithic auxiliary theory to solve the
original problem. Instead of solving the problem $\phi$ of theory $T$
directly, the formula $\phi$ is \emph{lifted} to the formula $\aux
\phi$ of the approximation theory $\aux T$.  The formula $\aux \phi$
is solved using a decision procedure for $\aux T$. $\aux T$ enables
approximation of the original problem and controlling the degree of
approximation. The search for the model is guided by a search through
the space of approximations expressible in $\aux T$.  Lifting the
formula $\phi$ to $\aux T$ introduces \emph{precision} as the means of
characterizing the degree of approximation. For different values of
precision, different approximations of the original formula are
obtained. The overall goal is to find a model of an approximation
$\aux m$ that can be translated back to a model $m$ of the original
formula $\phi$.

\Comment{This paragraph needs another go. PR: yes; the main point
  should we that we deviate from \cite{DBLP:journals/jar/ZeljicWR17}
  and don't use a monolithic theory? that has to be clarified} The
solving process can be seen as a two-tier search, in which the search
for a sufficiently precise approximation guides the actual model
search. Search for the approximation tries to capture essential
properties of the model and is performed by the \emph{abstract
  solver}. The low-level search is entirely unaware of the high-level
aspects. It is performed by the \emph{back-end solver} which seeks the
approximate model. The two tiers of search guide each other in turns,
until a solution is found or the search space of approximations has
been exhausted. In practice there is no need to implement a solver for
the monolithic approximation theory $\aux T$. Instead, \uppsat uses an
\emph{off-the-shelf} SMT solver as the back-end procedure to solve
lifted formulas.

\Comment{AZ: Find another place for this paragraph}
The overall goal for the approximation is to produce constraints that are
`easier' to solve than the input constraints. One example, that we consider
later, is approximating the theory of \fpa using the theory of reals, which
is considered `simpler' because it ignores the rounding behavior and special
values of \fpa semantics.\Comment{PR: the quotes look strange}

\paragraph{A bird's eye view on \uppsat.}
This paper focuses on the theory of \fpa and presents several
approximations suitable for solving \fpa formulas. We first discuss
the general structure of the approximations from the perspective of
\uppsat.

An \emph{approximation context} contains the following components:
1.~an \emph{input theory} $T$, the language of the problem to solve; 2.~an
\emph{output theory} $\aux{T}$, the language of in which we solve lifted
formulas; 3.~a \emph{precision domain,} the parameters used to indicate
degree of approximation; and 4.~a \emph{precision ordering,} defining 
an order among different approximations.

Given an approximation and a back-end solver, \uppsat takes a set of constraints
of the input theory $T$ and produces constraints of the output theory for the
back-end solver. Precision regulates the encoding from the input to the output
theory, and its domain and ordering are of consequence for encoding,
approximation refinement and termination.

The approximation context only determines the setting for the approximation, but does
not give the complete picture. For example, fixing an input and an
output theory does not uniquely determine the encoding, also
the choice of precision
domain and the precision ordering are essential for the expressiveness of the
approximation.
\Comment{AZ: High-level description of approximations}
Given an approximation context, to fully define an approximation we also
need to define the following components:
1.~\emph{encoding} of the formula based on precision;
2.~\emph{decoding} the values of the approximate model;
3.~\emph{model reconstruction strategy;}
4.~\emph{model-based refinement strategy;} and
5.~\emph{proof-based refinement strategy.}
The flow of data between these components can be seen in
Fig~\ref{fig:dataflow}.

Encoding of the formula and decoding of the approximate model are the core of
the approximation. These operations describe the two directions of moving
between the input and the output theory. The encoding aims to retain the
essential properties of the problem while making it easier to solve. The goal
of decoding is to translate a model for the approximate constraints to an
assignment of the input theory. These two operations are of course closely related (and
implemented by the \code{Codec} trait).

The purpose of model reconstruction strategy is to transform the
decoded model into a model of the original constraints. Sometimes the
model of the lifted formula will also be a model for the original
formula. However, often this is not the case, and a reconstruction
strategy is used to repair the assignment in an attempt to
find a satisfying assignment. Reconstruction strategies can range from
simple re-evaluation of the constraints to using a constraint solver
or an optimization procedure.

The goal of the model-based and proof-based refinement strategies is
to select the approximation for the next iteration based on the
available information. For example, given an approximate model and a
failed model we can infer which parts of the formula to refine. This
is expressed as a new precision value, specifically a precision value
which is greater than the previous one according to the precision
ordering.

In order to preserve completeness and termination, for the case of
decidable theories, we assume that every precision domain contains a
top element~$\top$, and that precision domains satisfy the ascending
chain condition (every ascending chain is
finite)~\cite{DBLP:journals/jar/ZeljicWR17}. By convention,
approximation in top precision~$\top$ corresponds to solving the
original, un-approximated constraint with the help of a fall-back
solver.

Implementation of these operations can be separated into two layers, a general
layer and a theory-specific layer. For example, significant parts of the
encoding and decoding are specific to theories involved in the approximation,
while the various strategies are mostly theory-independent, except for a few
details. Theory-independent layers are abstracted into templates, that provide
\emph{hook} functions for theory-specific details. \uppsat is designed around
the mix-and-match principle, to provide a sandbox for testing different approximations with little implementation effort.

Fig. \ref{fig:trait-dia} shows the traits (i.e., interfaces) that have
to be implemented by approximations in \uppsat. The
\emph{approximation} class takes an object implementing all four
traits, and combines them into an approximation to be used by the
abstract solver.

\begin{figure}
  \begin{center}
    \begin{lstlisting}[language=Scala,numbers=left]
trait AppContext {
  val inputTheory : Theory
  val outputTheory : Theory

  type Prec
  val pOrdering : POrdering[Prec]
}

trait Codec extends AppContext {
  def encodeFormula(
        ast : AST,
        pmap : PrecMap[Prec]) : AST
  def decodeModel(
        ast : AST,
        appMode : Model,
        pmap : PrecMap[Prec]) : Model
}

trait ModelReconstruction extends AppContext {
  def reconstruct(
        ast : AST,
        decodedModel : Model) : Model
}

trait ModelGuidedRefStrategy extends AppContext {
  def satRefine(
        ast : AST,
        decodedModel : Model,
        failedModel : Model,
        pmap : PrecMap[Prec]) : PrecMap[Prec]
}

trait ProofGuidedRefStrategy extends AppContext {
  def unsatRefine(
        ast : AST,
        core : List[AST],
        pmap : PrecMap[Prec]) : PrecMap[Prec]
}
\end{lstlisting}
   \end{center} 

    \caption{The basic traits neccessary to specify an
      approximation in \uppsat}
    \label{fig:trait-dia}
\end{figure}

Consider an implementation of the reduced
precision approximation of \fpa in a concise and compact manner.
This approximation comprises of dropping \fpa-specific elements,
such as the rounding modes, and replacing \fpa operations by the
corresponding real arithmetic operations. in certain cases, a
combination of operations may be necessary, e.g., in the case of the
\emph{fused-multiply-add} operation. In the case of the \fpa theory,
the approximation could hard-code one rounding mode for all
operations, change the variables and operations to have reduced
precision, or just omit some of the constraints.

\Comment{open source, extensible, collection of strategies and approximations, dsl for describing new approximations, etc.}

 \section{Specifying Approximations in \uppsat}
\label{sec:smallfloats}

In this section we show how to specify approximations in
\uppsat,\footnote{\url{https://github.com/uuverifiers/uppsat}} using
the example of reduced-precision
\fpa~\cite{DBLP:journals/jar/ZeljicWR17} from
Section~\ref{sec:reduced}. It should be remarked that one of the
design goals of \uppsat is the ability to define approximations in a
convenient, high-level way; the code we show in this section is mostly
identical to the actual implementation in \uppsat, modulo a small number of
 simplifications for the purpose of presentation.
We will first give
an intuition for this particular approximation, before breaking
it down into the elements that \uppsat requires.

\subsection{Approximation using Reduced-Precision \fpa}

Floating-point numbers are a two-parameter data type, denoted \fp{e,s}. The
parameters $e$ and $s$
are the number of bits used to store the exponent and the
significand in memory, respectively.
\Standard specifies several distinct combinations of $e$ and
$s$, for example, single precision \fp{8,24} and double precision \fp{11,53}
floating-point numbers. And indeed, these are the most commonly used data types
to represent real-valued data. Solving \fpa constraints typically involves
encoding them into bit-vector arithmetic and subsequently into propositional logic,
via a procedure called \emph{flattening} or \emph{bit-blasting}. The size and
complexity of the propositional formula depends on the size of floating-point
numbers in memory. Such an encoding of \fpa constraints can become prohibitively
large very quickly. However, many key values, e.g., special values, one, powers
of two, can be represented compactly and exist in floating-point representations
that contain very few bits. Therefore, reasoning over single- or
double-precision
floating-point numbers, for models that involve mostly (or only) these values can
be wasteful. Instead, we solve a \emph{reduced-precision} version of the
formula, i.e., we work with Reduced Precision Floating Points (RPFP). Reducing the precision does not affect the structure of the formula,
and only changes the sorts of floating-point variables, predicates and operations. Bit-blasting
reduced-precision constraints results in significantly smaller propositional
formulas, that are still expressive enough to find an approximate solution.

\Comment{PB: We probably should refer to fig:smallfloats}

\begin{figure}
  \begin{center}
    \begin{tikzpicture}[node distance=0]
    \node[draw=none, minimum height=0.4cm, minimum width=0.25 cm] (top) at (0,0){$n \,:$ };
    \node[draw=none, minimum height=0.4cm, minimum width=0.25 cm, below = 1cm of top] (mid) { $$ };
    \node[draw=none, minimum height=0.4cm, minimum width=0.25 cm, below= 1cm of mid] (bot) {$0 \,:$ };
    
    \node[draw=yellow!50, fill=yellow!50, minimum height=0.4cm, minimum width=0.25 cm, right = 0.5cm of top] (nSign) {1};     
    \node[draw=orange!50, fill=orange!50,  minimum height=0.4cm, minimum width = 2cm, right = of nSign] (nExp) {8};     
    \node[draw=red!50, fill=red!50,  minimum height=0.4cm, minimum width = 5.75cm,  right = of nExp] (nMan) {23};

    \node[draw=yellow!50, fill=yellow!50, minimum height=0.4cm, minimum width=0.25cm, right = 2.5cm of mid] (mSign) {1};
    \node[draw=orange!50, fill=orange!50,  minimum height=0.4cm, minimum width = 1.25cm, right = of mSign] (mExp) {5};  
    \node[draw=red!50, fill=red!50,  minimum height=0.4cm, minimum width = 2.5cm,  right = of mExp] (mMan) {10};

    \node[draw=yellow!50, fill=yellow!50, minimum height=0.4cm, minimum width=0.25cm, right= 3.75cm of bot ] (bSign) {1};
    \node[draw=orange!50, fill=orange!50,  minimum height=0.4cm, minimum width = 0.75cm, right = of bSign] (bExp) {3};
    \node[draw=red!50, fill=red!50,  minimum height=0.4cm, minimum width = 0.5cm,  right = of bExp] (bMan) {2};

    \node[draw=none, minimum height=0.4cm,, above= 0.2cm of bExp] {$\vdots$};
    \node[draw=none, minimum height=0.4cm,, above= 1.6cm of bExp] {$\vdots$};

    \path[->, draw] ($(bot.north) - (0.14,0)$) -- ($(top.south)- (0.14,0)$);

  \end{tikzpicture}

    \caption{Example of scaling a single precision floating-point sort}
    \label{fig:smallfloats}
  \end{center}
\end{figure}
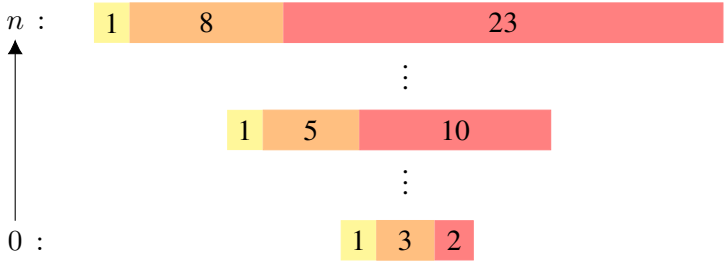

\subsection{Reduced-Precision FPA Approximation in \uppsat}

An approximation in \uppsat consists of several parts: an
approximation context (the ``approximation core''),
a codec, a model reconstruction strategy, and a
 refinement strategy for model- and proof-guided refinement.
Fig.~\ref{fig:rpfpapp} shows the
object~\code{RPFPApp} implementing the reduced-precision
floating-point approximation. The approximation object is implemented
using \scala\ \emph{mix-in traits} (shown in Fig.~\ref{fig:trait-dia}), which enable the modular
mix-and-match approximation design. In the following paragraphs, we
show the key points of reduced precision floating-point approximation
through its component traits.

\begin{figure}
  \begin{center}
    \begin{lstlisting}[language=Scala,numbers=left]
object RPFPApp extends RPFPContext 
                  with RPFPCodec
                  with EAAReconstruction
                  with RPFPModelRefinement
                  with RPFPProofRefinement
\end{lstlisting}
   \end{center}
  \caption{Specification of the reduced precision floating-point
    approximation as a \scala\ object.}
  \label{fig:rpfpapp}
\end{figure}

\paragraph{Approximation context.} An approximation context specifies input and
output theory, a precision domain and a precision ordering. The
reduced-precision floating-point approximation encodes floating-point constraints as
scaled-down floating-point constraints. Therefore, both the input and the output
theory are the quantifier-free floating-point theory (\code{FPTheory}). The
precision uniformly affects both the significand and the exponent, so a scalar
data type \lstinline!Prec = Int!
is sufficient to represent precision. In particular, we choose
integers in the range $[0,5]$ as the precision domain with the usual ordering.
Fig.~\ref{fig:smallfloatscontext} shows the specification of \code{RPFPContext} the
approximation context object for the reduced precision floating-point
approximation.

\begin{figure}
  \begin{center}
    \begin{lstlisting}[language=Scala,numbers=left]
trait RPFPContext extends AppContext {
  val  inputTheory  = FPTheory
  val  outputTheory = FPTheory   
  type Prec         = Int
  val  pOrdering    = new IntPOrder(0, 5)
}
\end{lstlisting}   \end{center}

  \caption{Specifying the approximation context for the reduced precision
    floating-point approximation.}
  \label{fig:smallfloatscontext}    
\end{figure}

\paragraph{Codec.} The essence of approximation takes place in the encoding of
the formula, and conversely how the approximate model is decoded.
These two operations are implemented by the \code{RPFPCodec} trait,
shown in Fig.~\ref{fig:encodeNode} for the case of the
reduced-precision \fpa. Reduced-precision floating-point approximation
scales-down the sort of floating-point variables and operations, while
keeping the high-level structure of the formula. Scaling for
operations and variables are performed based on precision values,
while predicate nodes are scaled to the largest sort among their
children. Constant literals and rounding modes remain unaffected by
encoding. The discrepancy in sorts due to individual precisions is
removed by inserting \emph{fp.toFP} casts where necessary. The \emph{fp.toFP} declaration is an SMT-lib function which casts a Floating Point value to a given sort. To ensure internal consistency
of the approximate models, all occurences of a variable share the same
precision. Predicate scaling requires that the sorts of the arguments
are known, i.e., arguments are already encoded when their parent node
is encoded. Therefore, we consider a formula as an abstract syntax
tree (AST) and use a post-order visit pattern over the formula.
\uppsat provides a template trait for such an encoding called
\code{PostOrderCodec}. To implement it, the user needs to define two
hook functions: \code{encodeNode} and \code{decodeNode}.

\Comment{PB: Here we use the words ``pad the arguments'', do we mean encode the children? They should be already encoded due to the post-order pattern?}
To encode a node, we scale the sort, pad the arguments, re-instantiate the
symbol to the new sort and bundle the new symbol with the padded children. These
steps are implemented in  the \code{encodeNode} hook function, shown in Fig.~\ref{fig:encodeNode}. The details of scaling
the sort are shown in the \code{scaleSort} auxiliary function. The sort scaling
is linear and consists of 6 sorts, starting with the \fp{3,3} up to (and
including) the original sort. The \code{cast} function adds a floating-point cast
\emph{fp.toFP} between the parent and the child node where necessary. Implementation
of the functions \code{cast} and \code{encodeSymbol} is straightforward and omitted
in the interest of brevity.

\begin{figure}
  \begin{center}
    \scriptsize
    \begin{lstlisting}[language=Scala,numbers=left]
trait RPFPCodec extends RPFPContext
                   with PostOrderCodec {
  def scaleSort(
        ast : AST,
        p : Int,
        encodedChildren : List[AST]) = {
    ast.symbol match {
      case _ : FloatingPointPredicateSymbol => {
        val childrenSorts =
          encodedChildren.filterNot(_.isLiteral).map(_.symbol.sort)
        childrenSorts.foldLeft(childrenSorts.head)(fpsortMaximum)
      }
   
      case _ : FloatingPointFunctionSymbol => {
        val FPSort(eBitWidth, sBitWidth) = sort
        val eBits = 3 + ((eBitWidth - 3) * p)/pOrdering.maxPrecision
        val sBits = 3 + ((sBitWidth - 3) * p)/pOrdering.maxPrecision
        FPSort(eBits, sBits)
      }
      
      case _ => sort 
    }
  }

  def encodeNode(
        ast : AST,
        encodedChildren : List[AST],
        precision : Int) : AST = {
    val sort = scaleSort(ast, precision, encodedChildren)
    val children = encodedChildren.map(cast(_, sort))
    val symbol = encodeSymbol(ast.symbol, sort, children)
    AST(symbol, ast.label, children)
  }

  // ...
}

\end{lstlisting}

-
   \end{center}
  \caption{Implementation of the reduced-precision encoding. \comment{
this code has to be polished a bit more. identifiers in the code are
inconsistent with other listings? In particular, it still contains
the word ``smallfloat''}}
  \label{fig:encodeNode}
\end{figure}

\begin{figure}
  \begin{center}
    \scriptsize
    \begin{lstlisting}[language=Scala,numbers=left]
trait RPFPCodec extends RPFPContext
                   with PostOrderCodec {

  // ...                     

  def decodeFPValue(
        symbol : ConcreteFunctionSymbol,
        value : AST,
        p : Int) : ConcreteFunctionSymbol = {
    (symbol.sort, value.symbol) match {
      case (FPSort(e, s), fp : FloatingPointLiteral) => {
        fp.getFactory match {
          case _ : FPSpecialValuesFactory => fp(FPSort(e, s))
  
          // If exponent bits are all zeros it is a special value
          case _ if !fp.eBits.contains(1) => {
            val sPrefix = fp.sBits.dropWhile(_ == 0)
            val eUnderflow = fp.sBits.length - sPrefix.length
            val sBits = sPrefix.tail ::: List.fill(s-sPrefix.length)(0)  
            val exp = - bias(fp.eBits.length) - eUnderflow
            val eBits = intToBits(biasExp(exp, e), e)                       
            FloatingPointLiteral(fp.sign, eBits, sBits, FPSort(e,s))
          }
          
          case _ => {
            val exp = unbiasExp(fp.eBits, fp.eBits.length)
            val eBits = intToBits(biasExp(exp, e), e)
            val missing = (s - 1) - fp.sBits.length
            val sBits = fp.sBits ::: List.fill(missing)(0)
            FloatingPointLiteral(fp.sign, eBits, sBits, FPSort(e, s))
          }
        }
      }      
      case _ => value.symbol
    }
  }
    
  def decodeNode(
        args : (Model, PrecMap[Prec]),
        decodedModel : Model,
        ast : AST) : Model = {
    val (appModel, pmap) = args
    
    val decodedValue =
          decodeFPValue(ast.symbol, appModel(ast), pmap(ast.label))
    
    decodedModel.set(ast, Leaf(decodedValue))
    decodedModel
  }  
}
\end{lstlisting}
   \end{center}
  \caption{Implementation of the reduced-precision decoding (part of the trait in Fig.~\ref{fig:encodeNode}.) \comment{
this code has to be polished a bit more. identifiers in the code are
inconsistent with other listings? In particular, it still contains
the word ``smallfloat''}}
  \label{fig:decodeNode}
\end{figure}

After the back-end solver returns a model of the approximate
constraints, it needs to be decoded. Decoding is essentially casting
variable assignments to their sort in the original formula. For
example, suppose a formula $\phi$ over variables $x$ and $y$ of sort
\fp{8,24} is encoded to the formula~$\hat\phi_{3,3}$ (as in
Ex.~\ref{ex:rpfp}), yielding a model $\aux m = \{x \mapsto 0_{3,3}; y
\mapsto 1_{3,3}\}$. \Comment{PB: Maybe we just remove this explanation of how change sort of floating points?} Decoding will cast these values from the model of
approximate constraints and translate them to the same values, but in
their full-precison sort, resulting in a variable assignment $m = \{x
\mapsto 0_{8,24}; y \mapsto 1_{8,24}\}$. Special values are also
decoded by re-instantiating them in the original sort. Other values
are decoded by adding the missing bits to their representation. The
missing bits in the encoded formula are implicitly set to zero. To
decode the significant, the missing zero bits are simply
re-inserted. Padding the exponent requires some attention due to the
details of \standard. The values of the exponent are stored with an
added \emph{bias value}, which is dependent on the exponent bit-width.
To pad the exponent, we first remove the bias of the exponent in
reduced precision, and then add the bias of full-precison
FP. (Subnormal floating-point values require more attention.)

\code{PostOrderCodec} implements the \code{decodeModel} function through the
\code{decodeNode} hook function. The hook function is applied to all the values
in the model of the approximate constraints. The decoding of the values is
performed by the \code{decodeFPValue} function, all shown in Fig.
\ref{fig:encodeNode}.

\paragraph{Model reconstruction strategy} specifies how to obtain a model of the
input constraints starting from the decoded model. A simple strategy to obtain a
reconstructed model is to satisfy the same Boolean constraints (constraints which are true or false, e.g., equalities, inequalities, predicate) as the approximate
model, i.e., to try and satisfy the Boolean structure in the same way. \Comment{PB: I changed atomic constraints to Boolean constraints} We call those constraints \emph{critical atoms}. However, due to the
difference in semantics, values of the decoded model are not guaranteed to
satisfy them. Typically, the rounding error, significantly larger in reduced
precision \fpa, accumulates and changes the value of critical atoms under the
original semantics. Therefore, evaluation of critical atoms under the original
semantics is necessary to ensure that the model satisfies the original formula.
In fact, rather than evaluating the critical atoms simply as a verification
step, evaluation can be used to infer the error-free values under the original
semantics. Starting from an empty candidate partial model, the constraints are
evaluated in a bottom-up fashion. Thus, the reconstruction can be defined by
defining the reconstruction of a single node in the \code{reconstructNode} hook
function, shown in Fig. \ref{fig:reconstructNode}.

The key to a good reconstruction strategy is propagation. Certain
constraints allow more information to be propagated than others. For
example, equality $x = y + z$ uniquely determines the value of $x$ if
the values of $y$ and $z$ are known and the equality is known to
hold. Whereas, an inequality for example, allows for less propagation. The decoded
model contains the information which critical atoms need to be
satisfied. The critical atoms combined with a bottom-up evaluation,
allow propagation to take place, by applying
\emph{equality-as-assignment}; if the following conditions are satisfied: 1.~the equality
is true in the approximate model, 2.~its left- or right-hand side is a
variable that is currently unassigned in the candidate model, and
3.~the value of the other side is defined in the candidate model, then
the variable can be assigned the value of the other side in the
candidate model. Equality-as-assignment is crucial for elimination of
rounding errors due to the RPFP encoding. Note that this
reconstruction strategy can fail if cyclic dependencies exist among
the constraints.

An important aspect of the reconstruction strategy is the order of
evaluation.  Bottom-up evaluation, bar equality-as-assignment,
requires that all the sub-expressions have a value in the candidate
model. The base case are variables which might be undefined in the
candidate model. If they are undefined in the candidate model when
they are needed for evaluation, they are assigned the value from the
decoded model. This means, that evaluation of inqualities ahead of
equalities might prevent equality-as-assignment to take
place. Therefore, we wish to evaluate predicates in an order such that
equality-as-assignment enabled critical atoms are evaluated
first. Therefore we separate all predicates of the form $x = y$ or $x
= f(\ldots)$, where $x, y$ var variables and $f$ is some operation or
predicate. We call these equations \emph{definitional}. In order to
maximise the propagation during the reconstruction,
definitional equalities are prioritised over the remaining
predicates.

Furthermore, the definitional equalities are sorted based on a
topological order of the variables in a graph defined by viewing
definitional equalities as directed edges in a graph. An equation of
the form $x = f(\ldots)$ generates edges from every variable on the
right hands side to $x$, and an equation of the form $x = y$ generates
an edge from $x$ to $y$ and one edge from $y$ to $x$. The topological
sorting of variables starts with the varibles occuring in definitional
equalities, that have the lowest input degree. Their values can be
safely copied from the decoded model. The resulting order of variables
corresponds to a bottom-up propagation through the formula, that
maximises applicatoins of equality-as-assignment in the
reconstruction. Any cyclic dependencies will be broken, with algorithm
picking any variable arbitrarily (we leave it to future work to design
a reasonable heuristic). After a topological order of the variables
have been established, the equalities ordered according to the
variable-ordering.

The model reconstruction performs a bottom-up reconstruction of critical atoms
in the topological order of the equalites followed by the remaining predicates.
Ordering the predicates in the described manner increases the likelihood of
propagation fixing rounding errors introduced by the FPFP encoding.

\begin{figure}
  \begin{center}
    \scriptsize
    \begin{lstlisting}
  def reconstructNode(decodedModel  : Model,
                      candidateModel : Model,
                      ast : AST) : Model = {
    val AST(symbol, label, children) = ast

    if (children.length > 0 &&
        !equalityAsAssignment(ast, decodedModel, candidateModel)) {
      val newChildren = for ( c <- children) yield {
        getCurrentValue(c, decodedModel, candidateModel)
      }

      val newAST = AST(symbol, label, newChildren.toList)
      val newValue = ModelReconstructor.evalAST(newAST, inputTheory)
      candidateModel.set(ast, newValue)
    }
    candidateModel
  }
  \end{lstlisting}
   \end{center}

  \caption{An example of post-order reconstruction, using equality-as-assignment}
  \label{fig:reconstructNode}
\end{figure}

\paragraph{Model-guided refinement strategy} takes place when model
reconstruction fails to obtain a model. Model-guided refinement increases the
precision of the formula, based on the decoded model and the failed candidate
model. The refinement increases the precision of operations, but only so far
that a more precise model is obtained in the next iteration. Comparison of the
evaluation of the formula under the two assignments, highlights critical atoms
that should be refined. These atoms evaluate to true in the approximate model
and to false in the failed candidate model. Since \fpa is a numerical domain, it
is possible to apply some notion of \emph{error} to determine which nodes
contribute the most to the discrepancies in evaluation and use them to rank the
sub-expressions. After ranking, only a portion of them is refined, say 30\%.
Refinement amounts to increasing precision by some amount, in this case a
constant. In general, one could use the error to determine by how much to
increase the precision. Since error-based refinement can be applied to any
numerical domain, \uppsat implements an \emph{error-based refinement strategy},
which is instantiated by providing an implementation of the \code{nodeError}
hook function, shown in Fig.~\ref{fig:modelRefinement}.

\begin{figure}
  \begin{center}
    \scriptsize
    \begin{lstlisting}[language=Scala,numbers=left]
trait RPFPMGRefinementStrategy
  extends RPFPContext 
     with ErrorBasedRefinementStrategy {

  def defaultRefinePrecision(p : Int) : Int = {
    p + 1
  }
  
  def nodeError(
        decodedModel : Model,
        failedModel : Model
        accu : Map[AST, Double],
        ast : AST) : Map[AST, Double] = {
    ast.symbol match {
      case literal : FloatingPointLiteral => accu      
      case fpfs : FloatingPointFunctionSymbol => {
        val Some(outErr) =
             computeRelativeError(ast, decodedModel, failedModel)
        val computedErrors =
             ast.children.map{
               computeRelativeError(_, decodedModel, failedModel)
             }
        val inErrors = computedErrors.collect{case Some(x) => x}
        val sumInErrors = inErrors.fold(0.0){(x,y) => x + y}
        val avgInErr = sumInErrors /  inErrors.length
        accu + (ast -> outErr / (1 + avgInErr))        
      }
      case _ => accu
    }
  }
}
\end{lstlisting}
   \end{center}

  \caption{Implementation of a model-guided refinement strategy based on relative errors}
  \label{fig:modelRefinement}
\end{figure}

\paragraph{Proof-guided refinement strategy} uses proofs of unsatisfiability to
refine the formula. Formula can be refined using unsatisfiable cores,
when an approximate model is not available. At the moment \uppsat has
no support for obtaining cores or proofs from the back-end
solvers. Instead, a na\"ive refinement strategy is used, which
increases all the precisions by a constant, shown in
Fig.~\ref{fig:proofRefinement}.

\begin{figure}
  \begin{center}
    \scriptsize
    \begin{lstlisting}[language=Scala,numbers=left]
trait RPFPPGRefinementStrategy
  extends UniformPGRefinementStrategy {
  
  def unsatRefinePrecision(p : Int) : Int = {
    p + 1
  }
}
\end{lstlisting}
   \end{center}

  \caption{A naive proof-guided refinement strategy uniformly increasing precision}
  \label{fig:proofRefinement}
\end{figure}

 \section{Approximations in \uppsat}
In this section, we discuss some more general aspects of approximations within
the \uppsat framework. In addition to listing alternatives to the components of
the RPFP approximation, some implementation details are discussed.

\paragraph{Precision domains} are crucial for both the expressiveness of the
encoding and the subtlety of the refinement. Precision can be \emph{uniform} or
\emph{compositional} in terms of their relationship with the formula. Uniform
precision assigns a single precision value to the entire formula, whereas
compositional precision associates different values with some or all parts of
the formula. As we have seen, the RPFP approximation uses a compositional
precision, which is associated with variable and function nodes. Uniform
precision is used in the BV and RA approximations, which are presented in the
next section.

From the perspective of encoding expressivity, precision can be a
scalar value or a vector. While in most cases scalar precision
suffices, vectors (or tuples) can be used to elegantly encode more
expressive approximations. For instance, a pair of precisions
associated with an \fpa node allow the significand and the exponent to
have independent bit-widths. Choosing a suitable precision domain is
important, both for the compactness of the definition of approximation
in \uppsat and for the performance of the resulting approximating
solver. Too crude a precision domain might yield a negligible
improvement of performance, while too fine a precision domain might
spend too much time wandering through the different approximations.

\paragraph{Encoding and decoding} are the heart of the
approximation. The two translations are intertwined, a simple elegant
encoding is useless if the model cannot be translated back in a
meaningful way. In fact, the encoding often suggests a natural way of
implementing the decoding, since the translations are in a sense
inverse. In general, an encoding is just an arbitrary translation of a
formula of the input theory to a formula of the output theory; in
practice, like for the RPFP approximation, the encoding does not
change the overall structure of the formula, but merely adjust the
sorts involved.  Other approximations might add global constraints in
the encoding, e.g., definitional equalities or impose ranges, or they
might add or remove nodes in the formula. For instance, the real-arithmetic
approximation RA of \fpa will not encode the rounding modes, since they
do not have an equivalent in real arithmetic. The decoding of a real
model needs to produce some reasonable values for the rounding modes
somehow. This can, for instance, be done by choosing a pre-selected default
value.

To maintain information of the relationship between the original and
the encoded formula, \uppsat uses \emph{labeled abstract syntax
  trees}. During the encoding, the result of the encoding is assigned
the label of the source node in the original formula that it encodes.
The labels offer a way to keep track of the translation, since the
encoding can be ambiguous to decode. All the approximations presented
in this paper are context-independent and node-based, i.e., it is
sufficient to specify the translation at the node level. \uppsat
offers a pattern for this kind of codec, called
\code{PostOrderCodec}. Overall, \uppsat can handle a broad range of
encodings, that can be specified succinctly within the framework.

\paragraph{Model reconstruction strategies} take place entirely in the
input theory, and as such can be combined with a number of different
encodings (they are independent of the chosen output theory). The
reconstruction strategy used by the RPFP approximation is simple in
the sense that it only evaluates expressions, and it does not pose
satisfiability queries to a solver. A different strategy, along
similar lines, might start the reconstruction from the difficult
(e.g., non-linear) constraints and then evaluate the remainder of the
formula. More complex strategies might use a solver during the
reconstruction to search for a model within some $\epsilon$ distance
of a decoded (failing) model. A \emph{numeric model lifting} strategy
was proposed by Ramachandran and
Wahl~\cite{DBLP:conf/fmcad/RamachandranW16}. Their method identifies a
subset of the model to be tweaked, and instantiates the formula as a
univariate satisfiability check. Except for a chosen variable, all the
variables in the formula are substituted by their value in the failing
decoded model. This approach often quickly patches the candidate
model. \uppsat can express these more advanced strategies, but
implementation and experiments in this direction have been left for
future work.

\paragraph{Refinement strategies} use the information obtained either
from the models or the proof of unsatisfiability to find a better
approximation. In cases when information is scarce (e.g., no proofs
are available in case of unsatisfiability), or the approximation is
very coarse and no useful information can be extracted from a decoded
model, a uniform refinement strategy can increase precision of the
entire formula. This is the case with fixed-point approximation BV and the
proof-guided refinement of the RPFP approximation.  In case of numeric
domains, a notion of \emph{error} can be used to determine which terms
to refine and by how much~\cite{DBLP:journals/jar/ZeljicWR17}. This is
the strategy used by the RPFP approximation. In the case of a
precision vector for each node in a formula, the error between the
decoded and candidate model can be used to refine either the exponent,
if the magnitude of the error is large, or the significant if the
error is very small.

\section{Other Approximations of \fpa}
\label{sec:approximations}
We have shown in detail the RPFP approximation of \fpa, and discussed
different components that can be used in general. In this section we
outline two further approximations of \fpa that have been implemented
in \uppsat: the fixed-point approximation BV
(Section~\ref{sec:fixedpointexample}), encoded as bit-vectors, and the
real-arithmetic approximation RA. Both approximations are currently
implemented in a more experimental and less refined way than the RPFP
approximation, but encouragingly, even simple approximations can give
rise to speed-ups compared to their back-end solvers (as shown in
Section~\ref{sec:experiments}).

\subsection{BV --- The Fixed-Point Approximation of \fpa}
\label{sec:fixedpoint}
The idea behind the BV approximation is to avoid the overhead of the rounding
semantics and special values of the \fpa, by encoding all the \fpa values and
variables and operations as values and operations of the fixed-point arithmetic.

\paragraph{The BV context.} The input theory is the theory of \fpa,
and the intended output theory is the theory of fixed-point
arithmetic. However, since fixed-point arithmetic is not commonly
supported by SMT solvers, we can encode fixed-point constraints in the
theory of fixed-width bit-vectors.  The precision determines the
number of integer and fractional binary digits in the fixed-point
representation of a number.  For simplicity, at this point we do not
mix multiple fixed-point formats in one formula, but instead apply
uniform precision in the BV approximation; as a result, all operations
in a constraint are encoded using the same fixed-point sort. As a
proof of concept, the precision domain is two-dimensional, with the
first component~$p_i$ in a pair~$(p_i, p_f)$ denoting the number of
integral, and the second component~$p_f$ the number of fractional bits
in the encoding, respectively. The precision domain ranges from
$(5,5)$ to $(25,25)$, with the maximum element~$(25,25) = \top$ being
interpreted as sending the original, unapproximated \fpa constraint to
Z3 as a fall-back solver.

\begin{Example}
  Given a variable of precision $(4, 5)$, we will have a domain of
  numbers between $1000.00000$ and $0111.11111$, which when interpreted
  in two's-complement notation are numbers between $-8$ and $7.96875$.
\end{Example}

\paragraph{The BV codec.}
A codec describes how values can be converted from the input theory to
the output theory, and vice versa. The floating-point operations are
in BV encoded as their fixed-point equivalents, which in turn are
encoded as bit-vector operations. This process is fairly
straightforward, with the exception of the rounding modes and special
\fpa values. The rounding modes and not-a-number values are omitted by
the encoding, while the remaining special values are encoded, with
respect to the current precision, either as zero or as the largest or
smallest value (in case of infinities). Translation of literal
floating-point constants amounts to a representation as the closest
value in the chosen fixed-point sort.
The decoding consists of converting a fixed-point number to a rational
number, followed by conversion to the closest floating-point number,
with some care taken for the special values.

\paragraph{BV reconstruction and refinement.}
The BV approximation uses the same model reconstruction strategy as
the RPFP approximation. In contrast, the chosen refinement strategy in
the BV approximation is currently very simple: since the precision is
uniform, the refinement is also uniform, regardless of whether an
approximate model is available or not. At each iteration, the
precision is increased by 4 in both dimensions, resulting in addition
of 4~bits to both the integral and fractional part of
numbers.\footnote{This means that the approximation does not really
  leverage the two-dimensional precision, and that maximal precision
  of the encoding is reached after at most 5 iterations.}

\subsection{RA --- The Real Arithmetic Approximation of \fpa}
\label{sec:real}

The third and possibly most obvious approach to approximate \fpa is by
encoding into real arithmetic constraints.  We present a comparatively
simplistic implementation of this kind of approximation, due to the
difficulty to refine approximations in real arithmetic in a meaningful
way (real arithmetic already represents to infinite-precision
arithmetic). Ramachandran and
Wahl~\cite{DBLP:conf/fmcad/RamachandranW16} describe a topological
notion of refinement, that requires a back-end solver that handles the
combined theory of real arithmetic and \fpa. However, solving
constraints over this combination of theories is challenging in
itself, and efficient SMT solvers are not publicly available, to the
best of our knowledge.

\paragraph{RA context.}
In the RA approximation, the \fpa is the input theory, and the output
theory is the theory of (non-linear) real arithmetic. The precision
domain is a uniform binary domain~$\{\bot, \top\}$, deciding whether
approximation is taking place at all ($\bot$), or whether the original
\fpa constraint is sent to a back-end solver (for $\top$; again, the
fall-back solver in this case is Z3). Essentially, this is a
hit-or-miss approximation, which either will work right away or
directly resort to the fall-back solver.

\paragraph{RA codec.}
The encoding is fairly straightforward, the \fpa operations are
translated as their real counter-parts, omitting the rounding modes in
the process. While the special values can be encoded, currently they
are not supported by the RA approximation. \fpa\ numerals are
converted to reals, i.e., in the case of normal \fpa numbers the
resulting real number is $(-1)^{\mathit{sign}}\cdot
\mathit{significand} \cdot 2^{\mathit{exponent}}$. Decoding will
translate a real number to the closest \fpa numeral.

\paragraph{RA reconstruction} coincides with the RPFP reconstruction.

\paragraph{RA refinement} is achieved by uniform refinement, and
results in the full precision~$\top$ after a single iteration. In the
case of the topological refinement proposed by Ramachandran and
Wahl~\cite{DBLP:conf/fmcad/RamachandranW16}, the precision domain
would be the same, but the precision itself would be compositional,
i.e., a precision would be associated with each node of the
formula. Essentially, the precision would represent a switch, deciding
whether a node should be encoded in real arithmetic or floating-point
arithmetic.

\Comment{PB: ``Essentially....'', is this the same thing said in the first paragraph of 5.2? Because after reading this paragraph I am thinking ``So why are we not doing that?'', but I guess it is because of no back-end}

\section{Related Work}
\label{sec:related}

\subsection{Approximations in General}

The concept of abstraction (and approximation) is central to software
engineering and program verification, and is increasingly employed in
general mathematical reasoning and in decision procedures as
well. Frequently only under- and over-approximations are considered,
i.e., the formula that is solved either implies or is implied by an
approximate formula. Counter-example guided abstraction
refinement~\cite{Clarke:2000:CAR:647769.734089} is a general concept
that is applied in many verification tools and decision procedures,
even on a relatively low level as in QBF
solvers~\cite{Janota:2012:SQC:2352219.2352233}, or in model-based
quantifier instantiation for SMT~\cite{DBLP:conf/cav/GeM09}. 

Specific instantiations of abstraction schemes in related areas also
include the bit-vector abstractions by Bryant et
al.~\cite{DBLP:conf/tacas/BryantKOSSB07} and Brummayer and
Biere~\cite{DBLP:conf/eurocast/BrummayerB09}, as well as the (mixed)
floating-point abstractions by Brillout et al.~\cite{bkw2009-fmcad}.

Van Khanh and Ogawa present approximations for solving polynomials
over reals~\cite{VanKhanh201227}. Gao et
al.~\cite{DBLP:conf/cade/GaoKC13} present a $\delta$-complete decision
procedure for non-linear reals, considering over-approximations of
constraints by means of $\delta$-weakening.

\subsection{Decision Procedures for Floating-Point Arithmetic}

The SMT solvers MathSAT~\cite{mathsat}, Z3~\cite{z3}, and
Sonolar~\cite{sonolar} feature bit-precise conversions from FPA to
bit-vector constraints, known as bit-blasting, and represent the
currently most commonly used solvers in program verification. As we
show in our experiments, bit-blasting can be boosted significantly
with the help of our approximation approach.

A general framework for decision procedures is Abstract CDCL,
introduced by D'Silva et al.~\cite{DBLP:conf/popl/DSilvaHK13}, which
was also instantiated for
FPA~\cite{DBLP:conf/tacas/DSilvaHKT12,Float-ACDCL}. This approach
relies on the definition of suitable abstract domains (as defined for
abstract interpretation~\cite{DBLP:conf/popl/CousotC77}) for
constraint propagation and learning. In our experimental evaluation
(Section~\ref{sec:experiments}), we compare to two decision procedures
for FPA that are implemented in MathSAT; instances of ACDCL and eager
translation to bit-vectors. ACDCL can seamlessly be integrated into
the \uppsat framework, for instance to solve approximations or to
derive an approximation based on abstract domains.

The work presented in this paper builds on previous research on the
use of approximations for solving \fpa
constraints~\cite{DBLP:conf/cade/ZeljicWR14,DBLP:journals/jar/ZeljicWR17}.
\uppsat is also close in spirit to the framework presented by
Ramachandran and Wahl~\cite{DBLP:conf/fmcad/RamachandranW16} for
efficiently solving FPA constraints based on the notion of `proxy'
theories, which correspond to our `output theories.' This framework
applies a relatively sophisticated method of reconstruction, by
applying a fall-back \fpa solver to a version of the input constraint
in which all but one variables have been substituted by their value in
a failing decoded model. Such reconstruction could also be realized in
\uppsat, and an implementation in \uppsat and experimental comparison
with other reconstruction methods is planned as future work.

A further recent approximation-based solver for \fpa is
XSat~\cite{DBLP:conf/cav/FuS16}. In XSat, reconstruction of models is
implemented with the help of randomized optimization, which results in
good performance, but does not give rise to a decision procedure
(incorrect sat/unsat results can be produced).

There is a long history of formalization and analysis of FPA concerns using
proof assistants, among others in Coq by
Melquiond~\cite{DBLP:journals/iandc/Melquiond12} and in HOL Light by
Harrison~\cite{harrison-floats}. Coq has also been integrated with a dedicated
floating-point prover called Gappa by Boldo et
al.~\cite{DBLP:conf/mkm/BoldoFM09}, which is based on interval reasoning and
forward error propagation to determine bounds on arithmetic expressions in
programs~\cite{Daumas:2010:CBE:1644001.1644003}. The ASTR\'EE static
analyzer~\cite{DBLP:conf/esop/CousotCFMMMR05} features abstract
interpretation-based analyses for FPA overflow and division-by-zero problems in
ANSI-C programs.

 \section{Experimental evaluation}
\label{sec:experiments}

 In this section we evaluate the effectiveness of the discussed
 approximations of
 \fpa, when combined with the bit-vectors, real and \fpa decision procedures implemented
 in MathSAT and Z3.

 \paragraph{Experimental setup.}
 The evaluation is done on the \emph{satisfiable} benchmarks of the
 \verb!QF_FP!  category of the SMT-LIB. Currently, \uppsat does not
 extract unsatisfiable cores from back-end, and none of the
 approximations have a meaningful proof-based refinement strategy, so
 that performance on unsatisfiable problems is guaranteed to be worse
 than that of the back-end solver. All experiments were done on an AMD
 Opteron 2220 SE machine, running 64-bit Linux, with memory limited to
 1.0gb, and with a timeout of one hour.

 \paragraph{\uppsat instances.} Table~\ref{tbl:configurations} shows
 combinations of approximation and back-end solver that we evaluate. The \uppsat
 instances are named in the form of \emph{APPROXIMATION}(\emph{back-end}). Note
 that the back-end needs to implement a decision procedure for the output theory
 of the approximation. As a consequence, we have three configurations for the
 RPFP approximation, by using bit-blasting procedures in Z3 and MathSAT and the
 ACDCL algorithm in MathSAT as decision-procedures for \fpa. \uppsat currently
 lacks support for the bit-vector theory in MathSAT, so for the BV
 approximation only the
 bit-vector solver in Z3 is used as the back-end. The back-end for the RA approximation
 is the nlsat tactic in Z3, since it is the only decision procedure in Z3 and
 MathSAT to support non-linear constraints over reals.
 \begin{table}[t] 
   \begin{center}
   \begin{tabular}{l|ccccc}
     & ACDCL & MathSAT& Z3& nlsat & \\ \hline
     RPFP & \Checkmark & \Checkmark & \Checkmark & & \\
     BV & & & \Checkmark & & \\
     RA & & & & \Checkmark
   \end{tabular}
   \end{center}

   \caption{Combinations of approximations, shown in rows, and back-end solvers,
     shown in columns, used to instantiate \uppsat are denoted by
     \smash{\Checkmark}. The
     instances of \uppsat are named in the format \emph{APPROXIMATION(back-end)} }
   \label{tbl:configurations}
 \end{table} 
 \paragraph{Investigated questions.}
 \Comment{Which implenetation? UppSat or the backend?}
 In previous work, we have observed that the RPFP
 approximation improves performance of bit-blasting implemented in the Z3 SMT
 solver~\cite{DBLP:journals/jar/ZeljicWR17}. Here we seek to reproduce those
 results, but also to see whether similar behavior can be observed with other
 implementations and algorithms. We were interested in answering the following
 research questions:

 \begin{itemize}
 \item Is the positive effect of the RPFP approximation on performance of the
   bit-blasting approach for \fpa independent of the implementation?
 \item Does the RPFP approximation have a positive effect on the ACDCL
   algorithm for \fpa?
 \item What is the impact of approximations on the state-of-the-art for the
   theory of \fpa?
 \end{itemize}

 \begin{table}[tb]
   \begin{center}
     \scriptsize
     \begin{tabular}[h!]{@{}lccc|ccccc@{}}
& acdcl & mathsat & z3 & BV & RPFP & RPFP & RPFP & RA\\
&  &  &  & (z3) & (acdcl) & (mathsat) & (z3) & (nlsat)\\
\hline
Solved & 86 & 99 & 97 & 91 & 78 & \textbf{101} & \textbf{101} & 90\\
Timeouts & 44 & 31 & 33 & 39 & 52 & 29 & 29 & 40\\
Best & 65 & 4 & 6 & 9 & 3 & 9 & 9 & 4\\
Average Iterations & - & - & - & 2.69 & 3.59 & 3.16 & 3.02 & 1.85\\
Max Precision & - & - & - & 23 & 2 & 1 & 2 & 110\\
Average Rank & 3.81 & 5.40 & 6.33 & 5.42 & 5.32 & 4.38 & 4.53 & 6.60\\
Total Time (s) & 10071 & 16748 & 34526 & 11979 & 8448 & 8279 & 14992 & 27169\\
Average Time (s) & 117.10 & 169.17 & 355.94 & 131.64 & 108.30 & 81.97 & 148.43 & 301.87\\
Only solver & 1 & 0 & 2 & 0 & 0 & 1 & 0 & 0\\
\end{tabular}
    \end{center}

   \caption{Comparison of solver performance in terms of  number of benchmarks
     solved and relative ranking of runtimes. The first three columns give
     numbers for the back-ends alone, while the other columns consider
     the configurations specified in Table~\ref{tbl:configurations}.}
  \label{tbl:all}
\end{table}

\Comment{PB: Changed from the 130 to 130 non-trivial, since there are a couple of benchmarks we don't support}
To answer these questions, we compare the performance of the back-ends
and the \uppsat instances on 130 non-trivial\footnote{The
  regression tests in the \texttt{wintersteiger} family were ignored
  for the evaluation.}  satisfiable benchmarks of the \verb!QF_FP!
category of the SMT-LIB benchmarks.  On each benchmark, solvers were
assigned a rank based on their solving time, i.e., if a solver had the
smallest solving time, it was assigned rank~1, the solver with the
next smallest solving time rank~2, etc.

\begin{figure}[tb]
  \begin{center}
    \makebox[\textwidth]{\includegraphics[width=0.75\paperwidth]{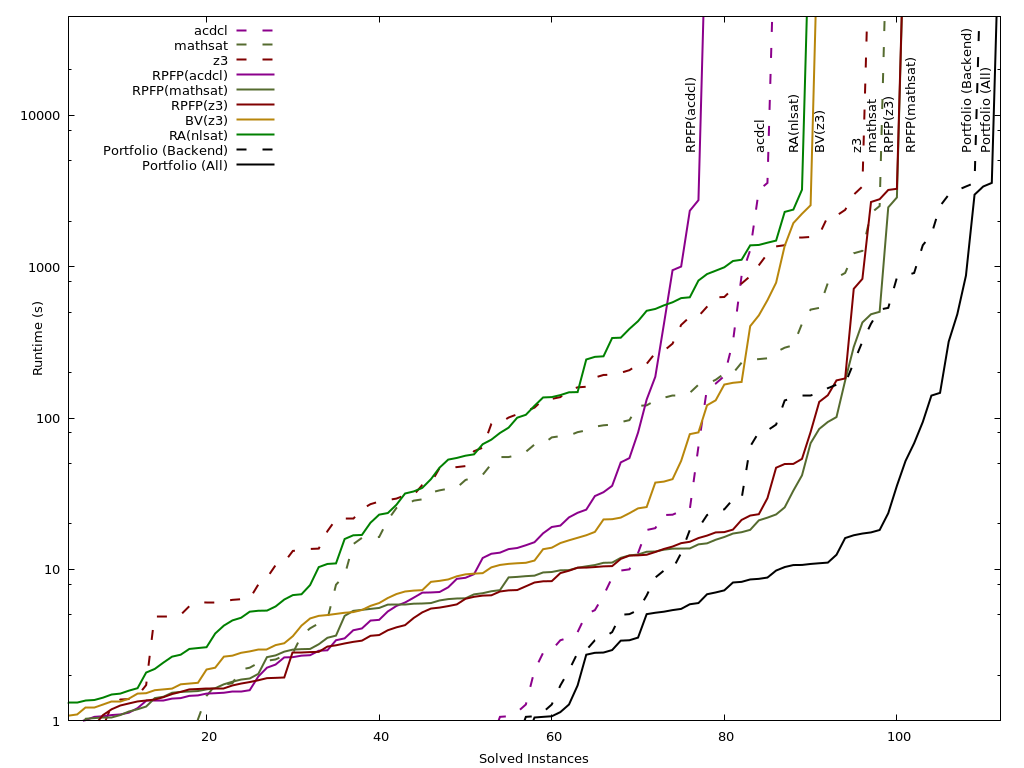}}
  \end{center}
  \caption{Cactus plot shows how many instances can be solved by a solver in a
    given amount of time. Comparison of all solvers on the satisfiable SMT-LIB benchmarks.}
  \label{fig:cactus}
\end{figure}

\Comment{PB: This is not refinement iterations}
The results are summarized in Table~\ref{tbl:all}, and a more detailed
view of runtimes is provided by the cactus plot shown in
Figure~\ref{fig:cactus}. Table~\ref{tbl:all} shows, for each solver,
the number of benchmarks solved within the 1 hour timeout, the number
of timeouts, the number of instances for which the solver was fastest,
the average number of refinement iterations on solved problems, the
number of benchmarks for which refinement reached maximum
precision~$\top$, the average rank, the total time needed to process
all benchmarks (excluding timeouts), the average solving time
(excluding timeouts), and the number of unique instances only solved by the respective
solver.

\paragraph{Discussion.}
We can observe that the RPFP approximation combined with bit-blasting,
either in Z3 or \mathsat, solves the largest number of instances. When
comparing the average rank, \mathsat\ comes out as the marginally
better choice of back-end. This is expected, based on the performance
on the back-ends themselves.  All the configurations shine on at least
a few benchmarks, indicating that the approximations do offer an
improvement. Furthermore, the ACDCL algorithm outperforms all the
other solvers on 65 benchmarks, which is also indicated by the lowest
average rank, but it solves fewer benchmarks that the bit-blasting
approaches in total.

Looking only at the approximations, we can see that on average the benchmarks
are solved using around three iterations. The notable exception is the RA
approximation, which performs at most two iterations,
the RA approximation and the
full \fpa semantics. This indicates that for many of the benchmarks,
full-precision encoding is not really necessary, since the RPFP approximation
rarely reaches maximum precision. However, the BV and RA approximations reach
maximal precision more often. In their defense, both BV and RA approximations
are presented as a proof of concept, since neither has tailored reconstruction
and refinement strategies.
\Comment{PB: Equality-as-assignment is arguably as tailored for RPFP as for BV?}

\paragraph{Virtual portfolios.}
To compare the impact of the approximations on the state-of-the-art,
we compare a virtual portfolio of the back-end solvers alone, and a
virtual portfolio of both the back-ends and the \uppsat
instances. Table~\ref{tbl:portfolio} shows the number of benchmarks
solved, the number of timeouts, and the total and average solving
time. The addition of the \uppsat instances allows only two more
benchmarks to be solved, compared to the back-end portfolio. However,
the total solving time is improved dramatically.

\begin{table}[h]
  \begin{center}
    \small
    \begin{tabular}[h!]{l cc}
 & Virtual Portfolio (Back-end) & Virtual Portfolio (All)\\
Solved & 110 & 112\\
Timeouts & 20 & 18\\
Total time & 25135 & 12516\\
Average time & 228.50 & 111.75\\
\end{tabular}
   \end{center}
  
  \caption{Comparison of virtual portfolio solver performance.}
  \label{tbl:portfolio}
\end{table}

\paragraph{Cactus plot.}
To complement the aggregated data, the cactus plot in Figure~\ref{fig:cactus}
shows on the X axis how many instances can be solved in the amount of time shown
on the Y axis, by each of the solvers and the portfolios. The \uppsat instances
are shown using full lines, while the back-ends are presented using dashed lines.
The colors denote the same back-end, e.g., mathsat and RPFP(mathsat) are both
colored green.

 It corroborates that the ACDCL algorithm is very efficient in
solving many benchmarks, solving as many as 68 in less than 10s, however,
eventually it gets overtaken by the other solvers. Looking more closely at the
RPFP approximation, we can conclude that it improves performance of bit-blasting
considerably, regardless of the implementation (\mathsat\ or Z3).
On the other hand, RPFP seems to
hinder, rather than help, the already very efficient ACDCL
algorithm.\footnote{Earlier experiments using the stable version 5.4.1 of
   MathSAT
  have shown similar effects of the RPFP approximation to those on the
  bit-blasting methods. However, overall the performance results were not
  consistent with performance of MathSAT in previous publications, and indicated
  a bug. We thank Alberto Griggio for promptly providing us with a
  corrected version of MathSAT, which we use in the evaluation.} Furthermore,
the virtual portfolios are also shown. While both portfolios solve more
instances than any individual solver, the portfolio based on the back-end solvers
and the \uppsat instances is a clear winner, showing the impact of presented
approximations on the state-of-the-art.

\newcommand{\trimFig}[1]{
       \hspace*{-17ex}
     \input{#1}
       \hspace*{-11.5ex}
}

\begin{figure}[p]
 \begin{minipage}{0.48\linewidth}
   \scalebox{0.7}{
     ~~~~~~
       \hspace*{-17ex}
     \input{figures/scatter_BV_z3__z3}
       \hspace*{-11.5ex}

   }
 \end{minipage}
 \hspace*{\fill}
 \begin{minipage}{0.48\linewidth}
   \scalebox{0.7}{
       
       \hspace*{-17ex}
     \input{figures/scatter_RPFP_z3__z3}
       \hspace*{-11.5ex}

   }
 \end{minipage}
 
 \begin{minipage}{0.48\linewidth}
   \scalebox{0.7}{
     
       \hspace*{-17ex}
     \input{figures/scatter_RA_nlsat__z3}
       \hspace*{-11.5ex}

   }
 \end{minipage}
 \hspace*{\fill}
 \begin{minipage}{0.48\linewidth}
   \scalebox{0.7}{
     
       \hspace*{-17ex}
     \input{figures/scatter_RPFP_mathsat__mathsat}
       \hspace*{-11.5ex}

   }
 \end{minipage}
 
 \begin{minipage}{0.48\linewidth}
   \hspace{0.5\linewidth}
 \end{minipage}
 \hspace*{\fill}
 \begin{minipage}{0.48\linewidth}
   \scalebox{0.7}{
     
       \hspace*{-17ex}
     \input{figures/scatter_RPFP_acdcl__acdcl}
       \hspace*{-11.5ex}

   }
 \end{minipage}
 
 \caption{Comparison of runtime performance of \uppsat instances against the
   corresponding back-end solver.}
 \label{fig:appvsback-end}
\end{figure}
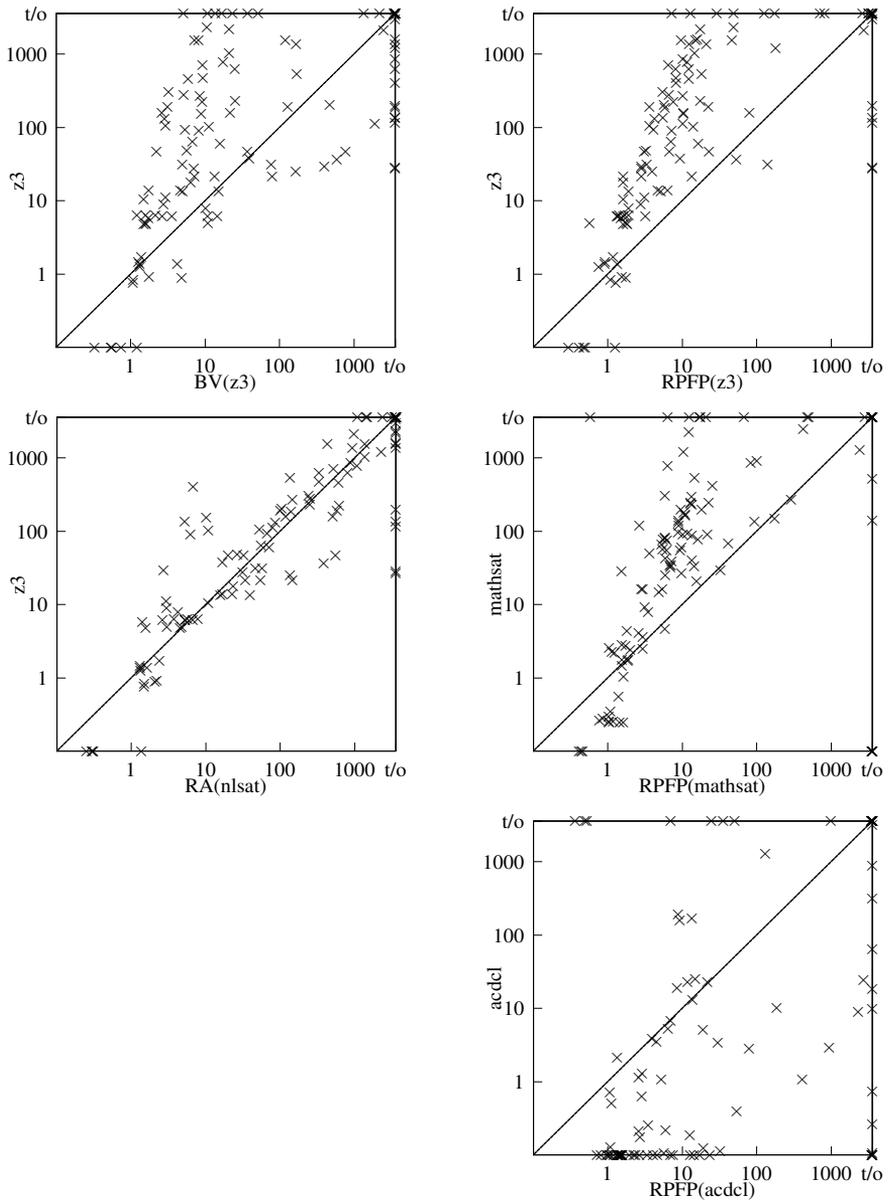

\paragraph{Detailed comparisons.}
To complete the picture, Figure~\ref{fig:appvsback-end} shows scatter plots of
each approximation against the corresponding back-end. The X axis denotes the
solving time of the \uppsat instance, while the Y axis denotes the solving time
of the back-end. Maximum value along either axes denotes a timeout. Benchmarks on
the diagonal show same solving time, above the diagonal the \uppsat instance is
faster, and below the diagonal the back-end is faster.
 In the case of BV and RA
approximations, we compare against Z3 as the baseline, since their respective
back-ends do not support the theory of \fpa.

The plots featuring Z3 and MathSAT as the back-end seem to show
super-linear speedup as a result of approximation, which is suggested
by the trend among the instances not being parallel to the diagonal.
The cause of this phenomenon is not entirely clear at this point, and
will need further investigation; one hypothesis is that approximation
is more effective on larger than on smaller \fpa problems.  Compared
to the experiments in \cite{DBLP:journals/jar/ZeljicWR17}, the
improved results are likely due to strengthening of the model
reconstruction strategy.

As evidenced by the previous data, the ACDCL algorithm does not
benefit from the RPFP approximation, in terms of performance. However, the
approximation does solve some instances without reach of the plain ACDCL
procedure.

In the cactus plot, the BV approximation shows improvement in performance over
the bit-blasting back-ends, but eventually falls behind in the number of solved
instances. Compared to the Z3 back-end, we can see traces of super-linear speed up,
but there are also cases where the BV approximation times out. Considering that
the BV approximation consists almost entirely of the encoding, and that it lacks
tailored reconstruction and refinement strategies, we find these results to be
quite promising.

The RA approximation also shows some improvement in performance compared to the
bit-blasting method. However, we can see from Table~\ref{tbl:all} that it
reaches maximum precision on all but 20 benchmarks, indicating its lack of
maturity. It is interesting, however, that it does outperform all the other
solvers on a few benchmarks. One approach would be to allow mixed constraints of
\fpa and RA, as proposed by Ramachandran et
al.~\cite{DBLP:conf/fmcad/RamachandranW16}, however none of the off-the-shelf SMT
solvers support this theory combination in a meaningful way yet.

Finally, a comparison of the three back-end solvers, and of the
corresponding combination of back-end solver and RPFP approximation,
is given in Fig.~\ref{fig:back-ends} in the Appendix.

 \section{Conclusion and Future Work}
\label{sec:conclusion}

We have presented a methodology and new framework, \uppsat, for
implementing approximating SMT solvers. The experimental evaluation
demonstrates, what has been showed earlier, the efficiency of
approximations as well as provides a simple implementaiton. The
presented framework does not only yield a significand preformance
improvement in comparison with the back-ends, it does so with a
modular code allowing for easy combinations of different strategies.

Approximation configurations presented here (RPFP(z3), RPFP(mathsat))
are shown to be state-of-the art in handling formulas in \fpa, where
they improve their performance of the respective back-end. For ACDCL
this is not the case, however, indicating that perhaps a different
method of approximation should be utilized. A strenght of the \uppsat
framework is that if a new SMT-solver would be improved on \fpa
formulas, it can easily be integrated into the approximation
framework, perhaps improving the underlying back-end even further.

The clear direction for improving \uppsat is to extend the general
framework with more abstract strategies, e.g., retrieve multiple
models from an approximate formula and/or apply multiple different
reconstruction strategies on approximate models. Currently, majority
of the time is spent on looking for models which means there is plenty
of room to make more sophisticated strategies in the framework.

Another big challenge is to extend \uppsat to be able to handle
unsatisfiable formulas efficiently. Currently, the proof refinement is
naive uniform refinement, but there is a potential to do much more
intelligent refinement. This research has been focused on finding
models, but it is very interesting to investigate how unsatisfiability
proofs from approximated formulas can be utilized for refinement.

The fixed point and real arithmetic approaches are presented here as a
proof of concept. They are simple and not much effort went into
instantiating the framework for these approximations. However, the
results shows that even uncomplicated approaches can be competitive;
this opens up the line of future work to design tailored refinement
and reconstruction strategies.

We present virtual portfolio solvers, which show some kind of optimal
choice among which solver to chose. It would be interesting to create
a real portfolio solver and see how well it can be tuned to pick the
correct solver (e.g., by machine learning algorithms) as well as
utilize parallelism.

\bibliographystyle{plain}
\bibliography{refs}

\appendix

\begin{figure}[p]
\begin{minipage}[h!]{0.48\linewidth}
  \scalebox{0.7}{
    
       \hspace*{-17ex}
     \input{figures/scatter_acdcl_z3}
       \hspace*{-11.5ex}

  }
\end{minipage}
 \hspace*{\fill}
\begin{minipage}{0.48\linewidth}
  \scalebox{0.7}{
    
       \hspace*{-17ex}
     \input{figures/scatter_RPFP_acdcl__RPFP_z3_}
       \hspace*{-11.5ex}

  }
\end{minipage}

\begin{minipage}{0.48\linewidth}
  \scalebox{0.7}{
    
       \hspace*{-17ex}
     \input{figures/scatter_mathsat_z3}
       \hspace*{-11.5ex}

  }
\end{minipage}
 \hspace*{\fill}
\begin{minipage}{0.48\linewidth}
  \scalebox{0.7}{
    
       \hspace*{-17ex}
     \input{figures/scatter_RPFP_mathsat__RPFP_z3_}
       \hspace*{-11.5ex}

  }
\end{minipage}

\begin{minipage}{0.48\linewidth}
  \scalebox{0.7}{
      
       \hspace*{-17ex}
     \input{figures/scatter_acdcl_mathsat}
       \hspace*{-11.5ex}

    }
  \end{minipage}
 \hspace*{\fill}
 \begin{minipage}{0.48\linewidth}
   \scalebox{0.7}{
      
       \hspace*{-17ex}
     \input{figures/scatter_RPFP_acdcl__RPFP_mathsat_}
       \hspace*{-11.5ex}

    }
  \end{minipage}

  \caption{Comparison of runtime performance the back-end solvers, without
    approaximation (left) and using RPFP approximation (right).}
  \label{fig:back-ends}
\end{figure}
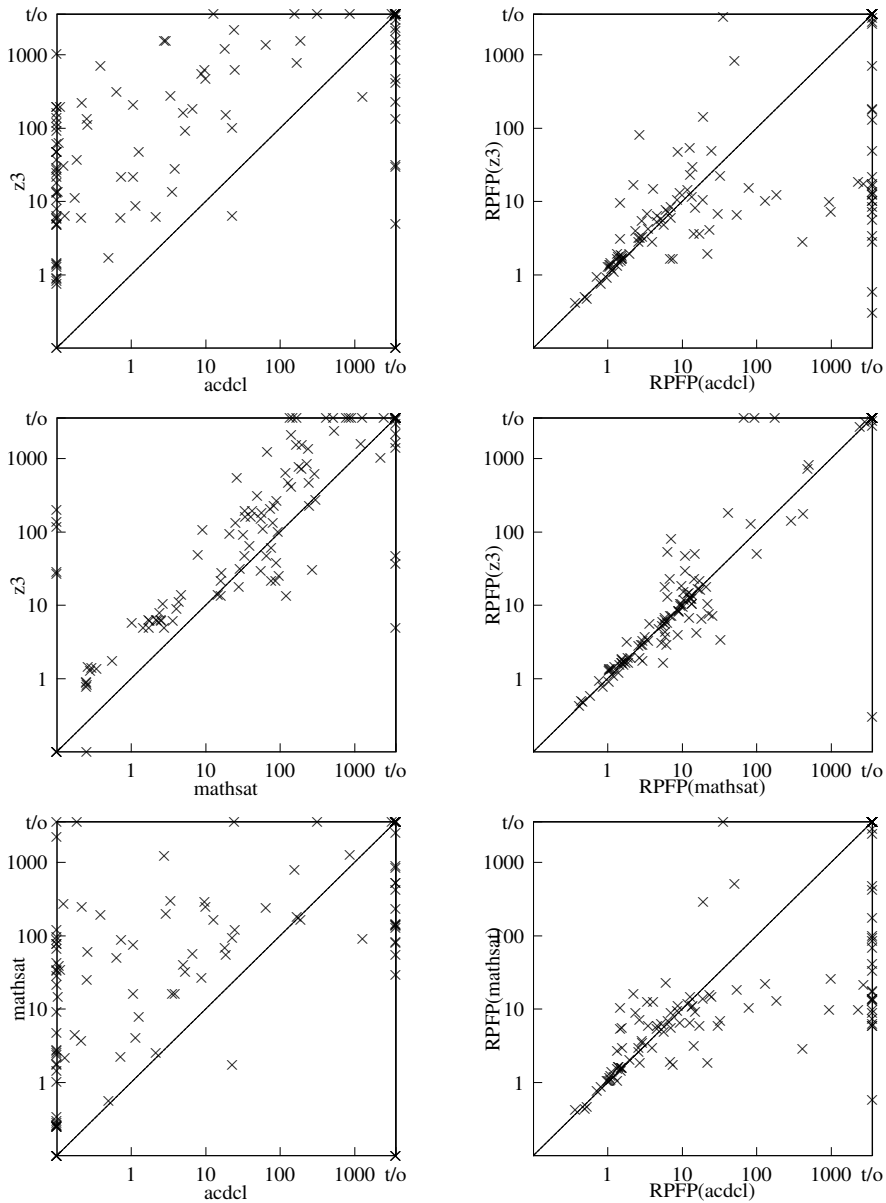

 \end{document}

%% file: figures/scatter_BV_z3__z3.tex
% GNUPLOT: LaTeX picture
\setlength{\unitlength}{0.240900pt}
\ifx\plotpoint\undefined\newsavebox{\plotpoint}\fi
\sbox{\plotpoint}{\rule[-0.200pt]{0.400pt}{0.400pt}}%
\begin{picture}(1500,900)(0,0)
\sbox{\plotpoint}{\rule[-0.200pt]{0.400pt}{0.400pt}}%
\put(390,274){\makebox(0,0)[r]{1}}
\put(410.0,274.0){\rule[-0.200pt]{4.818pt}{0.400pt}}
\put(390,439){\makebox(0,0)[r]{10}}
\put(410.0,439.0){\rule[-0.200pt]{4.818pt}{0.400pt}}
\put(390,603){\makebox(0,0)[r]{100}}
\put(410.0,603.0){\rule[-0.200pt]{4.818pt}{0.400pt}}
\put(390,768){\makebox(0,0)[r]{1000}}
\put(410.0,768.0){\rule[-0.200pt]{4.818pt}{0.400pt}}
\put(390,859){\makebox(0,0)[r]{t/o}}
\put(410.0,859.0){\rule[-0.200pt]{4.818pt}{0.400pt}}
\put(574,69){\makebox(0,0){1}}
\put(574.0,110.0){\rule[-0.200pt]{0.400pt}{4.818pt}}
\put(739,69){\makebox(0,0){10}}
\put(739.0,110.0){\rule[-0.200pt]{0.400pt}{4.818pt}}
\put(903,69){\makebox(0,0){100}}
\put(903.0,110.0){\rule[-0.200pt]{0.400pt}{4.818pt}}
\put(1068,69){\makebox(0,0){1000}}
\put(1068.0,110.0){\rule[-0.200pt]{0.400pt}{4.818pt}}
\put(1159,69){\makebox(0,0){t/o}}
\put(1159.0,110.0){\rule[-0.200pt]{0.400pt}{4.818pt}}
\put(410.0,110.0){\rule[-0.200pt]{0.400pt}{180.434pt}}
\put(410.0,110.0){\rule[-0.200pt]{180.434pt}{0.400pt}}
\put(1159.0,110.0){\rule[-0.200pt]{0.400pt}{180.434pt}}
\put(410.0,859.0){\rule[-0.200pt]{180.434pt}{0.400pt}}
\put(410,859){\line(1,0){749}}
\put(1159,110){\line(0,1){749}}
\put(324,484){\makebox(0,0){\rotatebox{90}{z3}}}
\put(784,29){\makebox(0,0){BV(z3)}}
\put(410,110){\usebox{\plotpoint}}
\multiput(410.00,110.59)(0.494,0.488){13}{\rule{0.500pt}{0.117pt}}
\multiput(410.00,109.17)(6.962,8.000){2}{\rule{0.250pt}{0.400pt}}
\multiput(418.00,118.59)(0.492,0.485){11}{\rule{0.500pt}{0.117pt}}
\multiput(418.00,117.17)(5.962,7.000){2}{\rule{0.250pt}{0.400pt}}
\multiput(425.00,125.59)(0.494,0.488){13}{\rule{0.500pt}{0.117pt}}
\multiput(425.00,124.17)(6.962,8.000){2}{\rule{0.250pt}{0.400pt}}
\multiput(433.00,133.59)(0.492,0.485){11}{\rule{0.500pt}{0.117pt}}
\multiput(433.00,132.17)(5.962,7.000){2}{\rule{0.250pt}{0.400pt}}
\multiput(440.00,140.59)(0.494,0.488){13}{\rule{0.500pt}{0.117pt}}
\multiput(440.00,139.17)(6.962,8.000){2}{\rule{0.250pt}{0.400pt}}
\multiput(448.00,148.59)(0.492,0.485){11}{\rule{0.500pt}{0.117pt}}
\multiput(448.00,147.17)(5.962,7.000){2}{\rule{0.250pt}{0.400pt}}
\multiput(455.00,155.59)(0.494,0.488){13}{\rule{0.500pt}{0.117pt}}
\multiput(455.00,154.17)(6.962,8.000){2}{\rule{0.250pt}{0.400pt}}
\multiput(463.00,163.59)(0.494,0.488){13}{\rule{0.500pt}{0.117pt}}
\multiput(463.00,162.17)(6.962,8.000){2}{\rule{0.250pt}{0.400pt}}
\multiput(471.00,171.59)(0.492,0.485){11}{\rule{0.500pt}{0.117pt}}
\multiput(471.00,170.17)(5.962,7.000){2}{\rule{0.250pt}{0.400pt}}
\multiput(478.00,178.59)(0.494,0.488){13}{\rule{0.500pt}{0.117pt}}
\multiput(478.00,177.17)(6.962,8.000){2}{\rule{0.250pt}{0.400pt}}
\multiput(486.00,186.59)(0.492,0.485){11}{\rule{0.500pt}{0.117pt}}
\multiput(486.00,185.17)(5.962,7.000){2}{\rule{0.250pt}{0.400pt}}
\multiput(493.00,193.59)(0.494,0.488){13}{\rule{0.500pt}{0.117pt}}
\multiput(493.00,192.17)(6.962,8.000){2}{\rule{0.250pt}{0.400pt}}
\multiput(501.00,201.59)(0.492,0.485){11}{\rule{0.500pt}{0.117pt}}
\multiput(501.00,200.17)(5.962,7.000){2}{\rule{0.250pt}{0.400pt}}
\multiput(508.00,208.59)(0.494,0.488){13}{\rule{0.500pt}{0.117pt}}
\multiput(508.00,207.17)(6.962,8.000){2}{\rule{0.250pt}{0.400pt}}
\multiput(516.00,216.59)(0.492,0.485){11}{\rule{0.500pt}{0.117pt}}
\multiput(516.00,215.17)(5.962,7.000){2}{\rule{0.250pt}{0.400pt}}
\multiput(523.00,223.59)(0.494,0.488){13}{\rule{0.500pt}{0.117pt}}
\multiput(523.00,222.17)(6.962,8.000){2}{\rule{0.250pt}{0.400pt}}
\multiput(531.00,231.59)(0.494,0.488){13}{\rule{0.500pt}{0.117pt}}
\multiput(531.00,230.17)(6.962,8.000){2}{\rule{0.250pt}{0.400pt}}
\multiput(539.00,239.59)(0.492,0.485){11}{\rule{0.500pt}{0.117pt}}
\multiput(539.00,238.17)(5.962,7.000){2}{\rule{0.250pt}{0.400pt}}
\multiput(546.00,246.59)(0.494,0.488){13}{\rule{0.500pt}{0.117pt}}
\multiput(546.00,245.17)(6.962,8.000){2}{\rule{0.250pt}{0.400pt}}
\multiput(554.00,254.59)(0.492,0.485){11}{\rule{0.500pt}{0.117pt}}
\multiput(554.00,253.17)(5.962,7.000){2}{\rule{0.250pt}{0.400pt}}
\multiput(561.00,261.59)(0.494,0.488){13}{\rule{0.500pt}{0.117pt}}
\multiput(561.00,260.17)(6.962,8.000){2}{\rule{0.250pt}{0.400pt}}
\multiput(569.00,269.59)(0.492,0.485){11}{\rule{0.500pt}{0.117pt}}
\multiput(569.00,268.17)(5.962,7.000){2}{\rule{0.250pt}{0.400pt}}
\multiput(576.00,276.59)(0.494,0.488){13}{\rule{0.500pt}{0.117pt}}
\multiput(576.00,275.17)(6.962,8.000){2}{\rule{0.250pt}{0.400pt}}
\multiput(584.00,284.59)(0.494,0.488){13}{\rule{0.500pt}{0.117pt}}
\multiput(584.00,283.17)(6.962,8.000){2}{\rule{0.250pt}{0.400pt}}
\multiput(592.00,292.59)(0.492,0.485){11}{\rule{0.500pt}{0.117pt}}
\multiput(592.00,291.17)(5.962,7.000){2}{\rule{0.250pt}{0.400pt}}
\multiput(599.00,299.59)(0.494,0.488){13}{\rule{0.500pt}{0.117pt}}
\multiput(599.00,298.17)(6.962,8.000){2}{\rule{0.250pt}{0.400pt}}
\multiput(607.00,307.59)(0.492,0.485){11}{\rule{0.500pt}{0.117pt}}
\multiput(607.00,306.17)(5.962,7.000){2}{\rule{0.250pt}{0.400pt}}
\multiput(614.00,314.59)(0.494,0.488){13}{\rule{0.500pt}{0.117pt}}
\multiput(614.00,313.17)(6.962,8.000){2}{\rule{0.250pt}{0.400pt}}
\multiput(622.00,322.59)(0.492,0.485){11}{\rule{0.500pt}{0.117pt}}
\multiput(622.00,321.17)(5.962,7.000){2}{\rule{0.250pt}{0.400pt}}
\multiput(629.00,329.59)(0.494,0.488){13}{\rule{0.500pt}{0.117pt}}
\multiput(629.00,328.17)(6.962,8.000){2}{\rule{0.250pt}{0.400pt}}
\multiput(637.00,337.59)(0.494,0.488){13}{\rule{0.500pt}{0.117pt}}
\multiput(637.00,336.17)(6.962,8.000){2}{\rule{0.250pt}{0.400pt}}
\multiput(645.00,345.59)(0.492,0.485){11}{\rule{0.500pt}{0.117pt}}
\multiput(645.00,344.17)(5.962,7.000){2}{\rule{0.250pt}{0.400pt}}
\multiput(652.00,352.59)(0.494,0.488){13}{\rule{0.500pt}{0.117pt}}
\multiput(652.00,351.17)(6.962,8.000){2}{\rule{0.250pt}{0.400pt}}
\multiput(660.00,360.59)(0.492,0.485){11}{\rule{0.500pt}{0.117pt}}
\multiput(660.00,359.17)(5.962,7.000){2}{\rule{0.250pt}{0.400pt}}
\multiput(667.00,367.59)(0.494,0.488){13}{\rule{0.500pt}{0.117pt}}
\multiput(667.00,366.17)(6.962,8.000){2}{\rule{0.250pt}{0.400pt}}
\multiput(675.00,375.59)(0.492,0.485){11}{\rule{0.500pt}{0.117pt}}
\multiput(675.00,374.17)(5.962,7.000){2}{\rule{0.250pt}{0.400pt}}
\multiput(682.00,382.59)(0.494,0.488){13}{\rule{0.500pt}{0.117pt}}
\multiput(682.00,381.17)(6.962,8.000){2}{\rule{0.250pt}{0.400pt}}
\multiput(690.00,390.59)(0.492,0.485){11}{\rule{0.500pt}{0.117pt}}
\multiput(690.00,389.17)(5.962,7.000){2}{\rule{0.250pt}{0.400pt}}
\multiput(697.00,397.59)(0.494,0.488){13}{\rule{0.500pt}{0.117pt}}
\multiput(697.00,396.17)(6.962,8.000){2}{\rule{0.250pt}{0.400pt}}
\multiput(705.00,405.59)(0.494,0.488){13}{\rule{0.500pt}{0.117pt}}
\multiput(705.00,404.17)(6.962,8.000){2}{\rule{0.250pt}{0.400pt}}
\multiput(713.00,413.59)(0.492,0.485){11}{\rule{0.500pt}{0.117pt}}
\multiput(713.00,412.17)(5.962,7.000){2}{\rule{0.250pt}{0.400pt}}
\multiput(720.00,420.59)(0.494,0.488){13}{\rule{0.500pt}{0.117pt}}
\multiput(720.00,419.17)(6.962,8.000){2}{\rule{0.250pt}{0.400pt}}
\multiput(728.00,428.59)(0.492,0.485){11}{\rule{0.500pt}{0.117pt}}
\multiput(728.00,427.17)(5.962,7.000){2}{\rule{0.250pt}{0.400pt}}
\multiput(735.00,435.59)(0.494,0.488){13}{\rule{0.500pt}{0.117pt}}
\multiput(735.00,434.17)(6.962,8.000){2}{\rule{0.250pt}{0.400pt}}
\multiput(743.00,443.59)(0.492,0.485){11}{\rule{0.500pt}{0.117pt}}
\multiput(743.00,442.17)(5.962,7.000){2}{\rule{0.250pt}{0.400pt}}
\multiput(750.00,450.59)(0.494,0.488){13}{\rule{0.500pt}{0.117pt}}
\multiput(750.00,449.17)(6.962,8.000){2}{\rule{0.250pt}{0.400pt}}
\multiput(758.00,458.59)(0.494,0.488){13}{\rule{0.500pt}{0.117pt}}
\multiput(758.00,457.17)(6.962,8.000){2}{\rule{0.250pt}{0.400pt}}
\multiput(766.00,466.59)(0.492,0.485){11}{\rule{0.500pt}{0.117pt}}
\multiput(766.00,465.17)(5.962,7.000){2}{\rule{0.250pt}{0.400pt}}
\multiput(773.00,473.59)(0.494,0.488){13}{\rule{0.500pt}{0.117pt}}
\multiput(773.00,472.17)(6.962,8.000){2}{\rule{0.250pt}{0.400pt}}
\multiput(781.00,481.59)(0.492,0.485){11}{\rule{0.500pt}{0.117pt}}
\multiput(781.00,480.17)(5.962,7.000){2}{\rule{0.250pt}{0.400pt}}
\multiput(788.00,488.59)(0.494,0.488){13}{\rule{0.500pt}{0.117pt}}
\multiput(788.00,487.17)(6.962,8.000){2}{\rule{0.250pt}{0.400pt}}
\multiput(796.00,496.59)(0.492,0.485){11}{\rule{0.500pt}{0.117pt}}
\multiput(796.00,495.17)(5.962,7.000){2}{\rule{0.250pt}{0.400pt}}
\multiput(803.00,503.59)(0.494,0.488){13}{\rule{0.500pt}{0.117pt}}
\multiput(803.00,502.17)(6.962,8.000){2}{\rule{0.250pt}{0.400pt}}
\multiput(811.00,511.59)(0.494,0.488){13}{\rule{0.500pt}{0.117pt}}
\multiput(811.00,510.17)(6.962,8.000){2}{\rule{0.250pt}{0.400pt}}
\multiput(819.00,519.59)(0.492,0.485){11}{\rule{0.500pt}{0.117pt}}
\multiput(819.00,518.17)(5.962,7.000){2}{\rule{0.250pt}{0.400pt}}
\multiput(826.00,526.59)(0.494,0.488){13}{\rule{0.500pt}{0.117pt}}
\multiput(826.00,525.17)(6.962,8.000){2}{\rule{0.250pt}{0.400pt}}
\multiput(834.00,534.59)(0.492,0.485){11}{\rule{0.500pt}{0.117pt}}
\multiput(834.00,533.17)(5.962,7.000){2}{\rule{0.250pt}{0.400pt}}
\multiput(841.00,541.59)(0.494,0.488){13}{\rule{0.500pt}{0.117pt}}
\multiput(841.00,540.17)(6.962,8.000){2}{\rule{0.250pt}{0.400pt}}
\multiput(849.00,549.59)(0.492,0.485){11}{\rule{0.500pt}{0.117pt}}
\multiput(849.00,548.17)(5.962,7.000){2}{\rule{0.250pt}{0.400pt}}
\multiput(856.00,556.59)(0.494,0.488){13}{\rule{0.500pt}{0.117pt}}
\multiput(856.00,555.17)(6.962,8.000){2}{\rule{0.250pt}{0.400pt}}
\multiput(864.00,564.59)(0.494,0.488){13}{\rule{0.500pt}{0.117pt}}
\multiput(864.00,563.17)(6.962,8.000){2}{\rule{0.250pt}{0.400pt}}
\multiput(872.00,572.59)(0.492,0.485){11}{\rule{0.500pt}{0.117pt}}
\multiput(872.00,571.17)(5.962,7.000){2}{\rule{0.250pt}{0.400pt}}
\multiput(879.00,579.59)(0.494,0.488){13}{\rule{0.500pt}{0.117pt}}
\multiput(879.00,578.17)(6.962,8.000){2}{\rule{0.250pt}{0.400pt}}
\multiput(887.00,587.59)(0.492,0.485){11}{\rule{0.500pt}{0.117pt}}
\multiput(887.00,586.17)(5.962,7.000){2}{\rule{0.250pt}{0.400pt}}
\multiput(894.00,594.59)(0.494,0.488){13}{\rule{0.500pt}{0.117pt}}
\multiput(894.00,593.17)(6.962,8.000){2}{\rule{0.250pt}{0.400pt}}
\multiput(902.00,602.59)(0.492,0.485){11}{\rule{0.500pt}{0.117pt}}
\multiput(902.00,601.17)(5.962,7.000){2}{\rule{0.250pt}{0.400pt}}
\multiput(909.00,609.59)(0.494,0.488){13}{\rule{0.500pt}{0.117pt}}
\multiput(909.00,608.17)(6.962,8.000){2}{\rule{0.250pt}{0.400pt}}
\multiput(917.00,617.59)(0.492,0.485){11}{\rule{0.500pt}{0.117pt}}
\multiput(917.00,616.17)(5.962,7.000){2}{\rule{0.250pt}{0.400pt}}
\multiput(924.00,624.59)(0.494,0.488){13}{\rule{0.500pt}{0.117pt}}
\multiput(924.00,623.17)(6.962,8.000){2}{\rule{0.250pt}{0.400pt}}
\multiput(932.00,632.59)(0.494,0.488){13}{\rule{0.500pt}{0.117pt}}
\multiput(932.00,631.17)(6.962,8.000){2}{\rule{0.250pt}{0.400pt}}
\multiput(940.00,640.59)(0.492,0.485){11}{\rule{0.500pt}{0.117pt}}
\multiput(940.00,639.17)(5.962,7.000){2}{\rule{0.250pt}{0.400pt}}
\multiput(947.00,647.59)(0.494,0.488){13}{\rule{0.500pt}{0.117pt}}
\multiput(947.00,646.17)(6.962,8.000){2}{\rule{0.250pt}{0.400pt}}
\multiput(955.00,655.59)(0.492,0.485){11}{\rule{0.500pt}{0.117pt}}
\multiput(955.00,654.17)(5.962,7.000){2}{\rule{0.250pt}{0.400pt}}
\multiput(962.00,662.59)(0.494,0.488){13}{\rule{0.500pt}{0.117pt}}
\multiput(962.00,661.17)(6.962,8.000){2}{\rule{0.250pt}{0.400pt}}
\multiput(970.00,670.59)(0.492,0.485){11}{\rule{0.500pt}{0.117pt}}
\multiput(970.00,669.17)(5.962,7.000){2}{\rule{0.250pt}{0.400pt}}
\multiput(977.00,677.59)(0.494,0.488){13}{\rule{0.500pt}{0.117pt}}
\multiput(977.00,676.17)(6.962,8.000){2}{\rule{0.250pt}{0.400pt}}
\multiput(985.00,685.59)(0.494,0.488){13}{\rule{0.500pt}{0.117pt}}
\multiput(985.00,684.17)(6.962,8.000){2}{\rule{0.250pt}{0.400pt}}
\multiput(993.00,693.59)(0.492,0.485){11}{\rule{0.500pt}{0.117pt}}
\multiput(993.00,692.17)(5.962,7.000){2}{\rule{0.250pt}{0.400pt}}
\multiput(1000.00,700.59)(0.494,0.488){13}{\rule{0.500pt}{0.117pt}}
\multiput(1000.00,699.17)(6.962,8.000){2}{\rule{0.250pt}{0.400pt}}
\multiput(1008.00,708.59)(0.492,0.485){11}{\rule{0.500pt}{0.117pt}}
\multiput(1008.00,707.17)(5.962,7.000){2}{\rule{0.250pt}{0.400pt}}
\multiput(1015.00,715.59)(0.494,0.488){13}{\rule{0.500pt}{0.117pt}}
\multiput(1015.00,714.17)(6.962,8.000){2}{\rule{0.250pt}{0.400pt}}
\multiput(1023.00,723.59)(0.492,0.485){11}{\rule{0.500pt}{0.117pt}}
\multiput(1023.00,722.17)(5.962,7.000){2}{\rule{0.250pt}{0.400pt}}
\multiput(1030.00,730.59)(0.494,0.488){13}{\rule{0.500pt}{0.117pt}}
\multiput(1030.00,729.17)(6.962,8.000){2}{\rule{0.250pt}{0.400pt}}
\multiput(1038.00,738.59)(0.494,0.488){13}{\rule{0.500pt}{0.117pt}}
\multiput(1038.00,737.17)(6.962,8.000){2}{\rule{0.250pt}{0.400pt}}
\multiput(1046.00,746.59)(0.492,0.485){11}{\rule{0.500pt}{0.117pt}}
\multiput(1046.00,745.17)(5.962,7.000){2}{\rule{0.250pt}{0.400pt}}
\multiput(1053.00,753.59)(0.494,0.488){13}{\rule{0.500pt}{0.117pt}}
\multiput(1053.00,752.17)(6.962,8.000){2}{\rule{0.250pt}{0.400pt}}
\multiput(1061.00,761.59)(0.492,0.485){11}{\rule{0.500pt}{0.117pt}}
\multiput(1061.00,760.17)(5.962,7.000){2}{\rule{0.250pt}{0.400pt}}
\multiput(1068.00,768.59)(0.494,0.488){13}{\rule{0.500pt}{0.117pt}}
\multiput(1068.00,767.17)(6.962,8.000){2}{\rule{0.250pt}{0.400pt}}
\multiput(1076.00,776.59)(0.492,0.485){11}{\rule{0.500pt}{0.117pt}}
\multiput(1076.00,775.17)(5.962,7.000){2}{\rule{0.250pt}{0.400pt}}
\multiput(1083.00,783.59)(0.494,0.488){13}{\rule{0.500pt}{0.117pt}}
\multiput(1083.00,782.17)(6.962,8.000){2}{\rule{0.250pt}{0.400pt}}
\multiput(1091.00,791.59)(0.492,0.485){11}{\rule{0.500pt}{0.117pt}}
\multiput(1091.00,790.17)(5.962,7.000){2}{\rule{0.250pt}{0.400pt}}
\multiput(1098.00,798.59)(0.494,0.488){13}{\rule{0.500pt}{0.117pt}}
\multiput(1098.00,797.17)(6.962,8.000){2}{\rule{0.250pt}{0.400pt}}
\multiput(1106.00,806.59)(0.494,0.488){13}{\rule{0.500pt}{0.117pt}}
\multiput(1106.00,805.17)(6.962,8.000){2}{\rule{0.250pt}{0.400pt}}
\multiput(1114.00,814.59)(0.492,0.485){11}{\rule{0.500pt}{0.117pt}}
\multiput(1114.00,813.17)(5.962,7.000){2}{\rule{0.250pt}{0.400pt}}
\multiput(1121.00,821.59)(0.494,0.488){13}{\rule{0.500pt}{0.117pt}}
\multiput(1121.00,820.17)(6.962,8.000){2}{\rule{0.250pt}{0.400pt}}
\multiput(1129.00,829.59)(0.492,0.485){11}{\rule{0.500pt}{0.117pt}}
\multiput(1129.00,828.17)(5.962,7.000){2}{\rule{0.250pt}{0.400pt}}
\multiput(1136.00,836.59)(0.494,0.488){13}{\rule{0.500pt}{0.117pt}}
\multiput(1136.00,835.17)(6.962,8.000){2}{\rule{0.250pt}{0.400pt}}
\multiput(1144.00,844.59)(0.492,0.485){11}{\rule{0.500pt}{0.117pt}}
\multiput(1144.00,843.17)(5.962,7.000){2}{\rule{0.250pt}{0.400pt}}
\multiput(1151.00,851.59)(0.494,0.488){13}{\rule{0.500pt}{0.117pt}}
\multiput(1151.00,850.17)(6.962,8.000){2}{\rule{0.250pt}{0.400pt}}
\put(532,110){\makebox(0,0){$\times$}}
\put(1159,510){\makebox(0,0){$\times$}}
\put(715,512){\makebox(0,0){$\times$}}
\put(1159,647){\makebox(0,0){$\times$}}
\put(666,403){\makebox(0,0){$\times$}}
\put(735,714){\makebox(0,0){$\times$}}
\put(941,790){\makebox(0,0){$\times$}}
\put(793,823){\makebox(0,0){$\times$}}
\put(1003,516){\makebox(0,0){$\times$}}
\put(1159,624){\makebox(0,0){$\times$}}
\put(767,404){\makebox(0,0){$\times$}}
\put(1159,704){\makebox(0,0){$\times$}}
\put(1159,757){\makebox(0,0){$\times$}}
\put(1159,802){\makebox(0,0){$\times$}}
\put(716,494){\makebox(0,0){$\times$}}
\put(1015,655){\makebox(0,0){$\times$}}
\put(589,407){\makebox(0,0){$\times$}}
\put(805,735){\makebox(0,0){$\times$}}
\put(1159,734){\makebox(0,0){$\times$}}
\put(1159,791){\makebox(0,0){$\times$}}
\put(678,298){\makebox(0,0){$\times$}}
\put(592,301){\makebox(0,0){$\times$}}
\put(596,291){\makebox(0,0){$\times$}}
\put(630,406){\makebox(0,0){$\times$}}
\put(614,399){\makebox(0,0){$\times$}}
\put(616,268){\makebox(0,0){$\times$}}
\put(688,265){\makebox(0,0){$\times$}}
\put(589,110){\makebox(0,0){$\times$}}
\put(580,255){\makebox(0,0){$\times$}}
\put(581,261){\makebox(0,0){$\times$}}
\put(741,422){\makebox(0,0){$\times$}}
\put(745,403){\makebox(0,0){$\times$}}
\put(1159,652){\makebox(0,0){$\times$}}
\put(1159,859){\makebox(0,0){$\times$}}
\put(1159,859){\makebox(0,0){$\times$}}
\put(1159,782){\makebox(0,0){$\times$}}
\put(1159,859){\makebox(0,0){$\times$}}
\put(1159,626){\makebox(0,0){$\times$}}
\put(746,389){\makebox(0,0){$\times$}}
\put(1159,859){\makebox(0,0){$\times$}}
\put(1159,859){\makebox(0,0){$\times$}}
\put(689,519){\makebox(0,0){$\times$}}
\put(645,403){\makebox(0,0){$\times$}}
\put(685,461){\makebox(0,0){$\times$}}
\put(595,297){\makebox(0,0){$\times$}}
\put(712,571){\makebox(0,0){$\times$}}
\put(1115,611){\makebox(0,0){$\times$}}
\put(761,494){\makebox(0,0){$\times$}}
\put(648,430){\makebox(0,0){$\times$}}
\put(690,460){\makebox(0,0){$\times$}}
\put(599,313){\makebox(0,0){$\times$}}
\put(726,597){\makebox(0,0){$\times$}}
\put(731,633){\makebox(0,0){$\times$}}
\put(748,604){\makebox(0,0){$\times$}}
\put(554,110){\makebox(0,0){$\times$}}
\put(533,110){\makebox(0,0){$\times$}}
\put(837,533){\makebox(0,0){$\times$}}
\put(833,549){\makebox(0,0){$\times$}}
\put(888,493){\makebox(0,0){$\times$}}
\put(940,504){\makebox(0,0){$\times$}}
\put(1050,550){\makebox(0,0){$\times$}}
\put(770,459){\makebox(0,0){$\times$}}
\put(773,567){\makebox(0,0){$\times$}}
\put(1031,531){\makebox(0,0){$\times$}}
\put(708,480){\makebox(0,0){$\times$}}
\put(632,549){\makebox(0,0){$\times$}}
\put(922,650){\makebox(0,0){$\times$}}
\put(793,769){\makebox(0,0){$\times$}}
\put(1159,859){\makebox(0,0){$\times$}}
\put(1159,859){\makebox(0,0){$\times$}}
\put(1159,859){\makebox(0,0){$\times$}}
\put(1159,859){\makebox(0,0){$\times$}}
\put(1159,859){\makebox(0,0){$\times$}}
\put(1159,859){\makebox(0,0){$\times$}}
\put(1159,859){\makebox(0,0){$\times$}}
\put(1159,859){\makebox(0,0){$\times$}}
\put(886,520){\makebox(0,0){$\times$}}
\put(1159,859){\makebox(0,0){$\times$}}
\put(1159,859){\makebox(0,0){$\times$}}
\put(1090,859){\makebox(0,0){$\times$}}
\put(1159,859){\makebox(0,0){$\times$}}
\put(615,461){\makebox(0,0){$\times$}}
\put(1159,859){\makebox(0,0){$\times$}}
\put(1159,859){\makebox(0,0){$\times$}}
\put(942,724){\makebox(0,0){$\times$}}
\put(1159,846){\makebox(0,0){$\times$}}
\put(1159,855){\makebox(0,0){$\times$}}
\put(1134,822){\makebox(0,0){$\times$}}
\put(495,110){\makebox(0,0){$\times$}}
\put(604,443){\makebox(0,0){$\times$}}
\put(608,388){\makebox(0,0){$\times$}}
\put(652,446){\makebox(0,0){$\times$}}
\put(604,387){\makebox(0,0){$\times$}}
\put(609,387){\makebox(0,0){$\times$}}
\put(608,406){\makebox(0,0){$\times$}}
\put(644,637){\makebox(0,0){$\times$}}
\put(692,676){\makebox(0,0){$\times$}}
\put(734,744){\makebox(0,0){$\times$}}
\put(779,749){\makebox(0,0){$\times$}}
\put(917,800){\makebox(0,0){$\times$}}
\put(652,607){\makebox(0,0){$\times$}}
\put(699,551){\makebox(0,0){$\times$}}
\put(728,674){\makebox(0,0){$\times$}}
\put(657,650){\makebox(0,0){$\times$}}
\put(695,598){\makebox(0,0){$\times$}}
\put(650,623){\makebox(0,0){$\times$}}
\put(795,637){\makebox(0,0){$\times$}}
\put(716,799){\makebox(0,0){$\times$}}
\put(659,684){\makebox(0,0){$\times$}}
\put(1125,859){\makebox(0,0){$\times$}}
\put(1159,859){\makebox(0,0){$\times$}}
\put(1159,859){\makebox(0,0){$\times$}}
\put(744,829){\makebox(0,0){$\times$}}
\put(745,859){\makebox(0,0){$\times$}}
\put(725,799){\makebox(0,0){$\times$}}
\put(1159,859){\makebox(0,0){$\times$}}
\put(1159,859){\makebox(0,0){$\times$}}
\put(1159,859){\makebox(0,0){$\times$}}
\put(800,859){\makebox(0,0){$\times$}}
\put(1159,859){\makebox(0,0){$\times$}}
\put(692,859){\makebox(0,0){$\times$}}
\put(806,662){\makebox(0,0){$\times$}}
\put(857,859){\makebox(0,0){$\times$}}
\put(762,859){\makebox(0,0){$\times$}}
\put(834,859){\makebox(0,0){$\times$}}
\put(702,713){\makebox(0,0){$\times$}}
\put(776,859){\makebox(0,0){$\times$}}
\put(733,661){\makebox(0,0){$\times$}}
\put(1159,614){\makebox(0,0){$\times$}}
\put(1159,514){\makebox(0,0){$\times$}}
\put(410.0,110.0){\rule[-0.200pt]{0.400pt}{180.434pt}}
\put(410.0,110.0){\rule[-0.200pt]{180.434pt}{0.400pt}}
\put(1159.0,110.0){\rule[-0.200pt]{0.400pt}{180.434pt}}
\put(410.0,859.0){\rule[-0.200pt]{180.434pt}{0.400pt}}
\end{picture}

%% file: figures/scatter_RPFP_z3__z3.tex
% GNUPLOT: LaTeX picture
\setlength{\unitlength}{0.240900pt}
\ifx\plotpoint\undefined\newsavebox{\plotpoint}\fi
\begin{picture}(1500,900)(0,0)
\sbox{\plotpoint}{\rule[-0.200pt]{0.400pt}{0.400pt}}%
\put(460,274){\makebox(0,0)[r]{1}}
\put(480.0,274.0){\rule[-0.200pt]{4.818pt}{0.400pt}}
\put(460,439){\makebox(0,0)[r]{10}}
\put(480.0,439.0){\rule[-0.200pt]{4.818pt}{0.400pt}}
\put(460,603){\makebox(0,0)[r]{100}}
\put(480.0,603.0){\rule[-0.200pt]{4.818pt}{0.400pt}}
\put(460,768){\makebox(0,0)[r]{1000}}
\put(480.0,768.0){\rule[-0.200pt]{4.818pt}{0.400pt}}
\put(460,859){\makebox(0,0)[r]{t/o}}
\put(480.0,859.0){\rule[-0.200pt]{4.818pt}{0.400pt}}
\put(644,69){\makebox(0,0){1}}
\put(644.0,110.0){\rule[-0.200pt]{0.400pt}{4.818pt}}
\put(809,69){\makebox(0,0){10}}
\put(809.0,110.0){\rule[-0.200pt]{0.400pt}{4.818pt}}
\put(973,69){\makebox(0,0){100}}
\put(973.0,110.0){\rule[-0.200pt]{0.400pt}{4.818pt}}
\put(1138,69){\makebox(0,0){1000}}
\put(1138.0,110.0){\rule[-0.200pt]{0.400pt}{4.818pt}}
\put(1229,69){\makebox(0,0){t/o}}
\put(1229.0,110.0){\rule[-0.200pt]{0.400pt}{4.818pt}}
\put(480.0,110.0){\rule[-0.200pt]{0.400pt}{180.434pt}}
\put(480.0,110.0){\rule[-0.200pt]{180.434pt}{0.400pt}}
\put(1229.0,110.0){\rule[-0.200pt]{0.400pt}{180.434pt}}
\put(480.0,859.0){\rule[-0.200pt]{180.434pt}{0.400pt}}
\put(480,859){\line(1,0){749}}
\put(1229,110){\line(0,1){749}}
\put(394,484){\makebox(0,0){\rotatebox{90}{z3}}}
\put(854,29){\makebox(0,0){RPFP(z3)}}
\put(480,110){\usebox{\plotpoint}}
\multiput(480.00,110.59)(0.494,0.488){13}{\rule{0.500pt}{0.117pt}}
\multiput(480.00,109.17)(6.962,8.000){2}{\rule{0.250pt}{0.400pt}}
\multiput(488.00,118.59)(0.492,0.485){11}{\rule{0.500pt}{0.117pt}}
\multiput(488.00,117.17)(5.962,7.000){2}{\rule{0.250pt}{0.400pt}}
\multiput(495.00,125.59)(0.494,0.488){13}{\rule{0.500pt}{0.117pt}}
\multiput(495.00,124.17)(6.962,8.000){2}{\rule{0.250pt}{0.400pt}}
\multiput(503.00,133.59)(0.492,0.485){11}{\rule{0.500pt}{0.117pt}}
\multiput(503.00,132.17)(5.962,7.000){2}{\rule{0.250pt}{0.400pt}}
\multiput(510.00,140.59)(0.494,0.488){13}{\rule{0.500pt}{0.117pt}}
\multiput(510.00,139.17)(6.962,8.000){2}{\rule{0.250pt}{0.400pt}}
\multiput(518.00,148.59)(0.492,0.485){11}{\rule{0.500pt}{0.117pt}}
\multiput(518.00,147.17)(5.962,7.000){2}{\rule{0.250pt}{0.400pt}}
\multiput(525.00,155.59)(0.494,0.488){13}{\rule{0.500pt}{0.117pt}}
\multiput(525.00,154.17)(6.962,8.000){2}{\rule{0.250pt}{0.400pt}}
\multiput(533.00,163.59)(0.494,0.488){13}{\rule{0.500pt}{0.117pt}}
\multiput(533.00,162.17)(6.962,8.000){2}{\rule{0.250pt}{0.400pt}}
\multiput(541.00,171.59)(0.492,0.485){11}{\rule{0.500pt}{0.117pt}}
\multiput(541.00,170.17)(5.962,7.000){2}{\rule{0.250pt}{0.400pt}}
\multiput(548.00,178.59)(0.494,0.488){13}{\rule{0.500pt}{0.117pt}}
\multiput(548.00,177.17)(6.962,8.000){2}{\rule{0.250pt}{0.400pt}}
\multiput(556.00,186.59)(0.492,0.485){11}{\rule{0.500pt}{0.117pt}}
\multiput(556.00,185.17)(5.962,7.000){2}{\rule{0.250pt}{0.400pt}}
\multiput(563.00,193.59)(0.494,0.488){13}{\rule{0.500pt}{0.117pt}}
\multiput(563.00,192.17)(6.962,8.000){2}{\rule{0.250pt}{0.400pt}}
\multiput(571.00,201.59)(0.492,0.485){11}{\rule{0.500pt}{0.117pt}}
\multiput(571.00,200.17)(5.962,7.000){2}{\rule{0.250pt}{0.400pt}}
\multiput(578.00,208.59)(0.494,0.488){13}{\rule{0.500pt}{0.117pt}}
\multiput(578.00,207.17)(6.962,8.000){2}{\rule{0.250pt}{0.400pt}}
\multiput(586.00,216.59)(0.492,0.485){11}{\rule{0.500pt}{0.117pt}}
\multiput(586.00,215.17)(5.962,7.000){2}{\rule{0.250pt}{0.400pt}}
\multiput(593.00,223.59)(0.494,0.488){13}{\rule{0.500pt}{0.117pt}}
\multiput(593.00,222.17)(6.962,8.000){2}{\rule{0.250pt}{0.400pt}}
\multiput(601.00,231.59)(0.494,0.488){13}{\rule{0.500pt}{0.117pt}}
\multiput(601.00,230.17)(6.962,8.000){2}{\rule{0.250pt}{0.400pt}}
\multiput(609.00,239.59)(0.492,0.485){11}{\rule{0.500pt}{0.117pt}}
\multiput(609.00,238.17)(5.962,7.000){2}{\rule{0.250pt}{0.400pt}}
\multiput(616.00,246.59)(0.494,0.488){13}{\rule{0.500pt}{0.117pt}}
\multiput(616.00,245.17)(6.962,8.000){2}{\rule{0.250pt}{0.400pt}}
\multiput(624.00,254.59)(0.492,0.485){11}{\rule{0.500pt}{0.117pt}}
\multiput(624.00,253.17)(5.962,7.000){2}{\rule{0.250pt}{0.400pt}}
\multiput(631.00,261.59)(0.494,0.488){13}{\rule{0.500pt}{0.117pt}}
\multiput(631.00,260.17)(6.962,8.000){2}{\rule{0.250pt}{0.400pt}}
\multiput(639.00,269.59)(0.492,0.485){11}{\rule{0.500pt}{0.117pt}}
\multiput(639.00,268.17)(5.962,7.000){2}{\rule{0.250pt}{0.400pt}}
\multiput(646.00,276.59)(0.494,0.488){13}{\rule{0.500pt}{0.117pt}}
\multiput(646.00,275.17)(6.962,8.000){2}{\rule{0.250pt}{0.400pt}}
\multiput(654.00,284.59)(0.494,0.488){13}{\rule{0.500pt}{0.117pt}}
\multiput(654.00,283.17)(6.962,8.000){2}{\rule{0.250pt}{0.400pt}}
\multiput(662.00,292.59)(0.492,0.485){11}{\rule{0.500pt}{0.117pt}}
\multiput(662.00,291.17)(5.962,7.000){2}{\rule{0.250pt}{0.400pt}}
\multiput(669.00,299.59)(0.494,0.488){13}{\rule{0.500pt}{0.117pt}}
\multiput(669.00,298.17)(6.962,8.000){2}{\rule{0.250pt}{0.400pt}}
\multiput(677.00,307.59)(0.492,0.485){11}{\rule{0.500pt}{0.117pt}}
\multiput(677.00,306.17)(5.962,7.000){2}{\rule{0.250pt}{0.400pt}}
\multiput(684.00,314.59)(0.494,0.488){13}{\rule{0.500pt}{0.117pt}}
\multiput(684.00,313.17)(6.962,8.000){2}{\rule{0.250pt}{0.400pt}}
\multiput(692.00,322.59)(0.492,0.485){11}{\rule{0.500pt}{0.117pt}}
\multiput(692.00,321.17)(5.962,7.000){2}{\rule{0.250pt}{0.400pt}}
\multiput(699.00,329.59)(0.494,0.488){13}{\rule{0.500pt}{0.117pt}}
\multiput(699.00,328.17)(6.962,8.000){2}{\rule{0.250pt}{0.400pt}}
\multiput(707.00,337.59)(0.494,0.488){13}{\rule{0.500pt}{0.117pt}}
\multiput(707.00,336.17)(6.962,8.000){2}{\rule{0.250pt}{0.400pt}}
\multiput(715.00,345.59)(0.492,0.485){11}{\rule{0.500pt}{0.117pt}}
\multiput(715.00,344.17)(5.962,7.000){2}{\rule{0.250pt}{0.400pt}}
\multiput(722.00,352.59)(0.494,0.488){13}{\rule{0.500pt}{0.117pt}}
\multiput(722.00,351.17)(6.962,8.000){2}{\rule{0.250pt}{0.400pt}}
\multiput(730.00,360.59)(0.492,0.485){11}{\rule{0.500pt}{0.117pt}}
\multiput(730.00,359.17)(5.962,7.000){2}{\rule{0.250pt}{0.400pt}}
\multiput(737.00,367.59)(0.494,0.488){13}{\rule{0.500pt}{0.117pt}}
\multiput(737.00,366.17)(6.962,8.000){2}{\rule{0.250pt}{0.400pt}}
\multiput(745.00,375.59)(0.492,0.485){11}{\rule{0.500pt}{0.117pt}}
\multiput(745.00,374.17)(5.962,7.000){2}{\rule{0.250pt}{0.400pt}}
\multiput(752.00,382.59)(0.494,0.488){13}{\rule{0.500pt}{0.117pt}}
\multiput(752.00,381.17)(6.962,8.000){2}{\rule{0.250pt}{0.400pt}}
\multiput(760.00,390.59)(0.492,0.485){11}{\rule{0.500pt}{0.117pt}}
\multiput(760.00,389.17)(5.962,7.000){2}{\rule{0.250pt}{0.400pt}}
\multiput(767.00,397.59)(0.494,0.488){13}{\rule{0.500pt}{0.117pt}}
\multiput(767.00,396.17)(6.962,8.000){2}{\rule{0.250pt}{0.400pt}}
\multiput(775.00,405.59)(0.494,0.488){13}{\rule{0.500pt}{0.117pt}}
\multiput(775.00,404.17)(6.962,8.000){2}{\rule{0.250pt}{0.400pt}}
\multiput(783.00,413.59)(0.492,0.485){11}{\rule{0.500pt}{0.117pt}}
\multiput(783.00,412.17)(5.962,7.000){2}{\rule{0.250pt}{0.400pt}}
\multiput(790.00,420.59)(0.494,0.488){13}{\rule{0.500pt}{0.117pt}}
\multiput(790.00,419.17)(6.962,8.000){2}{\rule{0.250pt}{0.400pt}}
\multiput(798.00,428.59)(0.492,0.485){11}{\rule{0.500pt}{0.117pt}}
\multiput(798.00,427.17)(5.962,7.000){2}{\rule{0.250pt}{0.400pt}}
\multiput(805.00,435.59)(0.494,0.488){13}{\rule{0.500pt}{0.117pt}}
\multiput(805.00,434.17)(6.962,8.000){2}{\rule{0.250pt}{0.400pt}}
\multiput(813.00,443.59)(0.492,0.485){11}{\rule{0.500pt}{0.117pt}}
\multiput(813.00,442.17)(5.962,7.000){2}{\rule{0.250pt}{0.400pt}}
\multiput(820.00,450.59)(0.494,0.488){13}{\rule{0.500pt}{0.117pt}}
\multiput(820.00,449.17)(6.962,8.000){2}{\rule{0.250pt}{0.400pt}}
\multiput(828.00,458.59)(0.494,0.488){13}{\rule{0.500pt}{0.117pt}}
\multiput(828.00,457.17)(6.962,8.000){2}{\rule{0.250pt}{0.400pt}}
\multiput(836.00,466.59)(0.492,0.485){11}{\rule{0.500pt}{0.117pt}}
\multiput(836.00,465.17)(5.962,7.000){2}{\rule{0.250pt}{0.400pt}}
\multiput(843.00,473.59)(0.494,0.488){13}{\rule{0.500pt}{0.117pt}}
\multiput(843.00,472.17)(6.962,8.000){2}{\rule{0.250pt}{0.400pt}}
\multiput(851.00,481.59)(0.492,0.485){11}{\rule{0.500pt}{0.117pt}}
\multiput(851.00,480.17)(5.962,7.000){2}{\rule{0.250pt}{0.400pt}}
\multiput(858.00,488.59)(0.494,0.488){13}{\rule{0.500pt}{0.117pt}}
\multiput(858.00,487.17)(6.962,8.000){2}{\rule{0.250pt}{0.400pt}}
\multiput(866.00,496.59)(0.492,0.485){11}{\rule{0.500pt}{0.117pt}}
\multiput(866.00,495.17)(5.962,7.000){2}{\rule{0.250pt}{0.400pt}}
\multiput(873.00,503.59)(0.494,0.488){13}{\rule{0.500pt}{0.117pt}}
\multiput(873.00,502.17)(6.962,8.000){2}{\rule{0.250pt}{0.400pt}}
\multiput(881.00,511.59)(0.494,0.488){13}{\rule{0.500pt}{0.117pt}}
\multiput(881.00,510.17)(6.962,8.000){2}{\rule{0.250pt}{0.400pt}}
\multiput(889.00,519.59)(0.492,0.485){11}{\rule{0.500pt}{0.117pt}}
\multiput(889.00,518.17)(5.962,7.000){2}{\rule{0.250pt}{0.400pt}}
\multiput(896.00,526.59)(0.494,0.488){13}{\rule{0.500pt}{0.117pt}}
\multiput(896.00,525.17)(6.962,8.000){2}{\rule{0.250pt}{0.400pt}}
\multiput(904.00,534.59)(0.492,0.485){11}{\rule{0.500pt}{0.117pt}}
\multiput(904.00,533.17)(5.962,7.000){2}{\rule{0.250pt}{0.400pt}}
\multiput(911.00,541.59)(0.494,0.488){13}{\rule{0.500pt}{0.117pt}}
\multiput(911.00,540.17)(6.962,8.000){2}{\rule{0.250pt}{0.400pt}}
\multiput(919.00,549.59)(0.492,0.485){11}{\rule{0.500pt}{0.117pt}}
\multiput(919.00,548.17)(5.962,7.000){2}{\rule{0.250pt}{0.400pt}}
\multiput(926.00,556.59)(0.494,0.488){13}{\rule{0.500pt}{0.117pt}}
\multiput(926.00,555.17)(6.962,8.000){2}{\rule{0.250pt}{0.400pt}}
\multiput(934.00,564.59)(0.494,0.488){13}{\rule{0.500pt}{0.117pt}}
\multiput(934.00,563.17)(6.962,8.000){2}{\rule{0.250pt}{0.400pt}}
\multiput(942.00,572.59)(0.492,0.485){11}{\rule{0.500pt}{0.117pt}}
\multiput(942.00,571.17)(5.962,7.000){2}{\rule{0.250pt}{0.400pt}}
\multiput(949.00,579.59)(0.494,0.488){13}{\rule{0.500pt}{0.117pt}}
\multiput(949.00,578.17)(6.962,8.000){2}{\rule{0.250pt}{0.400pt}}
\multiput(957.00,587.59)(0.492,0.485){11}{\rule{0.500pt}{0.117pt}}
\multiput(957.00,586.17)(5.962,7.000){2}{\rule{0.250pt}{0.400pt}}
\multiput(964.00,594.59)(0.494,0.488){13}{\rule{0.500pt}{0.117pt}}
\multiput(964.00,593.17)(6.962,8.000){2}{\rule{0.250pt}{0.400pt}}
\multiput(972.00,602.59)(0.492,0.485){11}{\rule{0.500pt}{0.117pt}}
\multiput(972.00,601.17)(5.962,7.000){2}{\rule{0.250pt}{0.400pt}}
\multiput(979.00,609.59)(0.494,0.488){13}{\rule{0.500pt}{0.117pt}}
\multiput(979.00,608.17)(6.962,8.000){2}{\rule{0.250pt}{0.400pt}}
\multiput(987.00,617.59)(0.492,0.485){11}{\rule{0.500pt}{0.117pt}}
\multiput(987.00,616.17)(5.962,7.000){2}{\rule{0.250pt}{0.400pt}}
\multiput(994.00,624.59)(0.494,0.488){13}{\rule{0.500pt}{0.117pt}}
\multiput(994.00,623.17)(6.962,8.000){2}{\rule{0.250pt}{0.400pt}}
\multiput(1002.00,632.59)(0.494,0.488){13}{\rule{0.500pt}{0.117pt}}
\multiput(1002.00,631.17)(6.962,8.000){2}{\rule{0.250pt}{0.400pt}}
\multiput(1010.00,640.59)(0.492,0.485){11}{\rule{0.500pt}{0.117pt}}
\multiput(1010.00,639.17)(5.962,7.000){2}{\rule{0.250pt}{0.400pt}}
\multiput(1017.00,647.59)(0.494,0.488){13}{\rule{0.500pt}{0.117pt}}
\multiput(1017.00,646.17)(6.962,8.000){2}{\rule{0.250pt}{0.400pt}}
\multiput(1025.00,655.59)(0.492,0.485){11}{\rule{0.500pt}{0.117pt}}
\multiput(1025.00,654.17)(5.962,7.000){2}{\rule{0.250pt}{0.400pt}}
\multiput(1032.00,662.59)(0.494,0.488){13}{\rule{0.500pt}{0.117pt}}
\multiput(1032.00,661.17)(6.962,8.000){2}{\rule{0.250pt}{0.400pt}}
\multiput(1040.00,670.59)(0.492,0.485){11}{\rule{0.500pt}{0.117pt}}
\multiput(1040.00,669.17)(5.962,7.000){2}{\rule{0.250pt}{0.400pt}}
\multiput(1047.00,677.59)(0.494,0.488){13}{\rule{0.500pt}{0.117pt}}
\multiput(1047.00,676.17)(6.962,8.000){2}{\rule{0.250pt}{0.400pt}}
\multiput(1055.00,685.59)(0.494,0.488){13}{\rule{0.500pt}{0.117pt}}
\multiput(1055.00,684.17)(6.962,8.000){2}{\rule{0.250pt}{0.400pt}}
\multiput(1063.00,693.59)(0.492,0.485){11}{\rule{0.500pt}{0.117pt}}
\multiput(1063.00,692.17)(5.962,7.000){2}{\rule{0.250pt}{0.400pt}}
\multiput(1070.00,700.59)(0.494,0.488){13}{\rule{0.500pt}{0.117pt}}
\multiput(1070.00,699.17)(6.962,8.000){2}{\rule{0.250pt}{0.400pt}}
\multiput(1078.00,708.59)(0.492,0.485){11}{\rule{0.500pt}{0.117pt}}
\multiput(1078.00,707.17)(5.962,7.000){2}{\rule{0.250pt}{0.400pt}}
\multiput(1085.00,715.59)(0.494,0.488){13}{\rule{0.500pt}{0.117pt}}
\multiput(1085.00,714.17)(6.962,8.000){2}{\rule{0.250pt}{0.400pt}}
\multiput(1093.00,723.59)(0.492,0.485){11}{\rule{0.500pt}{0.117pt}}
\multiput(1093.00,722.17)(5.962,7.000){2}{\rule{0.250pt}{0.400pt}}
\multiput(1100.00,730.59)(0.494,0.488){13}{\rule{0.500pt}{0.117pt}}
\multiput(1100.00,729.17)(6.962,8.000){2}{\rule{0.250pt}{0.400pt}}
\multiput(1108.00,738.59)(0.494,0.488){13}{\rule{0.500pt}{0.117pt}}
\multiput(1108.00,737.17)(6.962,8.000){2}{\rule{0.250pt}{0.400pt}}
\multiput(1116.00,746.59)(0.492,0.485){11}{\rule{0.500pt}{0.117pt}}
\multiput(1116.00,745.17)(5.962,7.000){2}{\rule{0.250pt}{0.400pt}}
\multiput(1123.00,753.59)(0.494,0.488){13}{\rule{0.500pt}{0.117pt}}
\multiput(1123.00,752.17)(6.962,8.000){2}{\rule{0.250pt}{0.400pt}}
\multiput(1131.00,761.59)(0.492,0.485){11}{\rule{0.500pt}{0.117pt}}
\multiput(1131.00,760.17)(5.962,7.000){2}{\rule{0.250pt}{0.400pt}}
\multiput(1138.00,768.59)(0.494,0.488){13}{\rule{0.500pt}{0.117pt}}
\multiput(1138.00,767.17)(6.962,8.000){2}{\rule{0.250pt}{0.400pt}}
\multiput(1146.00,776.59)(0.492,0.485){11}{\rule{0.500pt}{0.117pt}}
\multiput(1146.00,775.17)(5.962,7.000){2}{\rule{0.250pt}{0.400pt}}
\multiput(1153.00,783.59)(0.494,0.488){13}{\rule{0.500pt}{0.117pt}}
\multiput(1153.00,782.17)(6.962,8.000){2}{\rule{0.250pt}{0.400pt}}
\multiput(1161.00,791.59)(0.492,0.485){11}{\rule{0.500pt}{0.117pt}}
\multiput(1161.00,790.17)(5.962,7.000){2}{\rule{0.250pt}{0.400pt}}
\multiput(1168.00,798.59)(0.494,0.488){13}{\rule{0.500pt}{0.117pt}}
\multiput(1168.00,797.17)(6.962,8.000){2}{\rule{0.250pt}{0.400pt}}
\multiput(1176.00,806.59)(0.494,0.488){13}{\rule{0.500pt}{0.117pt}}
\multiput(1176.00,805.17)(6.962,8.000){2}{\rule{0.250pt}{0.400pt}}
\multiput(1184.00,814.59)(0.492,0.485){11}{\rule{0.500pt}{0.117pt}}
\multiput(1184.00,813.17)(5.962,7.000){2}{\rule{0.250pt}{0.400pt}}
\multiput(1191.00,821.59)(0.494,0.488){13}{\rule{0.500pt}{0.117pt}}
\multiput(1191.00,820.17)(6.962,8.000){2}{\rule{0.250pt}{0.400pt}}
\multiput(1199.00,829.59)(0.492,0.485){11}{\rule{0.500pt}{0.117pt}}
\multiput(1199.00,828.17)(5.962,7.000){2}{\rule{0.250pt}{0.400pt}}
\multiput(1206.00,836.59)(0.494,0.488){13}{\rule{0.500pt}{0.117pt}}
\multiput(1206.00,835.17)(6.962,8.000){2}{\rule{0.250pt}{0.400pt}}
\multiput(1214.00,844.59)(0.492,0.485){11}{\rule{0.500pt}{0.117pt}}
\multiput(1214.00,843.17)(5.962,7.000){2}{\rule{0.250pt}{0.400pt}}
\multiput(1221.00,851.59)(0.494,0.488){13}{\rule{0.500pt}{0.117pt}}
\multiput(1221.00,850.17)(6.962,8.000){2}{\rule{0.250pt}{0.400pt}}
\put(582,110){\makebox(0,0){$\times$}}
\put(1229,510){\makebox(0,0){$\times$}}
\put(719,512){\makebox(0,0){$\times$}}
\put(771,647){\makebox(0,0){$\times$}}
\put(667,403){\makebox(0,0){$\times$}}
\put(796,714){\makebox(0,0){$\times$}}
\put(825,790){\makebox(0,0){$\times$}}
\put(849,823){\makebox(0,0){$\times$}}
\put(719,516){\makebox(0,0){$\times$}}
\put(767,624){\makebox(0,0){$\times$}}
\put(665,404){\makebox(0,0){$\times$}}
\put(796,704){\makebox(0,0){$\times$}}
\put(811,757){\makebox(0,0){$\times$}}
\put(843,802){\makebox(0,0){$\times$}}
\put(719,494){\makebox(0,0){$\times$}}
\put(769,655){\makebox(0,0){$\times$}}
\put(670,407){\makebox(0,0){$\times$}}
\put(795,735){\makebox(0,0){$\times$}}
\put(824,734){\makebox(0,0){$\times$}}
\put(863,791){\makebox(0,0){$\times$}}
\put(638,298){\makebox(0,0){$\times$}}
\put(639,301){\makebox(0,0){$\times$}}
\put(625,291){\makebox(0,0){$\times$}}
\put(691,406){\makebox(0,0){$\times$}}
\put(674,399){\makebox(0,0){$\times$}}
\put(676,268){\makebox(0,0){$\times$}}
\put(685,265){\makebox(0,0){$\times$}}
\put(661,110){\makebox(0,0){$\times$}}
\put(663,255){\makebox(0,0){$\times$}}
\put(651,261){\makebox(0,0){$\times$}}
\put(691,422){\makebox(0,0){$\times$}}
\put(679,403){\makebox(0,0){$\times$}}
\put(1229,652){\makebox(0,0){$\times$}}
\put(1221,859){\makebox(0,0){$\times$}}
\put(1229,859){\makebox(0,0){$\times$}}
\put(1016,782){\makebox(0,0){$\times$}}
\put(1114,859){\makebox(0,0){$\times$}}
\put(1229,626){\makebox(0,0){$\times$}}
\put(605,389){\makebox(0,0){$\times$}}
\put(1229,859){\makebox(0,0){$\times$}}
\put(1229,859){\makebox(0,0){$\times$}}
\put(998,519){\makebox(0,0){$\times$}}
\put(728,403){\makebox(0,0){$\times$}}
\put(756,461){\makebox(0,0){$\times$}}
\put(666,297){\makebox(0,0){$\times$}}
\put(785,571){\makebox(0,0){$\times$}}
\put(811,611){\makebox(0,0){$\times$}}
\put(831,494){\makebox(0,0){$\times$}}
\put(718,430){\makebox(0,0){$\times$}}
\put(762,460){\makebox(0,0){$\times$}}
\put(657,313){\makebox(0,0){$\times$}}
\put(786,597){\makebox(0,0){$\times$}}
\put(812,633){\makebox(0,0){$\times$}}
\put(834,604){\makebox(0,0){$\times$}}
\put(591,110){\makebox(0,0){$\times$}}
\put(595,110){\makebox(0,0){$\times$}}
\put(805,533){\makebox(0,0){$\times$}}
\put(725,549){\makebox(0,0){$\times$}}
\put(679,493){\makebox(0,0){$\times$}}
\put(743,504){\makebox(0,0){$\times$}}
\put(781,550){\makebox(0,0){$\times$}}
\put(691,459){\makebox(0,0){$\times$}}
\put(845,567){\makebox(0,0){$\times$}}
\put(929,531){\makebox(0,0){$\times$}}
\put(679,480){\makebox(0,0){$\times$}}
\put(869,549){\makebox(0,0){$\times$}}
\put(737,650){\makebox(0,0){$\times$}}
\put(837,769){\makebox(0,0){$\times$}}
\put(1229,859){\makebox(0,0){$\times$}}
\put(1229,859){\makebox(0,0){$\times$}}
\put(1229,859){\makebox(0,0){$\times$}}
\put(1229,859){\makebox(0,0){$\times$}}
\put(1229,859){\makebox(0,0){$\times$}}
\put(1229,859){\makebox(0,0){$\times$}}
\put(1229,859){\makebox(0,0){$\times$}}
\put(1229,859){\makebox(0,0){$\times$}}
\put(731,520){\makebox(0,0){$\times$}}
\put(1229,859){\makebox(0,0){$\times$}}
\put(1229,859){\makebox(0,0){$\times$}}
\put(1222,859){\makebox(0,0){$\times$}}
\put(1229,859){\makebox(0,0){$\times$}}
\put(777,461){\makebox(0,0){$\times$}}
\put(1229,859){\makebox(0,0){$\times$}}
\put(1229,859){\makebox(0,0){$\times$}}
\put(852,724){\makebox(0,0){$\times$}}
\put(1229,846){\makebox(0,0){$\times$}}
\put(1229,855){\makebox(0,0){$\times$}}
\put(1211,822){\makebox(0,0){$\times$}}
\put(558,110){\makebox(0,0){$\times$}}
\put(679,443){\makebox(0,0){$\times$}}
\put(686,388){\makebox(0,0){$\times$}}
\put(726,446){\makebox(0,0){$\times$}}
\put(679,387){\makebox(0,0){$\times$}}
\put(688,387){\makebox(0,0){$\times$}}
\put(683,406){\makebox(0,0){$\times$}}
\put(958,637){\makebox(0,0){$\times$}}
\put(780,676){\makebox(0,0){$\times$}}
\put(779,744){\makebox(0,0){$\times$}}
\put(820,749){\makebox(0,0){$\times$}}
\put(839,800){\makebox(0,0){$\times$}}
\put(737,607){\makebox(0,0){$\times$}}
\put(730,551){\makebox(0,0){$\times$}}
\put(811,674){\makebox(0,0){$\times$}}
\put(867,650){\makebox(0,0){$\times$}}
\put(746,598){\makebox(0,0){$\times$}}
\put(748,623){\makebox(0,0){$\times$}}
\put(812,637){\makebox(0,0){$\times$}}
\put(807,799){\makebox(0,0){$\times$}}
\put(766,684){\makebox(0,0){$\times$}}
\put(1208,859){\makebox(0,0){$\times$}}
\put(1229,859){\makebox(0,0){$\times$}}
\put(1229,859){\makebox(0,0){$\times$}}
\put(923,829){\makebox(0,0){$\times$}}
\put(923,859){\makebox(0,0){$\times$}}
\put(919,799){\makebox(0,0){$\times$}}
\put(1229,859){\makebox(0,0){$\times$}}
\put(1229,859){\makebox(0,0){$\times$}}
\put(1229,859){\makebox(0,0){$\times$}}
\put(1124,859){\makebox(0,0){$\times$}}
\put(1229,859){\makebox(0,0){$\times$}}
\put(828,859){\makebox(0,0){$\times$}}
\put(849,662){\makebox(0,0){$\times$}}
\put(1014,859){\makebox(0,0){$\times$}}
\put(786,859){\makebox(0,0){$\times$}}
\put(886,859){\makebox(0,0){$\times$}}
\put(824,713){\makebox(0,0){$\times$}}
\put(991,859){\makebox(0,0){$\times$}}
\put(790,661){\makebox(0,0){$\times$}}
\put(1229,614){\makebox(0,0){$\times$}}
\put(1229,514){\makebox(0,0){$\times$}}
\put(480.0,110.0){\rule[-0.200pt]{0.400pt}{180.434pt}}
\put(480.0,110.0){\rule[-0.200pt]{180.434pt}{0.400pt}}
\put(1229.0,110.0){\rule[-0.200pt]{0.400pt}{180.434pt}}
\put(480.0,859.0){\rule[-0.200pt]{180.434pt}{0.400pt}}
\end{picture}

%% file: figures/scatter_RA_nlsat__z3.tex
% GNUPLOT: LaTeX picture
\setlength{\unitlength}{0.240900pt}
\ifx\plotpoint\undefined\newsavebox{\plotpoint}\fi
\begin{picture}(1500,900)(0,0)
\sbox{\plotpoint}{\rule[-0.200pt]{0.400pt}{0.400pt}}%
\put(460,274){\makebox(0,0)[r]{1}}
\put(480.0,274.0){\rule[-0.200pt]{4.818pt}{0.400pt}}
\put(460,439){\makebox(0,0)[r]{10}}
\put(480.0,439.0){\rule[-0.200pt]{4.818pt}{0.400pt}}
\put(460,603){\makebox(0,0)[r]{100}}
\put(480.0,603.0){\rule[-0.200pt]{4.818pt}{0.400pt}}
\put(460,768){\makebox(0,0)[r]{1000}}
\put(480.0,768.0){\rule[-0.200pt]{4.818pt}{0.400pt}}
\put(460,859){\makebox(0,0)[r]{t/o}}
\put(480.0,859.0){\rule[-0.200pt]{4.818pt}{0.400pt}}
\put(644,69){\makebox(0,0){1}}
\put(644.0,110.0){\rule[-0.200pt]{0.400pt}{4.818pt}}
\put(809,69){\makebox(0,0){10}}
\put(809.0,110.0){\rule[-0.200pt]{0.400pt}{4.818pt}}
\put(973,69){\makebox(0,0){100}}
\put(973.0,110.0){\rule[-0.200pt]{0.400pt}{4.818pt}}
\put(1138,69){\makebox(0,0){1000}}
\put(1138.0,110.0){\rule[-0.200pt]{0.400pt}{4.818pt}}
\put(1229,69){\makebox(0,0){t/o}}
\put(1229.0,110.0){\rule[-0.200pt]{0.400pt}{4.818pt}}
\put(480.0,110.0){\rule[-0.200pt]{0.400pt}{180.434pt}}
\put(480.0,110.0){\rule[-0.200pt]{180.434pt}{0.400pt}}
\put(1229.0,110.0){\rule[-0.200pt]{0.400pt}{180.434pt}}
\put(480.0,859.0){\rule[-0.200pt]{180.434pt}{0.400pt}}
\put(480,859){\line(1,0){749}}
\put(1229,110){\line(0,1){749}}
\put(394,484){\makebox(0,0){\rotatebox{90}{z3}}}
\put(854,29){\makebox(0,0){RA(nlsat)}}
\put(480,110){\usebox{\plotpoint}}
\multiput(480.00,110.59)(0.494,0.488){13}{\rule{0.500pt}{0.117pt}}
\multiput(480.00,109.17)(6.962,8.000){2}{\rule{0.250pt}{0.400pt}}
\multiput(488.00,118.59)(0.492,0.485){11}{\rule{0.500pt}{0.117pt}}
\multiput(488.00,117.17)(5.962,7.000){2}{\rule{0.250pt}{0.400pt}}
\multiput(495.00,125.59)(0.494,0.488){13}{\rule{0.500pt}{0.117pt}}
\multiput(495.00,124.17)(6.962,8.000){2}{\rule{0.250pt}{0.400pt}}
\multiput(503.00,133.59)(0.492,0.485){11}{\rule{0.500pt}{0.117pt}}
\multiput(503.00,132.17)(5.962,7.000){2}{\rule{0.250pt}{0.400pt}}
\multiput(510.00,140.59)(0.494,0.488){13}{\rule{0.500pt}{0.117pt}}
\multiput(510.00,139.17)(6.962,8.000){2}{\rule{0.250pt}{0.400pt}}
\multiput(518.00,148.59)(0.492,0.485){11}{\rule{0.500pt}{0.117pt}}
\multiput(518.00,147.17)(5.962,7.000){2}{\rule{0.250pt}{0.400pt}}
\multiput(525.00,155.59)(0.494,0.488){13}{\rule{0.500pt}{0.117pt}}
\multiput(525.00,154.17)(6.962,8.000){2}{\rule{0.250pt}{0.400pt}}
\multiput(533.00,163.59)(0.494,0.488){13}{\rule{0.500pt}{0.117pt}}
\multiput(533.00,162.17)(6.962,8.000){2}{\rule{0.250pt}{0.400pt}}
\multiput(541.00,171.59)(0.492,0.485){11}{\rule{0.500pt}{0.117pt}}
\multiput(541.00,170.17)(5.962,7.000){2}{\rule{0.250pt}{0.400pt}}
\multiput(548.00,178.59)(0.494,0.488){13}{\rule{0.500pt}{0.117pt}}
\multiput(548.00,177.17)(6.962,8.000){2}{\rule{0.250pt}{0.400pt}}
\multiput(556.00,186.59)(0.492,0.485){11}{\rule{0.500pt}{0.117pt}}
\multiput(556.00,185.17)(5.962,7.000){2}{\rule{0.250pt}{0.400pt}}
\multiput(563.00,193.59)(0.494,0.488){13}{\rule{0.500pt}{0.117pt}}
\multiput(563.00,192.17)(6.962,8.000){2}{\rule{0.250pt}{0.400pt}}
\multiput(571.00,201.59)(0.492,0.485){11}{\rule{0.500pt}{0.117pt}}
\multiput(571.00,200.17)(5.962,7.000){2}{\rule{0.250pt}{0.400pt}}
\multiput(578.00,208.59)(0.494,0.488){13}{\rule{0.500pt}{0.117pt}}
\multiput(578.00,207.17)(6.962,8.000){2}{\rule{0.250pt}{0.400pt}}
\multiput(586.00,216.59)(0.492,0.485){11}{\rule{0.500pt}{0.117pt}}
\multiput(586.00,215.17)(5.962,7.000){2}{\rule{0.250pt}{0.400pt}}
\multiput(593.00,223.59)(0.494,0.488){13}{\rule{0.500pt}{0.117pt}}
\multiput(593.00,222.17)(6.962,8.000){2}{\rule{0.250pt}{0.400pt}}
\multiput(601.00,231.59)(0.494,0.488){13}{\rule{0.500pt}{0.117pt}}
\multiput(601.00,230.17)(6.962,8.000){2}{\rule{0.250pt}{0.400pt}}
\multiput(609.00,239.59)(0.492,0.485){11}{\rule{0.500pt}{0.117pt}}
\multiput(609.00,238.17)(5.962,7.000){2}{\rule{0.250pt}{0.400pt}}
\multiput(616.00,246.59)(0.494,0.488){13}{\rule{0.500pt}{0.117pt}}
\multiput(616.00,245.17)(6.962,8.000){2}{\rule{0.250pt}{0.400pt}}
\multiput(624.00,254.59)(0.492,0.485){11}{\rule{0.500pt}{0.117pt}}
\multiput(624.00,253.17)(5.962,7.000){2}{\rule{0.250pt}{0.400pt}}
\multiput(631.00,261.59)(0.494,0.488){13}{\rule{0.500pt}{0.117pt}}
\multiput(631.00,260.17)(6.962,8.000){2}{\rule{0.250pt}{0.400pt}}
\multiput(639.00,269.59)(0.492,0.485){11}{\rule{0.500pt}{0.117pt}}
\multiput(639.00,268.17)(5.962,7.000){2}{\rule{0.250pt}{0.400pt}}
\multiput(646.00,276.59)(0.494,0.488){13}{\rule{0.500pt}{0.117pt}}
\multiput(646.00,275.17)(6.962,8.000){2}{\rule{0.250pt}{0.400pt}}
\multiput(654.00,284.59)(0.494,0.488){13}{\rule{0.500pt}{0.117pt}}
\multiput(654.00,283.17)(6.962,8.000){2}{\rule{0.250pt}{0.400pt}}
\multiput(662.00,292.59)(0.492,0.485){11}{\rule{0.500pt}{0.117pt}}
\multiput(662.00,291.17)(5.962,7.000){2}{\rule{0.250pt}{0.400pt}}
\multiput(669.00,299.59)(0.494,0.488){13}{\rule{0.500pt}{0.117pt}}
\multiput(669.00,298.17)(6.962,8.000){2}{\rule{0.250pt}{0.400pt}}
\multiput(677.00,307.59)(0.492,0.485){11}{\rule{0.500pt}{0.117pt}}
\multiput(677.00,306.17)(5.962,7.000){2}{\rule{0.250pt}{0.400pt}}
\multiput(684.00,314.59)(0.494,0.488){13}{\rule{0.500pt}{0.117pt}}
\multiput(684.00,313.17)(6.962,8.000){2}{\rule{0.250pt}{0.400pt}}
\multiput(692.00,322.59)(0.492,0.485){11}{\rule{0.500pt}{0.117pt}}
\multiput(692.00,321.17)(5.962,7.000){2}{\rule{0.250pt}{0.400pt}}
\multiput(699.00,329.59)(0.494,0.488){13}{\rule{0.500pt}{0.117pt}}
\multiput(699.00,328.17)(6.962,8.000){2}{\rule{0.250pt}{0.400pt}}
\multiput(707.00,337.59)(0.494,0.488){13}{\rule{0.500pt}{0.117pt}}
\multiput(707.00,336.17)(6.962,8.000){2}{\rule{0.250pt}{0.400pt}}
\multiput(715.00,345.59)(0.492,0.485){11}{\rule{0.500pt}{0.117pt}}
\multiput(715.00,344.17)(5.962,7.000){2}{\rule{0.250pt}{0.400pt}}
\multiput(722.00,352.59)(0.494,0.488){13}{\rule{0.500pt}{0.117pt}}
\multiput(722.00,351.17)(6.962,8.000){2}{\rule{0.250pt}{0.400pt}}
\multiput(730.00,360.59)(0.492,0.485){11}{\rule{0.500pt}{0.117pt}}
\multiput(730.00,359.17)(5.962,7.000){2}{\rule{0.250pt}{0.400pt}}
\multiput(737.00,367.59)(0.494,0.488){13}{\rule{0.500pt}{0.117pt}}
\multiput(737.00,366.17)(6.962,8.000){2}{\rule{0.250pt}{0.400pt}}
\multiput(745.00,375.59)(0.492,0.485){11}{\rule{0.500pt}{0.117pt}}
\multiput(745.00,374.17)(5.962,7.000){2}{\rule{0.250pt}{0.400pt}}
\multiput(752.00,382.59)(0.494,0.488){13}{\rule{0.500pt}{0.117pt}}
\multiput(752.00,381.17)(6.962,8.000){2}{\rule{0.250pt}{0.400pt}}
\multiput(760.00,390.59)(0.492,0.485){11}{\rule{0.500pt}{0.117pt}}
\multiput(760.00,389.17)(5.962,7.000){2}{\rule{0.250pt}{0.400pt}}
\multiput(767.00,397.59)(0.494,0.488){13}{\rule{0.500pt}{0.117pt}}
\multiput(767.00,396.17)(6.962,8.000){2}{\rule{0.250pt}{0.400pt}}
\multiput(775.00,405.59)(0.494,0.488){13}{\rule{0.500pt}{0.117pt}}
\multiput(775.00,404.17)(6.962,8.000){2}{\rule{0.250pt}{0.400pt}}
\multiput(783.00,413.59)(0.492,0.485){11}{\rule{0.500pt}{0.117pt}}
\multiput(783.00,412.17)(5.962,7.000){2}{\rule{0.250pt}{0.400pt}}
\multiput(790.00,420.59)(0.494,0.488){13}{\rule{0.500pt}{0.117pt}}
\multiput(790.00,419.17)(6.962,8.000){2}{\rule{0.250pt}{0.400pt}}
\multiput(798.00,428.59)(0.492,0.485){11}{\rule{0.500pt}{0.117pt}}
\multiput(798.00,427.17)(5.962,7.000){2}{\rule{0.250pt}{0.400pt}}
\multiput(805.00,435.59)(0.494,0.488){13}{\rule{0.500pt}{0.117pt}}
\multiput(805.00,434.17)(6.962,8.000){2}{\rule{0.250pt}{0.400pt}}
\multiput(813.00,443.59)(0.492,0.485){11}{\rule{0.500pt}{0.117pt}}
\multiput(813.00,442.17)(5.962,7.000){2}{\rule{0.250pt}{0.400pt}}
\multiput(820.00,450.59)(0.494,0.488){13}{\rule{0.500pt}{0.117pt}}
\multiput(820.00,449.17)(6.962,8.000){2}{\rule{0.250pt}{0.400pt}}
\multiput(828.00,458.59)(0.494,0.488){13}{\rule{0.500pt}{0.117pt}}
\multiput(828.00,457.17)(6.962,8.000){2}{\rule{0.250pt}{0.400pt}}
\multiput(836.00,466.59)(0.492,0.485){11}{\rule{0.500pt}{0.117pt}}
\multiput(836.00,465.17)(5.962,7.000){2}{\rule{0.250pt}{0.400pt}}
\multiput(843.00,473.59)(0.494,0.488){13}{\rule{0.500pt}{0.117pt}}
\multiput(843.00,472.17)(6.962,8.000){2}{\rule{0.250pt}{0.400pt}}
\multiput(851.00,481.59)(0.492,0.485){11}{\rule{0.500pt}{0.117pt}}
\multiput(851.00,480.17)(5.962,7.000){2}{\rule{0.250pt}{0.400pt}}
\multiput(858.00,488.59)(0.494,0.488){13}{\rule{0.500pt}{0.117pt}}
\multiput(858.00,487.17)(6.962,8.000){2}{\rule{0.250pt}{0.400pt}}
\multiput(866.00,496.59)(0.492,0.485){11}{\rule{0.500pt}{0.117pt}}
\multiput(866.00,495.17)(5.962,7.000){2}{\rule{0.250pt}{0.400pt}}
\multiput(873.00,503.59)(0.494,0.488){13}{\rule{0.500pt}{0.117pt}}
\multiput(873.00,502.17)(6.962,8.000){2}{\rule{0.250pt}{0.400pt}}
\multiput(881.00,511.59)(0.494,0.488){13}{\rule{0.500pt}{0.117pt}}
\multiput(881.00,510.17)(6.962,8.000){2}{\rule{0.250pt}{0.400pt}}
\multiput(889.00,519.59)(0.492,0.485){11}{\rule{0.500pt}{0.117pt}}
\multiput(889.00,518.17)(5.962,7.000){2}{\rule{0.250pt}{0.400pt}}
\multiput(896.00,526.59)(0.494,0.488){13}{\rule{0.500pt}{0.117pt}}
\multiput(896.00,525.17)(6.962,8.000){2}{\rule{0.250pt}{0.400pt}}
\multiput(904.00,534.59)(0.492,0.485){11}{\rule{0.500pt}{0.117pt}}
\multiput(904.00,533.17)(5.962,7.000){2}{\rule{0.250pt}{0.400pt}}
\multiput(911.00,541.59)(0.494,0.488){13}{\rule{0.500pt}{0.117pt}}
\multiput(911.00,540.17)(6.962,8.000){2}{\rule{0.250pt}{0.400pt}}
\multiput(919.00,549.59)(0.492,0.485){11}{\rule{0.500pt}{0.117pt}}
\multiput(919.00,548.17)(5.962,7.000){2}{\rule{0.250pt}{0.400pt}}
\multiput(926.00,556.59)(0.494,0.488){13}{\rule{0.500pt}{0.117pt}}
\multiput(926.00,555.17)(6.962,8.000){2}{\rule{0.250pt}{0.400pt}}
\multiput(934.00,564.59)(0.494,0.488){13}{\rule{0.500pt}{0.117pt}}
\multiput(934.00,563.17)(6.962,8.000){2}{\rule{0.250pt}{0.400pt}}
\multiput(942.00,572.59)(0.492,0.485){11}{\rule{0.500pt}{0.117pt}}
\multiput(942.00,571.17)(5.962,7.000){2}{\rule{0.250pt}{0.400pt}}
\multiput(949.00,579.59)(0.494,0.488){13}{\rule{0.500pt}{0.117pt}}
\multiput(949.00,578.17)(6.962,8.000){2}{\rule{0.250pt}{0.400pt}}
\multiput(957.00,587.59)(0.492,0.485){11}{\rule{0.500pt}{0.117pt}}
\multiput(957.00,586.17)(5.962,7.000){2}{\rule{0.250pt}{0.400pt}}
\multiput(964.00,594.59)(0.494,0.488){13}{\rule{0.500pt}{0.117pt}}
\multiput(964.00,593.17)(6.962,8.000){2}{\rule{0.250pt}{0.400pt}}
\multiput(972.00,602.59)(0.492,0.485){11}{\rule{0.500pt}{0.117pt}}
\multiput(972.00,601.17)(5.962,7.000){2}{\rule{0.250pt}{0.400pt}}
\multiput(979.00,609.59)(0.494,0.488){13}{\rule{0.500pt}{0.117pt}}
\multiput(979.00,608.17)(6.962,8.000){2}{\rule{0.250pt}{0.400pt}}
\multiput(987.00,617.59)(0.492,0.485){11}{\rule{0.500pt}{0.117pt}}
\multiput(987.00,616.17)(5.962,7.000){2}{\rule{0.250pt}{0.400pt}}
\multiput(994.00,624.59)(0.494,0.488){13}{\rule{0.500pt}{0.117pt}}
\multiput(994.00,623.17)(6.962,8.000){2}{\rule{0.250pt}{0.400pt}}
\multiput(1002.00,632.59)(0.494,0.488){13}{\rule{0.500pt}{0.117pt}}
\multiput(1002.00,631.17)(6.962,8.000){2}{\rule{0.250pt}{0.400pt}}
\multiput(1010.00,640.59)(0.492,0.485){11}{\rule{0.500pt}{0.117pt}}
\multiput(1010.00,639.17)(5.962,7.000){2}{\rule{0.250pt}{0.400pt}}
\multiput(1017.00,647.59)(0.494,0.488){13}{\rule{0.500pt}{0.117pt}}
\multiput(1017.00,646.17)(6.962,8.000){2}{\rule{0.250pt}{0.400pt}}
\multiput(1025.00,655.59)(0.492,0.485){11}{\rule{0.500pt}{0.117pt}}
\multiput(1025.00,654.17)(5.962,7.000){2}{\rule{0.250pt}{0.400pt}}
\multiput(1032.00,662.59)(0.494,0.488){13}{\rule{0.500pt}{0.117pt}}
\multiput(1032.00,661.17)(6.962,8.000){2}{\rule{0.250pt}{0.400pt}}
\multiput(1040.00,670.59)(0.492,0.485){11}{\rule{0.500pt}{0.117pt}}
\multiput(1040.00,669.17)(5.962,7.000){2}{\rule{0.250pt}{0.400pt}}
\multiput(1047.00,677.59)(0.494,0.488){13}{\rule{0.500pt}{0.117pt}}
\multiput(1047.00,676.17)(6.962,8.000){2}{\rule{0.250pt}{0.400pt}}
\multiput(1055.00,685.59)(0.494,0.488){13}{\rule{0.500pt}{0.117pt}}
\multiput(1055.00,684.17)(6.962,8.000){2}{\rule{0.250pt}{0.400pt}}
\multiput(1063.00,693.59)(0.492,0.485){11}{\rule{0.500pt}{0.117pt}}
\multiput(1063.00,692.17)(5.962,7.000){2}{\rule{0.250pt}{0.400pt}}
\multiput(1070.00,700.59)(0.494,0.488){13}{\rule{0.500pt}{0.117pt}}
\multiput(1070.00,699.17)(6.962,8.000){2}{\rule{0.250pt}{0.400pt}}
\multiput(1078.00,708.59)(0.492,0.485){11}{\rule{0.500pt}{0.117pt}}
\multiput(1078.00,707.17)(5.962,7.000){2}{\rule{0.250pt}{0.400pt}}
\multiput(1085.00,715.59)(0.494,0.488){13}{\rule{0.500pt}{0.117pt}}
\multiput(1085.00,714.17)(6.962,8.000){2}{\rule{0.250pt}{0.400pt}}
\multiput(1093.00,723.59)(0.492,0.485){11}{\rule{0.500pt}{0.117pt}}
\multiput(1093.00,722.17)(5.962,7.000){2}{\rule{0.250pt}{0.400pt}}
\multiput(1100.00,730.59)(0.494,0.488){13}{\rule{0.500pt}{0.117pt}}
\multiput(1100.00,729.17)(6.962,8.000){2}{\rule{0.250pt}{0.400pt}}
\multiput(1108.00,738.59)(0.494,0.488){13}{\rule{0.500pt}{0.117pt}}
\multiput(1108.00,737.17)(6.962,8.000){2}{\rule{0.250pt}{0.400pt}}
\multiput(1116.00,746.59)(0.492,0.485){11}{\rule{0.500pt}{0.117pt}}
\multiput(1116.00,745.17)(5.962,7.000){2}{\rule{0.250pt}{0.400pt}}
\multiput(1123.00,753.59)(0.494,0.488){13}{\rule{0.500pt}{0.117pt}}
\multiput(1123.00,752.17)(6.962,8.000){2}{\rule{0.250pt}{0.400pt}}
\multiput(1131.00,761.59)(0.492,0.485){11}{\rule{0.500pt}{0.117pt}}
\multiput(1131.00,760.17)(5.962,7.000){2}{\rule{0.250pt}{0.400pt}}
\multiput(1138.00,768.59)(0.494,0.488){13}{\rule{0.500pt}{0.117pt}}
\multiput(1138.00,767.17)(6.962,8.000){2}{\rule{0.250pt}{0.400pt}}
\multiput(1146.00,776.59)(0.492,0.485){11}{\rule{0.500pt}{0.117pt}}
\multiput(1146.00,775.17)(5.962,7.000){2}{\rule{0.250pt}{0.400pt}}
\multiput(1153.00,783.59)(0.494,0.488){13}{\rule{0.500pt}{0.117pt}}
\multiput(1153.00,782.17)(6.962,8.000){2}{\rule{0.250pt}{0.400pt}}
\multiput(1161.00,791.59)(0.492,0.485){11}{\rule{0.500pt}{0.117pt}}
\multiput(1161.00,790.17)(5.962,7.000){2}{\rule{0.250pt}{0.400pt}}
\multiput(1168.00,798.59)(0.494,0.488){13}{\rule{0.500pt}{0.117pt}}
\multiput(1168.00,797.17)(6.962,8.000){2}{\rule{0.250pt}{0.400pt}}
\multiput(1176.00,806.59)(0.494,0.488){13}{\rule{0.500pt}{0.117pt}}
\multiput(1176.00,805.17)(6.962,8.000){2}{\rule{0.250pt}{0.400pt}}
\multiput(1184.00,814.59)(0.492,0.485){11}{\rule{0.500pt}{0.117pt}}
\multiput(1184.00,813.17)(5.962,7.000){2}{\rule{0.250pt}{0.400pt}}
\multiput(1191.00,821.59)(0.494,0.488){13}{\rule{0.500pt}{0.117pt}}
\multiput(1191.00,820.17)(6.962,8.000){2}{\rule{0.250pt}{0.400pt}}
\multiput(1199.00,829.59)(0.492,0.485){11}{\rule{0.500pt}{0.117pt}}
\multiput(1199.00,828.17)(5.962,7.000){2}{\rule{0.250pt}{0.400pt}}
\multiput(1206.00,836.59)(0.494,0.488){13}{\rule{0.500pt}{0.117pt}}
\multiput(1206.00,835.17)(6.962,8.000){2}{\rule{0.250pt}{0.400pt}}
\multiput(1214.00,844.59)(0.492,0.485){11}{\rule{0.500pt}{0.117pt}}
\multiput(1214.00,843.17)(5.962,7.000){2}{\rule{0.250pt}{0.400pt}}
\multiput(1221.00,851.59)(0.494,0.488){13}{\rule{0.500pt}{0.117pt}}
\multiput(1221.00,850.17)(6.962,8.000){2}{\rule{0.250pt}{0.400pt}}
\put(558,110){\makebox(0,0){$\times$}}
\put(1229,510){\makebox(0,0){$\times$}}
\put(891,512){\makebox(0,0){$\times$}}
\put(998,647){\makebox(0,0){$\times$}}
\put(764,403){\makebox(0,0){$\times$}}
\put(1060,714){\makebox(0,0){$\times$}}
\put(1133,790){\makebox(0,0){$\times$}}
\put(1229,823){\makebox(0,0){$\times$}}
\put(716,516){\makebox(0,0){$\times$}}
\put(763,624){\makebox(0,0){$\times$}}
\put(781,404){\makebox(0,0){$\times$}}
\put(782,704){\makebox(0,0){$\times$}}
\put(1130,757){\makebox(0,0){$\times$}}
\put(1229,802){\makebox(0,0){$\times$}}
\put(897,494){\makebox(0,0){$\times$}}
\put(977,655){\makebox(0,0){$\times$}}
\put(769,407){\makebox(0,0){$\times$}}
\put(1060,735){\makebox(0,0){$\times$}}
\put(1123,734){\makebox(0,0){$\times$}}
\put(1229,791){\makebox(0,0){$\times$}}
\put(664,298){\makebox(0,0){$\times$}}
\put(664,301){\makebox(0,0){$\times$}}
\put(666,291){\makebox(0,0){$\times$}}
\put(739,406){\makebox(0,0){$\times$}}
\put(669,399){\makebox(0,0){$\times$}}
\put(701,268){\makebox(0,0){$\times$}}
\put(697,265){\makebox(0,0){$\times$}}
\put(667,110){\makebox(0,0){$\times$}}
\put(673,255){\makebox(0,0){$\times$}}
\put(674,261){\makebox(0,0){$\times$}}
\put(748,422){\makebox(0,0){$\times$}}
\put(764,403){\makebox(0,0){$\times$}}
\put(1229,652){\makebox(0,0){$\times$}}
\put(1229,859){\makebox(0,0){$\times$}}
\put(1229,859){\makebox(0,0){$\times$}}
\put(1197,782){\makebox(0,0){$\times$}}
\put(1229,859){\makebox(0,0){$\times$}}
\put(1229,626){\makebox(0,0){$\times$}}
\put(756,389){\makebox(0,0){$\times$}}
\put(1229,859){\makebox(0,0){$\times$}}
\put(1229,859){\makebox(0,0){$\times$}}
\put(919,519){\makebox(0,0){$\times$}}
\put(714,403){\makebox(0,0){$\times$}}
\put(868,461){\makebox(0,0){$\times$}}
\put(680,297){\makebox(0,0){$\times$}}
\put(932,571){\makebox(0,0){$\times$}}
\put(957,611){\makebox(0,0){$\times$}}
\put(1001,494){\makebox(0,0){$\times$}}
\put(723,430){\makebox(0,0){$\times$}}
\put(842,460){\makebox(0,0){$\times$}}
\put(707,313){\makebox(0,0){$\times$}}
\put(776,597){\makebox(0,0){$\times$}}
\put(811,633){\makebox(0,0){$\times$}}
\put(815,604){\makebox(0,0){$\times$}}
\put(546,110){\makebox(0,0){$\times$}}
\put(561,110){\makebox(0,0){$\times$}}
\put(846,533){\makebox(0,0){$\times$}}
\put(859,549){\makebox(0,0){$\times$}}
\put(930,493){\makebox(0,0){$\times$}}
\put(996,504){\makebox(0,0){$\times$}}
\put(1096,550){\makebox(0,0){$\times$}}
\put(907,459){\makebox(0,0){$\times$}}
\put(950,567){\makebox(0,0){$\times$}}
\put(1070,531){\makebox(0,0){$\times$}}
\put(870,480){\makebox(0,0){$\times$}}
\put(893,549){\makebox(0,0){$\times$}}
\put(974,650){\makebox(0,0){$\times$}}
\put(1161,769){\makebox(0,0){$\times$}}
\put(1229,859){\makebox(0,0){$\times$}}
\put(1229,859){\makebox(0,0){$\times$}}
\put(1229,859){\makebox(0,0){$\times$}}
\put(1229,859){\makebox(0,0){$\times$}}
\put(1229,859){\makebox(0,0){$\times$}}
\put(1229,859){\makebox(0,0){$\times$}}
\put(1229,859){\makebox(0,0){$\times$}}
\put(1229,859){\makebox(0,0){$\times$}}
\put(934,520){\makebox(0,0){$\times$}}
\put(1229,859){\makebox(0,0){$\times$}}
\put(1229,859){\makebox(0,0){$\times$}}
\put(1164,859){\makebox(0,0){$\times$}}
\put(1229,859){\makebox(0,0){$\times$}}
\put(846,461){\makebox(0,0){$\times$}}
\put(1229,859){\makebox(0,0){$\times$}}
\put(1229,859){\makebox(0,0){$\times$}}
\put(996,724){\makebox(0,0){$\times$}}
\put(1229,846){\makebox(0,0){$\times$}}
\put(1229,855){\makebox(0,0){$\times$}}
\put(1137,822){\makebox(0,0){$\times$}}
\put(560,110){\makebox(0,0){$\times$}}
\put(815,443){\makebox(0,0){$\times$}}
\put(724,388){\makebox(0,0){$\times$}}
\put(722,446){\makebox(0,0){$\times$}}
\put(753,387){\makebox(0,0){$\times$}}
\put(677,387){\makebox(0,0){$\times$}}
\put(792,406){\makebox(0,0){$\times$}}
\put(986,637){\makebox(0,0){$\times$}}
\put(1040,676){\makebox(0,0){$\times$}}
\put(1092,744){\makebox(0,0){$\times$}}
\put(1144,749){\makebox(0,0){$\times$}}
\put(1229,800){\makebox(0,0){$\times$}}
\put(928,607){\makebox(0,0){$\times$}}
\put(879,551){\makebox(0,0){$\times$}}
\put(1001,674){\makebox(0,0){$\times$}}
\put(1099,650){\makebox(0,0){$\times$}}
\put(945,598){\makebox(0,0){$\times$}}
\put(963,623){\makebox(0,0){$\times$}}
\put(1090,637){\makebox(0,0){$\times$}}
\put(1161,799){\makebox(0,0){$\times$}}
\put(1037,684){\makebox(0,0){$\times$}}
\put(1221,859){\makebox(0,0){$\times$}}
\put(1229,859){\makebox(0,0){$\times$}}
\put(1229,859){\makebox(0,0){$\times$}}
\put(1229,829){\makebox(0,0){$\times$}}
\put(1200,859){\makebox(0,0){$\times$}}
\put(1078,799){\makebox(0,0){$\times$}}
\put(1229,859){\makebox(0,0){$\times$}}
\put(1229,859){\makebox(0,0){$\times$}}
\put(1229,859){\makebox(0,0){$\times$}}
\put(1229,859){\makebox(0,0){$\times$}}
\put(1229,859){\makebox(0,0){$\times$}}
\put(1145,859){\makebox(0,0){$\times$}}
\put(1040,662){\makebox(0,0){$\times$}}
\put(1229,859){\makebox(0,0){$\times$}}
\put(1229,859){\makebox(0,0){$\times$}}
\put(1166,859){\makebox(0,0){$\times$}}
\put(1103,713){\makebox(0,0){$\times$}}
\put(1229,859){\makebox(0,0){$\times$}}
\put(1104,661){\makebox(0,0){$\times$}}
\put(1229,614){\makebox(0,0){$\times$}}
\put(1229,514){\makebox(0,0){$\times$}}
\put(480.0,110.0){\rule[-0.200pt]{0.400pt}{180.434pt}}
\put(480.0,110.0){\rule[-0.200pt]{180.434pt}{0.400pt}}
\put(1229.0,110.0){\rule[-0.200pt]{0.400pt}{180.434pt}}
\put(480.0,859.0){\rule[-0.200pt]{180.434pt}{0.400pt}}
\end{picture}

%% file: figures/scatter_RPFP_mathsat__mathsat.tex
% GNUPLOT: LaTeX picture
\setlength{\unitlength}{0.240900pt}
\ifx\plotpoint\undefined\newsavebox{\plotpoint}\fi
\begin{picture}(1500,900)(0,0)
\sbox{\plotpoint}{\rule[-0.200pt]{0.400pt}{0.400pt}}%
\put(460,274){\makebox(0,0)[r]{1}}
\put(480.0,274.0){\rule[-0.200pt]{4.818pt}{0.400pt}}
\put(460,439){\makebox(0,0)[r]{10}}
\put(480.0,439.0){\rule[-0.200pt]{4.818pt}{0.400pt}}
\put(460,603){\makebox(0,0)[r]{100}}
\put(480.0,603.0){\rule[-0.200pt]{4.818pt}{0.400pt}}
\put(460,768){\makebox(0,0)[r]{1000}}
\put(480.0,768.0){\rule[-0.200pt]{4.818pt}{0.400pt}}
\put(460,859){\makebox(0,0)[r]{t/o}}
\put(480.0,859.0){\rule[-0.200pt]{4.818pt}{0.400pt}}
\put(644,69){\makebox(0,0){1}}
\put(644.0,110.0){\rule[-0.200pt]{0.400pt}{4.818pt}}
\put(809,69){\makebox(0,0){10}}
\put(809.0,110.0){\rule[-0.200pt]{0.400pt}{4.818pt}}
\put(973,69){\makebox(0,0){100}}
\put(973.0,110.0){\rule[-0.200pt]{0.400pt}{4.818pt}}
\put(1138,69){\makebox(0,0){1000}}
\put(1138.0,110.0){\rule[-0.200pt]{0.400pt}{4.818pt}}
\put(1229,69){\makebox(0,0){t/o}}
\put(1229.0,110.0){\rule[-0.200pt]{0.400pt}{4.818pt}}
\put(480.0,110.0){\rule[-0.200pt]{0.400pt}{180.434pt}}
\put(480.0,110.0){\rule[-0.200pt]{180.434pt}{0.400pt}}
\put(1229.0,110.0){\rule[-0.200pt]{0.400pt}{180.434pt}}
\put(480.0,859.0){\rule[-0.200pt]{180.434pt}{0.400pt}}
\put(480,859){\line(1,0){749}}
\put(1229,110){\line(0,1){749}}
\put(394,484){\makebox(0,0){\rotatebox{90}{mathsat}}}
\put(854,29){\makebox(0,0){RPFP(mathsat)}}
\put(480,110){\usebox{\plotpoint}}
\multiput(480.00,110.59)(0.494,0.488){13}{\rule{0.500pt}{0.117pt}}
\multiput(480.00,109.17)(6.962,8.000){2}{\rule{0.250pt}{0.400pt}}
\multiput(488.00,118.59)(0.492,0.485){11}{\rule{0.500pt}{0.117pt}}
\multiput(488.00,117.17)(5.962,7.000){2}{\rule{0.250pt}{0.400pt}}
\multiput(495.00,125.59)(0.494,0.488){13}{\rule{0.500pt}{0.117pt}}
\multiput(495.00,124.17)(6.962,8.000){2}{\rule{0.250pt}{0.400pt}}
\multiput(503.00,133.59)(0.492,0.485){11}{\rule{0.500pt}{0.117pt}}
\multiput(503.00,132.17)(5.962,7.000){2}{\rule{0.250pt}{0.400pt}}
\multiput(510.00,140.59)(0.494,0.488){13}{\rule{0.500pt}{0.117pt}}
\multiput(510.00,139.17)(6.962,8.000){2}{\rule{0.250pt}{0.400pt}}
\multiput(518.00,148.59)(0.492,0.485){11}{\rule{0.500pt}{0.117pt}}
\multiput(518.00,147.17)(5.962,7.000){2}{\rule{0.250pt}{0.400pt}}
\multiput(525.00,155.59)(0.494,0.488){13}{\rule{0.500pt}{0.117pt}}
\multiput(525.00,154.17)(6.962,8.000){2}{\rule{0.250pt}{0.400pt}}
\multiput(533.00,163.59)(0.494,0.488){13}{\rule{0.500pt}{0.117pt}}
\multiput(533.00,162.17)(6.962,8.000){2}{\rule{0.250pt}{0.400pt}}
\multiput(541.00,171.59)(0.492,0.485){11}{\rule{0.500pt}{0.117pt}}
\multiput(541.00,170.17)(5.962,7.000){2}{\rule{0.250pt}{0.400pt}}
\multiput(548.00,178.59)(0.494,0.488){13}{\rule{0.500pt}{0.117pt}}
\multiput(548.00,177.17)(6.962,8.000){2}{\rule{0.250pt}{0.400pt}}
\multiput(556.00,186.59)(0.492,0.485){11}{\rule{0.500pt}{0.117pt}}
\multiput(556.00,185.17)(5.962,7.000){2}{\rule{0.250pt}{0.400pt}}
\multiput(563.00,193.59)(0.494,0.488){13}{\rule{0.500pt}{0.117pt}}
\multiput(563.00,192.17)(6.962,8.000){2}{\rule{0.250pt}{0.400pt}}
\multiput(571.00,201.59)(0.492,0.485){11}{\rule{0.500pt}{0.117pt}}
\multiput(571.00,200.17)(5.962,7.000){2}{\rule{0.250pt}{0.400pt}}
\multiput(578.00,208.59)(0.494,0.488){13}{\rule{0.500pt}{0.117pt}}
\multiput(578.00,207.17)(6.962,8.000){2}{\rule{0.250pt}{0.400pt}}
\multiput(586.00,216.59)(0.492,0.485){11}{\rule{0.500pt}{0.117pt}}
\multiput(586.00,215.17)(5.962,7.000){2}{\rule{0.250pt}{0.400pt}}
\multiput(593.00,223.59)(0.494,0.488){13}{\rule{0.500pt}{0.117pt}}
\multiput(593.00,222.17)(6.962,8.000){2}{\rule{0.250pt}{0.400pt}}
\multiput(601.00,231.59)(0.494,0.488){13}{\rule{0.500pt}{0.117pt}}
\multiput(601.00,230.17)(6.962,8.000){2}{\rule{0.250pt}{0.400pt}}
\multiput(609.00,239.59)(0.492,0.485){11}{\rule{0.500pt}{0.117pt}}
\multiput(609.00,238.17)(5.962,7.000){2}{\rule{0.250pt}{0.400pt}}
\multiput(616.00,246.59)(0.494,0.488){13}{\rule{0.500pt}{0.117pt}}
\multiput(616.00,245.17)(6.962,8.000){2}{\rule{0.250pt}{0.400pt}}
\multiput(624.00,254.59)(0.492,0.485){11}{\rule{0.500pt}{0.117pt}}
\multiput(624.00,253.17)(5.962,7.000){2}{\rule{0.250pt}{0.400pt}}
\multiput(631.00,261.59)(0.494,0.488){13}{\rule{0.500pt}{0.117pt}}
\multiput(631.00,260.17)(6.962,8.000){2}{\rule{0.250pt}{0.400pt}}
\multiput(639.00,269.59)(0.492,0.485){11}{\rule{0.500pt}{0.117pt}}
\multiput(639.00,268.17)(5.962,7.000){2}{\rule{0.250pt}{0.400pt}}
\multiput(646.00,276.59)(0.494,0.488){13}{\rule{0.500pt}{0.117pt}}
\multiput(646.00,275.17)(6.962,8.000){2}{\rule{0.250pt}{0.400pt}}
\multiput(654.00,284.59)(0.494,0.488){13}{\rule{0.500pt}{0.117pt}}
\multiput(654.00,283.17)(6.962,8.000){2}{\rule{0.250pt}{0.400pt}}
\multiput(662.00,292.59)(0.492,0.485){11}{\rule{0.500pt}{0.117pt}}
\multiput(662.00,291.17)(5.962,7.000){2}{\rule{0.250pt}{0.400pt}}
\multiput(669.00,299.59)(0.494,0.488){13}{\rule{0.500pt}{0.117pt}}
\multiput(669.00,298.17)(6.962,8.000){2}{\rule{0.250pt}{0.400pt}}
\multiput(677.00,307.59)(0.492,0.485){11}{\rule{0.500pt}{0.117pt}}
\multiput(677.00,306.17)(5.962,7.000){2}{\rule{0.250pt}{0.400pt}}
\multiput(684.00,314.59)(0.494,0.488){13}{\rule{0.500pt}{0.117pt}}
\multiput(684.00,313.17)(6.962,8.000){2}{\rule{0.250pt}{0.400pt}}
\multiput(692.00,322.59)(0.492,0.485){11}{\rule{0.500pt}{0.117pt}}
\multiput(692.00,321.17)(5.962,7.000){2}{\rule{0.250pt}{0.400pt}}
\multiput(699.00,329.59)(0.494,0.488){13}{\rule{0.500pt}{0.117pt}}
\multiput(699.00,328.17)(6.962,8.000){2}{\rule{0.250pt}{0.400pt}}
\multiput(707.00,337.59)(0.494,0.488){13}{\rule{0.500pt}{0.117pt}}
\multiput(707.00,336.17)(6.962,8.000){2}{\rule{0.250pt}{0.400pt}}
\multiput(715.00,345.59)(0.492,0.485){11}{\rule{0.500pt}{0.117pt}}
\multiput(715.00,344.17)(5.962,7.000){2}{\rule{0.250pt}{0.400pt}}
\multiput(722.00,352.59)(0.494,0.488){13}{\rule{0.500pt}{0.117pt}}
\multiput(722.00,351.17)(6.962,8.000){2}{\rule{0.250pt}{0.400pt}}
\multiput(730.00,360.59)(0.492,0.485){11}{\rule{0.500pt}{0.117pt}}
\multiput(730.00,359.17)(5.962,7.000){2}{\rule{0.250pt}{0.400pt}}
\multiput(737.00,367.59)(0.494,0.488){13}{\rule{0.500pt}{0.117pt}}
\multiput(737.00,366.17)(6.962,8.000){2}{\rule{0.250pt}{0.400pt}}
\multiput(745.00,375.59)(0.492,0.485){11}{\rule{0.500pt}{0.117pt}}
\multiput(745.00,374.17)(5.962,7.000){2}{\rule{0.250pt}{0.400pt}}
\multiput(752.00,382.59)(0.494,0.488){13}{\rule{0.500pt}{0.117pt}}
\multiput(752.00,381.17)(6.962,8.000){2}{\rule{0.250pt}{0.400pt}}
\multiput(760.00,390.59)(0.492,0.485){11}{\rule{0.500pt}{0.117pt}}
\multiput(760.00,389.17)(5.962,7.000){2}{\rule{0.250pt}{0.400pt}}
\multiput(767.00,397.59)(0.494,0.488){13}{\rule{0.500pt}{0.117pt}}
\multiput(767.00,396.17)(6.962,8.000){2}{\rule{0.250pt}{0.400pt}}
\multiput(775.00,405.59)(0.494,0.488){13}{\rule{0.500pt}{0.117pt}}
\multiput(775.00,404.17)(6.962,8.000){2}{\rule{0.250pt}{0.400pt}}
\multiput(783.00,413.59)(0.492,0.485){11}{\rule{0.500pt}{0.117pt}}
\multiput(783.00,412.17)(5.962,7.000){2}{\rule{0.250pt}{0.400pt}}
\multiput(790.00,420.59)(0.494,0.488){13}{\rule{0.500pt}{0.117pt}}
\multiput(790.00,419.17)(6.962,8.000){2}{\rule{0.250pt}{0.400pt}}
\multiput(798.00,428.59)(0.492,0.485){11}{\rule{0.500pt}{0.117pt}}
\multiput(798.00,427.17)(5.962,7.000){2}{\rule{0.250pt}{0.400pt}}
\multiput(805.00,435.59)(0.494,0.488){13}{\rule{0.500pt}{0.117pt}}
\multiput(805.00,434.17)(6.962,8.000){2}{\rule{0.250pt}{0.400pt}}
\multiput(813.00,443.59)(0.492,0.485){11}{\rule{0.500pt}{0.117pt}}
\multiput(813.00,442.17)(5.962,7.000){2}{\rule{0.250pt}{0.400pt}}
\multiput(820.00,450.59)(0.494,0.488){13}{\rule{0.500pt}{0.117pt}}
\multiput(820.00,449.17)(6.962,8.000){2}{\rule{0.250pt}{0.400pt}}
\multiput(828.00,458.59)(0.494,0.488){13}{\rule{0.500pt}{0.117pt}}
\multiput(828.00,457.17)(6.962,8.000){2}{\rule{0.250pt}{0.400pt}}
\multiput(836.00,466.59)(0.492,0.485){11}{\rule{0.500pt}{0.117pt}}
\multiput(836.00,465.17)(5.962,7.000){2}{\rule{0.250pt}{0.400pt}}
\multiput(843.00,473.59)(0.494,0.488){13}{\rule{0.500pt}{0.117pt}}
\multiput(843.00,472.17)(6.962,8.000){2}{\rule{0.250pt}{0.400pt}}
\multiput(851.00,481.59)(0.492,0.485){11}{\rule{0.500pt}{0.117pt}}
\multiput(851.00,480.17)(5.962,7.000){2}{\rule{0.250pt}{0.400pt}}
\multiput(858.00,488.59)(0.494,0.488){13}{\rule{0.500pt}{0.117pt}}
\multiput(858.00,487.17)(6.962,8.000){2}{\rule{0.250pt}{0.400pt}}
\multiput(866.00,496.59)(0.492,0.485){11}{\rule{0.500pt}{0.117pt}}
\multiput(866.00,495.17)(5.962,7.000){2}{\rule{0.250pt}{0.400pt}}
\multiput(873.00,503.59)(0.494,0.488){13}{\rule{0.500pt}{0.117pt}}
\multiput(873.00,502.17)(6.962,8.000){2}{\rule{0.250pt}{0.400pt}}
\multiput(881.00,511.59)(0.494,0.488){13}{\rule{0.500pt}{0.117pt}}
\multiput(881.00,510.17)(6.962,8.000){2}{\rule{0.250pt}{0.400pt}}
\multiput(889.00,519.59)(0.492,0.485){11}{\rule{0.500pt}{0.117pt}}
\multiput(889.00,518.17)(5.962,7.000){2}{\rule{0.250pt}{0.400pt}}
\multiput(896.00,526.59)(0.494,0.488){13}{\rule{0.500pt}{0.117pt}}
\multiput(896.00,525.17)(6.962,8.000){2}{\rule{0.250pt}{0.400pt}}
\multiput(904.00,534.59)(0.492,0.485){11}{\rule{0.500pt}{0.117pt}}
\multiput(904.00,533.17)(5.962,7.000){2}{\rule{0.250pt}{0.400pt}}
\multiput(911.00,541.59)(0.494,0.488){13}{\rule{0.500pt}{0.117pt}}
\multiput(911.00,540.17)(6.962,8.000){2}{\rule{0.250pt}{0.400pt}}
\multiput(919.00,549.59)(0.492,0.485){11}{\rule{0.500pt}{0.117pt}}
\multiput(919.00,548.17)(5.962,7.000){2}{\rule{0.250pt}{0.400pt}}
\multiput(926.00,556.59)(0.494,0.488){13}{\rule{0.500pt}{0.117pt}}
\multiput(926.00,555.17)(6.962,8.000){2}{\rule{0.250pt}{0.400pt}}
\multiput(934.00,564.59)(0.494,0.488){13}{\rule{0.500pt}{0.117pt}}
\multiput(934.00,563.17)(6.962,8.000){2}{\rule{0.250pt}{0.400pt}}
\multiput(942.00,572.59)(0.492,0.485){11}{\rule{0.500pt}{0.117pt}}
\multiput(942.00,571.17)(5.962,7.000){2}{\rule{0.250pt}{0.400pt}}
\multiput(949.00,579.59)(0.494,0.488){13}{\rule{0.500pt}{0.117pt}}
\multiput(949.00,578.17)(6.962,8.000){2}{\rule{0.250pt}{0.400pt}}
\multiput(957.00,587.59)(0.492,0.485){11}{\rule{0.500pt}{0.117pt}}
\multiput(957.00,586.17)(5.962,7.000){2}{\rule{0.250pt}{0.400pt}}
\multiput(964.00,594.59)(0.494,0.488){13}{\rule{0.500pt}{0.117pt}}
\multiput(964.00,593.17)(6.962,8.000){2}{\rule{0.250pt}{0.400pt}}
\multiput(972.00,602.59)(0.492,0.485){11}{\rule{0.500pt}{0.117pt}}
\multiput(972.00,601.17)(5.962,7.000){2}{\rule{0.250pt}{0.400pt}}
\multiput(979.00,609.59)(0.494,0.488){13}{\rule{0.500pt}{0.117pt}}
\multiput(979.00,608.17)(6.962,8.000){2}{\rule{0.250pt}{0.400pt}}
\multiput(987.00,617.59)(0.492,0.485){11}{\rule{0.500pt}{0.117pt}}
\multiput(987.00,616.17)(5.962,7.000){2}{\rule{0.250pt}{0.400pt}}
\multiput(994.00,624.59)(0.494,0.488){13}{\rule{0.500pt}{0.117pt}}
\multiput(994.00,623.17)(6.962,8.000){2}{\rule{0.250pt}{0.400pt}}
\multiput(1002.00,632.59)(0.494,0.488){13}{\rule{0.500pt}{0.117pt}}
\multiput(1002.00,631.17)(6.962,8.000){2}{\rule{0.250pt}{0.400pt}}
\multiput(1010.00,640.59)(0.492,0.485){11}{\rule{0.500pt}{0.117pt}}
\multiput(1010.00,639.17)(5.962,7.000){2}{\rule{0.250pt}{0.400pt}}
\multiput(1017.00,647.59)(0.494,0.488){13}{\rule{0.500pt}{0.117pt}}
\multiput(1017.00,646.17)(6.962,8.000){2}{\rule{0.250pt}{0.400pt}}
\multiput(1025.00,655.59)(0.492,0.485){11}{\rule{0.500pt}{0.117pt}}
\multiput(1025.00,654.17)(5.962,7.000){2}{\rule{0.250pt}{0.400pt}}
\multiput(1032.00,662.59)(0.494,0.488){13}{\rule{0.500pt}{0.117pt}}
\multiput(1032.00,661.17)(6.962,8.000){2}{\rule{0.250pt}{0.400pt}}
\multiput(1040.00,670.59)(0.492,0.485){11}{\rule{0.500pt}{0.117pt}}
\multiput(1040.00,669.17)(5.962,7.000){2}{\rule{0.250pt}{0.400pt}}
\multiput(1047.00,677.59)(0.494,0.488){13}{\rule{0.500pt}{0.117pt}}
\multiput(1047.00,676.17)(6.962,8.000){2}{\rule{0.250pt}{0.400pt}}
\multiput(1055.00,685.59)(0.494,0.488){13}{\rule{0.500pt}{0.117pt}}
\multiput(1055.00,684.17)(6.962,8.000){2}{\rule{0.250pt}{0.400pt}}
\multiput(1063.00,693.59)(0.492,0.485){11}{\rule{0.500pt}{0.117pt}}
\multiput(1063.00,692.17)(5.962,7.000){2}{\rule{0.250pt}{0.400pt}}
\multiput(1070.00,700.59)(0.494,0.488){13}{\rule{0.500pt}{0.117pt}}
\multiput(1070.00,699.17)(6.962,8.000){2}{\rule{0.250pt}{0.400pt}}
\multiput(1078.00,708.59)(0.492,0.485){11}{\rule{0.500pt}{0.117pt}}
\multiput(1078.00,707.17)(5.962,7.000){2}{\rule{0.250pt}{0.400pt}}
\multiput(1085.00,715.59)(0.494,0.488){13}{\rule{0.500pt}{0.117pt}}
\multiput(1085.00,714.17)(6.962,8.000){2}{\rule{0.250pt}{0.400pt}}
\multiput(1093.00,723.59)(0.492,0.485){11}{\rule{0.500pt}{0.117pt}}
\multiput(1093.00,722.17)(5.962,7.000){2}{\rule{0.250pt}{0.400pt}}
\multiput(1100.00,730.59)(0.494,0.488){13}{\rule{0.500pt}{0.117pt}}
\multiput(1100.00,729.17)(6.962,8.000){2}{\rule{0.250pt}{0.400pt}}
\multiput(1108.00,738.59)(0.494,0.488){13}{\rule{0.500pt}{0.117pt}}
\multiput(1108.00,737.17)(6.962,8.000){2}{\rule{0.250pt}{0.400pt}}
\multiput(1116.00,746.59)(0.492,0.485){11}{\rule{0.500pt}{0.117pt}}
\multiput(1116.00,745.17)(5.962,7.000){2}{\rule{0.250pt}{0.400pt}}
\multiput(1123.00,753.59)(0.494,0.488){13}{\rule{0.500pt}{0.117pt}}
\multiput(1123.00,752.17)(6.962,8.000){2}{\rule{0.250pt}{0.400pt}}
\multiput(1131.00,761.59)(0.492,0.485){11}{\rule{0.500pt}{0.117pt}}
\multiput(1131.00,760.17)(5.962,7.000){2}{\rule{0.250pt}{0.400pt}}
\multiput(1138.00,768.59)(0.494,0.488){13}{\rule{0.500pt}{0.117pt}}
\multiput(1138.00,767.17)(6.962,8.000){2}{\rule{0.250pt}{0.400pt}}
\multiput(1146.00,776.59)(0.492,0.485){11}{\rule{0.500pt}{0.117pt}}
\multiput(1146.00,775.17)(5.962,7.000){2}{\rule{0.250pt}{0.400pt}}
\multiput(1153.00,783.59)(0.494,0.488){13}{\rule{0.500pt}{0.117pt}}
\multiput(1153.00,782.17)(6.962,8.000){2}{\rule{0.250pt}{0.400pt}}
\multiput(1161.00,791.59)(0.492,0.485){11}{\rule{0.500pt}{0.117pt}}
\multiput(1161.00,790.17)(5.962,7.000){2}{\rule{0.250pt}{0.400pt}}
\multiput(1168.00,798.59)(0.494,0.488){13}{\rule{0.500pt}{0.117pt}}
\multiput(1168.00,797.17)(6.962,8.000){2}{\rule{0.250pt}{0.400pt}}
\multiput(1176.00,806.59)(0.494,0.488){13}{\rule{0.500pt}{0.117pt}}
\multiput(1176.00,805.17)(6.962,8.000){2}{\rule{0.250pt}{0.400pt}}
\multiput(1184.00,814.59)(0.492,0.485){11}{\rule{0.500pt}{0.117pt}}
\multiput(1184.00,813.17)(5.962,7.000){2}{\rule{0.250pt}{0.400pt}}
\multiput(1191.00,821.59)(0.494,0.488){13}{\rule{0.500pt}{0.117pt}}
\multiput(1191.00,820.17)(6.962,8.000){2}{\rule{0.250pt}{0.400pt}}
\multiput(1199.00,829.59)(0.492,0.485){11}{\rule{0.500pt}{0.117pt}}
\multiput(1199.00,828.17)(5.962,7.000){2}{\rule{0.250pt}{0.400pt}}
\multiput(1206.00,836.59)(0.494,0.488){13}{\rule{0.500pt}{0.117pt}}
\multiput(1206.00,835.17)(6.962,8.000){2}{\rule{0.250pt}{0.400pt}}
\multiput(1214.00,844.59)(0.492,0.485){11}{\rule{0.500pt}{0.117pt}}
\multiput(1214.00,843.17)(5.962,7.000){2}{\rule{0.250pt}{0.400pt}}
\multiput(1221.00,851.59)(0.494,0.488){13}{\rule{0.500pt}{0.117pt}}
\multiput(1221.00,850.17)(6.962,8.000){2}{\rule{0.250pt}{0.400pt}}
\put(582,110){\makebox(0,0){$\times$}}
\put(1229,110){\makebox(0,0){$\times$}}
\put(722,474){\makebox(0,0){$\times$}}
\put(766,563){\makebox(0,0){$\times$}}
\put(654,332){\makebox(0,0){$\times$}}
\put(801,622){\makebox(0,0){$\times$}}
\put(828,666){\makebox(0,0){$\times$}}
\put(862,859){\makebox(0,0){$\times$}}
\put(775,561){\makebox(0,0){$\times$}}
\put(771,588){\makebox(0,0){$\times$}}
\put(648,341){\makebox(0,0){$\times$}}
\put(801,628){\makebox(0,0){$\times$}}
\put(831,663){\makebox(0,0){$\times$}}
\put(849,859){\makebox(0,0){$\times$}}
\put(719,473){\makebox(0,0){$\times$}}
\put(776,582){\makebox(0,0){$\times$}}
\put(660,330){\makebox(0,0){$\times$}}
\put(802,616){\makebox(0,0){$\times$}}
\put(830,680){\makebox(0,0){$\times$}}
\put(848,859){\makebox(0,0){$\times$}}
\put(647,187){\makebox(0,0){$\times$}}
\put(626,179){\makebox(0,0){$\times$}}
\put(634,184){\makebox(0,0){$\times$}}
\put(689,313){\makebox(0,0){$\times$}}
\put(680,276){\makebox(0,0){$\times$}}
\put(671,175){\makebox(0,0){$\times$}}
\put(679,175){\makebox(0,0){$\times$}}
\put(647,176){\makebox(0,0){$\times$}}
\put(648,175){\makebox(0,0){$\times$}}
\put(657,177){\makebox(0,0){$\times$}}
\put(695,337){\makebox(0,0){$\times$}}
\put(676,315){\makebox(0,0){$\times$}}
\put(1229,110){\makebox(0,0){$\times$}}
\put(1213,859){\makebox(0,0){$\times$}}
\put(1229,859){\makebox(0,0){$\times$}}
\put(911,575){\makebox(0,0){$\times$}}
\put(1086,859){\makebox(0,0){$\times$}}
\put(1229,110){\makebox(0,0){$\times$}}
\put(606,859){\makebox(0,0){$\times$}}
\put(1229,859){\makebox(0,0){$\times$}}
\put(1229,859){\makebox(0,0){$\times$}}
\put(1050,675){\makebox(0,0){$\times$}}
\put(722,367){\makebox(0,0){$\times$}}
\put(758,466){\makebox(0,0){$\times$}}
\put(651,198){\makebox(0,0){$\times$}}
\put(786,536){\makebox(0,0){$\times$}}
\put(808,566){\makebox(0,0){$\times$}}
\put(831,594){\makebox(0,0){$\times$}}
\put(713,375){\makebox(0,0){$\times$}}
\put(765,473){\makebox(0,0){$\times$}}
\put(669,233){\makebox(0,0){$\times$}}
\put(783,522){\makebox(0,0){$\times$}}
\put(805,561){\makebox(0,0){$\times$}}
\put(821,599){\makebox(0,0){$\times$}}
\put(589,110){\makebox(0,0){$\times$}}
\put(586,110){\makebox(0,0){$\times$}}
\put(811,595){\makebox(0,0){$\times$}}
\put(764,574){\makebox(0,0){$\times$}}
\put(767,584){\makebox(0,0){$\times$}}
\put(800,601){\makebox(0,0){$\times$}}
\put(825,859){\makebox(0,0){$\times$}}
\put(715,617){\makebox(0,0){$\times$}}
\put(844,584){\makebox(0,0){$\times$}}
\put(777,859){\makebox(0,0){$\times$}}
\put(675,513){\makebox(0,0){$\times$}}
\put(836,525){\makebox(0,0){$\times$}}
\put(772,542){\makebox(0,0){$\times$}}
\put(824,826){\makebox(0,0){$\times$}}
\put(1229,859){\makebox(0,0){$\times$}}
\put(1229,859){\makebox(0,0){$\times$}}
\put(1229,859){\makebox(0,0){$\times$}}
\put(1229,859){\makebox(0,0){$\times$}}
\put(946,859){\makebox(0,0){$\times$}}
\put(1229,859){\makebox(0,0){$\times$}}
\put(1229,859){\makebox(0,0){$\times$}}
\put(1229,859){\makebox(0,0){$\times$}}
\put(894,515){\makebox(0,0){$\times$}}
\put(1229,859){\makebox(0,0){$\times$}}
\put(1229,859){\makebox(0,0){$\times$}}
\put(1229,721){\makebox(0,0){$\times$}}
\put(1229,859){\makebox(0,0){$\times$}}
\put(771,385){\makebox(0,0){$\times$}}
\put(969,625){\makebox(0,0){$\times$}}
\put(1014,631){\makebox(0,0){$\times$}}
\put(808,508){\makebox(0,0){$\times$}}
\put(1229,859){\makebox(0,0){$\times$}}
\put(1229,859){\makebox(0,0){$\times$}}
\put(1229,628){\makebox(0,0){$\times$}}
\put(1229,110){\makebox(0,0){$\times$}}
\put(684,345){\makebox(0,0){$\times$}}
\put(676,348){\makebox(0,0){$\times$}}
\put(687,380){\makebox(0,0){$\times$}}
\put(690,314){\makebox(0,0){$\times$}}
\put(676,301){\makebox(0,0){$\times$}}
\put(722,340){\makebox(0,0){$\times$}}
\put(785,527){\makebox(0,0){$\times$}}
\put(771,682){\makebox(0,0){$\times$}}
\put(852,651){\makebox(0,0){$\times$}}
\put(813,645){\makebox(0,0){$\times$}}
\put(812,782){\makebox(0,0){$\times$}}
\put(727,432){\makebox(0,0){$\times$}}
\put(734,422){\makebox(0,0){$\times$}}
\put(865,596){\makebox(0,0){$\times$}}
\put(782,526){\makebox(0,0){$\times$}}
\put(841,492){\makebox(0,0){$\times$}}
\put(772,505){\makebox(0,0){$\times$}}
\put(831,539){\makebox(0,0){$\times$}}
\put(806,652){\makebox(0,0){$\times$}}
\put(737,553){\makebox(0,0){$\times$}}
\put(1202,785){\makebox(0,0){$\times$}}
\put(1229,859){\makebox(0,0){$\times$}}
\put(1229,859){\makebox(0,0){$\times$}}
\put(837,723){\makebox(0,0){$\times$}}
\put(974,761){\makebox(0,0){$\times$}}
\put(816,639){\makebox(0,0){$\times$}}
\put(1229,859){\makebox(0,0){$\times$}}
\put(1229,859){\makebox(0,0){$\times$}}
\put(1229,859){\makebox(0,0){$\times$}}
\put(1089,859){\makebox(0,0){$\times$}}
\put(1229,859){\makebox(0,0){$\times$}}
\put(777,750){\makebox(0,0){$\times$}}
\put(771,590){\makebox(0,0){$\times$}}
\put(1077,833){\makebox(0,0){$\times$}}
\put(876,706){\makebox(0,0){$\times$}}
\put(816,640){\makebox(0,0){$\times$}}
\put(828,667){\makebox(0,0){$\times$}}
\put(961,756){\makebox(0,0){$\times$}}
\put(868,668){\makebox(0,0){$\times$}}
\put(1229,110){\makebox(0,0){$\times$}}
\put(1229,110){\makebox(0,0){$\times$}}
\put(480.0,110.0){\rule[-0.200pt]{0.400pt}{180.434pt}}
\put(480.0,110.0){\rule[-0.200pt]{180.434pt}{0.400pt}}
\put(1229.0,110.0){\rule[-0.200pt]{0.400pt}{180.434pt}}
\put(480.0,859.0){\rule[-0.200pt]{180.434pt}{0.400pt}}
\end{picture}

%% file: figures/scatter_RPFP_acdcl__acdcl.tex
% GNUPLOT: LaTeX picture
\setlength{\unitlength}{0.240900pt}
\ifx\plotpoint\undefined\newsavebox{\plotpoint}\fi
\begin{picture}(1500,900)(0,0)
\sbox{\plotpoint}{\rule[-0.200pt]{0.400pt}{0.400pt}}%
\put(460,274){\makebox(0,0)[r]{1}}
\put(480.0,274.0){\rule[-0.200pt]{4.818pt}{0.400pt}}
\put(460,439){\makebox(0,0)[r]{10}}
\put(480.0,439.0){\rule[-0.200pt]{4.818pt}{0.400pt}}
\put(460,603){\makebox(0,0)[r]{100}}
\put(480.0,603.0){\rule[-0.200pt]{4.818pt}{0.400pt}}
\put(460,768){\makebox(0,0)[r]{1000}}
\put(480.0,768.0){\rule[-0.200pt]{4.818pt}{0.400pt}}
\put(460,859){\makebox(0,0)[r]{t/o}}
\put(480.0,859.0){\rule[-0.200pt]{4.818pt}{0.400pt}}
\put(644,69){\makebox(0,0){1}}
\put(644.0,110.0){\rule[-0.200pt]{0.400pt}{4.818pt}}
\put(809,69){\makebox(0,0){10}}
\put(809.0,110.0){\rule[-0.200pt]{0.400pt}{4.818pt}}
\put(973,69){\makebox(0,0){100}}
\put(973.0,110.0){\rule[-0.200pt]{0.400pt}{4.818pt}}
\put(1138,69){\makebox(0,0){1000}}
\put(1138.0,110.0){\rule[-0.200pt]{0.400pt}{4.818pt}}
\put(1229,69){\makebox(0,0){t/o}}
\put(1229.0,110.0){\rule[-0.200pt]{0.400pt}{4.818pt}}
\put(480.0,110.0){\rule[-0.200pt]{0.400pt}{180.434pt}}
\put(480.0,110.0){\rule[-0.200pt]{180.434pt}{0.400pt}}
\put(1229.0,110.0){\rule[-0.200pt]{0.400pt}{180.434pt}}
\put(480.0,859.0){\rule[-0.200pt]{180.434pt}{0.400pt}}
\put(480,859){\line(1,0){749}}
\put(1229,110){\line(0,1){749}}
\put(394,484){\makebox(0,0){\rotatebox{90}{acdcl}}}
\put(854,29){\makebox(0,0){RPFP(acdcl)}}
\put(480,110){\usebox{\plotpoint}}
\multiput(480.00,110.59)(0.494,0.488){13}{\rule{0.500pt}{0.117pt}}
\multiput(480.00,109.17)(6.962,8.000){2}{\rule{0.250pt}{0.400pt}}
\multiput(488.00,118.59)(0.492,0.485){11}{\rule{0.500pt}{0.117pt}}
\multiput(488.00,117.17)(5.962,7.000){2}{\rule{0.250pt}{0.400pt}}
\multiput(495.00,125.59)(0.494,0.488){13}{\rule{0.500pt}{0.117pt}}
\multiput(495.00,124.17)(6.962,8.000){2}{\rule{0.250pt}{0.400pt}}
\multiput(503.00,133.59)(0.492,0.485){11}{\rule{0.500pt}{0.117pt}}
\multiput(503.00,132.17)(5.962,7.000){2}{\rule{0.250pt}{0.400pt}}
\multiput(510.00,140.59)(0.494,0.488){13}{\rule{0.500pt}{0.117pt}}
\multiput(510.00,139.17)(6.962,8.000){2}{\rule{0.250pt}{0.400pt}}
\multiput(518.00,148.59)(0.492,0.485){11}{\rule{0.500pt}{0.117pt}}
\multiput(518.00,147.17)(5.962,7.000){2}{\rule{0.250pt}{0.400pt}}
\multiput(525.00,155.59)(0.494,0.488){13}{\rule{0.500pt}{0.117pt}}
\multiput(525.00,154.17)(6.962,8.000){2}{\rule{0.250pt}{0.400pt}}
\multiput(533.00,163.59)(0.494,0.488){13}{\rule{0.500pt}{0.117pt}}
\multiput(533.00,162.17)(6.962,8.000){2}{\rule{0.250pt}{0.400pt}}
\multiput(541.00,171.59)(0.492,0.485){11}{\rule{0.500pt}{0.117pt}}
\multiput(541.00,170.17)(5.962,7.000){2}{\rule{0.250pt}{0.400pt}}
\multiput(548.00,178.59)(0.494,0.488){13}{\rule{0.500pt}{0.117pt}}
\multiput(548.00,177.17)(6.962,8.000){2}{\rule{0.250pt}{0.400pt}}
\multiput(556.00,186.59)(0.492,0.485){11}{\rule{0.500pt}{0.117pt}}
\multiput(556.00,185.17)(5.962,7.000){2}{\rule{0.250pt}{0.400pt}}
\multiput(563.00,193.59)(0.494,0.488){13}{\rule{0.500pt}{0.117pt}}
\multiput(563.00,192.17)(6.962,8.000){2}{\rule{0.250pt}{0.400pt}}
\multiput(571.00,201.59)(0.492,0.485){11}{\rule{0.500pt}{0.117pt}}
\multiput(571.00,200.17)(5.962,7.000){2}{\rule{0.250pt}{0.400pt}}
\multiput(578.00,208.59)(0.494,0.488){13}{\rule{0.500pt}{0.117pt}}
\multiput(578.00,207.17)(6.962,8.000){2}{\rule{0.250pt}{0.400pt}}
\multiput(586.00,216.59)(0.492,0.485){11}{\rule{0.500pt}{0.117pt}}
\multiput(586.00,215.17)(5.962,7.000){2}{\rule{0.250pt}{0.400pt}}
\multiput(593.00,223.59)(0.494,0.488){13}{\rule{0.500pt}{0.117pt}}
\multiput(593.00,222.17)(6.962,8.000){2}{\rule{0.250pt}{0.400pt}}
\multiput(601.00,231.59)(0.494,0.488){13}{\rule{0.500pt}{0.117pt}}
\multiput(601.00,230.17)(6.962,8.000){2}{\rule{0.250pt}{0.400pt}}
\multiput(609.00,239.59)(0.492,0.485){11}{\rule{0.500pt}{0.117pt}}
\multiput(609.00,238.17)(5.962,7.000){2}{\rule{0.250pt}{0.400pt}}
\multiput(616.00,246.59)(0.494,0.488){13}{\rule{0.500pt}{0.117pt}}
\multiput(616.00,245.17)(6.962,8.000){2}{\rule{0.250pt}{0.400pt}}
\multiput(624.00,254.59)(0.492,0.485){11}{\rule{0.500pt}{0.117pt}}
\multiput(624.00,253.17)(5.962,7.000){2}{\rule{0.250pt}{0.400pt}}
\multiput(631.00,261.59)(0.494,0.488){13}{\rule{0.500pt}{0.117pt}}
\multiput(631.00,260.17)(6.962,8.000){2}{\rule{0.250pt}{0.400pt}}
\multiput(639.00,269.59)(0.492,0.485){11}{\rule{0.500pt}{0.117pt}}
\multiput(639.00,268.17)(5.962,7.000){2}{\rule{0.250pt}{0.400pt}}
\multiput(646.00,276.59)(0.494,0.488){13}{\rule{0.500pt}{0.117pt}}
\multiput(646.00,275.17)(6.962,8.000){2}{\rule{0.250pt}{0.400pt}}
\multiput(654.00,284.59)(0.494,0.488){13}{\rule{0.500pt}{0.117pt}}
\multiput(654.00,283.17)(6.962,8.000){2}{\rule{0.250pt}{0.400pt}}
\multiput(662.00,292.59)(0.492,0.485){11}{\rule{0.500pt}{0.117pt}}
\multiput(662.00,291.17)(5.962,7.000){2}{\rule{0.250pt}{0.400pt}}
\multiput(669.00,299.59)(0.494,0.488){13}{\rule{0.500pt}{0.117pt}}
\multiput(669.00,298.17)(6.962,8.000){2}{\rule{0.250pt}{0.400pt}}
\multiput(677.00,307.59)(0.492,0.485){11}{\rule{0.500pt}{0.117pt}}
\multiput(677.00,306.17)(5.962,7.000){2}{\rule{0.250pt}{0.400pt}}
\multiput(684.00,314.59)(0.494,0.488){13}{\rule{0.500pt}{0.117pt}}
\multiput(684.00,313.17)(6.962,8.000){2}{\rule{0.250pt}{0.400pt}}
\multiput(692.00,322.59)(0.492,0.485){11}{\rule{0.500pt}{0.117pt}}
\multiput(692.00,321.17)(5.962,7.000){2}{\rule{0.250pt}{0.400pt}}
\multiput(699.00,329.59)(0.494,0.488){13}{\rule{0.500pt}{0.117pt}}
\multiput(699.00,328.17)(6.962,8.000){2}{\rule{0.250pt}{0.400pt}}
\multiput(707.00,337.59)(0.494,0.488){13}{\rule{0.500pt}{0.117pt}}
\multiput(707.00,336.17)(6.962,8.000){2}{\rule{0.250pt}{0.400pt}}
\multiput(715.00,345.59)(0.492,0.485){11}{\rule{0.500pt}{0.117pt}}
\multiput(715.00,344.17)(5.962,7.000){2}{\rule{0.250pt}{0.400pt}}
\multiput(722.00,352.59)(0.494,0.488){13}{\rule{0.500pt}{0.117pt}}
\multiput(722.00,351.17)(6.962,8.000){2}{\rule{0.250pt}{0.400pt}}
\multiput(730.00,360.59)(0.492,0.485){11}{\rule{0.500pt}{0.117pt}}
\multiput(730.00,359.17)(5.962,7.000){2}{\rule{0.250pt}{0.400pt}}
\multiput(737.00,367.59)(0.494,0.488){13}{\rule{0.500pt}{0.117pt}}
\multiput(737.00,366.17)(6.962,8.000){2}{\rule{0.250pt}{0.400pt}}
\multiput(745.00,375.59)(0.492,0.485){11}{\rule{0.500pt}{0.117pt}}
\multiput(745.00,374.17)(5.962,7.000){2}{\rule{0.250pt}{0.400pt}}
\multiput(752.00,382.59)(0.494,0.488){13}{\rule{0.500pt}{0.117pt}}
\multiput(752.00,381.17)(6.962,8.000){2}{\rule{0.250pt}{0.400pt}}
\multiput(760.00,390.59)(0.492,0.485){11}{\rule{0.500pt}{0.117pt}}
\multiput(760.00,389.17)(5.962,7.000){2}{\rule{0.250pt}{0.400pt}}
\multiput(767.00,397.59)(0.494,0.488){13}{\rule{0.500pt}{0.117pt}}
\multiput(767.00,396.17)(6.962,8.000){2}{\rule{0.250pt}{0.400pt}}
\multiput(775.00,405.59)(0.494,0.488){13}{\rule{0.500pt}{0.117pt}}
\multiput(775.00,404.17)(6.962,8.000){2}{\rule{0.250pt}{0.400pt}}
\multiput(783.00,413.59)(0.492,0.485){11}{\rule{0.500pt}{0.117pt}}
\multiput(783.00,412.17)(5.962,7.000){2}{\rule{0.250pt}{0.400pt}}
\multiput(790.00,420.59)(0.494,0.488){13}{\rule{0.500pt}{0.117pt}}
\multiput(790.00,419.17)(6.962,8.000){2}{\rule{0.250pt}{0.400pt}}
\multiput(798.00,428.59)(0.492,0.485){11}{\rule{0.500pt}{0.117pt}}
\multiput(798.00,427.17)(5.962,7.000){2}{\rule{0.250pt}{0.400pt}}
\multiput(805.00,435.59)(0.494,0.488){13}{\rule{0.500pt}{0.117pt}}
\multiput(805.00,434.17)(6.962,8.000){2}{\rule{0.250pt}{0.400pt}}
\multiput(813.00,443.59)(0.492,0.485){11}{\rule{0.500pt}{0.117pt}}
\multiput(813.00,442.17)(5.962,7.000){2}{\rule{0.250pt}{0.400pt}}
\multiput(820.00,450.59)(0.494,0.488){13}{\rule{0.500pt}{0.117pt}}
\multiput(820.00,449.17)(6.962,8.000){2}{\rule{0.250pt}{0.400pt}}
\multiput(828.00,458.59)(0.494,0.488){13}{\rule{0.500pt}{0.117pt}}
\multiput(828.00,457.17)(6.962,8.000){2}{\rule{0.250pt}{0.400pt}}
\multiput(836.00,466.59)(0.492,0.485){11}{\rule{0.500pt}{0.117pt}}
\multiput(836.00,465.17)(5.962,7.000){2}{\rule{0.250pt}{0.400pt}}
\multiput(843.00,473.59)(0.494,0.488){13}{\rule{0.500pt}{0.117pt}}
\multiput(843.00,472.17)(6.962,8.000){2}{\rule{0.250pt}{0.400pt}}
\multiput(851.00,481.59)(0.492,0.485){11}{\rule{0.500pt}{0.117pt}}
\multiput(851.00,480.17)(5.962,7.000){2}{\rule{0.250pt}{0.400pt}}
\multiput(858.00,488.59)(0.494,0.488){13}{\rule{0.500pt}{0.117pt}}
\multiput(858.00,487.17)(6.962,8.000){2}{\rule{0.250pt}{0.400pt}}
\multiput(866.00,496.59)(0.492,0.485){11}{\rule{0.500pt}{0.117pt}}
\multiput(866.00,495.17)(5.962,7.000){2}{\rule{0.250pt}{0.400pt}}
\multiput(873.00,503.59)(0.494,0.488){13}{\rule{0.500pt}{0.117pt}}
\multiput(873.00,502.17)(6.962,8.000){2}{\rule{0.250pt}{0.400pt}}
\multiput(881.00,511.59)(0.494,0.488){13}{\rule{0.500pt}{0.117pt}}
\multiput(881.00,510.17)(6.962,8.000){2}{\rule{0.250pt}{0.400pt}}
\multiput(889.00,519.59)(0.492,0.485){11}{\rule{0.500pt}{0.117pt}}
\multiput(889.00,518.17)(5.962,7.000){2}{\rule{0.250pt}{0.400pt}}
\multiput(896.00,526.59)(0.494,0.488){13}{\rule{0.500pt}{0.117pt}}
\multiput(896.00,525.17)(6.962,8.000){2}{\rule{0.250pt}{0.400pt}}
\multiput(904.00,534.59)(0.492,0.485){11}{\rule{0.500pt}{0.117pt}}
\multiput(904.00,533.17)(5.962,7.000){2}{\rule{0.250pt}{0.400pt}}
\multiput(911.00,541.59)(0.494,0.488){13}{\rule{0.500pt}{0.117pt}}
\multiput(911.00,540.17)(6.962,8.000){2}{\rule{0.250pt}{0.400pt}}
\multiput(919.00,549.59)(0.492,0.485){11}{\rule{0.500pt}{0.117pt}}
\multiput(919.00,548.17)(5.962,7.000){2}{\rule{0.250pt}{0.400pt}}
\multiput(926.00,556.59)(0.494,0.488){13}{\rule{0.500pt}{0.117pt}}
\multiput(926.00,555.17)(6.962,8.000){2}{\rule{0.250pt}{0.400pt}}
\multiput(934.00,564.59)(0.494,0.488){13}{\rule{0.500pt}{0.117pt}}
\multiput(934.00,563.17)(6.962,8.000){2}{\rule{0.250pt}{0.400pt}}
\multiput(942.00,572.59)(0.492,0.485){11}{\rule{0.500pt}{0.117pt}}
\multiput(942.00,571.17)(5.962,7.000){2}{\rule{0.250pt}{0.400pt}}
\multiput(949.00,579.59)(0.494,0.488){13}{\rule{0.500pt}{0.117pt}}
\multiput(949.00,578.17)(6.962,8.000){2}{\rule{0.250pt}{0.400pt}}
\multiput(957.00,587.59)(0.492,0.485){11}{\rule{0.500pt}{0.117pt}}
\multiput(957.00,586.17)(5.962,7.000){2}{\rule{0.250pt}{0.400pt}}
\multiput(964.00,594.59)(0.494,0.488){13}{\rule{0.500pt}{0.117pt}}
\multiput(964.00,593.17)(6.962,8.000){2}{\rule{0.250pt}{0.400pt}}
\multiput(972.00,602.59)(0.492,0.485){11}{\rule{0.500pt}{0.117pt}}
\multiput(972.00,601.17)(5.962,7.000){2}{\rule{0.250pt}{0.400pt}}
\multiput(979.00,609.59)(0.494,0.488){13}{\rule{0.500pt}{0.117pt}}
\multiput(979.00,608.17)(6.962,8.000){2}{\rule{0.250pt}{0.400pt}}
\multiput(987.00,617.59)(0.492,0.485){11}{\rule{0.500pt}{0.117pt}}
\multiput(987.00,616.17)(5.962,7.000){2}{\rule{0.250pt}{0.400pt}}
\multiput(994.00,624.59)(0.494,0.488){13}{\rule{0.500pt}{0.117pt}}
\multiput(994.00,623.17)(6.962,8.000){2}{\rule{0.250pt}{0.400pt}}
\multiput(1002.00,632.59)(0.494,0.488){13}{\rule{0.500pt}{0.117pt}}
\multiput(1002.00,631.17)(6.962,8.000){2}{\rule{0.250pt}{0.400pt}}
\multiput(1010.00,640.59)(0.492,0.485){11}{\rule{0.500pt}{0.117pt}}
\multiput(1010.00,639.17)(5.962,7.000){2}{\rule{0.250pt}{0.400pt}}
\multiput(1017.00,647.59)(0.494,0.488){13}{\rule{0.500pt}{0.117pt}}
\multiput(1017.00,646.17)(6.962,8.000){2}{\rule{0.250pt}{0.400pt}}
\multiput(1025.00,655.59)(0.492,0.485){11}{\rule{0.500pt}{0.117pt}}
\multiput(1025.00,654.17)(5.962,7.000){2}{\rule{0.250pt}{0.400pt}}
\multiput(1032.00,662.59)(0.494,0.488){13}{\rule{0.500pt}{0.117pt}}
\multiput(1032.00,661.17)(6.962,8.000){2}{\rule{0.250pt}{0.400pt}}
\multiput(1040.00,670.59)(0.492,0.485){11}{\rule{0.500pt}{0.117pt}}
\multiput(1040.00,669.17)(5.962,7.000){2}{\rule{0.250pt}{0.400pt}}
\multiput(1047.00,677.59)(0.494,0.488){13}{\rule{0.500pt}{0.117pt}}
\multiput(1047.00,676.17)(6.962,8.000){2}{\rule{0.250pt}{0.400pt}}
\multiput(1055.00,685.59)(0.494,0.488){13}{\rule{0.500pt}{0.117pt}}
\multiput(1055.00,684.17)(6.962,8.000){2}{\rule{0.250pt}{0.400pt}}
\multiput(1063.00,693.59)(0.492,0.485){11}{\rule{0.500pt}{0.117pt}}
\multiput(1063.00,692.17)(5.962,7.000){2}{\rule{0.250pt}{0.400pt}}
\multiput(1070.00,700.59)(0.494,0.488){13}{\rule{0.500pt}{0.117pt}}
\multiput(1070.00,699.17)(6.962,8.000){2}{\rule{0.250pt}{0.400pt}}
\multiput(1078.00,708.59)(0.492,0.485){11}{\rule{0.500pt}{0.117pt}}
\multiput(1078.00,707.17)(5.962,7.000){2}{\rule{0.250pt}{0.400pt}}
\multiput(1085.00,715.59)(0.494,0.488){13}{\rule{0.500pt}{0.117pt}}
\multiput(1085.00,714.17)(6.962,8.000){2}{\rule{0.250pt}{0.400pt}}
\multiput(1093.00,723.59)(0.492,0.485){11}{\rule{0.500pt}{0.117pt}}
\multiput(1093.00,722.17)(5.962,7.000){2}{\rule{0.250pt}{0.400pt}}
\multiput(1100.00,730.59)(0.494,0.488){13}{\rule{0.500pt}{0.117pt}}
\multiput(1100.00,729.17)(6.962,8.000){2}{\rule{0.250pt}{0.400pt}}
\multiput(1108.00,738.59)(0.494,0.488){13}{\rule{0.500pt}{0.117pt}}
\multiput(1108.00,737.17)(6.962,8.000){2}{\rule{0.250pt}{0.400pt}}
\multiput(1116.00,746.59)(0.492,0.485){11}{\rule{0.500pt}{0.117pt}}
\multiput(1116.00,745.17)(5.962,7.000){2}{\rule{0.250pt}{0.400pt}}
\multiput(1123.00,753.59)(0.494,0.488){13}{\rule{0.500pt}{0.117pt}}
\multiput(1123.00,752.17)(6.962,8.000){2}{\rule{0.250pt}{0.400pt}}
\multiput(1131.00,761.59)(0.492,0.485){11}{\rule{0.500pt}{0.117pt}}
\multiput(1131.00,760.17)(5.962,7.000){2}{\rule{0.250pt}{0.400pt}}
\multiput(1138.00,768.59)(0.494,0.488){13}{\rule{0.500pt}{0.117pt}}
\multiput(1138.00,767.17)(6.962,8.000){2}{\rule{0.250pt}{0.400pt}}
\multiput(1146.00,776.59)(0.492,0.485){11}{\rule{0.500pt}{0.117pt}}
\multiput(1146.00,775.17)(5.962,7.000){2}{\rule{0.250pt}{0.400pt}}
\multiput(1153.00,783.59)(0.494,0.488){13}{\rule{0.500pt}{0.117pt}}
\multiput(1153.00,782.17)(6.962,8.000){2}{\rule{0.250pt}{0.400pt}}
\multiput(1161.00,791.59)(0.492,0.485){11}{\rule{0.500pt}{0.117pt}}
\multiput(1161.00,790.17)(5.962,7.000){2}{\rule{0.250pt}{0.400pt}}
\multiput(1168.00,798.59)(0.494,0.488){13}{\rule{0.500pt}{0.117pt}}
\multiput(1168.00,797.17)(6.962,8.000){2}{\rule{0.250pt}{0.400pt}}
\multiput(1176.00,806.59)(0.494,0.488){13}{\rule{0.500pt}{0.117pt}}
\multiput(1176.00,805.17)(6.962,8.000){2}{\rule{0.250pt}{0.400pt}}
\multiput(1184.00,814.59)(0.492,0.485){11}{\rule{0.500pt}{0.117pt}}
\multiput(1184.00,813.17)(5.962,7.000){2}{\rule{0.250pt}{0.400pt}}
\multiput(1191.00,821.59)(0.494,0.488){13}{\rule{0.500pt}{0.117pt}}
\multiput(1191.00,820.17)(6.962,8.000){2}{\rule{0.250pt}{0.400pt}}
\multiput(1199.00,829.59)(0.492,0.485){11}{\rule{0.500pt}{0.117pt}}
\multiput(1199.00,828.17)(5.962,7.000){2}{\rule{0.250pt}{0.400pt}}
\multiput(1206.00,836.59)(0.494,0.488){13}{\rule{0.500pt}{0.117pt}}
\multiput(1206.00,835.17)(6.962,8.000){2}{\rule{0.250pt}{0.400pt}}
\multiput(1214.00,844.59)(0.492,0.485){11}{\rule{0.500pt}{0.117pt}}
\multiput(1214.00,843.17)(5.962,7.000){2}{\rule{0.250pt}{0.400pt}}
\multiput(1221.00,851.59)(0.494,0.488){13}{\rule{0.500pt}{0.117pt}}
\multiput(1221.00,850.17)(6.962,8.000){2}{\rule{0.250pt}{0.400pt}}
\put(573,859){\makebox(0,0){$\times$}}
\put(1229,110){\makebox(0,0){$\times$}}
\put(743,371){\makebox(0,0){$\times$}}
\put(784,410){\makebox(0,0){$\times$}}
\put(649,251){\makebox(0,0){$\times$}}
\put(784,859){\makebox(0,0){$\times$}}
\put(1229,572){\makebox(0,0){$\times$}}
\put(1210,502){\makebox(0,0){$\times$}}
\put(1229,859){\makebox(0,0){$\times$}}
\put(1229,859){\makebox(0,0){$\times$}}
\put(666,329){\makebox(0,0){$\times$}}
\put(1229,859){\makebox(0,0){$\times$}}
\put(1229,859){\makebox(0,0){$\times$}}
\put(1229,859){\makebox(0,0){$\times$}}
\put(1075,279){\makebox(0,0){$\times$}}
\put(763,279){\makebox(0,0){$\times$}}
\put(651,128){\makebox(0,0){$\times$}}
\put(838,504){\makebox(0,0){$\times$}}
\put(1229,437){\makebox(0,0){$\times$}}
\put(1229,859){\makebox(0,0){$\times$}}
\put(643,110){\makebox(0,0){$\times$}}
\put(621,110){\makebox(0,0){$\times$}}
\put(630,110){\makebox(0,0){$\times$}}
\put(865,498){\makebox(0,0){$\times$}}
\put(669,110){\makebox(0,0){$\times$}}
\put(674,110){\makebox(0,0){$\times$}}
\put(668,110){\makebox(0,0){$\times$}}
\put(645,110){\makebox(0,0){$\times$}}
\put(649,110){\makebox(0,0){$\times$}}
\put(658,110){\makebox(0,0){$\times$}}
\put(693,110){\makebox(0,0){$\times$}}
\put(675,110){\makebox(0,0){$\times$}}
\put(1229,110){\makebox(0,0){$\times$}}
\put(1229,686){\makebox(0,0){$\times$}}
\put(1229,859){\makebox(0,0){$\times$}}
\put(1229,482){\makebox(0,0){$\times$}}
\put(1229,851){\makebox(0,0){$\times$}}
\put(1229,110){\makebox(0,0){$\times$}}
\put(1229,859){\makebox(0,0){$\times$}}
\put(1229,859){\makebox(0,0){$\times$}}
\put(1229,859){\makebox(0,0){$\times$}}
\put(856,125){\makebox(0,0){$\times$}}
\put(713,164){\makebox(0,0){$\times$}}
\put(769,113){\makebox(0,0){$\times$}}
\put(651,110){\makebox(0,0){$\times$}}
\put(1229,115){\makebox(0,0){$\times$}}
\put(1229,178){\makebox(0,0){$\times$}}
\put(1229,253){\makebox(0,0){$\times$}}
\put(713,284){\makebox(0,0){$\times$}}
\put(753,365){\makebox(0,0){$\times$}}
\put(653,225){\makebox(0,0){$\times$}}
\put(778,394){\makebox(0,0){$\times$}}
\put(798,484){\makebox(0,0){$\times$}}
\put(821,498){\makebox(0,0){$\times$}}
\put(599,859){\makebox(0,0){$\times$}}
\put(594,859){\makebox(0,0){$\times$}}
\put(672,110){\makebox(0,0){$\times$}}
\put(672,110){\makebox(0,0){$\times$}}
\put(676,110){\makebox(0,0){$\times$}}
\put(706,110){\makebox(0,0){$\times$}}
\put(732,110){\makebox(0,0){$\times$}}
\put(666,110){\makebox(0,0){$\times$}}
\put(702,110){\makebox(0,0){$\times$}}
\put(826,155){\makebox(0,0){$\times$}}
\put(666,110){\makebox(0,0){$\times$}}
\put(827,110){\makebox(0,0){$\times$}}
\put(848,110){\makebox(0,0){$\times$}}
\put(745,110){\makebox(0,0){$\times$}}
\put(1229,859){\makebox(0,0){$\times$}}
\put(1229,859){\makebox(0,0){$\times$}}
\put(1229,859){\makebox(0,0){$\times$}}
\put(1229,859){\makebox(0,0){$\times$}}
\put(1229,859){\makebox(0,0){$\times$}}
\put(1229,859){\makebox(0,0){$\times$}}
\put(1229,859){\makebox(0,0){$\times$}}
\put(1229,859){\makebox(0,0){$\times$}}
\put(1229,859){\makebox(0,0){$\times$}}
\put(1229,859){\makebox(0,0){$\times$}}
\put(1229,859){\makebox(0,0){$\times$}}
\put(900,859){\makebox(0,0){$\times$}}
\put(1229,859){\makebox(0,0){$\times$}}
\put(754,110){\makebox(0,0){$\times$}}
\put(1229,859){\makebox(0,0){$\times$}}
\put(1229,859){\makebox(0,0){$\times$}}
\put(1198,430){\makebox(0,0){$\times$}}
\put(1229,859){\makebox(0,0){$\times$}}
\put(1229,859){\makebox(0,0){$\times$}}
\put(1229,859){\makebox(0,0){$\times$}}
\put(1229,110){\makebox(0,0){$\times$}}
\put(789,110){\makebox(0,0){$\times$}}
\put(676,110){\makebox(0,0){$\times$}}
\put(716,150){\makebox(0,0){$\times$}}
\put(783,110){\makebox(0,0){$\times$}}
\put(674,110){\makebox(0,0){$\times$}}
\put(677,110){\makebox(0,0){$\times$}}
\put(715,110){\makebox(0,0){$\times$}}
\put(888,362){\makebox(0,0){$\times$}}
\put(930,207){\makebox(0,0){$\times$}}
\put(831,641){\makebox(0,0){$\times$}}
\put(957,348){\makebox(0,0){$\times$}}
\put(835,110){\makebox(0,0){$\times$}}
\put(721,292){\makebox(0,0){$\times$}}
\put(993,786){\makebox(0,0){$\times$}}
\put(893,118){\makebox(0,0){$\times$}}
\put(870,110){\makebox(0,0){$\times$}}
\put(734,177){\makebox(0,0){$\times$}}
\put(855,390){\makebox(0,0){$\times$}}
\put(1134,351){\makebox(0,0){$\times$}}
\put(720,242){\makebox(0,0){$\times$}}
\put(1229,758){\makebox(0,0){$\times$}}
\put(1229,859){\makebox(0,0){$\times$}}
\put(1229,859){\makebox(0,0){$\times$}}
\put(874,859){\makebox(0,0){$\times$}}
\put(1229,859){\makebox(0,0){$\times$}}
\put(800,649){\makebox(0,0){$\times$}}
\put(1229,859){\makebox(0,0){$\times$}}
\put(1229,859){\makebox(0,0){$\times$}}
\put(1229,859){\makebox(0,0){$\times$}}
\put(925,859){\makebox(0,0){$\times$}}
\put(1229,859){\makebox(0,0){$\times$}}
\put(804,636){\makebox(0,0){$\times$}}
\put(1229,859){\makebox(0,0){$\times$}}
\put(1229,859){\makebox(0,0){$\times$}}
\put(1138,859){\makebox(0,0){$\times$}}
\put(832,457){\makebox(0,0){$\times$}}
\put(1018,439){\makebox(0,0){$\times$}}
\put(1229,859){\makebox(0,0){$\times$}}
\put(773,166){\makebox(0,0){$\times$}}
\put(1229,110){\makebox(0,0){$\times$}}
\put(1229,110){\makebox(0,0){$\times$}}
\put(480.0,110.0){\rule[-0.200pt]{0.400pt}{180.434pt}}
\put(480.0,110.0){\rule[-0.200pt]{180.434pt}{0.400pt}}
\put(1229.0,110.0){\rule[-0.200pt]{0.400pt}{180.434pt}}
\put(480.0,859.0){\rule[-0.200pt]{180.434pt}{0.400pt}}
\end{picture}

%% file: figures/scatter_acdcl_z3.tex
% GNUPLOT: LaTeX picture
\setlength{\unitlength}{0.240900pt}
\ifx\plotpoint\undefined\newsavebox{\plotpoint}\fi
\begin{picture}(1500,900)(0,0)
\sbox{\plotpoint}{\rule[-0.200pt]{0.400pt}{0.400pt}}%
\put(460,274){\makebox(0,0)[r]{1}}
\put(480.0,274.0){\rule[-0.200pt]{4.818pt}{0.400pt}}
\put(460,439){\makebox(0,0)[r]{10}}
\put(480.0,439.0){\rule[-0.200pt]{4.818pt}{0.400pt}}
\put(460,603){\makebox(0,0)[r]{100}}
\put(480.0,603.0){\rule[-0.200pt]{4.818pt}{0.400pt}}
\put(460,768){\makebox(0,0)[r]{1000}}
\put(480.0,768.0){\rule[-0.200pt]{4.818pt}{0.400pt}}
\put(460,859){\makebox(0,0)[r]{t/o}}
\put(480.0,859.0){\rule[-0.200pt]{4.818pt}{0.400pt}}
\put(644,69){\makebox(0,0){1}}
\put(644.0,110.0){\rule[-0.200pt]{0.400pt}{4.818pt}}
\put(809,69){\makebox(0,0){10}}
\put(809.0,110.0){\rule[-0.200pt]{0.400pt}{4.818pt}}
\put(973,69){\makebox(0,0){100}}
\put(973.0,110.0){\rule[-0.200pt]{0.400pt}{4.818pt}}
\put(1138,69){\makebox(0,0){1000}}
\put(1138.0,110.0){\rule[-0.200pt]{0.400pt}{4.818pt}}
\put(1229,69){\makebox(0,0){t/o}}
\put(1229.0,110.0){\rule[-0.200pt]{0.400pt}{4.818pt}}
\put(480.0,110.0){\rule[-0.200pt]{0.400pt}{180.434pt}}
\put(480.0,110.0){\rule[-0.200pt]{180.434pt}{0.400pt}}
\put(1229.0,110.0){\rule[-0.200pt]{0.400pt}{180.434pt}}
\put(480.0,859.0){\rule[-0.200pt]{180.434pt}{0.400pt}}
\put(480,859){\line(1,0){749}}
\put(1229,110){\line(0,1){749}}
\put(394,484){\makebox(0,0){\rotatebox{90}{z3}}}
\put(854,29){\makebox(0,0){acdcl}}
\put(480,110){\usebox{\plotpoint}}
\multiput(480.00,110.59)(0.494,0.488){13}{\rule{0.500pt}{0.117pt}}
\multiput(480.00,109.17)(6.962,8.000){2}{\rule{0.250pt}{0.400pt}}
\multiput(488.00,118.59)(0.492,0.485){11}{\rule{0.500pt}{0.117pt}}
\multiput(488.00,117.17)(5.962,7.000){2}{\rule{0.250pt}{0.400pt}}
\multiput(495.00,125.59)(0.494,0.488){13}{\rule{0.500pt}{0.117pt}}
\multiput(495.00,124.17)(6.962,8.000){2}{\rule{0.250pt}{0.400pt}}
\multiput(503.00,133.59)(0.492,0.485){11}{\rule{0.500pt}{0.117pt}}
\multiput(503.00,132.17)(5.962,7.000){2}{\rule{0.250pt}{0.400pt}}
\multiput(510.00,140.59)(0.494,0.488){13}{\rule{0.500pt}{0.117pt}}
\multiput(510.00,139.17)(6.962,8.000){2}{\rule{0.250pt}{0.400pt}}
\multiput(518.00,148.59)(0.492,0.485){11}{\rule{0.500pt}{0.117pt}}
\multiput(518.00,147.17)(5.962,7.000){2}{\rule{0.250pt}{0.400pt}}
\multiput(525.00,155.59)(0.494,0.488){13}{\rule{0.500pt}{0.117pt}}
\multiput(525.00,154.17)(6.962,8.000){2}{\rule{0.250pt}{0.400pt}}
\multiput(533.00,163.59)(0.494,0.488){13}{\rule{0.500pt}{0.117pt}}
\multiput(533.00,162.17)(6.962,8.000){2}{\rule{0.250pt}{0.400pt}}
\multiput(541.00,171.59)(0.492,0.485){11}{\rule{0.500pt}{0.117pt}}
\multiput(541.00,170.17)(5.962,7.000){2}{\rule{0.250pt}{0.400pt}}
\multiput(548.00,178.59)(0.494,0.488){13}{\rule{0.500pt}{0.117pt}}
\multiput(548.00,177.17)(6.962,8.000){2}{\rule{0.250pt}{0.400pt}}
\multiput(556.00,186.59)(0.492,0.485){11}{\rule{0.500pt}{0.117pt}}
\multiput(556.00,185.17)(5.962,7.000){2}{\rule{0.250pt}{0.400pt}}
\multiput(563.00,193.59)(0.494,0.488){13}{\rule{0.500pt}{0.117pt}}
\multiput(563.00,192.17)(6.962,8.000){2}{\rule{0.250pt}{0.400pt}}
\multiput(571.00,201.59)(0.492,0.485){11}{\rule{0.500pt}{0.117pt}}
\multiput(571.00,200.17)(5.962,7.000){2}{\rule{0.250pt}{0.400pt}}
\multiput(578.00,208.59)(0.494,0.488){13}{\rule{0.500pt}{0.117pt}}
\multiput(578.00,207.17)(6.962,8.000){2}{\rule{0.250pt}{0.400pt}}
\multiput(586.00,216.59)(0.492,0.485){11}{\rule{0.500pt}{0.117pt}}
\multiput(586.00,215.17)(5.962,7.000){2}{\rule{0.250pt}{0.400pt}}
\multiput(593.00,223.59)(0.494,0.488){13}{\rule{0.500pt}{0.117pt}}
\multiput(593.00,222.17)(6.962,8.000){2}{\rule{0.250pt}{0.400pt}}
\multiput(601.00,231.59)(0.494,0.488){13}{\rule{0.500pt}{0.117pt}}
\multiput(601.00,230.17)(6.962,8.000){2}{\rule{0.250pt}{0.400pt}}
\multiput(609.00,239.59)(0.492,0.485){11}{\rule{0.500pt}{0.117pt}}
\multiput(609.00,238.17)(5.962,7.000){2}{\rule{0.250pt}{0.400pt}}
\multiput(616.00,246.59)(0.494,0.488){13}{\rule{0.500pt}{0.117pt}}
\multiput(616.00,245.17)(6.962,8.000){2}{\rule{0.250pt}{0.400pt}}
\multiput(624.00,254.59)(0.492,0.485){11}{\rule{0.500pt}{0.117pt}}
\multiput(624.00,253.17)(5.962,7.000){2}{\rule{0.250pt}{0.400pt}}
\multiput(631.00,261.59)(0.494,0.488){13}{\rule{0.500pt}{0.117pt}}
\multiput(631.00,260.17)(6.962,8.000){2}{\rule{0.250pt}{0.400pt}}
\multiput(639.00,269.59)(0.492,0.485){11}{\rule{0.500pt}{0.117pt}}
\multiput(639.00,268.17)(5.962,7.000){2}{\rule{0.250pt}{0.400pt}}
\multiput(646.00,276.59)(0.494,0.488){13}{\rule{0.500pt}{0.117pt}}
\multiput(646.00,275.17)(6.962,8.000){2}{\rule{0.250pt}{0.400pt}}
\multiput(654.00,284.59)(0.494,0.488){13}{\rule{0.500pt}{0.117pt}}
\multiput(654.00,283.17)(6.962,8.000){2}{\rule{0.250pt}{0.400pt}}
\multiput(662.00,292.59)(0.492,0.485){11}{\rule{0.500pt}{0.117pt}}
\multiput(662.00,291.17)(5.962,7.000){2}{\rule{0.250pt}{0.400pt}}
\multiput(669.00,299.59)(0.494,0.488){13}{\rule{0.500pt}{0.117pt}}
\multiput(669.00,298.17)(6.962,8.000){2}{\rule{0.250pt}{0.400pt}}
\multiput(677.00,307.59)(0.492,0.485){11}{\rule{0.500pt}{0.117pt}}
\multiput(677.00,306.17)(5.962,7.000){2}{\rule{0.250pt}{0.400pt}}
\multiput(684.00,314.59)(0.494,0.488){13}{\rule{0.500pt}{0.117pt}}
\multiput(684.00,313.17)(6.962,8.000){2}{\rule{0.250pt}{0.400pt}}
\multiput(692.00,322.59)(0.492,0.485){11}{\rule{0.500pt}{0.117pt}}
\multiput(692.00,321.17)(5.962,7.000){2}{\rule{0.250pt}{0.400pt}}
\multiput(699.00,329.59)(0.494,0.488){13}{\rule{0.500pt}{0.117pt}}
\multiput(699.00,328.17)(6.962,8.000){2}{\rule{0.250pt}{0.400pt}}
\multiput(707.00,337.59)(0.494,0.488){13}{\rule{0.500pt}{0.117pt}}
\multiput(707.00,336.17)(6.962,8.000){2}{\rule{0.250pt}{0.400pt}}
\multiput(715.00,345.59)(0.492,0.485){11}{\rule{0.500pt}{0.117pt}}
\multiput(715.00,344.17)(5.962,7.000){2}{\rule{0.250pt}{0.400pt}}
\multiput(722.00,352.59)(0.494,0.488){13}{\rule{0.500pt}{0.117pt}}
\multiput(722.00,351.17)(6.962,8.000){2}{\rule{0.250pt}{0.400pt}}
\multiput(730.00,360.59)(0.492,0.485){11}{\rule{0.500pt}{0.117pt}}
\multiput(730.00,359.17)(5.962,7.000){2}{\rule{0.250pt}{0.400pt}}
\multiput(737.00,367.59)(0.494,0.488){13}{\rule{0.500pt}{0.117pt}}
\multiput(737.00,366.17)(6.962,8.000){2}{\rule{0.250pt}{0.400pt}}
\multiput(745.00,375.59)(0.492,0.485){11}{\rule{0.500pt}{0.117pt}}
\multiput(745.00,374.17)(5.962,7.000){2}{\rule{0.250pt}{0.400pt}}
\multiput(752.00,382.59)(0.494,0.488){13}{\rule{0.500pt}{0.117pt}}
\multiput(752.00,381.17)(6.962,8.000){2}{\rule{0.250pt}{0.400pt}}
\multiput(760.00,390.59)(0.492,0.485){11}{\rule{0.500pt}{0.117pt}}
\multiput(760.00,389.17)(5.962,7.000){2}{\rule{0.250pt}{0.400pt}}
\multiput(767.00,397.59)(0.494,0.488){13}{\rule{0.500pt}{0.117pt}}
\multiput(767.00,396.17)(6.962,8.000){2}{\rule{0.250pt}{0.400pt}}
\multiput(775.00,405.59)(0.494,0.488){13}{\rule{0.500pt}{0.117pt}}
\multiput(775.00,404.17)(6.962,8.000){2}{\rule{0.250pt}{0.400pt}}
\multiput(783.00,413.59)(0.492,0.485){11}{\rule{0.500pt}{0.117pt}}
\multiput(783.00,412.17)(5.962,7.000){2}{\rule{0.250pt}{0.400pt}}
\multiput(790.00,420.59)(0.494,0.488){13}{\rule{0.500pt}{0.117pt}}
\multiput(790.00,419.17)(6.962,8.000){2}{\rule{0.250pt}{0.400pt}}
\multiput(798.00,428.59)(0.492,0.485){11}{\rule{0.500pt}{0.117pt}}
\multiput(798.00,427.17)(5.962,7.000){2}{\rule{0.250pt}{0.400pt}}
\multiput(805.00,435.59)(0.494,0.488){13}{\rule{0.500pt}{0.117pt}}
\multiput(805.00,434.17)(6.962,8.000){2}{\rule{0.250pt}{0.400pt}}
\multiput(813.00,443.59)(0.492,0.485){11}{\rule{0.500pt}{0.117pt}}
\multiput(813.00,442.17)(5.962,7.000){2}{\rule{0.250pt}{0.400pt}}
\multiput(820.00,450.59)(0.494,0.488){13}{\rule{0.500pt}{0.117pt}}
\multiput(820.00,449.17)(6.962,8.000){2}{\rule{0.250pt}{0.400pt}}
\multiput(828.00,458.59)(0.494,0.488){13}{\rule{0.500pt}{0.117pt}}
\multiput(828.00,457.17)(6.962,8.000){2}{\rule{0.250pt}{0.400pt}}
\multiput(836.00,466.59)(0.492,0.485){11}{\rule{0.500pt}{0.117pt}}
\multiput(836.00,465.17)(5.962,7.000){2}{\rule{0.250pt}{0.400pt}}
\multiput(843.00,473.59)(0.494,0.488){13}{\rule{0.500pt}{0.117pt}}
\multiput(843.00,472.17)(6.962,8.000){2}{\rule{0.250pt}{0.400pt}}
\multiput(851.00,481.59)(0.492,0.485){11}{\rule{0.500pt}{0.117pt}}
\multiput(851.00,480.17)(5.962,7.000){2}{\rule{0.250pt}{0.400pt}}
\multiput(858.00,488.59)(0.494,0.488){13}{\rule{0.500pt}{0.117pt}}
\multiput(858.00,487.17)(6.962,8.000){2}{\rule{0.250pt}{0.400pt}}
\multiput(866.00,496.59)(0.492,0.485){11}{\rule{0.500pt}{0.117pt}}
\multiput(866.00,495.17)(5.962,7.000){2}{\rule{0.250pt}{0.400pt}}
\multiput(873.00,503.59)(0.494,0.488){13}{\rule{0.500pt}{0.117pt}}
\multiput(873.00,502.17)(6.962,8.000){2}{\rule{0.250pt}{0.400pt}}
\multiput(881.00,511.59)(0.494,0.488){13}{\rule{0.500pt}{0.117pt}}
\multiput(881.00,510.17)(6.962,8.000){2}{\rule{0.250pt}{0.400pt}}
\multiput(889.00,519.59)(0.492,0.485){11}{\rule{0.500pt}{0.117pt}}
\multiput(889.00,518.17)(5.962,7.000){2}{\rule{0.250pt}{0.400pt}}
\multiput(896.00,526.59)(0.494,0.488){13}{\rule{0.500pt}{0.117pt}}
\multiput(896.00,525.17)(6.962,8.000){2}{\rule{0.250pt}{0.400pt}}
\multiput(904.00,534.59)(0.492,0.485){11}{\rule{0.500pt}{0.117pt}}
\multiput(904.00,533.17)(5.962,7.000){2}{\rule{0.250pt}{0.400pt}}
\multiput(911.00,541.59)(0.494,0.488){13}{\rule{0.500pt}{0.117pt}}
\multiput(911.00,540.17)(6.962,8.000){2}{\rule{0.250pt}{0.400pt}}
\multiput(919.00,549.59)(0.492,0.485){11}{\rule{0.500pt}{0.117pt}}
\multiput(919.00,548.17)(5.962,7.000){2}{\rule{0.250pt}{0.400pt}}
\multiput(926.00,556.59)(0.494,0.488){13}{\rule{0.500pt}{0.117pt}}
\multiput(926.00,555.17)(6.962,8.000){2}{\rule{0.250pt}{0.400pt}}
\multiput(934.00,564.59)(0.494,0.488){13}{\rule{0.500pt}{0.117pt}}
\multiput(934.00,563.17)(6.962,8.000){2}{\rule{0.250pt}{0.400pt}}
\multiput(942.00,572.59)(0.492,0.485){11}{\rule{0.500pt}{0.117pt}}
\multiput(942.00,571.17)(5.962,7.000){2}{\rule{0.250pt}{0.400pt}}
\multiput(949.00,579.59)(0.494,0.488){13}{\rule{0.500pt}{0.117pt}}
\multiput(949.00,578.17)(6.962,8.000){2}{\rule{0.250pt}{0.400pt}}
\multiput(957.00,587.59)(0.492,0.485){11}{\rule{0.500pt}{0.117pt}}
\multiput(957.00,586.17)(5.962,7.000){2}{\rule{0.250pt}{0.400pt}}
\multiput(964.00,594.59)(0.494,0.488){13}{\rule{0.500pt}{0.117pt}}
\multiput(964.00,593.17)(6.962,8.000){2}{\rule{0.250pt}{0.400pt}}
\multiput(972.00,602.59)(0.492,0.485){11}{\rule{0.500pt}{0.117pt}}
\multiput(972.00,601.17)(5.962,7.000){2}{\rule{0.250pt}{0.400pt}}
\multiput(979.00,609.59)(0.494,0.488){13}{\rule{0.500pt}{0.117pt}}
\multiput(979.00,608.17)(6.962,8.000){2}{\rule{0.250pt}{0.400pt}}
\multiput(987.00,617.59)(0.492,0.485){11}{\rule{0.500pt}{0.117pt}}
\multiput(987.00,616.17)(5.962,7.000){2}{\rule{0.250pt}{0.400pt}}
\multiput(994.00,624.59)(0.494,0.488){13}{\rule{0.500pt}{0.117pt}}
\multiput(994.00,623.17)(6.962,8.000){2}{\rule{0.250pt}{0.400pt}}
\multiput(1002.00,632.59)(0.494,0.488){13}{\rule{0.500pt}{0.117pt}}
\multiput(1002.00,631.17)(6.962,8.000){2}{\rule{0.250pt}{0.400pt}}
\multiput(1010.00,640.59)(0.492,0.485){11}{\rule{0.500pt}{0.117pt}}
\multiput(1010.00,639.17)(5.962,7.000){2}{\rule{0.250pt}{0.400pt}}
\multiput(1017.00,647.59)(0.494,0.488){13}{\rule{0.500pt}{0.117pt}}
\multiput(1017.00,646.17)(6.962,8.000){2}{\rule{0.250pt}{0.400pt}}
\multiput(1025.00,655.59)(0.492,0.485){11}{\rule{0.500pt}{0.117pt}}
\multiput(1025.00,654.17)(5.962,7.000){2}{\rule{0.250pt}{0.400pt}}
\multiput(1032.00,662.59)(0.494,0.488){13}{\rule{0.500pt}{0.117pt}}
\multiput(1032.00,661.17)(6.962,8.000){2}{\rule{0.250pt}{0.400pt}}
\multiput(1040.00,670.59)(0.492,0.485){11}{\rule{0.500pt}{0.117pt}}
\multiput(1040.00,669.17)(5.962,7.000){2}{\rule{0.250pt}{0.400pt}}
\multiput(1047.00,677.59)(0.494,0.488){13}{\rule{0.500pt}{0.117pt}}
\multiput(1047.00,676.17)(6.962,8.000){2}{\rule{0.250pt}{0.400pt}}
\multiput(1055.00,685.59)(0.494,0.488){13}{\rule{0.500pt}{0.117pt}}
\multiput(1055.00,684.17)(6.962,8.000){2}{\rule{0.250pt}{0.400pt}}
\multiput(1063.00,693.59)(0.492,0.485){11}{\rule{0.500pt}{0.117pt}}
\multiput(1063.00,692.17)(5.962,7.000){2}{\rule{0.250pt}{0.400pt}}
\multiput(1070.00,700.59)(0.494,0.488){13}{\rule{0.500pt}{0.117pt}}
\multiput(1070.00,699.17)(6.962,8.000){2}{\rule{0.250pt}{0.400pt}}
\multiput(1078.00,708.59)(0.492,0.485){11}{\rule{0.500pt}{0.117pt}}
\multiput(1078.00,707.17)(5.962,7.000){2}{\rule{0.250pt}{0.400pt}}
\multiput(1085.00,715.59)(0.494,0.488){13}{\rule{0.500pt}{0.117pt}}
\multiput(1085.00,714.17)(6.962,8.000){2}{\rule{0.250pt}{0.400pt}}
\multiput(1093.00,723.59)(0.492,0.485){11}{\rule{0.500pt}{0.117pt}}
\multiput(1093.00,722.17)(5.962,7.000){2}{\rule{0.250pt}{0.400pt}}
\multiput(1100.00,730.59)(0.494,0.488){13}{\rule{0.500pt}{0.117pt}}
\multiput(1100.00,729.17)(6.962,8.000){2}{\rule{0.250pt}{0.400pt}}
\multiput(1108.00,738.59)(0.494,0.488){13}{\rule{0.500pt}{0.117pt}}
\multiput(1108.00,737.17)(6.962,8.000){2}{\rule{0.250pt}{0.400pt}}
\multiput(1116.00,746.59)(0.492,0.485){11}{\rule{0.500pt}{0.117pt}}
\multiput(1116.00,745.17)(5.962,7.000){2}{\rule{0.250pt}{0.400pt}}
\multiput(1123.00,753.59)(0.494,0.488){13}{\rule{0.500pt}{0.117pt}}
\multiput(1123.00,752.17)(6.962,8.000){2}{\rule{0.250pt}{0.400pt}}
\multiput(1131.00,761.59)(0.492,0.485){11}{\rule{0.500pt}{0.117pt}}
\multiput(1131.00,760.17)(5.962,7.000){2}{\rule{0.250pt}{0.400pt}}
\multiput(1138.00,768.59)(0.494,0.488){13}{\rule{0.500pt}{0.117pt}}
\multiput(1138.00,767.17)(6.962,8.000){2}{\rule{0.250pt}{0.400pt}}
\multiput(1146.00,776.59)(0.492,0.485){11}{\rule{0.500pt}{0.117pt}}
\multiput(1146.00,775.17)(5.962,7.000){2}{\rule{0.250pt}{0.400pt}}
\multiput(1153.00,783.59)(0.494,0.488){13}{\rule{0.500pt}{0.117pt}}
\multiput(1153.00,782.17)(6.962,8.000){2}{\rule{0.250pt}{0.400pt}}
\multiput(1161.00,791.59)(0.492,0.485){11}{\rule{0.500pt}{0.117pt}}
\multiput(1161.00,790.17)(5.962,7.000){2}{\rule{0.250pt}{0.400pt}}
\multiput(1168.00,798.59)(0.494,0.488){13}{\rule{0.500pt}{0.117pt}}
\multiput(1168.00,797.17)(6.962,8.000){2}{\rule{0.250pt}{0.400pt}}
\multiput(1176.00,806.59)(0.494,0.488){13}{\rule{0.500pt}{0.117pt}}
\multiput(1176.00,805.17)(6.962,8.000){2}{\rule{0.250pt}{0.400pt}}
\multiput(1184.00,814.59)(0.492,0.485){11}{\rule{0.500pt}{0.117pt}}
\multiput(1184.00,813.17)(5.962,7.000){2}{\rule{0.250pt}{0.400pt}}
\multiput(1191.00,821.59)(0.494,0.488){13}{\rule{0.500pt}{0.117pt}}
\multiput(1191.00,820.17)(6.962,8.000){2}{\rule{0.250pt}{0.400pt}}
\multiput(1199.00,829.59)(0.492,0.485){11}{\rule{0.500pt}{0.117pt}}
\multiput(1199.00,828.17)(5.962,7.000){2}{\rule{0.250pt}{0.400pt}}
\multiput(1206.00,836.59)(0.494,0.488){13}{\rule{0.500pt}{0.117pt}}
\multiput(1206.00,835.17)(6.962,8.000){2}{\rule{0.250pt}{0.400pt}}
\multiput(1214.00,844.59)(0.492,0.485){11}{\rule{0.500pt}{0.117pt}}
\multiput(1214.00,843.17)(5.962,7.000){2}{\rule{0.250pt}{0.400pt}}
\multiput(1221.00,851.59)(0.494,0.488){13}{\rule{0.500pt}{0.117pt}}
\multiput(1221.00,850.17)(6.962,8.000){2}{\rule{0.250pt}{0.400pt}}
\put(1229,110){\makebox(0,0){$\times$}}
\put(480,510){\makebox(0,0){$\times$}}
\put(741,512){\makebox(0,0){$\times$}}
\put(780,647){\makebox(0,0){$\times$}}
\put(621,403){\makebox(0,0){$\times$}}
\put(1229,714){\makebox(0,0){$\times$}}
\put(942,790){\makebox(0,0){$\times$}}
\put(872,823){\makebox(0,0){$\times$}}
\put(1229,516){\makebox(0,0){$\times$}}
\put(1229,624){\makebox(0,0){$\times$}}
\put(699,404){\makebox(0,0){$\times$}}
\put(1229,704){\makebox(0,0){$\times$}}
\put(1229,757){\makebox(0,0){$\times$}}
\put(1229,802){\makebox(0,0){$\times$}}
\put(649,494){\makebox(0,0){$\times$}}
\put(649,655){\makebox(0,0){$\times$}}
\put(498,407){\makebox(0,0){$\times$}}
\put(874,735){\makebox(0,0){$\times$}}
\put(807,734){\makebox(0,0){$\times$}}
\put(1229,791){\makebox(0,0){$\times$}}
\put(480,298){\makebox(0,0){$\times$}}
\put(480,301){\makebox(0,0){$\times$}}
\put(480,291){\makebox(0,0){$\times$}}
\put(868,406){\makebox(0,0){$\times$}}
\put(480,399){\makebox(0,0){$\times$}}
\put(480,268){\makebox(0,0){$\times$}}
\put(480,265){\makebox(0,0){$\times$}}
\put(480,110){\makebox(0,0){$\times$}}
\put(480,255){\makebox(0,0){$\times$}}
\put(480,261){\makebox(0,0){$\times$}}
\put(480,422){\makebox(0,0){$\times$}}
\put(480,403){\makebox(0,0){$\times$}}
\put(480,652){\makebox(0,0){$\times$}}
\put(1056,859){\makebox(0,0){$\times$}}
\put(1229,859){\makebox(0,0){$\times$}}
\put(852,782){\makebox(0,0){$\times$}}
\put(1221,859){\makebox(0,0){$\times$}}
\put(480,626){\makebox(0,0){$\times$}}
\put(1229,389){\makebox(0,0){$\times$}}
\put(1229,859){\makebox(0,0){$\times$}}
\put(1229,859){\makebox(0,0){$\times$}}
\put(495,519){\makebox(0,0){$\times$}}
\put(534,403){\makebox(0,0){$\times$}}
\put(483,461){\makebox(0,0){$\times$}}
\put(480,297){\makebox(0,0){$\times$}}
\put(485,571){\makebox(0,0){$\times$}}
\put(548,611){\makebox(0,0){$\times$}}
\put(623,494){\makebox(0,0){$\times$}}
\put(654,430){\makebox(0,0){$\times$}}
\put(735,460){\makebox(0,0){$\times$}}
\put(595,313){\makebox(0,0){$\times$}}
\put(764,597){\makebox(0,0){$\times$}}
\put(854,633){\makebox(0,0){$\times$}}
\put(868,604){\makebox(0,0){$\times$}}
\put(1229,110){\makebox(0,0){$\times$}}
\put(1229,110){\makebox(0,0){$\times$}}
\put(480,533){\makebox(0,0){$\times$}}
\put(480,549){\makebox(0,0){$\times$}}
\put(480,493){\makebox(0,0){$\times$}}
\put(480,504){\makebox(0,0){$\times$}}
\put(480,550){\makebox(0,0){$\times$}}
\put(480,459){\makebox(0,0){$\times$}}
\put(480,567){\makebox(0,0){$\times$}}
\put(525,531){\makebox(0,0){$\times$}}
\put(480,480){\makebox(0,0){$\times$}}
\put(480,549){\makebox(0,0){$\times$}}
\put(480,650){\makebox(0,0){$\times$}}
\put(480,769){\makebox(0,0){$\times$}}
\put(1229,859){\makebox(0,0){$\times$}}
\put(1229,859){\makebox(0,0){$\times$}}
\put(1229,859){\makebox(0,0){$\times$}}
\put(1229,859){\makebox(0,0){$\times$}}
\put(1229,859){\makebox(0,0){$\times$}}
\put(1229,859){\makebox(0,0){$\times$}}
\put(1229,859){\makebox(0,0){$\times$}}
\put(1229,859){\makebox(0,0){$\times$}}
\put(1229,520){\makebox(0,0){$\times$}}
\put(1229,859){\makebox(0,0){$\times$}}
\put(1229,859){\makebox(0,0){$\times$}}
\put(1229,859){\makebox(0,0){$\times$}}
\put(1229,859){\makebox(0,0){$\times$}}
\put(480,461){\makebox(0,0){$\times$}}
\put(1229,859){\makebox(0,0){$\times$}}
\put(1229,859){\makebox(0,0){$\times$}}
\put(800,724){\makebox(0,0){$\times$}}
\put(1229,846){\makebox(0,0){$\times$}}
\put(1229,855){\makebox(0,0){$\times$}}
\put(1229,822){\makebox(0,0){$\times$}}
\put(480,110){\makebox(0,0){$\times$}}
\put(480,443){\makebox(0,0){$\times$}}
\put(480,388){\makebox(0,0){$\times$}}
\put(520,446){\makebox(0,0){$\times$}}
\put(480,387){\makebox(0,0){$\times$}}
\put(480,387){\makebox(0,0){$\times$}}
\put(480,406){\makebox(0,0){$\times$}}
\put(480,637){\makebox(0,0){$\times$}}
\put(732,676){\makebox(0,0){$\times$}}
\put(577,744){\makebox(0,0){$\times$}}
\put(1011,749){\makebox(0,0){$\times$}}
\put(718,800){\makebox(0,0){$\times$}}
\put(480,607){\makebox(0,0){$\times$}}
\put(662,551){\makebox(0,0){$\times$}}
\put(1156,674){\makebox(0,0){$\times$}}
\put(488,650){\makebox(0,0){$\times$}}
\put(480,598){\makebox(0,0){$\times$}}
\put(547,623){\makebox(0,0){$\times$}}
\put(760,637){\makebox(0,0){$\times$}}
\put(721,799){\makebox(0,0){$\times$}}
\put(612,684){\makebox(0,0){$\times$}}
\put(1128,859){\makebox(0,0){$\times$}}
\put(1229,859){\makebox(0,0){$\times$}}
\put(1229,859){\makebox(0,0){$\times$}}
\put(1229,829){\makebox(0,0){$\times$}}
\put(1229,859){\makebox(0,0){$\times$}}
\put(1019,799){\makebox(0,0){$\times$}}
\put(1229,859){\makebox(0,0){$\times$}}
\put(1229,859){\makebox(0,0){$\times$}}
\put(1229,859){\makebox(0,0){$\times$}}
\put(1229,859){\makebox(0,0){$\times$}}
\put(1229,859){\makebox(0,0){$\times$}}
\put(1006,859){\makebox(0,0){$\times$}}
\put(1229,662){\makebox(0,0){$\times$}}
\put(1229,859){\makebox(0,0){$\times$}}
\put(1229,859){\makebox(0,0){$\times$}}
\put(827,859){\makebox(0,0){$\times$}}
\put(809,713){\makebox(0,0){$\times$}}
\put(1229,859){\makebox(0,0){$\times$}}
\put(536,661){\makebox(0,0){$\times$}}
\put(480,614){\makebox(0,0){$\times$}}
\put(480,514){\makebox(0,0){$\times$}}
\put(480.0,110.0){\rule[-0.200pt]{0.400pt}{180.434pt}}
\put(480.0,110.0){\rule[-0.200pt]{180.434pt}{0.400pt}}
\put(1229.0,110.0){\rule[-0.200pt]{0.400pt}{180.434pt}}
\put(480.0,859.0){\rule[-0.200pt]{180.434pt}{0.400pt}}
\end{picture}

%% file: figures/scatter_RPFP_acdcl__RPFP_z3_.tex
% GNUPLOT: LaTeX picture
\setlength{\unitlength}{0.240900pt}
\ifx\plotpoint\undefined\newsavebox{\plotpoint}\fi
\begin{picture}(1500,900)(0,0)
\sbox{\plotpoint}{\rule[-0.200pt]{0.400pt}{0.400pt}}%
\put(460,274){\makebox(0,0)[r]{1}}
\put(480.0,274.0){\rule[-0.200pt]{4.818pt}{0.400pt}}
\put(460,439){\makebox(0,0)[r]{10}}
\put(480.0,439.0){\rule[-0.200pt]{4.818pt}{0.400pt}}
\put(460,603){\makebox(0,0)[r]{100}}
\put(480.0,603.0){\rule[-0.200pt]{4.818pt}{0.400pt}}
\put(460,768){\makebox(0,0)[r]{1000}}
\put(480.0,768.0){\rule[-0.200pt]{4.818pt}{0.400pt}}
\put(460,859){\makebox(0,0)[r]{t/o}}
\put(480.0,859.0){\rule[-0.200pt]{4.818pt}{0.400pt}}
\put(644,69){\makebox(0,0){1}}
\put(644.0,110.0){\rule[-0.200pt]{0.400pt}{4.818pt}}
\put(809,69){\makebox(0,0){10}}
\put(809.0,110.0){\rule[-0.200pt]{0.400pt}{4.818pt}}
\put(973,69){\makebox(0,0){100}}
\put(973.0,110.0){\rule[-0.200pt]{0.400pt}{4.818pt}}
\put(1138,69){\makebox(0,0){1000}}
\put(1138.0,110.0){\rule[-0.200pt]{0.400pt}{4.818pt}}
\put(1229,69){\makebox(0,0){t/o}}
\put(1229.0,110.0){\rule[-0.200pt]{0.400pt}{4.818pt}}
\put(480.0,110.0){\rule[-0.200pt]{0.400pt}{180.434pt}}
\put(480.0,110.0){\rule[-0.200pt]{180.434pt}{0.400pt}}
\put(1229.0,110.0){\rule[-0.200pt]{0.400pt}{180.434pt}}
\put(480.0,859.0){\rule[-0.200pt]{180.434pt}{0.400pt}}
\put(480,859){\line(1,0){749}}
\put(1229,110){\line(0,1){749}}
\put(394,484){\makebox(0,0){\rotatebox{90}{RPFP(z3)}}}
\put(854,29){\makebox(0,0){RPFP(acdcl)}}
\put(480,110){\usebox{\plotpoint}}
\multiput(480.00,110.59)(0.494,0.488){13}{\rule{0.500pt}{0.117pt}}
\multiput(480.00,109.17)(6.962,8.000){2}{\rule{0.250pt}{0.400pt}}
\multiput(488.00,118.59)(0.492,0.485){11}{\rule{0.500pt}{0.117pt}}
\multiput(488.00,117.17)(5.962,7.000){2}{\rule{0.250pt}{0.400pt}}
\multiput(495.00,125.59)(0.494,0.488){13}{\rule{0.500pt}{0.117pt}}
\multiput(495.00,124.17)(6.962,8.000){2}{\rule{0.250pt}{0.400pt}}
\multiput(503.00,133.59)(0.492,0.485){11}{\rule{0.500pt}{0.117pt}}
\multiput(503.00,132.17)(5.962,7.000){2}{\rule{0.250pt}{0.400pt}}
\multiput(510.00,140.59)(0.494,0.488){13}{\rule{0.500pt}{0.117pt}}
\multiput(510.00,139.17)(6.962,8.000){2}{\rule{0.250pt}{0.400pt}}
\multiput(518.00,148.59)(0.492,0.485){11}{\rule{0.500pt}{0.117pt}}
\multiput(518.00,147.17)(5.962,7.000){2}{\rule{0.250pt}{0.400pt}}
\multiput(525.00,155.59)(0.494,0.488){13}{\rule{0.500pt}{0.117pt}}
\multiput(525.00,154.17)(6.962,8.000){2}{\rule{0.250pt}{0.400pt}}
\multiput(533.00,163.59)(0.494,0.488){13}{\rule{0.500pt}{0.117pt}}
\multiput(533.00,162.17)(6.962,8.000){2}{\rule{0.250pt}{0.400pt}}
\multiput(541.00,171.59)(0.492,0.485){11}{\rule{0.500pt}{0.117pt}}
\multiput(541.00,170.17)(5.962,7.000){2}{\rule{0.250pt}{0.400pt}}
\multiput(548.00,178.59)(0.494,0.488){13}{\rule{0.500pt}{0.117pt}}
\multiput(548.00,177.17)(6.962,8.000){2}{\rule{0.250pt}{0.400pt}}
\multiput(556.00,186.59)(0.492,0.485){11}{\rule{0.500pt}{0.117pt}}
\multiput(556.00,185.17)(5.962,7.000){2}{\rule{0.250pt}{0.400pt}}
\multiput(563.00,193.59)(0.494,0.488){13}{\rule{0.500pt}{0.117pt}}
\multiput(563.00,192.17)(6.962,8.000){2}{\rule{0.250pt}{0.400pt}}
\multiput(571.00,201.59)(0.492,0.485){11}{\rule{0.500pt}{0.117pt}}
\multiput(571.00,200.17)(5.962,7.000){2}{\rule{0.250pt}{0.400pt}}
\multiput(578.00,208.59)(0.494,0.488){13}{\rule{0.500pt}{0.117pt}}
\multiput(578.00,207.17)(6.962,8.000){2}{\rule{0.250pt}{0.400pt}}
\multiput(586.00,216.59)(0.492,0.485){11}{\rule{0.500pt}{0.117pt}}
\multiput(586.00,215.17)(5.962,7.000){2}{\rule{0.250pt}{0.400pt}}
\multiput(593.00,223.59)(0.494,0.488){13}{\rule{0.500pt}{0.117pt}}
\multiput(593.00,222.17)(6.962,8.000){2}{\rule{0.250pt}{0.400pt}}
\multiput(601.00,231.59)(0.494,0.488){13}{\rule{0.500pt}{0.117pt}}
\multiput(601.00,230.17)(6.962,8.000){2}{\rule{0.250pt}{0.400pt}}
\multiput(609.00,239.59)(0.492,0.485){11}{\rule{0.500pt}{0.117pt}}
\multiput(609.00,238.17)(5.962,7.000){2}{\rule{0.250pt}{0.400pt}}
\multiput(616.00,246.59)(0.494,0.488){13}{\rule{0.500pt}{0.117pt}}
\multiput(616.00,245.17)(6.962,8.000){2}{\rule{0.250pt}{0.400pt}}
\multiput(624.00,254.59)(0.492,0.485){11}{\rule{0.500pt}{0.117pt}}
\multiput(624.00,253.17)(5.962,7.000){2}{\rule{0.250pt}{0.400pt}}
\multiput(631.00,261.59)(0.494,0.488){13}{\rule{0.500pt}{0.117pt}}
\multiput(631.00,260.17)(6.962,8.000){2}{\rule{0.250pt}{0.400pt}}
\multiput(639.00,269.59)(0.492,0.485){11}{\rule{0.500pt}{0.117pt}}
\multiput(639.00,268.17)(5.962,7.000){2}{\rule{0.250pt}{0.400pt}}
\multiput(646.00,276.59)(0.494,0.488){13}{\rule{0.500pt}{0.117pt}}
\multiput(646.00,275.17)(6.962,8.000){2}{\rule{0.250pt}{0.400pt}}
\multiput(654.00,284.59)(0.494,0.488){13}{\rule{0.500pt}{0.117pt}}
\multiput(654.00,283.17)(6.962,8.000){2}{\rule{0.250pt}{0.400pt}}
\multiput(662.00,292.59)(0.492,0.485){11}{\rule{0.500pt}{0.117pt}}
\multiput(662.00,291.17)(5.962,7.000){2}{\rule{0.250pt}{0.400pt}}
\multiput(669.00,299.59)(0.494,0.488){13}{\rule{0.500pt}{0.117pt}}
\multiput(669.00,298.17)(6.962,8.000){2}{\rule{0.250pt}{0.400pt}}
\multiput(677.00,307.59)(0.492,0.485){11}{\rule{0.500pt}{0.117pt}}
\multiput(677.00,306.17)(5.962,7.000){2}{\rule{0.250pt}{0.400pt}}
\multiput(684.00,314.59)(0.494,0.488){13}{\rule{0.500pt}{0.117pt}}
\multiput(684.00,313.17)(6.962,8.000){2}{\rule{0.250pt}{0.400pt}}
\multiput(692.00,322.59)(0.492,0.485){11}{\rule{0.500pt}{0.117pt}}
\multiput(692.00,321.17)(5.962,7.000){2}{\rule{0.250pt}{0.400pt}}
\multiput(699.00,329.59)(0.494,0.488){13}{\rule{0.500pt}{0.117pt}}
\multiput(699.00,328.17)(6.962,8.000){2}{\rule{0.250pt}{0.400pt}}
\multiput(707.00,337.59)(0.494,0.488){13}{\rule{0.500pt}{0.117pt}}
\multiput(707.00,336.17)(6.962,8.000){2}{\rule{0.250pt}{0.400pt}}
\multiput(715.00,345.59)(0.492,0.485){11}{\rule{0.500pt}{0.117pt}}
\multiput(715.00,344.17)(5.962,7.000){2}{\rule{0.250pt}{0.400pt}}
\multiput(722.00,352.59)(0.494,0.488){13}{\rule{0.500pt}{0.117pt}}
\multiput(722.00,351.17)(6.962,8.000){2}{\rule{0.250pt}{0.400pt}}
\multiput(730.00,360.59)(0.492,0.485){11}{\rule{0.500pt}{0.117pt}}
\multiput(730.00,359.17)(5.962,7.000){2}{\rule{0.250pt}{0.400pt}}
\multiput(737.00,367.59)(0.494,0.488){13}{\rule{0.500pt}{0.117pt}}
\multiput(737.00,366.17)(6.962,8.000){2}{\rule{0.250pt}{0.400pt}}
\multiput(745.00,375.59)(0.492,0.485){11}{\rule{0.500pt}{0.117pt}}
\multiput(745.00,374.17)(5.962,7.000){2}{\rule{0.250pt}{0.400pt}}
\multiput(752.00,382.59)(0.494,0.488){13}{\rule{0.500pt}{0.117pt}}
\multiput(752.00,381.17)(6.962,8.000){2}{\rule{0.250pt}{0.400pt}}
\multiput(760.00,390.59)(0.492,0.485){11}{\rule{0.500pt}{0.117pt}}
\multiput(760.00,389.17)(5.962,7.000){2}{\rule{0.250pt}{0.400pt}}
\multiput(767.00,397.59)(0.494,0.488){13}{\rule{0.500pt}{0.117pt}}
\multiput(767.00,396.17)(6.962,8.000){2}{\rule{0.250pt}{0.400pt}}
\multiput(775.00,405.59)(0.494,0.488){13}{\rule{0.500pt}{0.117pt}}
\multiput(775.00,404.17)(6.962,8.000){2}{\rule{0.250pt}{0.400pt}}
\multiput(783.00,413.59)(0.492,0.485){11}{\rule{0.500pt}{0.117pt}}
\multiput(783.00,412.17)(5.962,7.000){2}{\rule{0.250pt}{0.400pt}}
\multiput(790.00,420.59)(0.494,0.488){13}{\rule{0.500pt}{0.117pt}}
\multiput(790.00,419.17)(6.962,8.000){2}{\rule{0.250pt}{0.400pt}}
\multiput(798.00,428.59)(0.492,0.485){11}{\rule{0.500pt}{0.117pt}}
\multiput(798.00,427.17)(5.962,7.000){2}{\rule{0.250pt}{0.400pt}}
\multiput(805.00,435.59)(0.494,0.488){13}{\rule{0.500pt}{0.117pt}}
\multiput(805.00,434.17)(6.962,8.000){2}{\rule{0.250pt}{0.400pt}}
\multiput(813.00,443.59)(0.492,0.485){11}{\rule{0.500pt}{0.117pt}}
\multiput(813.00,442.17)(5.962,7.000){2}{\rule{0.250pt}{0.400pt}}
\multiput(820.00,450.59)(0.494,0.488){13}{\rule{0.500pt}{0.117pt}}
\multiput(820.00,449.17)(6.962,8.000){2}{\rule{0.250pt}{0.400pt}}
\multiput(828.00,458.59)(0.494,0.488){13}{\rule{0.500pt}{0.117pt}}
\multiput(828.00,457.17)(6.962,8.000){2}{\rule{0.250pt}{0.400pt}}
\multiput(836.00,466.59)(0.492,0.485){11}{\rule{0.500pt}{0.117pt}}
\multiput(836.00,465.17)(5.962,7.000){2}{\rule{0.250pt}{0.400pt}}
\multiput(843.00,473.59)(0.494,0.488){13}{\rule{0.500pt}{0.117pt}}
\multiput(843.00,472.17)(6.962,8.000){2}{\rule{0.250pt}{0.400pt}}
\multiput(851.00,481.59)(0.492,0.485){11}{\rule{0.500pt}{0.117pt}}
\multiput(851.00,480.17)(5.962,7.000){2}{\rule{0.250pt}{0.400pt}}
\multiput(858.00,488.59)(0.494,0.488){13}{\rule{0.500pt}{0.117pt}}
\multiput(858.00,487.17)(6.962,8.000){2}{\rule{0.250pt}{0.400pt}}
\multiput(866.00,496.59)(0.492,0.485){11}{\rule{0.500pt}{0.117pt}}
\multiput(866.00,495.17)(5.962,7.000){2}{\rule{0.250pt}{0.400pt}}
\multiput(873.00,503.59)(0.494,0.488){13}{\rule{0.500pt}{0.117pt}}
\multiput(873.00,502.17)(6.962,8.000){2}{\rule{0.250pt}{0.400pt}}
\multiput(881.00,511.59)(0.494,0.488){13}{\rule{0.500pt}{0.117pt}}
\multiput(881.00,510.17)(6.962,8.000){2}{\rule{0.250pt}{0.400pt}}
\multiput(889.00,519.59)(0.492,0.485){11}{\rule{0.500pt}{0.117pt}}
\multiput(889.00,518.17)(5.962,7.000){2}{\rule{0.250pt}{0.400pt}}
\multiput(896.00,526.59)(0.494,0.488){13}{\rule{0.500pt}{0.117pt}}
\multiput(896.00,525.17)(6.962,8.000){2}{\rule{0.250pt}{0.400pt}}
\multiput(904.00,534.59)(0.492,0.485){11}{\rule{0.500pt}{0.117pt}}
\multiput(904.00,533.17)(5.962,7.000){2}{\rule{0.250pt}{0.400pt}}
\multiput(911.00,541.59)(0.494,0.488){13}{\rule{0.500pt}{0.117pt}}
\multiput(911.00,540.17)(6.962,8.000){2}{\rule{0.250pt}{0.400pt}}
\multiput(919.00,549.59)(0.492,0.485){11}{\rule{0.500pt}{0.117pt}}
\multiput(919.00,548.17)(5.962,7.000){2}{\rule{0.250pt}{0.400pt}}
\multiput(926.00,556.59)(0.494,0.488){13}{\rule{0.500pt}{0.117pt}}
\multiput(926.00,555.17)(6.962,8.000){2}{\rule{0.250pt}{0.400pt}}
\multiput(934.00,564.59)(0.494,0.488){13}{\rule{0.500pt}{0.117pt}}
\multiput(934.00,563.17)(6.962,8.000){2}{\rule{0.250pt}{0.400pt}}
\multiput(942.00,572.59)(0.492,0.485){11}{\rule{0.500pt}{0.117pt}}
\multiput(942.00,571.17)(5.962,7.000){2}{\rule{0.250pt}{0.400pt}}
\multiput(949.00,579.59)(0.494,0.488){13}{\rule{0.500pt}{0.117pt}}
\multiput(949.00,578.17)(6.962,8.000){2}{\rule{0.250pt}{0.400pt}}
\multiput(957.00,587.59)(0.492,0.485){11}{\rule{0.500pt}{0.117pt}}
\multiput(957.00,586.17)(5.962,7.000){2}{\rule{0.250pt}{0.400pt}}
\multiput(964.00,594.59)(0.494,0.488){13}{\rule{0.500pt}{0.117pt}}
\multiput(964.00,593.17)(6.962,8.000){2}{\rule{0.250pt}{0.400pt}}
\multiput(972.00,602.59)(0.492,0.485){11}{\rule{0.500pt}{0.117pt}}
\multiput(972.00,601.17)(5.962,7.000){2}{\rule{0.250pt}{0.400pt}}
\multiput(979.00,609.59)(0.494,0.488){13}{\rule{0.500pt}{0.117pt}}
\multiput(979.00,608.17)(6.962,8.000){2}{\rule{0.250pt}{0.400pt}}
\multiput(987.00,617.59)(0.492,0.485){11}{\rule{0.500pt}{0.117pt}}
\multiput(987.00,616.17)(5.962,7.000){2}{\rule{0.250pt}{0.400pt}}
\multiput(994.00,624.59)(0.494,0.488){13}{\rule{0.500pt}{0.117pt}}
\multiput(994.00,623.17)(6.962,8.000){2}{\rule{0.250pt}{0.400pt}}
\multiput(1002.00,632.59)(0.494,0.488){13}{\rule{0.500pt}{0.117pt}}
\multiput(1002.00,631.17)(6.962,8.000){2}{\rule{0.250pt}{0.400pt}}
\multiput(1010.00,640.59)(0.492,0.485){11}{\rule{0.500pt}{0.117pt}}
\multiput(1010.00,639.17)(5.962,7.000){2}{\rule{0.250pt}{0.400pt}}
\multiput(1017.00,647.59)(0.494,0.488){13}{\rule{0.500pt}{0.117pt}}
\multiput(1017.00,646.17)(6.962,8.000){2}{\rule{0.250pt}{0.400pt}}
\multiput(1025.00,655.59)(0.492,0.485){11}{\rule{0.500pt}{0.117pt}}
\multiput(1025.00,654.17)(5.962,7.000){2}{\rule{0.250pt}{0.400pt}}
\multiput(1032.00,662.59)(0.494,0.488){13}{\rule{0.500pt}{0.117pt}}
\multiput(1032.00,661.17)(6.962,8.000){2}{\rule{0.250pt}{0.400pt}}
\multiput(1040.00,670.59)(0.492,0.485){11}{\rule{0.500pt}{0.117pt}}
\multiput(1040.00,669.17)(5.962,7.000){2}{\rule{0.250pt}{0.400pt}}
\multiput(1047.00,677.59)(0.494,0.488){13}{\rule{0.500pt}{0.117pt}}
\multiput(1047.00,676.17)(6.962,8.000){2}{\rule{0.250pt}{0.400pt}}
\multiput(1055.00,685.59)(0.494,0.488){13}{\rule{0.500pt}{0.117pt}}
\multiput(1055.00,684.17)(6.962,8.000){2}{\rule{0.250pt}{0.400pt}}
\multiput(1063.00,693.59)(0.492,0.485){11}{\rule{0.500pt}{0.117pt}}
\multiput(1063.00,692.17)(5.962,7.000){2}{\rule{0.250pt}{0.400pt}}
\multiput(1070.00,700.59)(0.494,0.488){13}{\rule{0.500pt}{0.117pt}}
\multiput(1070.00,699.17)(6.962,8.000){2}{\rule{0.250pt}{0.400pt}}
\multiput(1078.00,708.59)(0.492,0.485){11}{\rule{0.500pt}{0.117pt}}
\multiput(1078.00,707.17)(5.962,7.000){2}{\rule{0.250pt}{0.400pt}}
\multiput(1085.00,715.59)(0.494,0.488){13}{\rule{0.500pt}{0.117pt}}
\multiput(1085.00,714.17)(6.962,8.000){2}{\rule{0.250pt}{0.400pt}}
\multiput(1093.00,723.59)(0.492,0.485){11}{\rule{0.500pt}{0.117pt}}
\multiput(1093.00,722.17)(5.962,7.000){2}{\rule{0.250pt}{0.400pt}}
\multiput(1100.00,730.59)(0.494,0.488){13}{\rule{0.500pt}{0.117pt}}
\multiput(1100.00,729.17)(6.962,8.000){2}{\rule{0.250pt}{0.400pt}}
\multiput(1108.00,738.59)(0.494,0.488){13}{\rule{0.500pt}{0.117pt}}
\multiput(1108.00,737.17)(6.962,8.000){2}{\rule{0.250pt}{0.400pt}}
\multiput(1116.00,746.59)(0.492,0.485){11}{\rule{0.500pt}{0.117pt}}
\multiput(1116.00,745.17)(5.962,7.000){2}{\rule{0.250pt}{0.400pt}}
\multiput(1123.00,753.59)(0.494,0.488){13}{\rule{0.500pt}{0.117pt}}
\multiput(1123.00,752.17)(6.962,8.000){2}{\rule{0.250pt}{0.400pt}}
\multiput(1131.00,761.59)(0.492,0.485){11}{\rule{0.500pt}{0.117pt}}
\multiput(1131.00,760.17)(5.962,7.000){2}{\rule{0.250pt}{0.400pt}}
\multiput(1138.00,768.59)(0.494,0.488){13}{\rule{0.500pt}{0.117pt}}
\multiput(1138.00,767.17)(6.962,8.000){2}{\rule{0.250pt}{0.400pt}}
\multiput(1146.00,776.59)(0.492,0.485){11}{\rule{0.500pt}{0.117pt}}
\multiput(1146.00,775.17)(5.962,7.000){2}{\rule{0.250pt}{0.400pt}}
\multiput(1153.00,783.59)(0.494,0.488){13}{\rule{0.500pt}{0.117pt}}
\multiput(1153.00,782.17)(6.962,8.000){2}{\rule{0.250pt}{0.400pt}}
\multiput(1161.00,791.59)(0.492,0.485){11}{\rule{0.500pt}{0.117pt}}
\multiput(1161.00,790.17)(5.962,7.000){2}{\rule{0.250pt}{0.400pt}}
\multiput(1168.00,798.59)(0.494,0.488){13}{\rule{0.500pt}{0.117pt}}
\multiput(1168.00,797.17)(6.962,8.000){2}{\rule{0.250pt}{0.400pt}}
\multiput(1176.00,806.59)(0.494,0.488){13}{\rule{0.500pt}{0.117pt}}
\multiput(1176.00,805.17)(6.962,8.000){2}{\rule{0.250pt}{0.400pt}}
\multiput(1184.00,814.59)(0.492,0.485){11}{\rule{0.500pt}{0.117pt}}
\multiput(1184.00,813.17)(5.962,7.000){2}{\rule{0.250pt}{0.400pt}}
\multiput(1191.00,821.59)(0.494,0.488){13}{\rule{0.500pt}{0.117pt}}
\multiput(1191.00,820.17)(6.962,8.000){2}{\rule{0.250pt}{0.400pt}}
\multiput(1199.00,829.59)(0.492,0.485){11}{\rule{0.500pt}{0.117pt}}
\multiput(1199.00,828.17)(5.962,7.000){2}{\rule{0.250pt}{0.400pt}}
\multiput(1206.00,836.59)(0.494,0.488){13}{\rule{0.500pt}{0.117pt}}
\multiput(1206.00,835.17)(6.962,8.000){2}{\rule{0.250pt}{0.400pt}}
\multiput(1214.00,844.59)(0.492,0.485){11}{\rule{0.500pt}{0.117pt}}
\multiput(1214.00,843.17)(5.962,7.000){2}{\rule{0.250pt}{0.400pt}}
\multiput(1221.00,851.59)(0.494,0.488){13}{\rule{0.500pt}{0.117pt}}
\multiput(1221.00,850.17)(6.962,8.000){2}{\rule{0.250pt}{0.400pt}}
\put(573,212){\makebox(0,0){$\times$}}
\put(1229,859){\makebox(0,0){$\times$}}
\put(743,349){\makebox(0,0){$\times$}}
\put(784,401){\makebox(0,0){$\times$}}
\put(649,297){\makebox(0,0){$\times$}}
\put(784,426){\makebox(0,0){$\times$}}
\put(1229,455){\makebox(0,0){$\times$}}
\put(1210,479){\makebox(0,0){$\times$}}
\put(1229,349){\makebox(0,0){$\times$}}
\put(1229,397){\makebox(0,0){$\times$}}
\put(666,295){\makebox(0,0){$\times$}}
\put(1229,426){\makebox(0,0){$\times$}}
\put(1229,441){\makebox(0,0){$\times$}}
\put(1229,473){\makebox(0,0){$\times$}}
\put(1075,349){\makebox(0,0){$\times$}}
\put(763,399){\makebox(0,0){$\times$}}
\put(651,300){\makebox(0,0){$\times$}}
\put(838,425){\makebox(0,0){$\times$}}
\put(1229,454){\makebox(0,0){$\times$}}
\put(1229,493){\makebox(0,0){$\times$}}
\put(643,268){\makebox(0,0){$\times$}}
\put(621,269){\makebox(0,0){$\times$}}
\put(630,255){\makebox(0,0){$\times$}}
\put(865,321){\makebox(0,0){$\times$}}
\put(669,304){\makebox(0,0){$\times$}}
\put(674,306){\makebox(0,0){$\times$}}
\put(668,315){\makebox(0,0){$\times$}}
\put(645,291){\makebox(0,0){$\times$}}
\put(649,293){\makebox(0,0){$\times$}}
\put(658,281){\makebox(0,0){$\times$}}
\put(693,321){\makebox(0,0){$\times$}}
\put(675,309){\makebox(0,0){$\times$}}
\put(1229,859){\makebox(0,0){$\times$}}
\put(1229,851){\makebox(0,0){$\times$}}
\put(1229,859){\makebox(0,0){$\times$}}
\put(1229,646){\makebox(0,0){$\times$}}
\put(1229,744){\makebox(0,0){$\times$}}
\put(1229,859){\makebox(0,0){$\times$}}
\put(1229,235){\makebox(0,0){$\times$}}
\put(1229,859){\makebox(0,0){$\times$}}
\put(1229,859){\makebox(0,0){$\times$}}
\put(856,628){\makebox(0,0){$\times$}}
\put(713,358){\makebox(0,0){$\times$}}
\put(769,386){\makebox(0,0){$\times$}}
\put(651,296){\makebox(0,0){$\times$}}
\put(1229,415){\makebox(0,0){$\times$}}
\put(1229,441){\makebox(0,0){$\times$}}
\put(1229,461){\makebox(0,0){$\times$}}
\put(713,348){\makebox(0,0){$\times$}}
\put(753,392){\makebox(0,0){$\times$}}
\put(653,287){\makebox(0,0){$\times$}}
\put(778,416){\makebox(0,0){$\times$}}
\put(798,442){\makebox(0,0){$\times$}}
\put(821,464){\makebox(0,0){$\times$}}
\put(599,221){\makebox(0,0){$\times$}}
\put(594,225){\makebox(0,0){$\times$}}
\put(672,435){\makebox(0,0){$\times$}}
\put(672,355){\makebox(0,0){$\times$}}
\put(676,309){\makebox(0,0){$\times$}}
\put(706,373){\makebox(0,0){$\times$}}
\put(732,411){\makebox(0,0){$\times$}}
\put(666,321){\makebox(0,0){$\times$}}
\put(702,475){\makebox(0,0){$\times$}}
\put(826,559){\makebox(0,0){$\times$}}
\put(666,309){\makebox(0,0){$\times$}}
\put(827,499){\makebox(0,0){$\times$}}
\put(848,367){\makebox(0,0){$\times$}}
\put(745,467){\makebox(0,0){$\times$}}
\put(1229,859){\makebox(0,0){$\times$}}
\put(1229,859){\makebox(0,0){$\times$}}
\put(1229,859){\makebox(0,0){$\times$}}
\put(1229,859){\makebox(0,0){$\times$}}
\put(1229,859){\makebox(0,0){$\times$}}
\put(1229,859){\makebox(0,0){$\times$}}
\put(1229,859){\makebox(0,0){$\times$}}
\put(1229,859){\makebox(0,0){$\times$}}
\put(1229,361){\makebox(0,0){$\times$}}
\put(1229,859){\makebox(0,0){$\times$}}
\put(1229,859){\makebox(0,0){$\times$}}
\put(900,852){\makebox(0,0){$\times$}}
\put(1229,859){\makebox(0,0){$\times$}}
\put(754,407){\makebox(0,0){$\times$}}
\put(1229,859){\makebox(0,0){$\times$}}
\put(1229,859){\makebox(0,0){$\times$}}
\put(1198,482){\makebox(0,0){$\times$}}
\put(1229,859){\makebox(0,0){$\times$}}
\put(1229,859){\makebox(0,0){$\times$}}
\put(1229,841){\makebox(0,0){$\times$}}
\put(1229,188){\makebox(0,0){$\times$}}
\put(789,309){\makebox(0,0){$\times$}}
\put(676,316){\makebox(0,0){$\times$}}
\put(716,356){\makebox(0,0){$\times$}}
\put(783,309){\makebox(0,0){$\times$}}
\put(674,318){\makebox(0,0){$\times$}}
\put(677,313){\makebox(0,0){$\times$}}
\put(715,588){\makebox(0,0){$\times$}}
\put(888,410){\makebox(0,0){$\times$}}
\put(930,409){\makebox(0,0){$\times$}}
\put(831,450){\makebox(0,0){$\times$}}
\put(957,469){\makebox(0,0){$\times$}}
\put(835,367){\makebox(0,0){$\times$}}
\put(721,360){\makebox(0,0){$\times$}}
\put(993,441){\makebox(0,0){$\times$}}
\put(893,497){\makebox(0,0){$\times$}}
\put(870,376){\makebox(0,0){$\times$}}
\put(734,378){\makebox(0,0){$\times$}}
\put(855,442){\makebox(0,0){$\times$}}
\put(1134,437){\makebox(0,0){$\times$}}
\put(720,396){\makebox(0,0){$\times$}}
\put(1229,838){\makebox(0,0){$\times$}}
\put(1229,859){\makebox(0,0){$\times$}}
\put(1229,859){\makebox(0,0){$\times$}}
\put(874,553){\makebox(0,0){$\times$}}
\put(1229,553){\makebox(0,0){$\times$}}
\put(800,549){\makebox(0,0){$\times$}}
\put(1229,859){\makebox(0,0){$\times$}}
\put(1229,859){\makebox(0,0){$\times$}}
\put(1229,859){\makebox(0,0){$\times$}}
\put(925,754){\makebox(0,0){$\times$}}
\put(1229,859){\makebox(0,0){$\times$}}
\put(804,458){\makebox(0,0){$\times$}}
\put(1229,479){\makebox(0,0){$\times$}}
\put(1229,644){\makebox(0,0){$\times$}}
\put(1138,416){\makebox(0,0){$\times$}}
\put(832,516){\makebox(0,0){$\times$}}
\put(1018,454){\makebox(0,0){$\times$}}
\put(1229,621){\makebox(0,0){$\times$}}
\put(773,420){\makebox(0,0){$\times$}}
\put(1229,859){\makebox(0,0){$\times$}}
\put(1229,859){\makebox(0,0){$\times$}}
\put(480.0,110.0){\rule[-0.200pt]{0.400pt}{180.434pt}}
\put(480.0,110.0){\rule[-0.200pt]{180.434pt}{0.400pt}}
\put(1229.0,110.0){\rule[-0.200pt]{0.400pt}{180.434pt}}
\put(480.0,859.0){\rule[-0.200pt]{180.434pt}{0.400pt}}
\end{picture}

%% file: figures/scatter_mathsat_z3.tex
% GNUPLOT: LaTeX picture
\setlength{\unitlength}{0.240900pt}
\ifx\plotpoint\undefined\newsavebox{\plotpoint}\fi
\begin{picture}(1500,900)(0,0)
\sbox{\plotpoint}{\rule[-0.200pt]{0.400pt}{0.400pt}}%
\put(460,274){\makebox(0,0)[r]{1}}
\put(480.0,274.0){\rule[-0.200pt]{4.818pt}{0.400pt}}
\put(460,439){\makebox(0,0)[r]{10}}
\put(480.0,439.0){\rule[-0.200pt]{4.818pt}{0.400pt}}
\put(460,603){\makebox(0,0)[r]{100}}
\put(480.0,603.0){\rule[-0.200pt]{4.818pt}{0.400pt}}
\put(460,768){\makebox(0,0)[r]{1000}}
\put(480.0,768.0){\rule[-0.200pt]{4.818pt}{0.400pt}}
\put(460,859){\makebox(0,0)[r]{t/o}}
\put(480.0,859.0){\rule[-0.200pt]{4.818pt}{0.400pt}}
\put(644,69){\makebox(0,0){1}}
\put(644.0,110.0){\rule[-0.200pt]{0.400pt}{4.818pt}}
\put(809,69){\makebox(0,0){10}}
\put(809.0,110.0){\rule[-0.200pt]{0.400pt}{4.818pt}}
\put(973,69){\makebox(0,0){100}}
\put(973.0,110.0){\rule[-0.200pt]{0.400pt}{4.818pt}}
\put(1138,69){\makebox(0,0){1000}}
\put(1138.0,110.0){\rule[-0.200pt]{0.400pt}{4.818pt}}
\put(1229,69){\makebox(0,0){t/o}}
\put(1229.0,110.0){\rule[-0.200pt]{0.400pt}{4.818pt}}
\put(480.0,110.0){\rule[-0.200pt]{0.400pt}{180.434pt}}
\put(480.0,110.0){\rule[-0.200pt]{180.434pt}{0.400pt}}
\put(1229.0,110.0){\rule[-0.200pt]{0.400pt}{180.434pt}}
\put(480.0,859.0){\rule[-0.200pt]{180.434pt}{0.400pt}}
\put(480,859){\line(1,0){749}}
\put(1229,110){\line(0,1){749}}
\put(394,484){\makebox(0,0){\rotatebox{90}{z3}}}
\put(854,29){\makebox(0,0){mathsat}}
\put(480,110){\usebox{\plotpoint}}
\multiput(480.00,110.59)(0.494,0.488){13}{\rule{0.500pt}{0.117pt}}
\multiput(480.00,109.17)(6.962,8.000){2}{\rule{0.250pt}{0.400pt}}
\multiput(488.00,118.59)(0.492,0.485){11}{\rule{0.500pt}{0.117pt}}
\multiput(488.00,117.17)(5.962,7.000){2}{\rule{0.250pt}{0.400pt}}
\multiput(495.00,125.59)(0.494,0.488){13}{\rule{0.500pt}{0.117pt}}
\multiput(495.00,124.17)(6.962,8.000){2}{\rule{0.250pt}{0.400pt}}
\multiput(503.00,133.59)(0.492,0.485){11}{\rule{0.500pt}{0.117pt}}
\multiput(503.00,132.17)(5.962,7.000){2}{\rule{0.250pt}{0.400pt}}
\multiput(510.00,140.59)(0.494,0.488){13}{\rule{0.500pt}{0.117pt}}
\multiput(510.00,139.17)(6.962,8.000){2}{\rule{0.250pt}{0.400pt}}
\multiput(518.00,148.59)(0.492,0.485){11}{\rule{0.500pt}{0.117pt}}
\multiput(518.00,147.17)(5.962,7.000){2}{\rule{0.250pt}{0.400pt}}
\multiput(525.00,155.59)(0.494,0.488){13}{\rule{0.500pt}{0.117pt}}
\multiput(525.00,154.17)(6.962,8.000){2}{\rule{0.250pt}{0.400pt}}
\multiput(533.00,163.59)(0.494,0.488){13}{\rule{0.500pt}{0.117pt}}
\multiput(533.00,162.17)(6.962,8.000){2}{\rule{0.250pt}{0.400pt}}
\multiput(541.00,171.59)(0.492,0.485){11}{\rule{0.500pt}{0.117pt}}
\multiput(541.00,170.17)(5.962,7.000){2}{\rule{0.250pt}{0.400pt}}
\multiput(548.00,178.59)(0.494,0.488){13}{\rule{0.500pt}{0.117pt}}
\multiput(548.00,177.17)(6.962,8.000){2}{\rule{0.250pt}{0.400pt}}
\multiput(556.00,186.59)(0.492,0.485){11}{\rule{0.500pt}{0.117pt}}
\multiput(556.00,185.17)(5.962,7.000){2}{\rule{0.250pt}{0.400pt}}
\multiput(563.00,193.59)(0.494,0.488){13}{\rule{0.500pt}{0.117pt}}
\multiput(563.00,192.17)(6.962,8.000){2}{\rule{0.250pt}{0.400pt}}
\multiput(571.00,201.59)(0.492,0.485){11}{\rule{0.500pt}{0.117pt}}
\multiput(571.00,200.17)(5.962,7.000){2}{\rule{0.250pt}{0.400pt}}
\multiput(578.00,208.59)(0.494,0.488){13}{\rule{0.500pt}{0.117pt}}
\multiput(578.00,207.17)(6.962,8.000){2}{\rule{0.250pt}{0.400pt}}
\multiput(586.00,216.59)(0.492,0.485){11}{\rule{0.500pt}{0.117pt}}
\multiput(586.00,215.17)(5.962,7.000){2}{\rule{0.250pt}{0.400pt}}
\multiput(593.00,223.59)(0.494,0.488){13}{\rule{0.500pt}{0.117pt}}
\multiput(593.00,222.17)(6.962,8.000){2}{\rule{0.250pt}{0.400pt}}
\multiput(601.00,231.59)(0.494,0.488){13}{\rule{0.500pt}{0.117pt}}
\multiput(601.00,230.17)(6.962,8.000){2}{\rule{0.250pt}{0.400pt}}
\multiput(609.00,239.59)(0.492,0.485){11}{\rule{0.500pt}{0.117pt}}
\multiput(609.00,238.17)(5.962,7.000){2}{\rule{0.250pt}{0.400pt}}
\multiput(616.00,246.59)(0.494,0.488){13}{\rule{0.500pt}{0.117pt}}
\multiput(616.00,245.17)(6.962,8.000){2}{\rule{0.250pt}{0.400pt}}
\multiput(624.00,254.59)(0.492,0.485){11}{\rule{0.500pt}{0.117pt}}
\multiput(624.00,253.17)(5.962,7.000){2}{\rule{0.250pt}{0.400pt}}
\multiput(631.00,261.59)(0.494,0.488){13}{\rule{0.500pt}{0.117pt}}
\multiput(631.00,260.17)(6.962,8.000){2}{\rule{0.250pt}{0.400pt}}
\multiput(639.00,269.59)(0.492,0.485){11}{\rule{0.500pt}{0.117pt}}
\multiput(639.00,268.17)(5.962,7.000){2}{\rule{0.250pt}{0.400pt}}
\multiput(646.00,276.59)(0.494,0.488){13}{\rule{0.500pt}{0.117pt}}
\multiput(646.00,275.17)(6.962,8.000){2}{\rule{0.250pt}{0.400pt}}
\multiput(654.00,284.59)(0.494,0.488){13}{\rule{0.500pt}{0.117pt}}
\multiput(654.00,283.17)(6.962,8.000){2}{\rule{0.250pt}{0.400pt}}
\multiput(662.00,292.59)(0.492,0.485){11}{\rule{0.500pt}{0.117pt}}
\multiput(662.00,291.17)(5.962,7.000){2}{\rule{0.250pt}{0.400pt}}
\multiput(669.00,299.59)(0.494,0.488){13}{\rule{0.500pt}{0.117pt}}
\multiput(669.00,298.17)(6.962,8.000){2}{\rule{0.250pt}{0.400pt}}
\multiput(677.00,307.59)(0.492,0.485){11}{\rule{0.500pt}{0.117pt}}
\multiput(677.00,306.17)(5.962,7.000){2}{\rule{0.250pt}{0.400pt}}
\multiput(684.00,314.59)(0.494,0.488){13}{\rule{0.500pt}{0.117pt}}
\multiput(684.00,313.17)(6.962,8.000){2}{\rule{0.250pt}{0.400pt}}
\multiput(692.00,322.59)(0.492,0.485){11}{\rule{0.500pt}{0.117pt}}
\multiput(692.00,321.17)(5.962,7.000){2}{\rule{0.250pt}{0.400pt}}
\multiput(699.00,329.59)(0.494,0.488){13}{\rule{0.500pt}{0.117pt}}
\multiput(699.00,328.17)(6.962,8.000){2}{\rule{0.250pt}{0.400pt}}
\multiput(707.00,337.59)(0.494,0.488){13}{\rule{0.500pt}{0.117pt}}
\multiput(707.00,336.17)(6.962,8.000){2}{\rule{0.250pt}{0.400pt}}
\multiput(715.00,345.59)(0.492,0.485){11}{\rule{0.500pt}{0.117pt}}
\multiput(715.00,344.17)(5.962,7.000){2}{\rule{0.250pt}{0.400pt}}
\multiput(722.00,352.59)(0.494,0.488){13}{\rule{0.500pt}{0.117pt}}
\multiput(722.00,351.17)(6.962,8.000){2}{\rule{0.250pt}{0.400pt}}
\multiput(730.00,360.59)(0.492,0.485){11}{\rule{0.500pt}{0.117pt}}
\multiput(730.00,359.17)(5.962,7.000){2}{\rule{0.250pt}{0.400pt}}
\multiput(737.00,367.59)(0.494,0.488){13}{\rule{0.500pt}{0.117pt}}
\multiput(737.00,366.17)(6.962,8.000){2}{\rule{0.250pt}{0.400pt}}
\multiput(745.00,375.59)(0.492,0.485){11}{\rule{0.500pt}{0.117pt}}
\multiput(745.00,374.17)(5.962,7.000){2}{\rule{0.250pt}{0.400pt}}
\multiput(752.00,382.59)(0.494,0.488){13}{\rule{0.500pt}{0.117pt}}
\multiput(752.00,381.17)(6.962,8.000){2}{\rule{0.250pt}{0.400pt}}
\multiput(760.00,390.59)(0.492,0.485){11}{\rule{0.500pt}{0.117pt}}
\multiput(760.00,389.17)(5.962,7.000){2}{\rule{0.250pt}{0.400pt}}
\multiput(767.00,397.59)(0.494,0.488){13}{\rule{0.500pt}{0.117pt}}
\multiput(767.00,396.17)(6.962,8.000){2}{\rule{0.250pt}{0.400pt}}
\multiput(775.00,405.59)(0.494,0.488){13}{\rule{0.500pt}{0.117pt}}
\multiput(775.00,404.17)(6.962,8.000){2}{\rule{0.250pt}{0.400pt}}
\multiput(783.00,413.59)(0.492,0.485){11}{\rule{0.500pt}{0.117pt}}
\multiput(783.00,412.17)(5.962,7.000){2}{\rule{0.250pt}{0.400pt}}
\multiput(790.00,420.59)(0.494,0.488){13}{\rule{0.500pt}{0.117pt}}
\multiput(790.00,419.17)(6.962,8.000){2}{\rule{0.250pt}{0.400pt}}
\multiput(798.00,428.59)(0.492,0.485){11}{\rule{0.500pt}{0.117pt}}
\multiput(798.00,427.17)(5.962,7.000){2}{\rule{0.250pt}{0.400pt}}
\multiput(805.00,435.59)(0.494,0.488){13}{\rule{0.500pt}{0.117pt}}
\multiput(805.00,434.17)(6.962,8.000){2}{\rule{0.250pt}{0.400pt}}
\multiput(813.00,443.59)(0.492,0.485){11}{\rule{0.500pt}{0.117pt}}
\multiput(813.00,442.17)(5.962,7.000){2}{\rule{0.250pt}{0.400pt}}
\multiput(820.00,450.59)(0.494,0.488){13}{\rule{0.500pt}{0.117pt}}
\multiput(820.00,449.17)(6.962,8.000){2}{\rule{0.250pt}{0.400pt}}
\multiput(828.00,458.59)(0.494,0.488){13}{\rule{0.500pt}{0.117pt}}
\multiput(828.00,457.17)(6.962,8.000){2}{\rule{0.250pt}{0.400pt}}
\multiput(836.00,466.59)(0.492,0.485){11}{\rule{0.500pt}{0.117pt}}
\multiput(836.00,465.17)(5.962,7.000){2}{\rule{0.250pt}{0.400pt}}
\multiput(843.00,473.59)(0.494,0.488){13}{\rule{0.500pt}{0.117pt}}
\multiput(843.00,472.17)(6.962,8.000){2}{\rule{0.250pt}{0.400pt}}
\multiput(851.00,481.59)(0.492,0.485){11}{\rule{0.500pt}{0.117pt}}
\multiput(851.00,480.17)(5.962,7.000){2}{\rule{0.250pt}{0.400pt}}
\multiput(858.00,488.59)(0.494,0.488){13}{\rule{0.500pt}{0.117pt}}
\multiput(858.00,487.17)(6.962,8.000){2}{\rule{0.250pt}{0.400pt}}
\multiput(866.00,496.59)(0.492,0.485){11}{\rule{0.500pt}{0.117pt}}
\multiput(866.00,495.17)(5.962,7.000){2}{\rule{0.250pt}{0.400pt}}
\multiput(873.00,503.59)(0.494,0.488){13}{\rule{0.500pt}{0.117pt}}
\multiput(873.00,502.17)(6.962,8.000){2}{\rule{0.250pt}{0.400pt}}
\multiput(881.00,511.59)(0.494,0.488){13}{\rule{0.500pt}{0.117pt}}
\multiput(881.00,510.17)(6.962,8.000){2}{\rule{0.250pt}{0.400pt}}
\multiput(889.00,519.59)(0.492,0.485){11}{\rule{0.500pt}{0.117pt}}
\multiput(889.00,518.17)(5.962,7.000){2}{\rule{0.250pt}{0.400pt}}
\multiput(896.00,526.59)(0.494,0.488){13}{\rule{0.500pt}{0.117pt}}
\multiput(896.00,525.17)(6.962,8.000){2}{\rule{0.250pt}{0.400pt}}
\multiput(904.00,534.59)(0.492,0.485){11}{\rule{0.500pt}{0.117pt}}
\multiput(904.00,533.17)(5.962,7.000){2}{\rule{0.250pt}{0.400pt}}
\multiput(911.00,541.59)(0.494,0.488){13}{\rule{0.500pt}{0.117pt}}
\multiput(911.00,540.17)(6.962,8.000){2}{\rule{0.250pt}{0.400pt}}
\multiput(919.00,549.59)(0.492,0.485){11}{\rule{0.500pt}{0.117pt}}
\multiput(919.00,548.17)(5.962,7.000){2}{\rule{0.250pt}{0.400pt}}
\multiput(926.00,556.59)(0.494,0.488){13}{\rule{0.500pt}{0.117pt}}
\multiput(926.00,555.17)(6.962,8.000){2}{\rule{0.250pt}{0.400pt}}
\multiput(934.00,564.59)(0.494,0.488){13}{\rule{0.500pt}{0.117pt}}
\multiput(934.00,563.17)(6.962,8.000){2}{\rule{0.250pt}{0.400pt}}
\multiput(942.00,572.59)(0.492,0.485){11}{\rule{0.500pt}{0.117pt}}
\multiput(942.00,571.17)(5.962,7.000){2}{\rule{0.250pt}{0.400pt}}
\multiput(949.00,579.59)(0.494,0.488){13}{\rule{0.500pt}{0.117pt}}
\multiput(949.00,578.17)(6.962,8.000){2}{\rule{0.250pt}{0.400pt}}
\multiput(957.00,587.59)(0.492,0.485){11}{\rule{0.500pt}{0.117pt}}
\multiput(957.00,586.17)(5.962,7.000){2}{\rule{0.250pt}{0.400pt}}
\multiput(964.00,594.59)(0.494,0.488){13}{\rule{0.500pt}{0.117pt}}
\multiput(964.00,593.17)(6.962,8.000){2}{\rule{0.250pt}{0.400pt}}
\multiput(972.00,602.59)(0.492,0.485){11}{\rule{0.500pt}{0.117pt}}
\multiput(972.00,601.17)(5.962,7.000){2}{\rule{0.250pt}{0.400pt}}
\multiput(979.00,609.59)(0.494,0.488){13}{\rule{0.500pt}{0.117pt}}
\multiput(979.00,608.17)(6.962,8.000){2}{\rule{0.250pt}{0.400pt}}
\multiput(987.00,617.59)(0.492,0.485){11}{\rule{0.500pt}{0.117pt}}
\multiput(987.00,616.17)(5.962,7.000){2}{\rule{0.250pt}{0.400pt}}
\multiput(994.00,624.59)(0.494,0.488){13}{\rule{0.500pt}{0.117pt}}
\multiput(994.00,623.17)(6.962,8.000){2}{\rule{0.250pt}{0.400pt}}
\multiput(1002.00,632.59)(0.494,0.488){13}{\rule{0.500pt}{0.117pt}}
\multiput(1002.00,631.17)(6.962,8.000){2}{\rule{0.250pt}{0.400pt}}
\multiput(1010.00,640.59)(0.492,0.485){11}{\rule{0.500pt}{0.117pt}}
\multiput(1010.00,639.17)(5.962,7.000){2}{\rule{0.250pt}{0.400pt}}
\multiput(1017.00,647.59)(0.494,0.488){13}{\rule{0.500pt}{0.117pt}}
\multiput(1017.00,646.17)(6.962,8.000){2}{\rule{0.250pt}{0.400pt}}
\multiput(1025.00,655.59)(0.492,0.485){11}{\rule{0.500pt}{0.117pt}}
\multiput(1025.00,654.17)(5.962,7.000){2}{\rule{0.250pt}{0.400pt}}
\multiput(1032.00,662.59)(0.494,0.488){13}{\rule{0.500pt}{0.117pt}}
\multiput(1032.00,661.17)(6.962,8.000){2}{\rule{0.250pt}{0.400pt}}
\multiput(1040.00,670.59)(0.492,0.485){11}{\rule{0.500pt}{0.117pt}}
\multiput(1040.00,669.17)(5.962,7.000){2}{\rule{0.250pt}{0.400pt}}
\multiput(1047.00,677.59)(0.494,0.488){13}{\rule{0.500pt}{0.117pt}}
\multiput(1047.00,676.17)(6.962,8.000){2}{\rule{0.250pt}{0.400pt}}
\multiput(1055.00,685.59)(0.494,0.488){13}{\rule{0.500pt}{0.117pt}}
\multiput(1055.00,684.17)(6.962,8.000){2}{\rule{0.250pt}{0.400pt}}
\multiput(1063.00,693.59)(0.492,0.485){11}{\rule{0.500pt}{0.117pt}}
\multiput(1063.00,692.17)(5.962,7.000){2}{\rule{0.250pt}{0.400pt}}
\multiput(1070.00,700.59)(0.494,0.488){13}{\rule{0.500pt}{0.117pt}}
\multiput(1070.00,699.17)(6.962,8.000){2}{\rule{0.250pt}{0.400pt}}
\multiput(1078.00,708.59)(0.492,0.485){11}{\rule{0.500pt}{0.117pt}}
\multiput(1078.00,707.17)(5.962,7.000){2}{\rule{0.250pt}{0.400pt}}
\multiput(1085.00,715.59)(0.494,0.488){13}{\rule{0.500pt}{0.117pt}}
\multiput(1085.00,714.17)(6.962,8.000){2}{\rule{0.250pt}{0.400pt}}
\multiput(1093.00,723.59)(0.492,0.485){11}{\rule{0.500pt}{0.117pt}}
\multiput(1093.00,722.17)(5.962,7.000){2}{\rule{0.250pt}{0.400pt}}
\multiput(1100.00,730.59)(0.494,0.488){13}{\rule{0.500pt}{0.117pt}}
\multiput(1100.00,729.17)(6.962,8.000){2}{\rule{0.250pt}{0.400pt}}
\multiput(1108.00,738.59)(0.494,0.488){13}{\rule{0.500pt}{0.117pt}}
\multiput(1108.00,737.17)(6.962,8.000){2}{\rule{0.250pt}{0.400pt}}
\multiput(1116.00,746.59)(0.492,0.485){11}{\rule{0.500pt}{0.117pt}}
\multiput(1116.00,745.17)(5.962,7.000){2}{\rule{0.250pt}{0.400pt}}
\multiput(1123.00,753.59)(0.494,0.488){13}{\rule{0.500pt}{0.117pt}}
\multiput(1123.00,752.17)(6.962,8.000){2}{\rule{0.250pt}{0.400pt}}
\multiput(1131.00,761.59)(0.492,0.485){11}{\rule{0.500pt}{0.117pt}}
\multiput(1131.00,760.17)(5.962,7.000){2}{\rule{0.250pt}{0.400pt}}
\multiput(1138.00,768.59)(0.494,0.488){13}{\rule{0.500pt}{0.117pt}}
\multiput(1138.00,767.17)(6.962,8.000){2}{\rule{0.250pt}{0.400pt}}
\multiput(1146.00,776.59)(0.492,0.485){11}{\rule{0.500pt}{0.117pt}}
\multiput(1146.00,775.17)(5.962,7.000){2}{\rule{0.250pt}{0.400pt}}
\multiput(1153.00,783.59)(0.494,0.488){13}{\rule{0.500pt}{0.117pt}}
\multiput(1153.00,782.17)(6.962,8.000){2}{\rule{0.250pt}{0.400pt}}
\multiput(1161.00,791.59)(0.492,0.485){11}{\rule{0.500pt}{0.117pt}}
\multiput(1161.00,790.17)(5.962,7.000){2}{\rule{0.250pt}{0.400pt}}
\multiput(1168.00,798.59)(0.494,0.488){13}{\rule{0.500pt}{0.117pt}}
\multiput(1168.00,797.17)(6.962,8.000){2}{\rule{0.250pt}{0.400pt}}
\multiput(1176.00,806.59)(0.494,0.488){13}{\rule{0.500pt}{0.117pt}}
\multiput(1176.00,805.17)(6.962,8.000){2}{\rule{0.250pt}{0.400pt}}
\multiput(1184.00,814.59)(0.492,0.485){11}{\rule{0.500pt}{0.117pt}}
\multiput(1184.00,813.17)(5.962,7.000){2}{\rule{0.250pt}{0.400pt}}
\multiput(1191.00,821.59)(0.494,0.488){13}{\rule{0.500pt}{0.117pt}}
\multiput(1191.00,820.17)(6.962,8.000){2}{\rule{0.250pt}{0.400pt}}
\multiput(1199.00,829.59)(0.492,0.485){11}{\rule{0.500pt}{0.117pt}}
\multiput(1199.00,828.17)(5.962,7.000){2}{\rule{0.250pt}{0.400pt}}
\multiput(1206.00,836.59)(0.494,0.488){13}{\rule{0.500pt}{0.117pt}}
\multiput(1206.00,835.17)(6.962,8.000){2}{\rule{0.250pt}{0.400pt}}
\multiput(1214.00,844.59)(0.492,0.485){11}{\rule{0.500pt}{0.117pt}}
\multiput(1214.00,843.17)(5.962,7.000){2}{\rule{0.250pt}{0.400pt}}
\multiput(1221.00,851.59)(0.494,0.488){13}{\rule{0.500pt}{0.117pt}}
\multiput(1221.00,850.17)(6.962,8.000){2}{\rule{0.250pt}{0.400pt}}
\put(480,110){\makebox(0,0){$\times$}}
\put(480,510){\makebox(0,0){$\times$}}
\put(844,512){\makebox(0,0){$\times$}}
\put(933,647){\makebox(0,0){$\times$}}
\put(702,403){\makebox(0,0){$\times$}}
\put(992,714){\makebox(0,0){$\times$}}
\put(1036,790){\makebox(0,0){$\times$}}
\put(1229,823){\makebox(0,0){$\times$}}
\put(931,516){\makebox(0,0){$\times$}}
\put(958,624){\makebox(0,0){$\times$}}
\put(711,404){\makebox(0,0){$\times$}}
\put(998,704){\makebox(0,0){$\times$}}
\put(1033,757){\makebox(0,0){$\times$}}
\put(1229,802){\makebox(0,0){$\times$}}
\put(843,494){\makebox(0,0){$\times$}}
\put(952,655){\makebox(0,0){$\times$}}
\put(700,407){\makebox(0,0){$\times$}}
\put(986,735){\makebox(0,0){$\times$}}
\put(1050,734){\makebox(0,0){$\times$}}
\put(1229,791){\makebox(0,0){$\times$}}
\put(557,298){\makebox(0,0){$\times$}}
\put(549,301){\makebox(0,0){$\times$}}
\put(554,291){\makebox(0,0){$\times$}}
\put(683,406){\makebox(0,0){$\times$}}
\put(646,399){\makebox(0,0){$\times$}}
\put(545,268){\makebox(0,0){$\times$}}
\put(545,265){\makebox(0,0){$\times$}}
\put(546,110){\makebox(0,0){$\times$}}
\put(545,255){\makebox(0,0){$\times$}}
\put(547,261){\makebox(0,0){$\times$}}
\put(707,422){\makebox(0,0){$\times$}}
\put(685,403){\makebox(0,0){$\times$}}
\put(480,652){\makebox(0,0){$\times$}}
\put(1229,859){\makebox(0,0){$\times$}}
\put(1229,859){\makebox(0,0){$\times$}}
\put(945,782){\makebox(0,0){$\times$}}
\put(1229,859){\makebox(0,0){$\times$}}
\put(480,626){\makebox(0,0){$\times$}}
\put(1229,389){\makebox(0,0){$\times$}}
\put(1229,859){\makebox(0,0){$\times$}}
\put(1229,859){\makebox(0,0){$\times$}}
\put(1045,519){\makebox(0,0){$\times$}}
\put(737,403){\makebox(0,0){$\times$}}
\put(836,461){\makebox(0,0){$\times$}}
\put(568,297){\makebox(0,0){$\times$}}
\put(906,571){\makebox(0,0){$\times$}}
\put(936,611){\makebox(0,0){$\times$}}
\put(964,494){\makebox(0,0){$\times$}}
\put(745,430){\makebox(0,0){$\times$}}
\put(843,460){\makebox(0,0){$\times$}}
\put(603,313){\makebox(0,0){$\times$}}
\put(892,597){\makebox(0,0){$\times$}}
\put(931,633){\makebox(0,0){$\times$}}
\put(969,604){\makebox(0,0){$\times$}}
\put(480,110){\makebox(0,0){$\times$}}
\put(480,110){\makebox(0,0){$\times$}}
\put(965,533){\makebox(0,0){$\times$}}
\put(944,549){\makebox(0,0){$\times$}}
\put(954,493){\makebox(0,0){$\times$}}
\put(971,504){\makebox(0,0){$\times$}}
\put(1229,550){\makebox(0,0){$\times$}}
\put(987,459){\makebox(0,0){$\times$}}
\put(954,567){\makebox(0,0){$\times$}}
\put(1229,531){\makebox(0,0){$\times$}}
\put(883,480){\makebox(0,0){$\times$}}
\put(895,549){\makebox(0,0){$\times$}}
\put(912,650){\makebox(0,0){$\times$}}
\put(1196,769){\makebox(0,0){$\times$}}
\put(1229,859){\makebox(0,0){$\times$}}
\put(1229,859){\makebox(0,0){$\times$}}
\put(1229,859){\makebox(0,0){$\times$}}
\put(1229,859){\makebox(0,0){$\times$}}
\put(1229,859){\makebox(0,0){$\times$}}
\put(1229,859){\makebox(0,0){$\times$}}
\put(1229,859){\makebox(0,0){$\times$}}
\put(1229,859){\makebox(0,0){$\times$}}
\put(885,520){\makebox(0,0){$\times$}}
\put(1229,859){\makebox(0,0){$\times$}}
\put(1229,859){\makebox(0,0){$\times$}}
\put(1091,859){\makebox(0,0){$\times$}}
\put(1229,859){\makebox(0,0){$\times$}}
\put(755,461){\makebox(0,0){$\times$}}
\put(995,859){\makebox(0,0){$\times$}}
\put(1001,859){\makebox(0,0){$\times$}}
\put(878,724){\makebox(0,0){$\times$}}
\put(1229,846){\makebox(0,0){$\times$}}
\put(1229,855){\makebox(0,0){$\times$}}
\put(998,822){\makebox(0,0){$\times$}}
\put(480,110){\makebox(0,0){$\times$}}
\put(715,443){\makebox(0,0){$\times$}}
\put(718,388){\makebox(0,0){$\times$}}
\put(750,446){\makebox(0,0){$\times$}}
\put(684,387){\makebox(0,0){$\times$}}
\put(671,387){\makebox(0,0){$\times$}}
\put(710,406){\makebox(0,0){$\times$}}
\put(897,637){\makebox(0,0){$\times$}}
\put(1052,676){\makebox(0,0){$\times$}}
\put(1021,744){\makebox(0,0){$\times$}}
\put(1015,749){\makebox(0,0){$\times$}}
\put(1152,800){\makebox(0,0){$\times$}}
\put(802,607){\makebox(0,0){$\times$}}
\put(792,551){\makebox(0,0){$\times$}}
\put(966,674){\makebox(0,0){$\times$}}
\put(896,650){\makebox(0,0){$\times$}}
\put(862,598){\makebox(0,0){$\times$}}
\put(875,623){\makebox(0,0){$\times$}}
\put(909,637){\makebox(0,0){$\times$}}
\put(1022,799){\makebox(0,0){$\times$}}
\put(923,684){\makebox(0,0){$\times$}}
\put(1155,859){\makebox(0,0){$\times$}}
\put(1229,859){\makebox(0,0){$\times$}}
\put(1229,859){\makebox(0,0){$\times$}}
\put(1093,829){\makebox(0,0){$\times$}}
\put(1131,859){\makebox(0,0){$\times$}}
\put(1009,799){\makebox(0,0){$\times$}}
\put(1229,859){\makebox(0,0){$\times$}}
\put(1229,859){\makebox(0,0){$\times$}}
\put(1229,859){\makebox(0,0){$\times$}}
\put(1229,859){\makebox(0,0){$\times$}}
\put(1229,859){\makebox(0,0){$\times$}}
\put(1120,859){\makebox(0,0){$\times$}}
\put(960,662){\makebox(0,0){$\times$}}
\put(1203,859){\makebox(0,0){$\times$}}
\put(1076,859){\makebox(0,0){$\times$}}
\put(1010,859){\makebox(0,0){$\times$}}
\put(1037,713){\makebox(0,0){$\times$}}
\put(1126,859){\makebox(0,0){$\times$}}
\put(1038,661){\makebox(0,0){$\times$}}
\put(480,614){\makebox(0,0){$\times$}}
\put(480,514){\makebox(0,0){$\times$}}
\put(480.0,110.0){\rule[-0.200pt]{0.400pt}{180.434pt}}
\put(480.0,110.0){\rule[-0.200pt]{180.434pt}{0.400pt}}
\put(1229.0,110.0){\rule[-0.200pt]{0.400pt}{180.434pt}}
\put(480.0,859.0){\rule[-0.200pt]{180.434pt}{0.400pt}}
\end{picture}

%% file: figures/scatter_RPFP_mathsat__RPFP_z3_.tex
% GNUPLOT: LaTeX picture
\setlength{\unitlength}{0.240900pt}
\ifx\plotpoint\undefined\newsavebox{\plotpoint}\fi
\begin{picture}(1500,900)(0,0)
\sbox{\plotpoint}{\rule[-0.200pt]{0.400pt}{0.400pt}}%
\put(460,274){\makebox(0,0)[r]{1}}
\put(480.0,274.0){\rule[-0.200pt]{4.818pt}{0.400pt}}
\put(460,439){\makebox(0,0)[r]{10}}
\put(480.0,439.0){\rule[-0.200pt]{4.818pt}{0.400pt}}
\put(460,603){\makebox(0,0)[r]{100}}
\put(480.0,603.0){\rule[-0.200pt]{4.818pt}{0.400pt}}
\put(460,768){\makebox(0,0)[r]{1000}}
\put(480.0,768.0){\rule[-0.200pt]{4.818pt}{0.400pt}}
\put(460,859){\makebox(0,0)[r]{t/o}}
\put(480.0,859.0){\rule[-0.200pt]{4.818pt}{0.400pt}}
\put(644,69){\makebox(0,0){1}}
\put(644.0,110.0){\rule[-0.200pt]{0.400pt}{4.818pt}}
\put(809,69){\makebox(0,0){10}}
\put(809.0,110.0){\rule[-0.200pt]{0.400pt}{4.818pt}}
\put(973,69){\makebox(0,0){100}}
\put(973.0,110.0){\rule[-0.200pt]{0.400pt}{4.818pt}}
\put(1138,69){\makebox(0,0){1000}}
\put(1138.0,110.0){\rule[-0.200pt]{0.400pt}{4.818pt}}
\put(1229,69){\makebox(0,0){t/o}}
\put(1229.0,110.0){\rule[-0.200pt]{0.400pt}{4.818pt}}
\put(480.0,110.0){\rule[-0.200pt]{0.400pt}{180.434pt}}
\put(480.0,110.0){\rule[-0.200pt]{180.434pt}{0.400pt}}
\put(1229.0,110.0){\rule[-0.200pt]{0.400pt}{180.434pt}}
\put(480.0,859.0){\rule[-0.200pt]{180.434pt}{0.400pt}}
\put(480,859){\line(1,0){749}}
\put(1229,110){\line(0,1){749}}
\put(394,484){\makebox(0,0){\rotatebox{90}{RPFP(z3)}}}
\put(854,29){\makebox(0,0){RPFP(mathsat)}}
\put(480,110){\usebox{\plotpoint}}
\multiput(480.00,110.59)(0.494,0.488){13}{\rule{0.500pt}{0.117pt}}
\multiput(480.00,109.17)(6.962,8.000){2}{\rule{0.250pt}{0.400pt}}
\multiput(488.00,118.59)(0.492,0.485){11}{\rule{0.500pt}{0.117pt}}
\multiput(488.00,117.17)(5.962,7.000){2}{\rule{0.250pt}{0.400pt}}
\multiput(495.00,125.59)(0.494,0.488){13}{\rule{0.500pt}{0.117pt}}
\multiput(495.00,124.17)(6.962,8.000){2}{\rule{0.250pt}{0.400pt}}
\multiput(503.00,133.59)(0.492,0.485){11}{\rule{0.500pt}{0.117pt}}
\multiput(503.00,132.17)(5.962,7.000){2}{\rule{0.250pt}{0.400pt}}
\multiput(510.00,140.59)(0.494,0.488){13}{\rule{0.500pt}{0.117pt}}
\multiput(510.00,139.17)(6.962,8.000){2}{\rule{0.250pt}{0.400pt}}
\multiput(518.00,148.59)(0.492,0.485){11}{\rule{0.500pt}{0.117pt}}
\multiput(518.00,147.17)(5.962,7.000){2}{\rule{0.250pt}{0.400pt}}
\multiput(525.00,155.59)(0.494,0.488){13}{\rule{0.500pt}{0.117pt}}
\multiput(525.00,154.17)(6.962,8.000){2}{\rule{0.250pt}{0.400pt}}
\multiput(533.00,163.59)(0.494,0.488){13}{\rule{0.500pt}{0.117pt}}
\multiput(533.00,162.17)(6.962,8.000){2}{\rule{0.250pt}{0.400pt}}
\multiput(541.00,171.59)(0.492,0.485){11}{\rule{0.500pt}{0.117pt}}
\multiput(541.00,170.17)(5.962,7.000){2}{\rule{0.250pt}{0.400pt}}
\multiput(548.00,178.59)(0.494,0.488){13}{\rule{0.500pt}{0.117pt}}
\multiput(548.00,177.17)(6.962,8.000){2}{\rule{0.250pt}{0.400pt}}
\multiput(556.00,186.59)(0.492,0.485){11}{\rule{0.500pt}{0.117pt}}
\multiput(556.00,185.17)(5.962,7.000){2}{\rule{0.250pt}{0.400pt}}
\multiput(563.00,193.59)(0.494,0.488){13}{\rule{0.500pt}{0.117pt}}
\multiput(563.00,192.17)(6.962,8.000){2}{\rule{0.250pt}{0.400pt}}
\multiput(571.00,201.59)(0.492,0.485){11}{\rule{0.500pt}{0.117pt}}
\multiput(571.00,200.17)(5.962,7.000){2}{\rule{0.250pt}{0.400pt}}
\multiput(578.00,208.59)(0.494,0.488){13}{\rule{0.500pt}{0.117pt}}
\multiput(578.00,207.17)(6.962,8.000){2}{\rule{0.250pt}{0.400pt}}
\multiput(586.00,216.59)(0.492,0.485){11}{\rule{0.500pt}{0.117pt}}
\multiput(586.00,215.17)(5.962,7.000){2}{\rule{0.250pt}{0.400pt}}
\multiput(593.00,223.59)(0.494,0.488){13}{\rule{0.500pt}{0.117pt}}
\multiput(593.00,222.17)(6.962,8.000){2}{\rule{0.250pt}{0.400pt}}
\multiput(601.00,231.59)(0.494,0.488){13}{\rule{0.500pt}{0.117pt}}
\multiput(601.00,230.17)(6.962,8.000){2}{\rule{0.250pt}{0.400pt}}
\multiput(609.00,239.59)(0.492,0.485){11}{\rule{0.500pt}{0.117pt}}
\multiput(609.00,238.17)(5.962,7.000){2}{\rule{0.250pt}{0.400pt}}
\multiput(616.00,246.59)(0.494,0.488){13}{\rule{0.500pt}{0.117pt}}
\multiput(616.00,245.17)(6.962,8.000){2}{\rule{0.250pt}{0.400pt}}
\multiput(624.00,254.59)(0.492,0.485){11}{\rule{0.500pt}{0.117pt}}
\multiput(624.00,253.17)(5.962,7.000){2}{\rule{0.250pt}{0.400pt}}
\multiput(631.00,261.59)(0.494,0.488){13}{\rule{0.500pt}{0.117pt}}
\multiput(631.00,260.17)(6.962,8.000){2}{\rule{0.250pt}{0.400pt}}
\multiput(639.00,269.59)(0.492,0.485){11}{\rule{0.500pt}{0.117pt}}
\multiput(639.00,268.17)(5.962,7.000){2}{\rule{0.250pt}{0.400pt}}
\multiput(646.00,276.59)(0.494,0.488){13}{\rule{0.500pt}{0.117pt}}
\multiput(646.00,275.17)(6.962,8.000){2}{\rule{0.250pt}{0.400pt}}
\multiput(654.00,284.59)(0.494,0.488){13}{\rule{0.500pt}{0.117pt}}
\multiput(654.00,283.17)(6.962,8.000){2}{\rule{0.250pt}{0.400pt}}
\multiput(662.00,292.59)(0.492,0.485){11}{\rule{0.500pt}{0.117pt}}
\multiput(662.00,291.17)(5.962,7.000){2}{\rule{0.250pt}{0.400pt}}
\multiput(669.00,299.59)(0.494,0.488){13}{\rule{0.500pt}{0.117pt}}
\multiput(669.00,298.17)(6.962,8.000){2}{\rule{0.250pt}{0.400pt}}
\multiput(677.00,307.59)(0.492,0.485){11}{\rule{0.500pt}{0.117pt}}
\multiput(677.00,306.17)(5.962,7.000){2}{\rule{0.250pt}{0.400pt}}
\multiput(684.00,314.59)(0.494,0.488){13}{\rule{0.500pt}{0.117pt}}
\multiput(684.00,313.17)(6.962,8.000){2}{\rule{0.250pt}{0.400pt}}
\multiput(692.00,322.59)(0.492,0.485){11}{\rule{0.500pt}{0.117pt}}
\multiput(692.00,321.17)(5.962,7.000){2}{\rule{0.250pt}{0.400pt}}
\multiput(699.00,329.59)(0.494,0.488){13}{\rule{0.500pt}{0.117pt}}
\multiput(699.00,328.17)(6.962,8.000){2}{\rule{0.250pt}{0.400pt}}
\multiput(707.00,337.59)(0.494,0.488){13}{\rule{0.500pt}{0.117pt}}
\multiput(707.00,336.17)(6.962,8.000){2}{\rule{0.250pt}{0.400pt}}
\multiput(715.00,345.59)(0.492,0.485){11}{\rule{0.500pt}{0.117pt}}
\multiput(715.00,344.17)(5.962,7.000){2}{\rule{0.250pt}{0.400pt}}
\multiput(722.00,352.59)(0.494,0.488){13}{\rule{0.500pt}{0.117pt}}
\multiput(722.00,351.17)(6.962,8.000){2}{\rule{0.250pt}{0.400pt}}
\multiput(730.00,360.59)(0.492,0.485){11}{\rule{0.500pt}{0.117pt}}
\multiput(730.00,359.17)(5.962,7.000){2}{\rule{0.250pt}{0.400pt}}
\multiput(737.00,367.59)(0.494,0.488){13}{\rule{0.500pt}{0.117pt}}
\multiput(737.00,366.17)(6.962,8.000){2}{\rule{0.250pt}{0.400pt}}
\multiput(745.00,375.59)(0.492,0.485){11}{\rule{0.500pt}{0.117pt}}
\multiput(745.00,374.17)(5.962,7.000){2}{\rule{0.250pt}{0.400pt}}
\multiput(752.00,382.59)(0.494,0.488){13}{\rule{0.500pt}{0.117pt}}
\multiput(752.00,381.17)(6.962,8.000){2}{\rule{0.250pt}{0.400pt}}
\multiput(760.00,390.59)(0.492,0.485){11}{\rule{0.500pt}{0.117pt}}
\multiput(760.00,389.17)(5.962,7.000){2}{\rule{0.250pt}{0.400pt}}
\multiput(767.00,397.59)(0.494,0.488){13}{\rule{0.500pt}{0.117pt}}
\multiput(767.00,396.17)(6.962,8.000){2}{\rule{0.250pt}{0.400pt}}
\multiput(775.00,405.59)(0.494,0.488){13}{\rule{0.500pt}{0.117pt}}
\multiput(775.00,404.17)(6.962,8.000){2}{\rule{0.250pt}{0.400pt}}
\multiput(783.00,413.59)(0.492,0.485){11}{\rule{0.500pt}{0.117pt}}
\multiput(783.00,412.17)(5.962,7.000){2}{\rule{0.250pt}{0.400pt}}
\multiput(790.00,420.59)(0.494,0.488){13}{\rule{0.500pt}{0.117pt}}
\multiput(790.00,419.17)(6.962,8.000){2}{\rule{0.250pt}{0.400pt}}
\multiput(798.00,428.59)(0.492,0.485){11}{\rule{0.500pt}{0.117pt}}
\multiput(798.00,427.17)(5.962,7.000){2}{\rule{0.250pt}{0.400pt}}
\multiput(805.00,435.59)(0.494,0.488){13}{\rule{0.500pt}{0.117pt}}
\multiput(805.00,434.17)(6.962,8.000){2}{\rule{0.250pt}{0.400pt}}
\multiput(813.00,443.59)(0.492,0.485){11}{\rule{0.500pt}{0.117pt}}
\multiput(813.00,442.17)(5.962,7.000){2}{\rule{0.250pt}{0.400pt}}
\multiput(820.00,450.59)(0.494,0.488){13}{\rule{0.500pt}{0.117pt}}
\multiput(820.00,449.17)(6.962,8.000){2}{\rule{0.250pt}{0.400pt}}
\multiput(828.00,458.59)(0.494,0.488){13}{\rule{0.500pt}{0.117pt}}
\multiput(828.00,457.17)(6.962,8.000){2}{\rule{0.250pt}{0.400pt}}
\multiput(836.00,466.59)(0.492,0.485){11}{\rule{0.500pt}{0.117pt}}
\multiput(836.00,465.17)(5.962,7.000){2}{\rule{0.250pt}{0.400pt}}
\multiput(843.00,473.59)(0.494,0.488){13}{\rule{0.500pt}{0.117pt}}
\multiput(843.00,472.17)(6.962,8.000){2}{\rule{0.250pt}{0.400pt}}
\multiput(851.00,481.59)(0.492,0.485){11}{\rule{0.500pt}{0.117pt}}
\multiput(851.00,480.17)(5.962,7.000){2}{\rule{0.250pt}{0.400pt}}
\multiput(858.00,488.59)(0.494,0.488){13}{\rule{0.500pt}{0.117pt}}
\multiput(858.00,487.17)(6.962,8.000){2}{\rule{0.250pt}{0.400pt}}
\multiput(866.00,496.59)(0.492,0.485){11}{\rule{0.500pt}{0.117pt}}
\multiput(866.00,495.17)(5.962,7.000){2}{\rule{0.250pt}{0.400pt}}
\multiput(873.00,503.59)(0.494,0.488){13}{\rule{0.500pt}{0.117pt}}
\multiput(873.00,502.17)(6.962,8.000){2}{\rule{0.250pt}{0.400pt}}
\multiput(881.00,511.59)(0.494,0.488){13}{\rule{0.500pt}{0.117pt}}
\multiput(881.00,510.17)(6.962,8.000){2}{\rule{0.250pt}{0.400pt}}
\multiput(889.00,519.59)(0.492,0.485){11}{\rule{0.500pt}{0.117pt}}
\multiput(889.00,518.17)(5.962,7.000){2}{\rule{0.250pt}{0.400pt}}
\multiput(896.00,526.59)(0.494,0.488){13}{\rule{0.500pt}{0.117pt}}
\multiput(896.00,525.17)(6.962,8.000){2}{\rule{0.250pt}{0.400pt}}
\multiput(904.00,534.59)(0.492,0.485){11}{\rule{0.500pt}{0.117pt}}
\multiput(904.00,533.17)(5.962,7.000){2}{\rule{0.250pt}{0.400pt}}
\multiput(911.00,541.59)(0.494,0.488){13}{\rule{0.500pt}{0.117pt}}
\multiput(911.00,540.17)(6.962,8.000){2}{\rule{0.250pt}{0.400pt}}
\multiput(919.00,549.59)(0.492,0.485){11}{\rule{0.500pt}{0.117pt}}
\multiput(919.00,548.17)(5.962,7.000){2}{\rule{0.250pt}{0.400pt}}
\multiput(926.00,556.59)(0.494,0.488){13}{\rule{0.500pt}{0.117pt}}
\multiput(926.00,555.17)(6.962,8.000){2}{\rule{0.250pt}{0.400pt}}
\multiput(934.00,564.59)(0.494,0.488){13}{\rule{0.500pt}{0.117pt}}
\multiput(934.00,563.17)(6.962,8.000){2}{\rule{0.250pt}{0.400pt}}
\multiput(942.00,572.59)(0.492,0.485){11}{\rule{0.500pt}{0.117pt}}
\multiput(942.00,571.17)(5.962,7.000){2}{\rule{0.250pt}{0.400pt}}
\multiput(949.00,579.59)(0.494,0.488){13}{\rule{0.500pt}{0.117pt}}
\multiput(949.00,578.17)(6.962,8.000){2}{\rule{0.250pt}{0.400pt}}
\multiput(957.00,587.59)(0.492,0.485){11}{\rule{0.500pt}{0.117pt}}
\multiput(957.00,586.17)(5.962,7.000){2}{\rule{0.250pt}{0.400pt}}
\multiput(964.00,594.59)(0.494,0.488){13}{\rule{0.500pt}{0.117pt}}
\multiput(964.00,593.17)(6.962,8.000){2}{\rule{0.250pt}{0.400pt}}
\multiput(972.00,602.59)(0.492,0.485){11}{\rule{0.500pt}{0.117pt}}
\multiput(972.00,601.17)(5.962,7.000){2}{\rule{0.250pt}{0.400pt}}
\multiput(979.00,609.59)(0.494,0.488){13}{\rule{0.500pt}{0.117pt}}
\multiput(979.00,608.17)(6.962,8.000){2}{\rule{0.250pt}{0.400pt}}
\multiput(987.00,617.59)(0.492,0.485){11}{\rule{0.500pt}{0.117pt}}
\multiput(987.00,616.17)(5.962,7.000){2}{\rule{0.250pt}{0.400pt}}
\multiput(994.00,624.59)(0.494,0.488){13}{\rule{0.500pt}{0.117pt}}
\multiput(994.00,623.17)(6.962,8.000){2}{\rule{0.250pt}{0.400pt}}
\multiput(1002.00,632.59)(0.494,0.488){13}{\rule{0.500pt}{0.117pt}}
\multiput(1002.00,631.17)(6.962,8.000){2}{\rule{0.250pt}{0.400pt}}
\multiput(1010.00,640.59)(0.492,0.485){11}{\rule{0.500pt}{0.117pt}}
\multiput(1010.00,639.17)(5.962,7.000){2}{\rule{0.250pt}{0.400pt}}
\multiput(1017.00,647.59)(0.494,0.488){13}{\rule{0.500pt}{0.117pt}}
\multiput(1017.00,646.17)(6.962,8.000){2}{\rule{0.250pt}{0.400pt}}
\multiput(1025.00,655.59)(0.492,0.485){11}{\rule{0.500pt}{0.117pt}}
\multiput(1025.00,654.17)(5.962,7.000){2}{\rule{0.250pt}{0.400pt}}
\multiput(1032.00,662.59)(0.494,0.488){13}{\rule{0.500pt}{0.117pt}}
\multiput(1032.00,661.17)(6.962,8.000){2}{\rule{0.250pt}{0.400pt}}
\multiput(1040.00,670.59)(0.492,0.485){11}{\rule{0.500pt}{0.117pt}}
\multiput(1040.00,669.17)(5.962,7.000){2}{\rule{0.250pt}{0.400pt}}
\multiput(1047.00,677.59)(0.494,0.488){13}{\rule{0.500pt}{0.117pt}}
\multiput(1047.00,676.17)(6.962,8.000){2}{\rule{0.250pt}{0.400pt}}
\multiput(1055.00,685.59)(0.494,0.488){13}{\rule{0.500pt}{0.117pt}}
\multiput(1055.00,684.17)(6.962,8.000){2}{\rule{0.250pt}{0.400pt}}
\multiput(1063.00,693.59)(0.492,0.485){11}{\rule{0.500pt}{0.117pt}}
\multiput(1063.00,692.17)(5.962,7.000){2}{\rule{0.250pt}{0.400pt}}
\multiput(1070.00,700.59)(0.494,0.488){13}{\rule{0.500pt}{0.117pt}}
\multiput(1070.00,699.17)(6.962,8.000){2}{\rule{0.250pt}{0.400pt}}
\multiput(1078.00,708.59)(0.492,0.485){11}{\rule{0.500pt}{0.117pt}}
\multiput(1078.00,707.17)(5.962,7.000){2}{\rule{0.250pt}{0.400pt}}
\multiput(1085.00,715.59)(0.494,0.488){13}{\rule{0.500pt}{0.117pt}}
\multiput(1085.00,714.17)(6.962,8.000){2}{\rule{0.250pt}{0.400pt}}
\multiput(1093.00,723.59)(0.492,0.485){11}{\rule{0.500pt}{0.117pt}}
\multiput(1093.00,722.17)(5.962,7.000){2}{\rule{0.250pt}{0.400pt}}
\multiput(1100.00,730.59)(0.494,0.488){13}{\rule{0.500pt}{0.117pt}}
\multiput(1100.00,729.17)(6.962,8.000){2}{\rule{0.250pt}{0.400pt}}
\multiput(1108.00,738.59)(0.494,0.488){13}{\rule{0.500pt}{0.117pt}}
\multiput(1108.00,737.17)(6.962,8.000){2}{\rule{0.250pt}{0.400pt}}
\multiput(1116.00,746.59)(0.492,0.485){11}{\rule{0.500pt}{0.117pt}}
\multiput(1116.00,745.17)(5.962,7.000){2}{\rule{0.250pt}{0.400pt}}
\multiput(1123.00,753.59)(0.494,0.488){13}{\rule{0.500pt}{0.117pt}}
\multiput(1123.00,752.17)(6.962,8.000){2}{\rule{0.250pt}{0.400pt}}
\multiput(1131.00,761.59)(0.492,0.485){11}{\rule{0.500pt}{0.117pt}}
\multiput(1131.00,760.17)(5.962,7.000){2}{\rule{0.250pt}{0.400pt}}
\multiput(1138.00,768.59)(0.494,0.488){13}{\rule{0.500pt}{0.117pt}}
\multiput(1138.00,767.17)(6.962,8.000){2}{\rule{0.250pt}{0.400pt}}
\multiput(1146.00,776.59)(0.492,0.485){11}{\rule{0.500pt}{0.117pt}}
\multiput(1146.00,775.17)(5.962,7.000){2}{\rule{0.250pt}{0.400pt}}
\multiput(1153.00,783.59)(0.494,0.488){13}{\rule{0.500pt}{0.117pt}}
\multiput(1153.00,782.17)(6.962,8.000){2}{\rule{0.250pt}{0.400pt}}
\multiput(1161.00,791.59)(0.492,0.485){11}{\rule{0.500pt}{0.117pt}}
\multiput(1161.00,790.17)(5.962,7.000){2}{\rule{0.250pt}{0.400pt}}
\multiput(1168.00,798.59)(0.494,0.488){13}{\rule{0.500pt}{0.117pt}}
\multiput(1168.00,797.17)(6.962,8.000){2}{\rule{0.250pt}{0.400pt}}
\multiput(1176.00,806.59)(0.494,0.488){13}{\rule{0.500pt}{0.117pt}}
\multiput(1176.00,805.17)(6.962,8.000){2}{\rule{0.250pt}{0.400pt}}
\multiput(1184.00,814.59)(0.492,0.485){11}{\rule{0.500pt}{0.117pt}}
\multiput(1184.00,813.17)(5.962,7.000){2}{\rule{0.250pt}{0.400pt}}
\multiput(1191.00,821.59)(0.494,0.488){13}{\rule{0.500pt}{0.117pt}}
\multiput(1191.00,820.17)(6.962,8.000){2}{\rule{0.250pt}{0.400pt}}
\multiput(1199.00,829.59)(0.492,0.485){11}{\rule{0.500pt}{0.117pt}}
\multiput(1199.00,828.17)(5.962,7.000){2}{\rule{0.250pt}{0.400pt}}
\multiput(1206.00,836.59)(0.494,0.488){13}{\rule{0.500pt}{0.117pt}}
\multiput(1206.00,835.17)(6.962,8.000){2}{\rule{0.250pt}{0.400pt}}
\multiput(1214.00,844.59)(0.492,0.485){11}{\rule{0.500pt}{0.117pt}}
\multiput(1214.00,843.17)(5.962,7.000){2}{\rule{0.250pt}{0.400pt}}
\multiput(1221.00,851.59)(0.494,0.488){13}{\rule{0.500pt}{0.117pt}}
\multiput(1221.00,850.17)(6.962,8.000){2}{\rule{0.250pt}{0.400pt}}
\put(582,212){\makebox(0,0){$\times$}}
\put(1229,859){\makebox(0,0){$\times$}}
\put(722,349){\makebox(0,0){$\times$}}
\put(766,401){\makebox(0,0){$\times$}}
\put(654,297){\makebox(0,0){$\times$}}
\put(801,426){\makebox(0,0){$\times$}}
\put(828,455){\makebox(0,0){$\times$}}
\put(862,479){\makebox(0,0){$\times$}}
\put(775,349){\makebox(0,0){$\times$}}
\put(771,397){\makebox(0,0){$\times$}}
\put(648,295){\makebox(0,0){$\times$}}
\put(801,426){\makebox(0,0){$\times$}}
\put(831,441){\makebox(0,0){$\times$}}
\put(849,473){\makebox(0,0){$\times$}}
\put(719,349){\makebox(0,0){$\times$}}
\put(776,399){\makebox(0,0){$\times$}}
\put(660,300){\makebox(0,0){$\times$}}
\put(802,425){\makebox(0,0){$\times$}}
\put(830,454){\makebox(0,0){$\times$}}
\put(848,493){\makebox(0,0){$\times$}}
\put(647,268){\makebox(0,0){$\times$}}
\put(626,269){\makebox(0,0){$\times$}}
\put(634,255){\makebox(0,0){$\times$}}
\put(689,321){\makebox(0,0){$\times$}}
\put(680,304){\makebox(0,0){$\times$}}
\put(671,306){\makebox(0,0){$\times$}}
\put(679,315){\makebox(0,0){$\times$}}
\put(647,291){\makebox(0,0){$\times$}}
\put(648,293){\makebox(0,0){$\times$}}
\put(657,281){\makebox(0,0){$\times$}}
\put(695,321){\makebox(0,0){$\times$}}
\put(676,309){\makebox(0,0){$\times$}}
\put(1229,859){\makebox(0,0){$\times$}}
\put(1213,851){\makebox(0,0){$\times$}}
\put(1229,859){\makebox(0,0){$\times$}}
\put(911,646){\makebox(0,0){$\times$}}
\put(1086,744){\makebox(0,0){$\times$}}
\put(1229,859){\makebox(0,0){$\times$}}
\put(606,235){\makebox(0,0){$\times$}}
\put(1229,859){\makebox(0,0){$\times$}}
\put(1229,859){\makebox(0,0){$\times$}}
\put(1050,628){\makebox(0,0){$\times$}}
\put(722,358){\makebox(0,0){$\times$}}
\put(758,386){\makebox(0,0){$\times$}}
\put(651,296){\makebox(0,0){$\times$}}
\put(786,415){\makebox(0,0){$\times$}}
\put(808,441){\makebox(0,0){$\times$}}
\put(831,461){\makebox(0,0){$\times$}}
\put(713,348){\makebox(0,0){$\times$}}
\put(765,392){\makebox(0,0){$\times$}}
\put(669,287){\makebox(0,0){$\times$}}
\put(783,416){\makebox(0,0){$\times$}}
\put(805,442){\makebox(0,0){$\times$}}
\put(821,464){\makebox(0,0){$\times$}}
\put(589,221){\makebox(0,0){$\times$}}
\put(586,225){\makebox(0,0){$\times$}}
\put(811,435){\makebox(0,0){$\times$}}
\put(764,355){\makebox(0,0){$\times$}}
\put(767,309){\makebox(0,0){$\times$}}
\put(800,373){\makebox(0,0){$\times$}}
\put(825,411){\makebox(0,0){$\times$}}
\put(715,321){\makebox(0,0){$\times$}}
\put(844,475){\makebox(0,0){$\times$}}
\put(777,559){\makebox(0,0){$\times$}}
\put(675,309){\makebox(0,0){$\times$}}
\put(836,499){\makebox(0,0){$\times$}}
\put(772,367){\makebox(0,0){$\times$}}
\put(824,467){\makebox(0,0){$\times$}}
\put(1229,859){\makebox(0,0){$\times$}}
\put(1229,859){\makebox(0,0){$\times$}}
\put(1229,859){\makebox(0,0){$\times$}}
\put(1229,859){\makebox(0,0){$\times$}}
\put(946,859){\makebox(0,0){$\times$}}
\put(1229,859){\makebox(0,0){$\times$}}
\put(1229,859){\makebox(0,0){$\times$}}
\put(1229,859){\makebox(0,0){$\times$}}
\put(894,361){\makebox(0,0){$\times$}}
\put(1229,859){\makebox(0,0){$\times$}}
\put(1229,859){\makebox(0,0){$\times$}}
\put(1229,852){\makebox(0,0){$\times$}}
\put(1229,859){\makebox(0,0){$\times$}}
\put(771,407){\makebox(0,0){$\times$}}
\put(969,859){\makebox(0,0){$\times$}}
\put(1014,859){\makebox(0,0){$\times$}}
\put(808,482){\makebox(0,0){$\times$}}
\put(1229,859){\makebox(0,0){$\times$}}
\put(1229,859){\makebox(0,0){$\times$}}
\put(1229,841){\makebox(0,0){$\times$}}
\put(1229,188){\makebox(0,0){$\times$}}
\put(684,309){\makebox(0,0){$\times$}}
\put(676,316){\makebox(0,0){$\times$}}
\put(687,356){\makebox(0,0){$\times$}}
\put(690,309){\makebox(0,0){$\times$}}
\put(676,318){\makebox(0,0){$\times$}}
\put(722,313){\makebox(0,0){$\times$}}
\put(785,588){\makebox(0,0){$\times$}}
\put(771,410){\makebox(0,0){$\times$}}
\put(852,409){\makebox(0,0){$\times$}}
\put(813,450){\makebox(0,0){$\times$}}
\put(812,469){\makebox(0,0){$\times$}}
\put(727,367){\makebox(0,0){$\times$}}
\put(734,360){\makebox(0,0){$\times$}}
\put(865,441){\makebox(0,0){$\times$}}
\put(782,497){\makebox(0,0){$\times$}}
\put(841,376){\makebox(0,0){$\times$}}
\put(772,378){\makebox(0,0){$\times$}}
\put(831,442){\makebox(0,0){$\times$}}
\put(806,437){\makebox(0,0){$\times$}}
\put(737,396){\makebox(0,0){$\times$}}
\put(1202,838){\makebox(0,0){$\times$}}
\put(1229,859){\makebox(0,0){$\times$}}
\put(1229,859){\makebox(0,0){$\times$}}
\put(837,553){\makebox(0,0){$\times$}}
\put(974,553){\makebox(0,0){$\times$}}
\put(816,549){\makebox(0,0){$\times$}}
\put(1229,859){\makebox(0,0){$\times$}}
\put(1229,859){\makebox(0,0){$\times$}}
\put(1229,859){\makebox(0,0){$\times$}}
\put(1089,754){\makebox(0,0){$\times$}}
\put(1229,859){\makebox(0,0){$\times$}}
\put(777,458){\makebox(0,0){$\times$}}
\put(771,479){\makebox(0,0){$\times$}}
\put(1077,644){\makebox(0,0){$\times$}}
\put(876,416){\makebox(0,0){$\times$}}
\put(816,516){\makebox(0,0){$\times$}}
\put(828,454){\makebox(0,0){$\times$}}
\put(961,621){\makebox(0,0){$\times$}}
\put(868,420){\makebox(0,0){$\times$}}
\put(1229,859){\makebox(0,0){$\times$}}
\put(1229,859){\makebox(0,0){$\times$}}
\put(480.0,110.0){\rule[-0.200pt]{0.400pt}{180.434pt}}
\put(480.0,110.0){\rule[-0.200pt]{180.434pt}{0.400pt}}
\put(1229.0,110.0){\rule[-0.200pt]{0.400pt}{180.434pt}}
\put(480.0,859.0){\rule[-0.200pt]{180.434pt}{0.400pt}}
\end{picture}

%% file: figures/scatter_acdcl_mathsat.tex
% GNUPLOT: LaTeX picture
\setlength{\unitlength}{0.240900pt}
\ifx\plotpoint\undefined\newsavebox{\plotpoint}\fi
\begin{picture}(1500,900)(0,0)
\sbox{\plotpoint}{\rule[-0.200pt]{0.400pt}{0.400pt}}%
\put(460,274){\makebox(0,0)[r]{1}}
\put(480.0,274.0){\rule[-0.200pt]{4.818pt}{0.400pt}}
\put(460,439){\makebox(0,0)[r]{10}}
\put(480.0,439.0){\rule[-0.200pt]{4.818pt}{0.400pt}}
\put(460,603){\makebox(0,0)[r]{100}}
\put(480.0,603.0){\rule[-0.200pt]{4.818pt}{0.400pt}}
\put(460,768){\makebox(0,0)[r]{1000}}
\put(480.0,768.0){\rule[-0.200pt]{4.818pt}{0.400pt}}
\put(460,859){\makebox(0,0)[r]{t/o}}
\put(480.0,859.0){\rule[-0.200pt]{4.818pt}{0.400pt}}
\put(644,69){\makebox(0,0){1}}
\put(644.0,110.0){\rule[-0.200pt]{0.400pt}{4.818pt}}
\put(809,69){\makebox(0,0){10}}
\put(809.0,110.0){\rule[-0.200pt]{0.400pt}{4.818pt}}
\put(973,69){\makebox(0,0){100}}
\put(973.0,110.0){\rule[-0.200pt]{0.400pt}{4.818pt}}
\put(1138,69){\makebox(0,0){1000}}
\put(1138.0,110.0){\rule[-0.200pt]{0.400pt}{4.818pt}}
\put(1229,69){\makebox(0,0){t/o}}
\put(1229.0,110.0){\rule[-0.200pt]{0.400pt}{4.818pt}}
\put(480.0,110.0){\rule[-0.200pt]{0.400pt}{180.434pt}}
\put(480.0,110.0){\rule[-0.200pt]{180.434pt}{0.400pt}}
\put(1229.0,110.0){\rule[-0.200pt]{0.400pt}{180.434pt}}
\put(480.0,859.0){\rule[-0.200pt]{180.434pt}{0.400pt}}
\put(480,859){\line(1,0){749}}
\put(1229,110){\line(0,1){749}}
\put(394,484){\makebox(0,0){\rotatebox{90}{mathsat}}}
\put(854,29){\makebox(0,0){acdcl}}
\put(480,110){\usebox{\plotpoint}}
\multiput(480.00,110.59)(0.494,0.488){13}{\rule{0.500pt}{0.117pt}}
\multiput(480.00,109.17)(6.962,8.000){2}{\rule{0.250pt}{0.400pt}}
\multiput(488.00,118.59)(0.492,0.485){11}{\rule{0.500pt}{0.117pt}}
\multiput(488.00,117.17)(5.962,7.000){2}{\rule{0.250pt}{0.400pt}}
\multiput(495.00,125.59)(0.494,0.488){13}{\rule{0.500pt}{0.117pt}}
\multiput(495.00,124.17)(6.962,8.000){2}{\rule{0.250pt}{0.400pt}}
\multiput(503.00,133.59)(0.492,0.485){11}{\rule{0.500pt}{0.117pt}}
\multiput(503.00,132.17)(5.962,7.000){2}{\rule{0.250pt}{0.400pt}}
\multiput(510.00,140.59)(0.494,0.488){13}{\rule{0.500pt}{0.117pt}}
\multiput(510.00,139.17)(6.962,8.000){2}{\rule{0.250pt}{0.400pt}}
\multiput(518.00,148.59)(0.492,0.485){11}{\rule{0.500pt}{0.117pt}}
\multiput(518.00,147.17)(5.962,7.000){2}{\rule{0.250pt}{0.400pt}}
\multiput(525.00,155.59)(0.494,0.488){13}{\rule{0.500pt}{0.117pt}}
\multiput(525.00,154.17)(6.962,8.000){2}{\rule{0.250pt}{0.400pt}}
\multiput(533.00,163.59)(0.494,0.488){13}{\rule{0.500pt}{0.117pt}}
\multiput(533.00,162.17)(6.962,8.000){2}{\rule{0.250pt}{0.400pt}}
\multiput(541.00,171.59)(0.492,0.485){11}{\rule{0.500pt}{0.117pt}}
\multiput(541.00,170.17)(5.962,7.000){2}{\rule{0.250pt}{0.400pt}}
\multiput(548.00,178.59)(0.494,0.488){13}{\rule{0.500pt}{0.117pt}}
\multiput(548.00,177.17)(6.962,8.000){2}{\rule{0.250pt}{0.400pt}}
\multiput(556.00,186.59)(0.492,0.485){11}{\rule{0.500pt}{0.117pt}}
\multiput(556.00,185.17)(5.962,7.000){2}{\rule{0.250pt}{0.400pt}}
\multiput(563.00,193.59)(0.494,0.488){13}{\rule{0.500pt}{0.117pt}}
\multiput(563.00,192.17)(6.962,8.000){2}{\rule{0.250pt}{0.400pt}}
\multiput(571.00,201.59)(0.492,0.485){11}{\rule{0.500pt}{0.117pt}}
\multiput(571.00,200.17)(5.962,7.000){2}{\rule{0.250pt}{0.400pt}}
\multiput(578.00,208.59)(0.494,0.488){13}{\rule{0.500pt}{0.117pt}}
\multiput(578.00,207.17)(6.962,8.000){2}{\rule{0.250pt}{0.400pt}}
\multiput(586.00,216.59)(0.492,0.485){11}{\rule{0.500pt}{0.117pt}}
\multiput(586.00,215.17)(5.962,7.000){2}{\rule{0.250pt}{0.400pt}}
\multiput(593.00,223.59)(0.494,0.488){13}{\rule{0.500pt}{0.117pt}}
\multiput(593.00,222.17)(6.962,8.000){2}{\rule{0.250pt}{0.400pt}}
\multiput(601.00,231.59)(0.494,0.488){13}{\rule{0.500pt}{0.117pt}}
\multiput(601.00,230.17)(6.962,8.000){2}{\rule{0.250pt}{0.400pt}}
\multiput(609.00,239.59)(0.492,0.485){11}{\rule{0.500pt}{0.117pt}}
\multiput(609.00,238.17)(5.962,7.000){2}{\rule{0.250pt}{0.400pt}}
\multiput(616.00,246.59)(0.494,0.488){13}{\rule{0.500pt}{0.117pt}}
\multiput(616.00,245.17)(6.962,8.000){2}{\rule{0.250pt}{0.400pt}}
\multiput(624.00,254.59)(0.492,0.485){11}{\rule{0.500pt}{0.117pt}}
\multiput(624.00,253.17)(5.962,7.000){2}{\rule{0.250pt}{0.400pt}}
\multiput(631.00,261.59)(0.494,0.488){13}{\rule{0.500pt}{0.117pt}}
\multiput(631.00,260.17)(6.962,8.000){2}{\rule{0.250pt}{0.400pt}}
\multiput(639.00,269.59)(0.492,0.485){11}{\rule{0.500pt}{0.117pt}}
\multiput(639.00,268.17)(5.962,7.000){2}{\rule{0.250pt}{0.400pt}}
\multiput(646.00,276.59)(0.494,0.488){13}{\rule{0.500pt}{0.117pt}}
\multiput(646.00,275.17)(6.962,8.000){2}{\rule{0.250pt}{0.400pt}}
\multiput(654.00,284.59)(0.494,0.488){13}{\rule{0.500pt}{0.117pt}}
\multiput(654.00,283.17)(6.962,8.000){2}{\rule{0.250pt}{0.400pt}}
\multiput(662.00,292.59)(0.492,0.485){11}{\rule{0.500pt}{0.117pt}}
\multiput(662.00,291.17)(5.962,7.000){2}{\rule{0.250pt}{0.400pt}}
\multiput(669.00,299.59)(0.494,0.488){13}{\rule{0.500pt}{0.117pt}}
\multiput(669.00,298.17)(6.962,8.000){2}{\rule{0.250pt}{0.400pt}}
\multiput(677.00,307.59)(0.492,0.485){11}{\rule{0.500pt}{0.117pt}}
\multiput(677.00,306.17)(5.962,7.000){2}{\rule{0.250pt}{0.400pt}}
\multiput(684.00,314.59)(0.494,0.488){13}{\rule{0.500pt}{0.117pt}}
\multiput(684.00,313.17)(6.962,8.000){2}{\rule{0.250pt}{0.400pt}}
\multiput(692.00,322.59)(0.492,0.485){11}{\rule{0.500pt}{0.117pt}}
\multiput(692.00,321.17)(5.962,7.000){2}{\rule{0.250pt}{0.400pt}}
\multiput(699.00,329.59)(0.494,0.488){13}{\rule{0.500pt}{0.117pt}}
\multiput(699.00,328.17)(6.962,8.000){2}{\rule{0.250pt}{0.400pt}}
\multiput(707.00,337.59)(0.494,0.488){13}{\rule{0.500pt}{0.117pt}}
\multiput(707.00,336.17)(6.962,8.000){2}{\rule{0.250pt}{0.400pt}}
\multiput(715.00,345.59)(0.492,0.485){11}{\rule{0.500pt}{0.117pt}}
\multiput(715.00,344.17)(5.962,7.000){2}{\rule{0.250pt}{0.400pt}}
\multiput(722.00,352.59)(0.494,0.488){13}{\rule{0.500pt}{0.117pt}}
\multiput(722.00,351.17)(6.962,8.000){2}{\rule{0.250pt}{0.400pt}}
\multiput(730.00,360.59)(0.492,0.485){11}{\rule{0.500pt}{0.117pt}}
\multiput(730.00,359.17)(5.962,7.000){2}{\rule{0.250pt}{0.400pt}}
\multiput(737.00,367.59)(0.494,0.488){13}{\rule{0.500pt}{0.117pt}}
\multiput(737.00,366.17)(6.962,8.000){2}{\rule{0.250pt}{0.400pt}}
\multiput(745.00,375.59)(0.492,0.485){11}{\rule{0.500pt}{0.117pt}}
\multiput(745.00,374.17)(5.962,7.000){2}{\rule{0.250pt}{0.400pt}}
\multiput(752.00,382.59)(0.494,0.488){13}{\rule{0.500pt}{0.117pt}}
\multiput(752.00,381.17)(6.962,8.000){2}{\rule{0.250pt}{0.400pt}}
\multiput(760.00,390.59)(0.492,0.485){11}{\rule{0.500pt}{0.117pt}}
\multiput(760.00,389.17)(5.962,7.000){2}{\rule{0.250pt}{0.400pt}}
\multiput(767.00,397.59)(0.494,0.488){13}{\rule{0.500pt}{0.117pt}}
\multiput(767.00,396.17)(6.962,8.000){2}{\rule{0.250pt}{0.400pt}}
\multiput(775.00,405.59)(0.494,0.488){13}{\rule{0.500pt}{0.117pt}}
\multiput(775.00,404.17)(6.962,8.000){2}{\rule{0.250pt}{0.400pt}}
\multiput(783.00,413.59)(0.492,0.485){11}{\rule{0.500pt}{0.117pt}}
\multiput(783.00,412.17)(5.962,7.000){2}{\rule{0.250pt}{0.400pt}}
\multiput(790.00,420.59)(0.494,0.488){13}{\rule{0.500pt}{0.117pt}}
\multiput(790.00,419.17)(6.962,8.000){2}{\rule{0.250pt}{0.400pt}}
\multiput(798.00,428.59)(0.492,0.485){11}{\rule{0.500pt}{0.117pt}}
\multiput(798.00,427.17)(5.962,7.000){2}{\rule{0.250pt}{0.400pt}}
\multiput(805.00,435.59)(0.494,0.488){13}{\rule{0.500pt}{0.117pt}}
\multiput(805.00,434.17)(6.962,8.000){2}{\rule{0.250pt}{0.400pt}}
\multiput(813.00,443.59)(0.492,0.485){11}{\rule{0.500pt}{0.117pt}}
\multiput(813.00,442.17)(5.962,7.000){2}{\rule{0.250pt}{0.400pt}}
\multiput(820.00,450.59)(0.494,0.488){13}{\rule{0.500pt}{0.117pt}}
\multiput(820.00,449.17)(6.962,8.000){2}{\rule{0.250pt}{0.400pt}}
\multiput(828.00,458.59)(0.494,0.488){13}{\rule{0.500pt}{0.117pt}}
\multiput(828.00,457.17)(6.962,8.000){2}{\rule{0.250pt}{0.400pt}}
\multiput(836.00,466.59)(0.492,0.485){11}{\rule{0.500pt}{0.117pt}}
\multiput(836.00,465.17)(5.962,7.000){2}{\rule{0.250pt}{0.400pt}}
\multiput(843.00,473.59)(0.494,0.488){13}{\rule{0.500pt}{0.117pt}}
\multiput(843.00,472.17)(6.962,8.000){2}{\rule{0.250pt}{0.400pt}}
\multiput(851.00,481.59)(0.492,0.485){11}{\rule{0.500pt}{0.117pt}}
\multiput(851.00,480.17)(5.962,7.000){2}{\rule{0.250pt}{0.400pt}}
\multiput(858.00,488.59)(0.494,0.488){13}{\rule{0.500pt}{0.117pt}}
\multiput(858.00,487.17)(6.962,8.000){2}{\rule{0.250pt}{0.400pt}}
\multiput(866.00,496.59)(0.492,0.485){11}{\rule{0.500pt}{0.117pt}}
\multiput(866.00,495.17)(5.962,7.000){2}{\rule{0.250pt}{0.400pt}}
\multiput(873.00,503.59)(0.494,0.488){13}{\rule{0.500pt}{0.117pt}}
\multiput(873.00,502.17)(6.962,8.000){2}{\rule{0.250pt}{0.400pt}}
\multiput(881.00,511.59)(0.494,0.488){13}{\rule{0.500pt}{0.117pt}}
\multiput(881.00,510.17)(6.962,8.000){2}{\rule{0.250pt}{0.400pt}}
\multiput(889.00,519.59)(0.492,0.485){11}{\rule{0.500pt}{0.117pt}}
\multiput(889.00,518.17)(5.962,7.000){2}{\rule{0.250pt}{0.400pt}}
\multiput(896.00,526.59)(0.494,0.488){13}{\rule{0.500pt}{0.117pt}}
\multiput(896.00,525.17)(6.962,8.000){2}{\rule{0.250pt}{0.400pt}}
\multiput(904.00,534.59)(0.492,0.485){11}{\rule{0.500pt}{0.117pt}}
\multiput(904.00,533.17)(5.962,7.000){2}{\rule{0.250pt}{0.400pt}}
\multiput(911.00,541.59)(0.494,0.488){13}{\rule{0.500pt}{0.117pt}}
\multiput(911.00,540.17)(6.962,8.000){2}{\rule{0.250pt}{0.400pt}}
\multiput(919.00,549.59)(0.492,0.485){11}{\rule{0.500pt}{0.117pt}}
\multiput(919.00,548.17)(5.962,7.000){2}{\rule{0.250pt}{0.400pt}}
\multiput(926.00,556.59)(0.494,0.488){13}{\rule{0.500pt}{0.117pt}}
\multiput(926.00,555.17)(6.962,8.000){2}{\rule{0.250pt}{0.400pt}}
\multiput(934.00,564.59)(0.494,0.488){13}{\rule{0.500pt}{0.117pt}}
\multiput(934.00,563.17)(6.962,8.000){2}{\rule{0.250pt}{0.400pt}}
\multiput(942.00,572.59)(0.492,0.485){11}{\rule{0.500pt}{0.117pt}}
\multiput(942.00,571.17)(5.962,7.000){2}{\rule{0.250pt}{0.400pt}}
\multiput(949.00,579.59)(0.494,0.488){13}{\rule{0.500pt}{0.117pt}}
\multiput(949.00,578.17)(6.962,8.000){2}{\rule{0.250pt}{0.400pt}}
\multiput(957.00,587.59)(0.492,0.485){11}{\rule{0.500pt}{0.117pt}}
\multiput(957.00,586.17)(5.962,7.000){2}{\rule{0.250pt}{0.400pt}}
\multiput(964.00,594.59)(0.494,0.488){13}{\rule{0.500pt}{0.117pt}}
\multiput(964.00,593.17)(6.962,8.000){2}{\rule{0.250pt}{0.400pt}}
\multiput(972.00,602.59)(0.492,0.485){11}{\rule{0.500pt}{0.117pt}}
\multiput(972.00,601.17)(5.962,7.000){2}{\rule{0.250pt}{0.400pt}}
\multiput(979.00,609.59)(0.494,0.488){13}{\rule{0.500pt}{0.117pt}}
\multiput(979.00,608.17)(6.962,8.000){2}{\rule{0.250pt}{0.400pt}}
\multiput(987.00,617.59)(0.492,0.485){11}{\rule{0.500pt}{0.117pt}}
\multiput(987.00,616.17)(5.962,7.000){2}{\rule{0.250pt}{0.400pt}}
\multiput(994.00,624.59)(0.494,0.488){13}{\rule{0.500pt}{0.117pt}}
\multiput(994.00,623.17)(6.962,8.000){2}{\rule{0.250pt}{0.400pt}}
\multiput(1002.00,632.59)(0.494,0.488){13}{\rule{0.500pt}{0.117pt}}
\multiput(1002.00,631.17)(6.962,8.000){2}{\rule{0.250pt}{0.400pt}}
\multiput(1010.00,640.59)(0.492,0.485){11}{\rule{0.500pt}{0.117pt}}
\multiput(1010.00,639.17)(5.962,7.000){2}{\rule{0.250pt}{0.400pt}}
\multiput(1017.00,647.59)(0.494,0.488){13}{\rule{0.500pt}{0.117pt}}
\multiput(1017.00,646.17)(6.962,8.000){2}{\rule{0.250pt}{0.400pt}}
\multiput(1025.00,655.59)(0.492,0.485){11}{\rule{0.500pt}{0.117pt}}
\multiput(1025.00,654.17)(5.962,7.000){2}{\rule{0.250pt}{0.400pt}}
\multiput(1032.00,662.59)(0.494,0.488){13}{\rule{0.500pt}{0.117pt}}
\multiput(1032.00,661.17)(6.962,8.000){2}{\rule{0.250pt}{0.400pt}}
\multiput(1040.00,670.59)(0.492,0.485){11}{\rule{0.500pt}{0.117pt}}
\multiput(1040.00,669.17)(5.962,7.000){2}{\rule{0.250pt}{0.400pt}}
\multiput(1047.00,677.59)(0.494,0.488){13}{\rule{0.500pt}{0.117pt}}
\multiput(1047.00,676.17)(6.962,8.000){2}{\rule{0.250pt}{0.400pt}}
\multiput(1055.00,685.59)(0.494,0.488){13}{\rule{0.500pt}{0.117pt}}
\multiput(1055.00,684.17)(6.962,8.000){2}{\rule{0.250pt}{0.400pt}}
\multiput(1063.00,693.59)(0.492,0.485){11}{\rule{0.500pt}{0.117pt}}
\multiput(1063.00,692.17)(5.962,7.000){2}{\rule{0.250pt}{0.400pt}}
\multiput(1070.00,700.59)(0.494,0.488){13}{\rule{0.500pt}{0.117pt}}
\multiput(1070.00,699.17)(6.962,8.000){2}{\rule{0.250pt}{0.400pt}}
\multiput(1078.00,708.59)(0.492,0.485){11}{\rule{0.500pt}{0.117pt}}
\multiput(1078.00,707.17)(5.962,7.000){2}{\rule{0.250pt}{0.400pt}}
\multiput(1085.00,715.59)(0.494,0.488){13}{\rule{0.500pt}{0.117pt}}
\multiput(1085.00,714.17)(6.962,8.000){2}{\rule{0.250pt}{0.400pt}}
\multiput(1093.00,723.59)(0.492,0.485){11}{\rule{0.500pt}{0.117pt}}
\multiput(1093.00,722.17)(5.962,7.000){2}{\rule{0.250pt}{0.400pt}}
\multiput(1100.00,730.59)(0.494,0.488){13}{\rule{0.500pt}{0.117pt}}
\multiput(1100.00,729.17)(6.962,8.000){2}{\rule{0.250pt}{0.400pt}}
\multiput(1108.00,738.59)(0.494,0.488){13}{\rule{0.500pt}{0.117pt}}
\multiput(1108.00,737.17)(6.962,8.000){2}{\rule{0.250pt}{0.400pt}}
\multiput(1116.00,746.59)(0.492,0.485){11}{\rule{0.500pt}{0.117pt}}
\multiput(1116.00,745.17)(5.962,7.000){2}{\rule{0.250pt}{0.400pt}}
\multiput(1123.00,753.59)(0.494,0.488){13}{\rule{0.500pt}{0.117pt}}
\multiput(1123.00,752.17)(6.962,8.000){2}{\rule{0.250pt}{0.400pt}}
\multiput(1131.00,761.59)(0.492,0.485){11}{\rule{0.500pt}{0.117pt}}
\multiput(1131.00,760.17)(5.962,7.000){2}{\rule{0.250pt}{0.400pt}}
\multiput(1138.00,768.59)(0.494,0.488){13}{\rule{0.500pt}{0.117pt}}
\multiput(1138.00,767.17)(6.962,8.000){2}{\rule{0.250pt}{0.400pt}}
\multiput(1146.00,776.59)(0.492,0.485){11}{\rule{0.500pt}{0.117pt}}
\multiput(1146.00,775.17)(5.962,7.000){2}{\rule{0.250pt}{0.400pt}}
\multiput(1153.00,783.59)(0.494,0.488){13}{\rule{0.500pt}{0.117pt}}
\multiput(1153.00,782.17)(6.962,8.000){2}{\rule{0.250pt}{0.400pt}}
\multiput(1161.00,791.59)(0.492,0.485){11}{\rule{0.500pt}{0.117pt}}
\multiput(1161.00,790.17)(5.962,7.000){2}{\rule{0.250pt}{0.400pt}}
\multiput(1168.00,798.59)(0.494,0.488){13}{\rule{0.500pt}{0.117pt}}
\multiput(1168.00,797.17)(6.962,8.000){2}{\rule{0.250pt}{0.400pt}}
\multiput(1176.00,806.59)(0.494,0.488){13}{\rule{0.500pt}{0.117pt}}
\multiput(1176.00,805.17)(6.962,8.000){2}{\rule{0.250pt}{0.400pt}}
\multiput(1184.00,814.59)(0.492,0.485){11}{\rule{0.500pt}{0.117pt}}
\multiput(1184.00,813.17)(5.962,7.000){2}{\rule{0.250pt}{0.400pt}}
\multiput(1191.00,821.59)(0.494,0.488){13}{\rule{0.500pt}{0.117pt}}
\multiput(1191.00,820.17)(6.962,8.000){2}{\rule{0.250pt}{0.400pt}}
\multiput(1199.00,829.59)(0.492,0.485){11}{\rule{0.500pt}{0.117pt}}
\multiput(1199.00,828.17)(5.962,7.000){2}{\rule{0.250pt}{0.400pt}}
\multiput(1206.00,836.59)(0.494,0.488){13}{\rule{0.500pt}{0.117pt}}
\multiput(1206.00,835.17)(6.962,8.000){2}{\rule{0.250pt}{0.400pt}}
\multiput(1214.00,844.59)(0.492,0.485){11}{\rule{0.500pt}{0.117pt}}
\multiput(1214.00,843.17)(5.962,7.000){2}{\rule{0.250pt}{0.400pt}}
\multiput(1221.00,851.59)(0.494,0.488){13}{\rule{0.500pt}{0.117pt}}
\multiput(1221.00,850.17)(6.962,8.000){2}{\rule{0.250pt}{0.400pt}}
\put(1229,110){\makebox(0,0){$\times$}}
\put(480,110){\makebox(0,0){$\times$}}
\put(741,474){\makebox(0,0){$\times$}}
\put(780,563){\makebox(0,0){$\times$}}
\put(621,332){\makebox(0,0){$\times$}}
\put(1229,622){\makebox(0,0){$\times$}}
\put(942,666){\makebox(0,0){$\times$}}
\put(872,859){\makebox(0,0){$\times$}}
\put(1229,561){\makebox(0,0){$\times$}}
\put(1229,588){\makebox(0,0){$\times$}}
\put(699,341){\makebox(0,0){$\times$}}
\put(1229,628){\makebox(0,0){$\times$}}
\put(1229,663){\makebox(0,0){$\times$}}
\put(1229,859){\makebox(0,0){$\times$}}
\put(649,473){\makebox(0,0){$\times$}}
\put(649,582){\makebox(0,0){$\times$}}
\put(498,330){\makebox(0,0){$\times$}}
\put(874,616){\makebox(0,0){$\times$}}
\put(807,680){\makebox(0,0){$\times$}}
\put(1229,859){\makebox(0,0){$\times$}}
\put(480,187){\makebox(0,0){$\times$}}
\put(480,179){\makebox(0,0){$\times$}}
\put(480,184){\makebox(0,0){$\times$}}
\put(868,313){\makebox(0,0){$\times$}}
\put(480,276){\makebox(0,0){$\times$}}
\put(480,175){\makebox(0,0){$\times$}}
\put(480,175){\makebox(0,0){$\times$}}
\put(480,176){\makebox(0,0){$\times$}}
\put(480,175){\makebox(0,0){$\times$}}
\put(480,177){\makebox(0,0){$\times$}}
\put(480,337){\makebox(0,0){$\times$}}
\put(480,315){\makebox(0,0){$\times$}}
\put(480,110){\makebox(0,0){$\times$}}
\put(1056,859){\makebox(0,0){$\times$}}
\put(1229,859){\makebox(0,0){$\times$}}
\put(852,575){\makebox(0,0){$\times$}}
\put(1221,859){\makebox(0,0){$\times$}}
\put(480,110){\makebox(0,0){$\times$}}
\put(1229,859){\makebox(0,0){$\times$}}
\put(1229,859){\makebox(0,0){$\times$}}
\put(1229,859){\makebox(0,0){$\times$}}
\put(495,675){\makebox(0,0){$\times$}}
\put(534,367){\makebox(0,0){$\times$}}
\put(483,466){\makebox(0,0){$\times$}}
\put(480,198){\makebox(0,0){$\times$}}
\put(485,536){\makebox(0,0){$\times$}}
\put(548,566){\makebox(0,0){$\times$}}
\put(623,594){\makebox(0,0){$\times$}}
\put(654,375){\makebox(0,0){$\times$}}
\put(735,473){\makebox(0,0){$\times$}}
\put(595,233){\makebox(0,0){$\times$}}
\put(764,522){\makebox(0,0){$\times$}}
\put(854,561){\makebox(0,0){$\times$}}
\put(868,599){\makebox(0,0){$\times$}}
\put(1229,110){\makebox(0,0){$\times$}}
\put(1229,110){\makebox(0,0){$\times$}}
\put(480,595){\makebox(0,0){$\times$}}
\put(480,574){\makebox(0,0){$\times$}}
\put(480,584){\makebox(0,0){$\times$}}
\put(480,601){\makebox(0,0){$\times$}}
\put(480,859){\makebox(0,0){$\times$}}
\put(480,617){\makebox(0,0){$\times$}}
\put(480,584){\makebox(0,0){$\times$}}
\put(525,859){\makebox(0,0){$\times$}}
\put(480,513){\makebox(0,0){$\times$}}
\put(480,525){\makebox(0,0){$\times$}}
\put(480,542){\makebox(0,0){$\times$}}
\put(480,826){\makebox(0,0){$\times$}}
\put(1229,859){\makebox(0,0){$\times$}}
\put(1229,859){\makebox(0,0){$\times$}}
\put(1229,859){\makebox(0,0){$\times$}}
\put(1229,859){\makebox(0,0){$\times$}}
\put(1229,859){\makebox(0,0){$\times$}}
\put(1229,859){\makebox(0,0){$\times$}}
\put(1229,859){\makebox(0,0){$\times$}}
\put(1229,859){\makebox(0,0){$\times$}}
\put(1229,515){\makebox(0,0){$\times$}}
\put(1229,859){\makebox(0,0){$\times$}}
\put(1229,859){\makebox(0,0){$\times$}}
\put(1229,721){\makebox(0,0){$\times$}}
\put(1229,859){\makebox(0,0){$\times$}}
\put(480,385){\makebox(0,0){$\times$}}
\put(1229,625){\makebox(0,0){$\times$}}
\put(1229,631){\makebox(0,0){$\times$}}
\put(800,508){\makebox(0,0){$\times$}}
\put(1229,859){\makebox(0,0){$\times$}}
\put(1229,859){\makebox(0,0){$\times$}}
\put(1229,628){\makebox(0,0){$\times$}}
\put(480,110){\makebox(0,0){$\times$}}
\put(480,345){\makebox(0,0){$\times$}}
\put(480,348){\makebox(0,0){$\times$}}
\put(520,380){\makebox(0,0){$\times$}}
\put(480,314){\makebox(0,0){$\times$}}
\put(480,301){\makebox(0,0){$\times$}}
\put(480,340){\makebox(0,0){$\times$}}
\put(480,527){\makebox(0,0){$\times$}}
\put(732,682){\makebox(0,0){$\times$}}
\put(577,651){\makebox(0,0){$\times$}}
\put(1011,645){\makebox(0,0){$\times$}}
\put(718,782){\makebox(0,0){$\times$}}
\put(480,432){\makebox(0,0){$\times$}}
\put(662,422){\makebox(0,0){$\times$}}
\put(1156,596){\makebox(0,0){$\times$}}
\put(488,526){\makebox(0,0){$\times$}}
\put(480,492){\makebox(0,0){$\times$}}
\put(547,505){\makebox(0,0){$\times$}}
\put(760,539){\makebox(0,0){$\times$}}
\put(721,652){\makebox(0,0){$\times$}}
\put(612,553){\makebox(0,0){$\times$}}
\put(1128,785){\makebox(0,0){$\times$}}
\put(1229,859){\makebox(0,0){$\times$}}
\put(1229,859){\makebox(0,0){$\times$}}
\put(1229,723){\makebox(0,0){$\times$}}
\put(1229,761){\makebox(0,0){$\times$}}
\put(1019,639){\makebox(0,0){$\times$}}
\put(1229,859){\makebox(0,0){$\times$}}
\put(1229,859){\makebox(0,0){$\times$}}
\put(1229,859){\makebox(0,0){$\times$}}
\put(1229,859){\makebox(0,0){$\times$}}
\put(1229,859){\makebox(0,0){$\times$}}
\put(1006,750){\makebox(0,0){$\times$}}
\put(1229,590){\makebox(0,0){$\times$}}
\put(1229,833){\makebox(0,0){$\times$}}
\put(1229,706){\makebox(0,0){$\times$}}
\put(827,640){\makebox(0,0){$\times$}}
\put(809,667){\makebox(0,0){$\times$}}
\put(1229,756){\makebox(0,0){$\times$}}
\put(536,668){\makebox(0,0){$\times$}}
\put(480,110){\makebox(0,0){$\times$}}
\put(480,110){\makebox(0,0){$\times$}}
\put(480.0,110.0){\rule[-0.200pt]{0.400pt}{180.434pt}}
\put(480.0,110.0){\rule[-0.200pt]{180.434pt}{0.400pt}}
\put(1229.0,110.0){\rule[-0.200pt]{0.400pt}{180.434pt}}
\put(480.0,859.0){\rule[-0.200pt]{180.434pt}{0.400pt}}
\end{picture}

%% file: figures/scatter_RPFP_acdcl__RPFP_mathsat_.tex
% GNUPLOT: LaTeX picture
\setlength{\unitlength}{0.240900pt}
\ifx\plotpoint\undefined\newsavebox{\plotpoint}\fi
\begin{picture}(1500,900)(0,0)
\sbox{\plotpoint}{\rule[-0.200pt]{0.400pt}{0.400pt}}%
\put(460,274){\makebox(0,0)[r]{1}}
\put(480.0,274.0){\rule[-0.200pt]{4.818pt}{0.400pt}}
\put(460,439){\makebox(0,0)[r]{10}}
\put(480.0,439.0){\rule[-0.200pt]{4.818pt}{0.400pt}}
\put(460,603){\makebox(0,0)[r]{100}}
\put(480.0,603.0){\rule[-0.200pt]{4.818pt}{0.400pt}}
\put(460,768){\makebox(0,0)[r]{1000}}
\put(480.0,768.0){\rule[-0.200pt]{4.818pt}{0.400pt}}
\put(460,859){\makebox(0,0)[r]{t/o}}
\put(480.0,859.0){\rule[-0.200pt]{4.818pt}{0.400pt}}
\put(644,69){\makebox(0,0){1}}
\put(644.0,110.0){\rule[-0.200pt]{0.400pt}{4.818pt}}
\put(809,69){\makebox(0,0){10}}
\put(809.0,110.0){\rule[-0.200pt]{0.400pt}{4.818pt}}
\put(973,69){\makebox(0,0){100}}
\put(973.0,110.0){\rule[-0.200pt]{0.400pt}{4.818pt}}
\put(1138,69){\makebox(0,0){1000}}
\put(1138.0,110.0){\rule[-0.200pt]{0.400pt}{4.818pt}}
\put(1229,69){\makebox(0,0){t/o}}
\put(1229.0,110.0){\rule[-0.200pt]{0.400pt}{4.818pt}}
\put(480.0,110.0){\rule[-0.200pt]{0.400pt}{180.434pt}}
\put(480.0,110.0){\rule[-0.200pt]{180.434pt}{0.400pt}}
\put(1229.0,110.0){\rule[-0.200pt]{0.400pt}{180.434pt}}
\put(480.0,859.0){\rule[-0.200pt]{180.434pt}{0.400pt}}
\put(480,859){\line(1,0){749}}
\put(1229,110){\line(0,1){749}}
\put(394,484){\makebox(0,0){\rotatebox{90}{RPFP(mathsat)}}}
\put(854,29){\makebox(0,0){RPFP(acdcl)}}
\put(480,110){\usebox{\plotpoint}}
\multiput(480.00,110.59)(0.494,0.488){13}{\rule{0.500pt}{0.117pt}}
\multiput(480.00,109.17)(6.962,8.000){2}{\rule{0.250pt}{0.400pt}}
\multiput(488.00,118.59)(0.492,0.485){11}{\rule{0.500pt}{0.117pt}}
\multiput(488.00,117.17)(5.962,7.000){2}{\rule{0.250pt}{0.400pt}}
\multiput(495.00,125.59)(0.494,0.488){13}{\rule{0.500pt}{0.117pt}}
\multiput(495.00,124.17)(6.962,8.000){2}{\rule{0.250pt}{0.400pt}}
\multiput(503.00,133.59)(0.492,0.485){11}{\rule{0.500pt}{0.117pt}}
\multiput(503.00,132.17)(5.962,7.000){2}{\rule{0.250pt}{0.400pt}}
\multiput(510.00,140.59)(0.494,0.488){13}{\rule{0.500pt}{0.117pt}}
\multiput(510.00,139.17)(6.962,8.000){2}{\rule{0.250pt}{0.400pt}}
\multiput(518.00,148.59)(0.492,0.485){11}{\rule{0.500pt}{0.117pt}}
\multiput(518.00,147.17)(5.962,7.000){2}{\rule{0.250pt}{0.400pt}}
\multiput(525.00,155.59)(0.494,0.488){13}{\rule{0.500pt}{0.117pt}}
\multiput(525.00,154.17)(6.962,8.000){2}{\rule{0.250pt}{0.400pt}}
\multiput(533.00,163.59)(0.494,0.488){13}{\rule{0.500pt}{0.117pt}}
\multiput(533.00,162.17)(6.962,8.000){2}{\rule{0.250pt}{0.400pt}}
\multiput(541.00,171.59)(0.492,0.485){11}{\rule{0.500pt}{0.117pt}}
\multiput(541.00,170.17)(5.962,7.000){2}{\rule{0.250pt}{0.400pt}}
\multiput(548.00,178.59)(0.494,0.488){13}{\rule{0.500pt}{0.117pt}}
\multiput(548.00,177.17)(6.962,8.000){2}{\rule{0.250pt}{0.400pt}}
\multiput(556.00,186.59)(0.492,0.485){11}{\rule{0.500pt}{0.117pt}}
\multiput(556.00,185.17)(5.962,7.000){2}{\rule{0.250pt}{0.400pt}}
\multiput(563.00,193.59)(0.494,0.488){13}{\rule{0.500pt}{0.117pt}}
\multiput(563.00,192.17)(6.962,8.000){2}{\rule{0.250pt}{0.400pt}}
\multiput(571.00,201.59)(0.492,0.485){11}{\rule{0.500pt}{0.117pt}}
\multiput(571.00,200.17)(5.962,7.000){2}{\rule{0.250pt}{0.400pt}}
\multiput(578.00,208.59)(0.494,0.488){13}{\rule{0.500pt}{0.117pt}}
\multiput(578.00,207.17)(6.962,8.000){2}{\rule{0.250pt}{0.400pt}}
\multiput(586.00,216.59)(0.492,0.485){11}{\rule{0.500pt}{0.117pt}}
\multiput(586.00,215.17)(5.962,7.000){2}{\rule{0.250pt}{0.400pt}}
\multiput(593.00,223.59)(0.494,0.488){13}{\rule{0.500pt}{0.117pt}}
\multiput(593.00,222.17)(6.962,8.000){2}{\rule{0.250pt}{0.400pt}}
\multiput(601.00,231.59)(0.494,0.488){13}{\rule{0.500pt}{0.117pt}}
\multiput(601.00,230.17)(6.962,8.000){2}{\rule{0.250pt}{0.400pt}}
\multiput(609.00,239.59)(0.492,0.485){11}{\rule{0.500pt}{0.117pt}}
\multiput(609.00,238.17)(5.962,7.000){2}{\rule{0.250pt}{0.400pt}}
\multiput(616.00,246.59)(0.494,0.488){13}{\rule{0.500pt}{0.117pt}}
\multiput(616.00,245.17)(6.962,8.000){2}{\rule{0.250pt}{0.400pt}}
\multiput(624.00,254.59)(0.492,0.485){11}{\rule{0.500pt}{0.117pt}}
\multiput(624.00,253.17)(5.962,7.000){2}{\rule{0.250pt}{0.400pt}}
\multiput(631.00,261.59)(0.494,0.488){13}{\rule{0.500pt}{0.117pt}}
\multiput(631.00,260.17)(6.962,8.000){2}{\rule{0.250pt}{0.400pt}}
\multiput(639.00,269.59)(0.492,0.485){11}{\rule{0.500pt}{0.117pt}}
\multiput(639.00,268.17)(5.962,7.000){2}{\rule{0.250pt}{0.400pt}}
\multiput(646.00,276.59)(0.494,0.488){13}{\rule{0.500pt}{0.117pt}}
\multiput(646.00,275.17)(6.962,8.000){2}{\rule{0.250pt}{0.400pt}}
\multiput(654.00,284.59)(0.494,0.488){13}{\rule{0.500pt}{0.117pt}}
\multiput(654.00,283.17)(6.962,8.000){2}{\rule{0.250pt}{0.400pt}}
\multiput(662.00,292.59)(0.492,0.485){11}{\rule{0.500pt}{0.117pt}}
\multiput(662.00,291.17)(5.962,7.000){2}{\rule{0.250pt}{0.400pt}}
\multiput(669.00,299.59)(0.494,0.488){13}{\rule{0.500pt}{0.117pt}}
\multiput(669.00,298.17)(6.962,8.000){2}{\rule{0.250pt}{0.400pt}}
\multiput(677.00,307.59)(0.492,0.485){11}{\rule{0.500pt}{0.117pt}}
\multiput(677.00,306.17)(5.962,7.000){2}{\rule{0.250pt}{0.400pt}}
\multiput(684.00,314.59)(0.494,0.488){13}{\rule{0.500pt}{0.117pt}}
\multiput(684.00,313.17)(6.962,8.000){2}{\rule{0.250pt}{0.400pt}}
\multiput(692.00,322.59)(0.492,0.485){11}{\rule{0.500pt}{0.117pt}}
\multiput(692.00,321.17)(5.962,7.000){2}{\rule{0.250pt}{0.400pt}}
\multiput(699.00,329.59)(0.494,0.488){13}{\rule{0.500pt}{0.117pt}}
\multiput(699.00,328.17)(6.962,8.000){2}{\rule{0.250pt}{0.400pt}}
\multiput(707.00,337.59)(0.494,0.488){13}{\rule{0.500pt}{0.117pt}}
\multiput(707.00,336.17)(6.962,8.000){2}{\rule{0.250pt}{0.400pt}}
\multiput(715.00,345.59)(0.492,0.485){11}{\rule{0.500pt}{0.117pt}}
\multiput(715.00,344.17)(5.962,7.000){2}{\rule{0.250pt}{0.400pt}}
\multiput(722.00,352.59)(0.494,0.488){13}{\rule{0.500pt}{0.117pt}}
\multiput(722.00,351.17)(6.962,8.000){2}{\rule{0.250pt}{0.400pt}}
\multiput(730.00,360.59)(0.492,0.485){11}{\rule{0.500pt}{0.117pt}}
\multiput(730.00,359.17)(5.962,7.000){2}{\rule{0.250pt}{0.400pt}}
\multiput(737.00,367.59)(0.494,0.488){13}{\rule{0.500pt}{0.117pt}}
\multiput(737.00,366.17)(6.962,8.000){2}{\rule{0.250pt}{0.400pt}}
\multiput(745.00,375.59)(0.492,0.485){11}{\rule{0.500pt}{0.117pt}}
\multiput(745.00,374.17)(5.962,7.000){2}{\rule{0.250pt}{0.400pt}}
\multiput(752.00,382.59)(0.494,0.488){13}{\rule{0.500pt}{0.117pt}}
\multiput(752.00,381.17)(6.962,8.000){2}{\rule{0.250pt}{0.400pt}}
\multiput(760.00,390.59)(0.492,0.485){11}{\rule{0.500pt}{0.117pt}}
\multiput(760.00,389.17)(5.962,7.000){2}{\rule{0.250pt}{0.400pt}}
\multiput(767.00,397.59)(0.494,0.488){13}{\rule{0.500pt}{0.117pt}}
\multiput(767.00,396.17)(6.962,8.000){2}{\rule{0.250pt}{0.400pt}}
\multiput(775.00,405.59)(0.494,0.488){13}{\rule{0.500pt}{0.117pt}}
\multiput(775.00,404.17)(6.962,8.000){2}{\rule{0.250pt}{0.400pt}}
\multiput(783.00,413.59)(0.492,0.485){11}{\rule{0.500pt}{0.117pt}}
\multiput(783.00,412.17)(5.962,7.000){2}{\rule{0.250pt}{0.400pt}}
\multiput(790.00,420.59)(0.494,0.488){13}{\rule{0.500pt}{0.117pt}}
\multiput(790.00,419.17)(6.962,8.000){2}{\rule{0.250pt}{0.400pt}}
\multiput(798.00,428.59)(0.492,0.485){11}{\rule{0.500pt}{0.117pt}}
\multiput(798.00,427.17)(5.962,7.000){2}{\rule{0.250pt}{0.400pt}}
\multiput(805.00,435.59)(0.494,0.488){13}{\rule{0.500pt}{0.117pt}}
\multiput(805.00,434.17)(6.962,8.000){2}{\rule{0.250pt}{0.400pt}}
\multiput(813.00,443.59)(0.492,0.485){11}{\rule{0.500pt}{0.117pt}}
\multiput(813.00,442.17)(5.962,7.000){2}{\rule{0.250pt}{0.400pt}}
\multiput(820.00,450.59)(0.494,0.488){13}{\rule{0.500pt}{0.117pt}}
\multiput(820.00,449.17)(6.962,8.000){2}{\rule{0.250pt}{0.400pt}}
\multiput(828.00,458.59)(0.494,0.488){13}{\rule{0.500pt}{0.117pt}}
\multiput(828.00,457.17)(6.962,8.000){2}{\rule{0.250pt}{0.400pt}}
\multiput(836.00,466.59)(0.492,0.485){11}{\rule{0.500pt}{0.117pt}}
\multiput(836.00,465.17)(5.962,7.000){2}{\rule{0.250pt}{0.400pt}}
\multiput(843.00,473.59)(0.494,0.488){13}{\rule{0.500pt}{0.117pt}}
\multiput(843.00,472.17)(6.962,8.000){2}{\rule{0.250pt}{0.400pt}}
\multiput(851.00,481.59)(0.492,0.485){11}{\rule{0.500pt}{0.117pt}}
\multiput(851.00,480.17)(5.962,7.000){2}{\rule{0.250pt}{0.400pt}}
\multiput(858.00,488.59)(0.494,0.488){13}{\rule{0.500pt}{0.117pt}}
\multiput(858.00,487.17)(6.962,8.000){2}{\rule{0.250pt}{0.400pt}}
\multiput(866.00,496.59)(0.492,0.485){11}{\rule{0.500pt}{0.117pt}}
\multiput(866.00,495.17)(5.962,7.000){2}{\rule{0.250pt}{0.400pt}}
\multiput(873.00,503.59)(0.494,0.488){13}{\rule{0.500pt}{0.117pt}}
\multiput(873.00,502.17)(6.962,8.000){2}{\rule{0.250pt}{0.400pt}}
\multiput(881.00,511.59)(0.494,0.488){13}{\rule{0.500pt}{0.117pt}}
\multiput(881.00,510.17)(6.962,8.000){2}{\rule{0.250pt}{0.400pt}}
\multiput(889.00,519.59)(0.492,0.485){11}{\rule{0.500pt}{0.117pt}}
\multiput(889.00,518.17)(5.962,7.000){2}{\rule{0.250pt}{0.400pt}}
\multiput(896.00,526.59)(0.494,0.488){13}{\rule{0.500pt}{0.117pt}}
\multiput(896.00,525.17)(6.962,8.000){2}{\rule{0.250pt}{0.400pt}}
\multiput(904.00,534.59)(0.492,0.485){11}{\rule{0.500pt}{0.117pt}}
\multiput(904.00,533.17)(5.962,7.000){2}{\rule{0.250pt}{0.400pt}}
\multiput(911.00,541.59)(0.494,0.488){13}{\rule{0.500pt}{0.117pt}}
\multiput(911.00,540.17)(6.962,8.000){2}{\rule{0.250pt}{0.400pt}}
\multiput(919.00,549.59)(0.492,0.485){11}{\rule{0.500pt}{0.117pt}}
\multiput(919.00,548.17)(5.962,7.000){2}{\rule{0.250pt}{0.400pt}}
\multiput(926.00,556.59)(0.494,0.488){13}{\rule{0.500pt}{0.117pt}}
\multiput(926.00,555.17)(6.962,8.000){2}{\rule{0.250pt}{0.400pt}}
\multiput(934.00,564.59)(0.494,0.488){13}{\rule{0.500pt}{0.117pt}}
\multiput(934.00,563.17)(6.962,8.000){2}{\rule{0.250pt}{0.400pt}}
\multiput(942.00,572.59)(0.492,0.485){11}{\rule{0.500pt}{0.117pt}}
\multiput(942.00,571.17)(5.962,7.000){2}{\rule{0.250pt}{0.400pt}}
\multiput(949.00,579.59)(0.494,0.488){13}{\rule{0.500pt}{0.117pt}}
\multiput(949.00,578.17)(6.962,8.000){2}{\rule{0.250pt}{0.400pt}}
\multiput(957.00,587.59)(0.492,0.485){11}{\rule{0.500pt}{0.117pt}}
\multiput(957.00,586.17)(5.962,7.000){2}{\rule{0.250pt}{0.400pt}}
\multiput(964.00,594.59)(0.494,0.488){13}{\rule{0.500pt}{0.117pt}}
\multiput(964.00,593.17)(6.962,8.000){2}{\rule{0.250pt}{0.400pt}}
\multiput(972.00,602.59)(0.492,0.485){11}{\rule{0.500pt}{0.117pt}}
\multiput(972.00,601.17)(5.962,7.000){2}{\rule{0.250pt}{0.400pt}}
\multiput(979.00,609.59)(0.494,0.488){13}{\rule{0.500pt}{0.117pt}}
\multiput(979.00,608.17)(6.962,8.000){2}{\rule{0.250pt}{0.400pt}}
\multiput(987.00,617.59)(0.492,0.485){11}{\rule{0.500pt}{0.117pt}}
\multiput(987.00,616.17)(5.962,7.000){2}{\rule{0.250pt}{0.400pt}}
\multiput(994.00,624.59)(0.494,0.488){13}{\rule{0.500pt}{0.117pt}}
\multiput(994.00,623.17)(6.962,8.000){2}{\rule{0.250pt}{0.400pt}}
\multiput(1002.00,632.59)(0.494,0.488){13}{\rule{0.500pt}{0.117pt}}
\multiput(1002.00,631.17)(6.962,8.000){2}{\rule{0.250pt}{0.400pt}}
\multiput(1010.00,640.59)(0.492,0.485){11}{\rule{0.500pt}{0.117pt}}
\multiput(1010.00,639.17)(5.962,7.000){2}{\rule{0.250pt}{0.400pt}}
\multiput(1017.00,647.59)(0.494,0.488){13}{\rule{0.500pt}{0.117pt}}
\multiput(1017.00,646.17)(6.962,8.000){2}{\rule{0.250pt}{0.400pt}}
\multiput(1025.00,655.59)(0.492,0.485){11}{\rule{0.500pt}{0.117pt}}
\multiput(1025.00,654.17)(5.962,7.000){2}{\rule{0.250pt}{0.400pt}}
\multiput(1032.00,662.59)(0.494,0.488){13}{\rule{0.500pt}{0.117pt}}
\multiput(1032.00,661.17)(6.962,8.000){2}{\rule{0.250pt}{0.400pt}}
\multiput(1040.00,670.59)(0.492,0.485){11}{\rule{0.500pt}{0.117pt}}
\multiput(1040.00,669.17)(5.962,7.000){2}{\rule{0.250pt}{0.400pt}}
\multiput(1047.00,677.59)(0.494,0.488){13}{\rule{0.500pt}{0.117pt}}
\multiput(1047.00,676.17)(6.962,8.000){2}{\rule{0.250pt}{0.400pt}}
\multiput(1055.00,685.59)(0.494,0.488){13}{\rule{0.500pt}{0.117pt}}
\multiput(1055.00,684.17)(6.962,8.000){2}{\rule{0.250pt}{0.400pt}}
\multiput(1063.00,693.59)(0.492,0.485){11}{\rule{0.500pt}{0.117pt}}
\multiput(1063.00,692.17)(5.962,7.000){2}{\rule{0.250pt}{0.400pt}}
\multiput(1070.00,700.59)(0.494,0.488){13}{\rule{0.500pt}{0.117pt}}
\multiput(1070.00,699.17)(6.962,8.000){2}{\rule{0.250pt}{0.400pt}}
\multiput(1078.00,708.59)(0.492,0.485){11}{\rule{0.500pt}{0.117pt}}
\multiput(1078.00,707.17)(5.962,7.000){2}{\rule{0.250pt}{0.400pt}}
\multiput(1085.00,715.59)(0.494,0.488){13}{\rule{0.500pt}{0.117pt}}
\multiput(1085.00,714.17)(6.962,8.000){2}{\rule{0.250pt}{0.400pt}}
\multiput(1093.00,723.59)(0.492,0.485){11}{\rule{0.500pt}{0.117pt}}
\multiput(1093.00,722.17)(5.962,7.000){2}{\rule{0.250pt}{0.400pt}}
\multiput(1100.00,730.59)(0.494,0.488){13}{\rule{0.500pt}{0.117pt}}
\multiput(1100.00,729.17)(6.962,8.000){2}{\rule{0.250pt}{0.400pt}}
\multiput(1108.00,738.59)(0.494,0.488){13}{\rule{0.500pt}{0.117pt}}
\multiput(1108.00,737.17)(6.962,8.000){2}{\rule{0.250pt}{0.400pt}}
\multiput(1116.00,746.59)(0.492,0.485){11}{\rule{0.500pt}{0.117pt}}
\multiput(1116.00,745.17)(5.962,7.000){2}{\rule{0.250pt}{0.400pt}}
\multiput(1123.00,753.59)(0.494,0.488){13}{\rule{0.500pt}{0.117pt}}
\multiput(1123.00,752.17)(6.962,8.000){2}{\rule{0.250pt}{0.400pt}}
\multiput(1131.00,761.59)(0.492,0.485){11}{\rule{0.500pt}{0.117pt}}
\multiput(1131.00,760.17)(5.962,7.000){2}{\rule{0.250pt}{0.400pt}}
\multiput(1138.00,768.59)(0.494,0.488){13}{\rule{0.500pt}{0.117pt}}
\multiput(1138.00,767.17)(6.962,8.000){2}{\rule{0.250pt}{0.400pt}}
\multiput(1146.00,776.59)(0.492,0.485){11}{\rule{0.500pt}{0.117pt}}
\multiput(1146.00,775.17)(5.962,7.000){2}{\rule{0.250pt}{0.400pt}}
\multiput(1153.00,783.59)(0.494,0.488){13}{\rule{0.500pt}{0.117pt}}
\multiput(1153.00,782.17)(6.962,8.000){2}{\rule{0.250pt}{0.400pt}}
\multiput(1161.00,791.59)(0.492,0.485){11}{\rule{0.500pt}{0.117pt}}
\multiput(1161.00,790.17)(5.962,7.000){2}{\rule{0.250pt}{0.400pt}}
\multiput(1168.00,798.59)(0.494,0.488){13}{\rule{0.500pt}{0.117pt}}
\multiput(1168.00,797.17)(6.962,8.000){2}{\rule{0.250pt}{0.400pt}}
\multiput(1176.00,806.59)(0.494,0.488){13}{\rule{0.500pt}{0.117pt}}
\multiput(1176.00,805.17)(6.962,8.000){2}{\rule{0.250pt}{0.400pt}}
\multiput(1184.00,814.59)(0.492,0.485){11}{\rule{0.500pt}{0.117pt}}
\multiput(1184.00,813.17)(5.962,7.000){2}{\rule{0.250pt}{0.400pt}}
\multiput(1191.00,821.59)(0.494,0.488){13}{\rule{0.500pt}{0.117pt}}
\multiput(1191.00,820.17)(6.962,8.000){2}{\rule{0.250pt}{0.400pt}}
\multiput(1199.00,829.59)(0.492,0.485){11}{\rule{0.500pt}{0.117pt}}
\multiput(1199.00,828.17)(5.962,7.000){2}{\rule{0.250pt}{0.400pt}}
\multiput(1206.00,836.59)(0.494,0.488){13}{\rule{0.500pt}{0.117pt}}
\multiput(1206.00,835.17)(6.962,8.000){2}{\rule{0.250pt}{0.400pt}}
\multiput(1214.00,844.59)(0.492,0.485){11}{\rule{0.500pt}{0.117pt}}
\multiput(1214.00,843.17)(5.962,7.000){2}{\rule{0.250pt}{0.400pt}}
\multiput(1221.00,851.59)(0.494,0.488){13}{\rule{0.500pt}{0.117pt}}
\multiput(1221.00,850.17)(6.962,8.000){2}{\rule{0.250pt}{0.400pt}}
\put(573,212){\makebox(0,0){$\times$}}
\put(1229,859){\makebox(0,0){$\times$}}
\put(743,352){\makebox(0,0){$\times$}}
\put(784,396){\makebox(0,0){$\times$}}
\put(649,284){\makebox(0,0){$\times$}}
\put(784,431){\makebox(0,0){$\times$}}
\put(1229,458){\makebox(0,0){$\times$}}
\put(1210,492){\makebox(0,0){$\times$}}
\put(1229,405){\makebox(0,0){$\times$}}
\put(1229,401){\makebox(0,0){$\times$}}
\put(666,278){\makebox(0,0){$\times$}}
\put(1229,431){\makebox(0,0){$\times$}}
\put(1229,461){\makebox(0,0){$\times$}}
\put(1229,479){\makebox(0,0){$\times$}}
\put(1075,349){\makebox(0,0){$\times$}}
\put(763,406){\makebox(0,0){$\times$}}
\put(651,290){\makebox(0,0){$\times$}}
\put(838,432){\makebox(0,0){$\times$}}
\put(1229,460){\makebox(0,0){$\times$}}
\put(1229,478){\makebox(0,0){$\times$}}
\put(643,277){\makebox(0,0){$\times$}}
\put(621,256){\makebox(0,0){$\times$}}
\put(630,264){\makebox(0,0){$\times$}}
\put(865,319){\makebox(0,0){$\times$}}
\put(669,310){\makebox(0,0){$\times$}}
\put(674,301){\makebox(0,0){$\times$}}
\put(668,309){\makebox(0,0){$\times$}}
\put(645,277){\makebox(0,0){$\times$}}
\put(649,278){\makebox(0,0){$\times$}}
\put(658,287){\makebox(0,0){$\times$}}
\put(693,325){\makebox(0,0){$\times$}}
\put(675,306){\makebox(0,0){$\times$}}
\put(1229,859){\makebox(0,0){$\times$}}
\put(1229,843){\makebox(0,0){$\times$}}
\put(1229,859){\makebox(0,0){$\times$}}
\put(1229,541){\makebox(0,0){$\times$}}
\put(1229,716){\makebox(0,0){$\times$}}
\put(1229,859){\makebox(0,0){$\times$}}
\put(1229,236){\makebox(0,0){$\times$}}
\put(1229,859){\makebox(0,0){$\times$}}
\put(1229,859){\makebox(0,0){$\times$}}
\put(856,680){\makebox(0,0){$\times$}}
\put(713,352){\makebox(0,0){$\times$}}
\put(769,388){\makebox(0,0){$\times$}}
\put(651,281){\makebox(0,0){$\times$}}
\put(1229,416){\makebox(0,0){$\times$}}
\put(1229,438){\makebox(0,0){$\times$}}
\put(1229,461){\makebox(0,0){$\times$}}
\put(713,343){\makebox(0,0){$\times$}}
\put(753,395){\makebox(0,0){$\times$}}
\put(653,299){\makebox(0,0){$\times$}}
\put(778,413){\makebox(0,0){$\times$}}
\put(798,435){\makebox(0,0){$\times$}}
\put(821,451){\makebox(0,0){$\times$}}
\put(599,219){\makebox(0,0){$\times$}}
\put(594,216){\makebox(0,0){$\times$}}
\put(672,441){\makebox(0,0){$\times$}}
\put(672,394){\makebox(0,0){$\times$}}
\put(676,397){\makebox(0,0){$\times$}}
\put(706,430){\makebox(0,0){$\times$}}
\put(732,455){\makebox(0,0){$\times$}}
\put(666,345){\makebox(0,0){$\times$}}
\put(702,474){\makebox(0,0){$\times$}}
\put(826,407){\makebox(0,0){$\times$}}
\put(666,305){\makebox(0,0){$\times$}}
\put(827,466){\makebox(0,0){$\times$}}
\put(848,402){\makebox(0,0){$\times$}}
\put(745,454){\makebox(0,0){$\times$}}
\put(1229,859){\makebox(0,0){$\times$}}
\put(1229,859){\makebox(0,0){$\times$}}
\put(1229,859){\makebox(0,0){$\times$}}
\put(1229,859){\makebox(0,0){$\times$}}
\put(1229,576){\makebox(0,0){$\times$}}
\put(1229,859){\makebox(0,0){$\times$}}
\put(1229,859){\makebox(0,0){$\times$}}
\put(1229,859){\makebox(0,0){$\times$}}
\put(1229,524){\makebox(0,0){$\times$}}
\put(1229,859){\makebox(0,0){$\times$}}
\put(1229,859){\makebox(0,0){$\times$}}
\put(900,859){\makebox(0,0){$\times$}}
\put(1229,859){\makebox(0,0){$\times$}}
\put(754,401){\makebox(0,0){$\times$}}
\put(1229,599){\makebox(0,0){$\times$}}
\put(1229,644){\makebox(0,0){$\times$}}
\put(1198,438){\makebox(0,0){$\times$}}
\put(1229,859){\makebox(0,0){$\times$}}
\put(1229,859){\makebox(0,0){$\times$}}
\put(1229,859){\makebox(0,0){$\times$}}
\put(1229,859){\makebox(0,0){$\times$}}
\put(789,314){\makebox(0,0){$\times$}}
\put(676,306){\makebox(0,0){$\times$}}
\put(716,317){\makebox(0,0){$\times$}}
\put(783,320){\makebox(0,0){$\times$}}
\put(674,306){\makebox(0,0){$\times$}}
\put(677,352){\makebox(0,0){$\times$}}
\put(715,415){\makebox(0,0){$\times$}}
\put(888,401){\makebox(0,0){$\times$}}
\put(930,482){\makebox(0,0){$\times$}}
\put(831,443){\makebox(0,0){$\times$}}
\put(957,442){\makebox(0,0){$\times$}}
\put(835,357){\makebox(0,0){$\times$}}
\put(721,364){\makebox(0,0){$\times$}}
\put(993,495){\makebox(0,0){$\times$}}
\put(893,412){\makebox(0,0){$\times$}}
\put(870,471){\makebox(0,0){$\times$}}
\put(734,402){\makebox(0,0){$\times$}}
\put(855,461){\makebox(0,0){$\times$}}
\put(1134,436){\makebox(0,0){$\times$}}
\put(720,367){\makebox(0,0){$\times$}}
\put(1229,832){\makebox(0,0){$\times$}}
\put(1229,859){\makebox(0,0){$\times$}}
\put(1229,859){\makebox(0,0){$\times$}}
\put(874,467){\makebox(0,0){$\times$}}
\put(1229,604){\makebox(0,0){$\times$}}
\put(800,446){\makebox(0,0){$\times$}}
\put(1229,859){\makebox(0,0){$\times$}}
\put(1229,859){\makebox(0,0){$\times$}}
\put(1229,859){\makebox(0,0){$\times$}}
\put(925,719){\makebox(0,0){$\times$}}
\put(1229,859){\makebox(0,0){$\times$}}
\put(804,407){\makebox(0,0){$\times$}}
\put(1229,401){\makebox(0,0){$\times$}}
\put(1229,707){\makebox(0,0){$\times$}}
\put(1138,506){\makebox(0,0){$\times$}}
\put(832,446){\makebox(0,0){$\times$}}
\put(1018,458){\makebox(0,0){$\times$}}
\put(1229,591){\makebox(0,0){$\times$}}
\put(773,498){\makebox(0,0){$\times$}}
\put(1229,859){\makebox(0,0){$\times$}}
\put(1229,859){\makebox(0,0){$\times$}}
\put(480.0,110.0){\rule[-0.200pt]{0.400pt}{180.434pt}}
\put(480.0,110.0){\rule[-0.200pt]{180.434pt}{0.400pt}}
\put(1229.0,110.0){\rule[-0.200pt]{0.400pt}{180.434pt}}
\put(480.0,859.0){\rule[-0.200pt]{180.434pt}{0.400pt}}
\end{picture}

%% file: fresh.bbl
\begin{thebibliography}{10}
\providecommand{\url}[1]{\texttt{#1}}
\providecommand{\urlprefix}{URL }

\bibitem{DBLP:conf/mkm/BoldoFM09}
Boldo, S., Filli{\^a}tre, J.C., Melquiond, G.: Combining {Coq} and {Gappa} for
  certifying floating-point programs. In: Intelligent Computer Mathematics
  (Calculemus); MKM/CICM. LNCS, vol. 5625. Springer (2009)

\bibitem{Float-ACDCL}
Brain, M., D'Silva, V., Griggio, A., Haller, L., Kroening, D.: Deciding
  floating-point logic with abstract conflict driven clause learning. FMSD
  (2013)

\bibitem{bkw2009-fmcad}
Brillout, A., Kroening, D., Wahl, T.: Mixed abstractions for floating-point
  arithmetic. In: FMCAD. IEEE (2009)

\bibitem{DBLP:conf/eurocast/BrummayerB09}
Brummayer, R., Biere, A.: Effective bit-width and under-approximation. In:
  EUROCAST. LNCS, vol. 5717. Springer (2009)

\bibitem{DBLP:conf/tacas/BryantKOSSB07}
Bryant, R.E., Kroening, D., Ouaknine, J., Seshia, S.A., Strichman, O., Brady,
  B.A.: Deciding bit-vector arithmetic with abstraction. In: TACAS. LNCS, vol.
  4424. Springer (2007)

\bibitem{mathsat}
Cimatti, A., Griggio, A., Schaafsma, B.J., Sebastiani, R.: The {MathSAT5} {SMT}
  solver. In: TACAS. LNCS, vol. 7795. Springer (2013)

\bibitem{Clarke:2000:CAR:647769.734089}
Clarke, E.M., Grumberg, O., Jha, S., Lu, Y., Veith, H.: Counterexample-guided
  abstraction refinement. In: CAV. LNCS, vol. 1855. Springer (2000)

\bibitem{DBLP:conf/popl/CousotC77}
Cousot, P., Cousot, R.: Abstract interpretation: A unified lattice model for
  static analysis of programs by construction or approximation of fixpoints.
  In: Proceedings, Fourth ACM Symposium on Principles of Programming Languages,
  Los Angeles, USA. pp. 238--252. ACM Press (1977)

\bibitem{DBLP:conf/esop/CousotCFMMMR05}
Cousot, P., Cousot, R., Feret, J., Mauborgne, L., Min{\'e}, A., Monniaux, D.,
  Rival, X.: The {ASTRE\'E} analyzer. In: ESOP. LNCS, vol. 3444. Springer
  (2005)

\bibitem{Daumas:2010:CBE:1644001.1644003}
Daumas, M., Melquiond, G.: Certification of bounds on expressions involving
  rounded operators. ACM Trans. Math. Softw.  37(1) (2010)

\bibitem{DBLP:conf/popl/DSilvaHK13}
D'Silva, V., Haller, L., Kroening, D.: Abstract conflict driven learning. In:
  POPL. ACM (2013)

\bibitem{DBLP:conf/tacas/DSilvaHKT12}
D'Silva, V., Haller, L., Kroening, D., Tautschnig, M.: Numeric bounds analysis
  with conflict-driven learning. In: TACAS. LNCS, vol. 7214. Springer (2012)

\bibitem{DBLP:conf/cav/FuS16}
Fu, Z., Su, Z.: {XSat}: {A} fast floating-point satisfiability solver. In:
  Chaudhuri, S., Farzan, A. (eds.) Computer Aided Verification - 28th
  International Conference, {CAV} 2016, Toronto, ON, Canada, July 17-23, 2016,
  Proceedings, Part {II}. Lecture Notes in Computer Science, vol. 9780, pp.
  187--209. Springer (2016), \url{https://doi.org/10.1007/978-3-319-41540-6}

\bibitem{DBLP:conf/cade/GaoKC13}
Gao, S., Kong, S., Clarke, E.M.: {dReal}: An {SMT} solver for nonlinear
  theories over the reals. In: CADE. LNCS, vol. 7898. Springer (2013)

\bibitem{DBLP:conf/cav/GeM09}
Ge, Y., de~Moura, L.: Complete instantiation for quantified formulas in
  satisfiability modulo theories. In: CAV. LNCS, vol. 5643. Springer (2009)

\bibitem{harrison-floats}
Harrison, J.: Floating point verification in {HOL} {L}ight: the exponential
  function. TR 428, University of Cambridge Computer Laboratory (1997),
  available on the Web as
  {\verb+http://www.cl.cam.ac.uk/~jrh13/papers/tang.html+}

\bibitem{Janota:2012:SQC:2352219.2352233}
Janota, M., Klieber, W., Marques-Silva, J., Clarke, E.: Solving {QBF} with
  counterexample guided refinement. In: SAT. LNCS, vol. 7317. Springer (2012)

\bibitem{VanKhanh201227}
Khanh, T.V., Ogawa, M.: {SMT} for polynomial constraints on real numbers. In:
  TAPAS. Electronic Notes in Theoretical Computer Science, vol. 289 (2012)

\bibitem{sonolar}
Lapschies, F., Peleska, J., Gorbachuk, E., Mangels, T.: {SONOLAR} {SMT}-solver.
  In: Satisfiability modulo theories competition; system description (2012)

\bibitem{DBLP:journals/iandc/Melquiond12}
Melquiond, G.: Floating-point arithmetic in the {Coq} system. In: Conf. on Real
  Numbers and Computers. Information \& Computation, vol. 216. Elsevier (2012)

\bibitem{z3}
de~Moura, L., Bj{\o}rner, N.: {Z3}: An efficient {SMT} solver. In: TACAS. LNCS,
  vol. 4963. Springer (2008)

\bibitem{DBLP:conf/fmcad/RamachandranW16}
Ramachandran, J., Wahl, T.: Integrating proxy theories and numeric model
  lifting for floating-point arithmetic. In: 2016 Formal Methods in
  Computer-Aided Design, {FMCAD} 2016, Mountain View, CA, USA, October 3-6,
  2016. {IEEE} (2016)

\bibitem{DBLP:conf/cade/ZeljicWR14}
Zeljic, A., Wintersteiger, C.M., R{\"{u}}mmer, P.: Approximations for model
  construction. In: Demri, S., Kapur, D., Weidenbach, C. (eds.) Automated
  Reasoning - 7th International Joint Conference, {IJCAR} 2014, Held as Part of
  the Vienna Summer of Logic, {VSL} 2014, Vienna, Austria, July 19-22, 2014.
  Proceedings. Lecture Notes in Computer Science, vol. 8562, pp. 344--359.
  Springer (2014), \url{https://doi.org/10.1007/978-3-319-08587-6}

\bibitem{DBLP:journals/jar/ZeljicWR17}
Zeljic, A., Wintersteiger, C.M., R{\"{u}}mmer, P.: An approximation framework
  for solvers and decision procedures. J. Autom. Reasoning  58(1),  127--147
  (2017), \url{https://doi.org/10.1007/s10817-016-9393-1}

\end{thebibliography}
